%% file: main.tex
\newif\ifllncs
\llncsfalse

\newif\ifieee
\ieeetrue

\newif\ifacm
\acmfalse

\ifllncs
	\documentclass{llncs}
\fi	
\ifieee
	\documentclass[conference]{IEEEtran}
\fi

\ifacm
	\documentclass[sigconf]{acmart}
\fi

\ifllncs
	\usepackage{geometry}
\fi
\usepackage{graphicx}
\usepackage{longtable}
\usepackage{color}
\usepackage{amsmath}
\usepackage{hyperref}
\ifllncs
	\usepackage{subfig}
	\usepackage{floatrow}
	\newfloatcommand{capbtabbox}{table}[][\FBwidth]
\else
\ifieee
	\usepackage[tight,footnotesize]{subfigure}
	\usepackage[bottom]{footmisc}
\else
	\usepackage[tight,footnotesize]{subfigure}
\fi
\fi

\usepackage{pgf}
\usepackage[strings]{underscore} 
\input{macros}

\newcommand{\figurewidth}{0.48}
\begin{document}
\title{Diversification Across Mining Pools: Optimal Mining
	Strategies under PoW}
\ifllncs
\else
\ifieee
\author{\IEEEauthorblockN{Panagiotis Chatzigiannis, Foteini Baldimtsi, Igor Griva and Jiasun Li}
	\IEEEauthorblockA{George Mason University \\
		Fairfax VA 22030\\
		Email: pchatzig@gmu.edu, foteini@gmu.edu, igriva@gmu.edu, jli29@gmu.edu}}
	
\else
\author{}
\fi
\fi
%
\ifieee
	\maketitle 
\fi         
%
\begin{abstract}
Mining is a central operation of all proof-of-work (PoW) based cryptocurrencies. The vast majority of miners today participate in ``mining pools'' instead of ``solo mining'' in order to lower risk and achieve a more steady income. However, this rise of participation in mining pools negatively affects the decentralization levels of most cryptocurrencies.

In this work, we look into mining pools from the point of view of a miner: We present an analytical model and implement a computational tool that allows miners to optimally distribute their computational power over multiple pools and PoW cryptocurrencies (i.e. build a mining portfolio), taking into account their risk aversion levels. Our tool allows miners to maximize their risk-adjusted earnings by diversifying across multiple mining pools which enhances PoW decentralization. Finally, we run an experiment in Bitcoin historical data and demonstrate that a miner diversifying over multiple pools, as instructed by our model/tool, receives a higher overall Sharpe ratio (i.e. average excess reward over its standard deviation/volatility).
\end{abstract}
\ifacm



	\maketitle
\fi

\input{intro}

\input{fees}

\input{active_miners}
\input{evaluation}
\input{experiment}
\input{conclusion}

\ifllncs
\else
\ifieee
	\bibliographystyle{IEEEtran}
\else
	\clearpage
	\clearpage
	\newpage
	\bibliographystyle{ACM-Reference-Format}
\fi
\fi

\bibliography{mybibliography,abbrev3,crypto}
%

%
%
%
%

\ifllncs
\else
\ifieee
	\begin{appendices}
	\include{appendix}	
	\end{appendices}
\else

\appendix

\include{appendix}
\fi
\fi

\end{document}

%% file: macros.tex
\newcommand{\fee}[1]{f_{#1}}
\newcommand{\totalhashrate}[1]{\poolhashrate{#1}}
\newcommand{\poolhashrate}[1]{\Lambda_{#1}}
\newcommand{\blockreward}[1]{R_{#1}}
\newcommand{\minerpoolhashrate}[1]{\lambda_{#1}}
\newcommand{\powalgnum}{\mathsf{A}}
\newcommand{\powalg}[1]{\alpha_{#1}}
\newcommand{\minertotalhashrate}{\minerpoolhashrate{\powalgnum}}
\newcommand{\minertotalalgrate}[1]{\minerpoolhashrate{#1}}
\newcommand{\cara}{\rho}
\newcommand{\poolsnum}{M}
\newcommand{\poolid}{m}
\newcommand{\cryptonum}{\mathsf{C}}
\newcommand{\cryptoid}{c}
\newcommand{\blocktime}[1]{D_{#1}}
\newcommand{\txfee}[1]{\mathsf{tx}_{#1}}
\newcommand{\interval}{t}
\newcommand{\blockday}{d}
\newcommand{\exchange}[1]{E_{#1}}
\newcommand{\difficulty}[1]{T_{#1}}
\newcommand{\blocksnum}[1]{B_{#1}}
\newcommand{\sharpe}{\mathsf{S}}
\newcommand{\pps}{\mathsf{PPS}}
\newcommand{\pplns}{\mathsf{PPLNS}}
\newcommand{\totalreward}{\mathsf{P}}
\newcommand{\rewardpercentage}{z}

\def\comments_on{0} 

\ifnum\comments_on=1

\newcommand{\foteini}[1]{{
		{\color{red}\textbf{Foteini:}
			\emph{#1}
		}\normalcolor}}

\newcommand{\panagiotis}[1]{{
		{\color{magenta}\textbf{Panagiotis:}
			#1
		}\normalcolor}}

\newcommand{\authnote}[1]{#1} 
\else

\newcommand{\foteini}[1]{}

\newcommand{\panagiotis}[1]{}

\newcommand{\authnote}[1]{}

\fi

%% file: intro.tex
\section{Introduction}
The majority of cryptocurrencies use some type of proof-of-work (PoW) based consensus mechanism to order and finalize transactions stored in the blockchain. At any given time, a set of users all over the world (called miners or maintainers) compete in solving a PoW puzzle that will allow them to post the next block in the blockchain and at the same time claim the ``coinbase'' reward and any relevant transaction fees. In the early years of cryptocurrencies solo mining was the norm, and a miner using his own hardware would attempt to solve the PoW puzzle himself, earning the reward. However, as the exchange rate of cryptocurrencies increased, the PoW competition become fiercer, specialized hardware was manufactured just for the purpose of mining particular types of PoW (e.g. Bitcoin or Ethereum mining ASICs~\cite{bitcoin-asics}), and eventually users formed coalitions for better chances of solving the puzzle. 

These coalitions known as \emph{mining pools}, where miners are all continuously trying to mine a block with the ``pool manager'' being the reward recipient, enabled participating users to reduce their mining risks\footnote{We measure a miner's ``risk'' by the variance of rewards over time.}. 
After the establishment of mining pools, it become nearly impossible for ``solo'' miners to compete on the mining game, even if they were using specialized hardware, or else they could risk not to earn any rewards at all during the hardware's lifetime.

The selection of a mining pool is not a trivial task. A large number of pools exist each offering different reward distribution methods and earning fees (as we further discuss in Section \ref{reward-methods}). At the same time different pools control a different ratio of the overall hashrate consumed by a cryptocurrency and larger pools (in terms of hashrate) offer lower risk, as they typically offer more frequent payouts to the miners. But how can a miner make an optimal decision about which mining pools to participate in and for which cryptocurrencies at any given time considering the variety of possible options?

\textbf{Our contributions.}
We present an analytical tool that allows risk-averse miners to optimally create a mining portfolio that maximizes their risk-adjusted rewards. We characterize miners by their total computational resources (i.e. hash power) and their risk aversion level, and mining pools by their total computational power (i.e. hash rate) and the reward mechanism they offer. 
We model the hash rate allocation  as an optimization problem that aims to maximize the miner's expected utility. In Section~\ref{sec:model}, we provide three different versions of our model. The first one, inspired by~\cite{jiasun}, concerns a miner that wishes to mine on a single cryptocurrency, while aiming to diversify among any number of mining pools (including the solo mining option). The second version captures miners which diversify across different cryptocurrencies that use the same PoW mining algorithm. In our third version, we model miners who also wish to diversify across cryptocurrencies with different PoW mining algorithms. Our modeling technique is based on standard utility maximization, and extends the  Markowitz Modern portfolio theory~\cite{doi:10.1111/j.1540-6261.1952.tb01525.x} to multiple mining pools, rather than multiple assets.

In Section~\ref{sec:evaluation} we present an implementation of our model. We develop a Python tool that uses the constrained optimization by linear approximation (COBYLA)~\cite{Powell1994} method to automate the pool distribution for an active miner. A miner can use our tool by providing as input its own mining power (for \emph{any} PoW type) as well as his risk aversion rate and \emph{any} number of pools he wishes to take into account when computing the optimal distribution of his mining power. As expected, we observe that for ``reasonable'' values of risk aversion level, the miner would generally allocate more of his resources to pools offering large hash power combined with small fees, without however neglecting other pools that are not as ``lucrative''.

Finally, to illustrate the usefulness of our tool, we run an experiment on Bitcoin historical data (Section~\ref{simulated-results}). We start by considering a Bitcoin miner who starts mining passively on a single chosen pool for 4 months and compute his earnings on a daily basis. 

Then, we consider a miner with the same hash power
who using our tool, would ``actively'' diversify every 3 days over 3 Bitcoin pools and reallocate his hash power accordingly.  In our experiment we observe  that the ``active'' miner improves his reward over risk ratio (Sharpe ratio~\cite{10.2307/2351741}) by 260\% compared to the ``passive'' miner. In our experiments, for the ``passive'' miner we picked a large and reputable pool (Slush pool) while for the ``active'' miner we added one more  pool of equivalent size and fee structure (ViaBTC), as well as a lower-fee smaller pool (DPool) for a better illustration. Note that there are several degrees of freedom in our experiments (i,e., time periods, set of pools selected etc.). Thus, we include replications with different parameters (time period, pools, risk aversion, miner’s power and frequency of diversification) to show how each one of them can affect the end result.

\vspace{0.2cm}
\noindent \textbf{Related Work.} 
There exist a few running tools  which on input a miner's hashing power suggest which cryptocurrency is currently most profitable. For instance, tools like Multipoolminer~\cite{multipoolminer}, Smartmine~\cite{smartmine} or Minergate~\cite{minergate} start by benchmarking the CPU/GPU of the miner, (which we consider an orthogonal service to the third version of our model) and then suggest a cryptocurrency which would offer the best reward at that time.  To make that decision, they look into various cryptocurrencies' parameters (i.e. block time, reward etc.) and the current difficulty (their exact model is unclear). Which pools the miner will use towards mining the suggested cryptocurrency is either hard-coded by the tool, or chosen by the miner.
Our method differs from such tools in various aspects. Most importantly, in our model we take the risk aversion rate of the miner into account, which is an important factor when making financial decisions. 
Moreover,  we focus on allocating a miner's power over \emph{multiple} pools, as opposed to just different cryptocurrencies, which can benefit the decentralization within a cryptocurrency.

Miner's risk aversion was taken into account by Fisch et al~\cite{10.1007/978-3-319-71924-5-15} who provided  a top-down (from  pool's point of view) analysis, i.e. focused on optimal pool operation strategies towards maximizing the pool's expected utility. 
Cong et al.~\cite{jiasun} also took risk aversion into account in their modeling,
however the focus of their work was different from ours. 
In particular, they focus on \emph{the interaction} between miners and pools. They first demonstrate the significant risk-diversification benefit offered by mining pools for individual miners, highlighting risk-sharing as a natural centralizing force.
Then, they demonstrate that the risk-sharing benefit within a large pool could be alternatively obtained
through miner diversification across multiple small pools. Finally, they present an equilibrium model where multiple pool managers compete in fees to attract
customer miners. In our work, we focus on the miners' side: we develop tools to help miners diversify among different pools and cryptocurrencies to maximize their risk-adjusted earnings.

Because we emphasize on hash rate allocations across multiple mining pools, either within the same or over different cryptocurrencies, our analysis distinguishes from contemporary works such as~\cite{DBLP:conf/esorics/BissiasLT18}, who present an economic model of hash power allocation over different cryptocurrencies sharing the same PoW algorithm in a Markowitz fashion.
This perspective also sets us apart from~\cite{DBLP:journals/corr/abs-1805-08979} who study mining across different currencies in a strategic fashion, without accounting for pooled mining.

Finally, some recent works examined the case where miners change the mining pool they mine with, in order to optimize their rewards from a \emph{network performance scope} (communications delay).
Y.Lewenberg et al. \cite{Lewenberg:2015:BMP:2772879.2773270} showed how network delays incentivize miners to switch among pools in order to optimize their payoffs due to the non-linearity introduced. 
X.Liu et al. \cite{8326513} discussed how to dynamically select a mining pool taking into account the pool's computation power (hash rate) and the network's propagation delay. 
Network performance is an important aspect when diversifying across pools, and we view \cite{Lewenberg:2015:BMP:2772879.2773270,8326513} as complementary to our work.

%% file: fees.tex
\section{Mining Background }
\label{sec:mining}

 As of today, the majority of blockchain-based cryptocurrencies use PoW for maintaining their ledger. 
 Miners listen the network for (a) pending transactions and (b) new blocks of transactions to be posted on the ledger. The role of a miner is to select a subset of pending transactions,  assemble them to a new block and perform computational work towards finding a random nonce $r$ that will make the block valid and allow the miner to append it on the blockchain and thus win the reward. 
 The brute-forcing process of finding a suitable nonce $r$ such that together with the rest of the block contents $b$ satisfies the property $H(b||r) < \difficulty{}$ for some target value $\difficulty{}$ is called ``mining''. For Bitcoin, this translates to finding a suitable hash pre-image using the double SHA256 hash function. 

Note that, although solving PoW puzzles was initially done using ordinary CPUs, the increasing prices of cryptocurrencies have resulted in a ``hardware race'' to develop the most efficient mining hardware using Application Specific Integrated Circuits (ASICs), designed to perform SHA256 hashing operations in many orders of magnitude faster than CPUs or GPUs.

\subsection{Mining Pools} Mining is a random process, in most cryptocurrencies the computational power devoted to solve the PoW  puzzle is very high, which implies a high variance on the miner's reward. In Bitcoin, even for a miner using state of the art ASICs, there is a good probability that he never gets a block mined during the hardware's lifetime. This led to the formation of mining pools, where coalitions of miners are all continuously trying to mine a block, with the ``pool manager'' being the reward recipient. If any of the participating miners finds a  solution to the PoW puzzle, the pool manager receives the block reward $\blockreward{}$ and distributes to the  participants, while possibly keeping a small cut (or fee $\fee{}$). The block reward $\blockreward{}$ distribution is based on how much work these miners performed. A method for the pool manager to measure how much effort each miner has put into the pool is by keeping a record of \emph{shares}, which are ``near-solutions'' to the PoW puzzle (or ``near-valid'' blocks), satisfying the property $\difficulty{s} < H(b||r) < \difficulty{}$ where $\difficulty{s}$ is the ``share difficulty''. There are several methods to distribute the reward $\blockreward{}$ to the miners, which we analyze below.

\ifllncs
\else
		\begin{table}[!t]		
		\label{formula-notation}
		\caption{Notation}
		\centering		
		\begin{tabular}{|l|l|l|}
			\hline
			Total number of pools & $M$\\
			Hashrate and fee of pool $\poolid$ &  $\poolhashrate{\poolid}$,$\fee{\poolid}$\\
			Transaction fee & $\txfee{}$\\
			Miner's hashing power allocated to pool $\poolid$ & \\(and cryptocurrency $\cryptoid$)& $\minerpoolhashrate{\poolid}$($\minerpoolhashrate{\poolid,\cryptoid}$)\\
			Miner's total hashing power & \\ (for mining algorithm $\powalg{}$) & $\minertotalhashrate$($\minertotalalgrate{\powalg{}}$)\\
			Constant Absolute Risk Aversion (CARA)& $\cara$\\
			Cryptocurrency $\cryptoid$ total hashrate & $\totalhashrate{c}$\\
			Block time and block reward of cryptocurrency $\cryptoid$ & $\blocktime{c}$,$\blockreward{c}$ \\
			Total number of cryprocurrencies & $\cryptonum$ \\
			Total number of PoW mining algorithms & $\powalgnum$\\
			(Average) network difficulty& $\difficulty{}$\\
			Cryptocurrency exchange rate & $\exchange{\cryptoid}$ \\ 
			Number of blocks found on a day $\blockday$ & $\blocksnum{\blockday}$\\
			Diversification interval (days) & $\interval$\\
			Sharpe ratio, Total accumulated reward/payoff & $\sharpe$,$\totalreward$ \\
			\hline
		\end{tabular}
		\end{table}	
\fi

\subsection{Reward Methods in Mining Pools}
\label{reward-methods}
Different pools offer slightly different reward methods (or a mix of them), with the most popular being: \textit{pay per share} (PPS), \emph{Proportional}, and \emph{pay per last $N$ shares} (PPLNS). 
In the PPS reward method, the miners are not immediately paid when a block is found, however each block reward is deposited into a ``central pool fund'' or ``bucket''. The miners are paid proportionally to the shares submitted throughout their participation in the pool, regardless of if and when the pool has found a solution to the PoW puzzle. PPS is generally considered to offer a steady, almost guaranteed income, independent of the pool's ``luck'' finding a block.
In the proportional reward method, whenever a pool solves the puzzle for a new block, the new block reward is distributed to the pool's miners proportionally to the number of shares each miner has submitted to the pool for that particular block. This method however was found to be vulnerable to the ``pool-hopping'' attack, where the miners could exploit the pool's expected earnings, variance and maturity time and ``hop away'' to another pool or solo-mining when the pool's attractiveness is low \cite{DBLP:journals/corr/abs-1112-4980}. 
The PPLNS method was implemented to counter this attack, where the miners' reward is distributed according to the ``recent'' number of submitted shares, thus invalidating shares submitted early. As shown in Appendix \ref{app:PoolData}, the PPLNS method in mining pools is the most popular reward method today. In addition, many other PPLNS variants are currently being used by mining pools, e.g the RBPPS (Round Based Pay Per Share) method where the pool only pays the reward after the block eventually gets confirmed by the network (thus excluding deprecated blocks).  We refer the reader to \cite{pool-comparison} and \cite{DBLP:journals/corr/abs-1112-4980} for thorough and complete analysis of pool reward methods.
In our setting, we mostly consider pools that offer a PPLNS reward method (or its variants). We generally do not consider pools that use the PPS method, as the miners' expected earnings do not depend on the variance of finding blocks.

\vspace{0.1cm}
\noindent \textit{Mining pools offering multiple reward methods.}
As mentioned above, some mining pools offer multiple reward systems (i.e. the Coinotron pool~\cite{coinotron} offers both PPS and PPLNS).
We study these types of pools separately, as some miners might opt for different fee contracts within the same pool.

Let a mining pool $\poolid$ with total hashrate $\poolhashrate{\poolid}$, offering both PPLNS and PPS reward systems to choose from, where $\rewardpercentage$  is the percentage of pool's hashing power paid using a PPLNS fee contract, and $(1-\rewardpercentage)$ is the percentage paid using a PPS fee contract. 
Also let $\minerpoolhashrate{m}$ be a miner's hashing power allocated to pool $m$ and $\blockreward{{\minerpoolhashrate{\poolid}}}$ the miner's reward when a block is eventually mined by the pool bringing a total reward $\blockreward{}$. 

The pool manager should make sure to keep paying its PPS miners at a steady rate, compensating for any pool ``luck'' fluctuations in any given period when trying to find a block. To achieve this, the manager needs to maintain a ``bucket'' containing an adequate amount of coins, and keep replenishing it with $\blockreward{\pps}$ (i.e. reward of PPS) each time a block is ``mined'' by the pool with block reward $\blockreward{}$, to keep paying PPS miners during periods of bad ``luck''. Consequently, when the pool collectively ``mines'' a block with reward $\blockreward{}$,  the pool manager can select one of the following three miner payment strategies, which also determine the exact PPLNS miners' reward:

\noindent \textbf{Strategy 1:} The pool manager splits $\blockreward{}$ into $\blockreward{\pps} = (1-\rewardpercentage)\blockreward{}$ and $\blockreward{\pplns} = \rewardpercentage\blockreward{}$. By this strategy, $\blockreward{\minerpoolhashrate{\poolid}} = \frac{\minerpoolhashrate{\poolid}}{\rewardpercentage \poolhashrate{\poolid}}\rewardpercentage \blockreward{} = \blockreward{} \frac{\minerpoolhashrate{\poolid}}{\poolhashrate{\poolid}}$. In other words, the miner having contributed a hashrate of $\minerpoolhashrate{\poolid}$ will get a reward based on the percent of the hashrate he contributed with respect to the total hashrate of PPLNS miners, multiplied by $\blockreward{\pplns}$ (i.e. reward of PPLNS).

\noindent \textbf{Strategy 2:} The pool manager pays the PPLNS miners based on the total hashrate of the pool $\poolhashrate{\poolid}$, then allocate the remainder of the rewards to the PPS bucket. By this strategy, $\blockreward{\minerpoolhashrate{\poolid}} = \blockreward{}\frac{\minerpoolhashrate{\poolid}}{\poolhashrate{\poolid}}$ which effectively results to the same paid amount as in Strategy 1.

\noindent \textbf{Strategy 3:} 
First replenish the PPS ``bucket'' based on the total amount $\tilde{R}$ was paid off to the PPS miners since the last block was found, then pay PPLNS based on what is left of the total reward. Following this strategy: $\blockreward{\pplns} = \blockreward \ - \tilde{\blockreward{}}$ and  miner's reward is $\blockreward{\minerpoolhashrate{\poolid}} = \blockreward{\pplns}\frac{\minerpoolhashrate{\poolid}}{\rewardpercentage\poolhashrate{\poolid}} = (\blockreward{} - \tilde{\blockreward{}})\frac{\minerpoolhashrate{\poolid}}{\rewardpercentage\poolhashrate{\poolid}}$. Effectively, the pool manager by this strategy transfers some of his risk to the PPLNS miners.

To our knowledge, no mining pool that offers both PPLNS and PPS reward systems specifies which strategy it follows.
Using public data to prove which strategy a mining pool follows is a non-trivial process.
We assume that pools offering both PPS and PPLNS reward mechanisms follow either Strategy 1 or 2 which are the most intuitive and produce the same end result for the miners.

%% file: active_miners.tex
\section{Active Miner's Problem}
\label{sec:model}

Our model builds on the recent study of \cite{jiasun}, which considers the problem of an ``active'' miner, who given a set of  $\poolsnum$ mining pools, where each pool $\poolid$ has a total hashrate $\poolhashrate{\poolid}$, and fee $\fee{\poolid}$, wishes to maximize his expected utility, or the weighted average of his expected earnings. 
In \cite{jiasun} the miner's utility is quantified based on the available pools' parameters and his Constant-Absolute-Risk-Aversion (CARA) $\rho$, as well as his own available hashing power and the expected reward value. We describe this model (which serves as the base of our schemes) in Section \ref{sec:single-crypto}.

\subsection{Mining on a Single Cryptocurrency}
\label{sec:single-crypto}
We consider an active miner, who owns mining hardware with PoW hashing power $\minertotalhashrate$ and is mining on a single cryptocurrency with total hashrate $\totalhashrate{}$. Ideally he would like to distribute $\minertotalhashrate$ among $\poolsnum$ mining pools of different sizes offering different fee structures (while possibly keeping a portion of his power for zero-fee solo mining) in a way that  maximizes his expected utility. Doing so is not trivial, as mining pools differ widely in their size and service fees, and in many cases there is no correlation between these two attributes.

A first attempt was done in~\cite{jiasun} where the authors introduced an analytical utility function to capture miners' risk aversion when considering mining pools offering a PPLNS reward method. 
The idea is to express the expected utility of pool-mining, in terms of a miner's PoW hash power $\minertotalhashrate$, pools' hashrates $\poolhashrate{\poolid}$, pools' fees $\fee{\poolid}$, a CARA value $\cara$ chosen by the miner and block reward $\blockreward{}$. 
The CARA value essentially quantifies how risk averse a miner is, where $\cara=0$ means that the miner is risk neutral.
The expected utility is calculated by Equation~\ref{formula}, where an active miner should find the vector ${\{\minerpoolhashrate{\poolid}\}}_{\poolid=1}^{\poolsnum}$ maximizing his utility value. This vector expresses an allocation of his mining power $\minertotalhashrate$ over $\poolsnum$ different pools, where $\minerpoolhashrate{\poolid}$ denotes the mining power allocated to pool $\poolid$:
\ifllncs
\else
	\begin{align}
	\label{formula}
	\sum_{\poolid=1}^{\poolsnum}(\minerpoolhashrate{\poolid} + \poolhashrate{\poolid})(1 - e^{-\cara\blockreward (1-\fee{\poolid})\frac{\minerpoolhashrate{\poolid}}{\minerpoolhashrate{\poolid} + \poolhashrate{\poolid}}}) + \nonumber\\ (\minertotalhashrate - \sum_{\poolid=1}^{\poolsnum}\minerpoolhashrate{\poolid})(1-e^{-\cara\blockreward{}})
	\end{align}
\fi
under constraints: \\
\begin{equation}
\notag
\sum_{\poolid=1}^{\poolsnum}\minerpoolhashrate{\poolid} \le \minertotalhashrate
\ \text{ and } \
\minerpoolhashrate{\poolid} \ge 0, \forall \poolid \in \poolsnum .
\end{equation}

By solving this optimization problem, we are given the optimal distribution of the miner's total hashpower $\minertotalhashrate$ to $\poolsnum$ pools. Note that the second term of Equation \ref{formula} expresses the left-over hashpower for the miner to mine ``solo''. If the miner is risk-neutral (i.e. $\cara=0$), solo mining (which has a zero fee) is the optimal solution.

The reasoning behind this model is that the total payoff $\totalreward$ of a miner, who has allocated his resources as above, is equal to the weighted sum of each pool's expected reward, with the weights being the miner's hash rate percentage in the pool, plus the expected reward from solo mining, which is expressed as: 
\begin{equation}\label{pplns-payoff}
\totalreward =	\sum_{m=1}^M \frac{\lambda_m}{\lambda_m + \Lambda_m} R \tilde{N}_{pool, m} + R \tilde{N}_{solo} 
\end{equation}
where $\tilde{N}_{pool, m} \sim {Poisson}(\frac{\lambda_m+\Lambda_m}{\Lambda})$ and $\tilde{N}_{solo} \sim Poisson(\frac{\lambda_{solo}}{\Lambda})$. The miner's expected utility $u(\cdot)$ is:
\begin{equation}\label{mgf}
u(\cdot) = {E} [ - e^{-\cara \totalreward} ]
\end{equation}
Notice that for a Poisson distributed variable $x$ with parameter $\lambda$, its moment generating function $E[e^{w x}]$ for any parameter $w$, is given by $e^{\lambda (e^w-1)}$. Now from Equations \ref{pplns-payoff} and \ref{mgf} we can derive Equation \ref{formula}. 
A similar reasoning applies to the rest of our models presented below.

\subsection{Allowing Selection of PPS pools}
The problem with the model of Eq.~\ref{formula} is that it restricts the miners to choose between pools that only offer the PPLNS reward method. Below, we provide a model that allows miners to  
 choose mining pools also offering PPS reward systems. In this case, a rational miner will choose to add only the PPS pool that offers the smaller fee, and disregard pools with higher PPS fees. Eq.~\ref{formula} is then transformed as follows (changes denoted in {\color{blue} blue} color):
\ifllncs
\else
	\begin{align}
	\label{SingleCryptPPS}	
	\sum_{\poolid=1}^{\poolsnum}(\minerpoolhashrate{\poolid} + \poolhashrate{\poolid})(1 - e^{-\cara \blockreward{} ( 1 - \fee{\poolid} ) \frac{\minerpoolhashrate{\poolid}}{\minerpoolhashrate{\poolid} + \poolhashrate{\poolid}}}) \nonumber
	\\ 
	+ (\minertotalhashrate {\color{blue} - \lambda_{\pps}} -  \sum_{\poolid=1}^{\poolsnum}\minerpoolhashrate{\poolid})(1-e^{-\cara\blockreward{}}) 
	{\color{blue} + \lambda_{\pps} (1 - \fee{\pps}) \cara \blockreward{}}
	\end{align}	
	under constraints: 
	\begin{align}
	\notag
	\sum_{\poolid=1}^{\poolsnum}\minerpoolhashrate{\poolid} {\color{blue}+\lambda_{\pps}} \le \minertotalhashrate, \ \ 
	\minerpoolhashrate{\poolid} \ge 0, \forall \poolid \in \poolsnum, \ \\ \text{ and }  \ 
	{\color{blue}\lambda_{\pps} \ge 0} . \nonumber
	\end{align}
	
\fi

%

The logic behind this model, similar to our basic single-cryptocurrency model, is that when assuming a miner who is willing to allocate his mining power towards PPS/PPLNS pools and solo-mining, we add to his total payoff $\totalreward$ a constant reward amount proportionate to the power allocated to the PPS pool over the total cryptocurrency's hashpower, minus the pool's fee, expressed as follows:
\ifllncs
\else
	\begin{align}\label{payoff}
	\totalreward =	\sum_{m=1}^M \frac{\lambda_m}{\lambda_m + \Lambda_m}(1-f_{\pps}) R \tilde{N}_{pool, m} + \nonumber
	\\R \tilde{N}_{solo} + R \frac{\lambda_{\pps} (1-f_{\pps})}{\Lambda} 
	\end{align}
\fi
From Equations \ref{payoff} and \ref{mgf} we derive Equation \ref{SingleCryptPPS}, where the last term $\lambda_{\pps} (1 - \fee{\pps}) \cara \blockreward{}$ expresses a steady income from the PPS pool.

\subsection{Mining Across Multiple Cryptocurrencies}
\label{multi-crypto-single-pow}
We now consider a miner who owns mining hardware with PoW hash power $\minertotalhashrate$ and wants to maximize his ``risk-sharing benefit'' value by mining over $\cryptonum$ different cryptocurrencies and $\poolsnum$ pools in total, provided that each cryptocurrency $\cryptoid \in \cryptonum$ uses the same PoW mining algorithm $\powalg{}$. The allocation of the miner's hashing power $\minertotalhashrate$ will  now be  ${\{\minerpoolhashrate{\poolid,\cryptoid}\}}_{\poolid=1}^{\poolsnum}, \cryptoid \in \cryptonum$ and the first constraint in Equation~\ref{formula} takes the form:
\ifllncs
\else
\begin{equation}
\notag
\sum_{\cryptoid \in \cryptonum}\sum_{\poolid=1}^{\poolsnum}\minerpoolhashrate{\poolid,\cryptoid} \le \minertotalhashrate.
\end{equation} 
\fi

However, each cryptocurrency $\cryptoid$ has its own block reward $\blockreward{\cryptoid}$\footnote{We should be careful to express $\blockreward{\cryptoid}$ in a fiat currency value (e.g. USD), as we do not take different cryptocurrency exchange rates into consideration.} and its own \emph{average} block time $\blocktime{\cryptoid}$. Thus, we also consider the \emph{reward over time} ratio  $\frac{\blockreward{\cryptoid}}{\blocktime{\cryptoid}}$, as it effectively normalizes $\blockreward{\cryptoid}$ over different cryptocurrencies. In addition, a miner's hashrate $\minerpoolhashrate{\poolid}$ allocated to pool $\poolid$ that mines cryptocurrency $\cryptoid$ should be normalized to each cryptocurrency's total  hashrate $\totalhashrate{\cryptoid}$. So, in this (more general) case, Equation \ref{formula} takes the following form, under the new constraints outlined above:
\ifllncs
\else
	\begin{align}
	\label{formula2}
	\sum_{\cryptoid \in \cryptonum}\Bigg(\sum_{\poolid=1}^{\poolsnum}&\frac{(\minerpoolhashrate{\poolid,\cryptoid} + \poolhashrate{\poolid,\cryptoid})}{\blocktime{\cryptoid}\totalhashrate{\cryptoid}} (1 - e^{-\cara\blockreward{\cryptoid}(1-\fee{\poolid,\cryptoid})\frac{\minerpoolhashrate{\poolid,\cryptoid}}{\minerpoolhashrate{\poolid,\cryptoid} + \poolhashrate{\poolid,\cryptoid}}}) \nonumber\\ &+ \frac{\minerpoolhashrate{0,\cryptoid}}{\blocktime{\cryptoid}\totalhashrate{\cryptoid}} (1-e^{-\cara\blockreward{\cryptoid}})\Bigg)
	\end{align}
\fi
where $\minerpoolhashrate{0,\cryptoid}$ denotes solo mining for cryptocurrency $\cryptoid$, and the first constraint of Equation \ref{formula} is more precisely expressed as

\begin{equation}
\notag
\sum_{\cryptoid \in \cryptonum}\Bigg(\sum_{\poolid=1}^{\poolsnum}\minerpoolhashrate{\poolid,\cryptoid} + \minerpoolhashrate{0,\cryptoid}\Bigg) \le \minertotalhashrate .
\end{equation}

The above constraint shows that the miner could choose to diversify his solo-mining (which was initially expressed by the second term in Equation \ref{formula}) over multiple currencies as well, in a similar fashion as he would do by diversifying across multiple mining pools.

\subsection{Mining Across Cryptocurrencies with Different PoW Algorithms}
\label{multi-crypto-multi-pow}
In the previous sections we assumed that the active miner's goal is to maximize his ``risk-sharing benefit'' value, given that he mines on one or more different cryptocurrencies using the same PoW algorithm. We now generalize our consideration, by allowing a miner to distribute his power across $\cryptonum$ different cryptocurrencies, and across $\powalgnum$ different PoW algorithms\footnote{This of course assumes that the miner owns CPU/GPU mining hardware, since ASICs are restricted to specific PoW algorithms.}.

Let $\powalgnum$ be the set of PoW algorithms. Since each algorithm solves a different version of the PoW puzzle, and uses a different set of hash functions, the miner's ``total'' hashrate $\minertotalalgrate{\powalg{i}}$ for each algorithm $\powalg{i}$ will change. However the miner might choose to allocate his hardware ``power'' among different mining PoW puzzles at the same time (in CPU mining, that would require setting priority levels to each Operating System process $i$). In this case, we can use Equation \ref{formula2}, but its first constraint will become

\begin{equation}
\notag
\sum_{\powalg{i} \in \powalgnum}\sum_{\cryptoid \in \cryptonum}\sum_{\poolid=1}^{\poolsnum}\frac{\minerpoolhashrate{\poolid,\cryptoid}}{\minertotalalgrate{\powalg{i}}} \le 1.
\end{equation}
\ifllncs
\else
\subsubsection*{Example}
\fi
Assume a miner who owns some amount of computational power (CPUs and/or GPUs). With his hardware, he could mine exclusively cryptocurrency $\cryptoid_{1}$ which uses PoW mining algorithm $\powalg{1}$, and his maximum hashrate would be $\minertotalalgrate{\powalg{1}}$. Alternatively, he could mine exclusively coin $\cryptoid_{2}$ which uses PoW mining algorithm $\powalg{2}$, at a hashing rate of $\minertotalalgrate{\powalg{2}}$. Now he wishes to diversify his risk among these two cryptocurrencies (for simplicity we assume that he chooses to mine them only on a single pool each). The resulting constraint would be
\begin{equation}
\notag
\frac{\minerpoolhashrate{1,1}}{\minertotalalgrate{\powalg{1}}} + \frac{\minerpoolhashrate{2,2}}{\minertotalalgrate{\powalg{2}}} \le 1.
\end{equation}
Each term represents the ``percentage'' of the miner's CPU (and/or GPU) power devoted to mining on a specific PoW algorithm. The sum of the ratios cannot exceed 1 which represents the total CPU and/or GPU power of the hardware\footnote{In our assumption we do not take a ``dual mining'' GPU setup into account.}.

\subsection{Transaction Fees in Mining Pools}
\label{pools-tx-fees}

In the early days of cryptocurrencies, transaction fees were negligible compared to block rewards. However during recent periods, transaction fees have risen to considerable amounts, especially in the case of Bitcoin~\cite{bitinfocharts}. It is up to the pool manager to decide if  the  transaction fees are distributed to the participating miners or is kept by the pool for profit. However, especially during  periods of high transaction fees, not including these fees to the miners' reward $\blockreward{}$ can be thought intuitively as a ``hidden fee''. Let $\txfee{\cryptoid}$ the \emph{average} transaction fee for cryptocurrency $\cryptoid$ observed during a recent period of time. Also we set $\txfee{\poolid,\cryptoid} = 1$ if the pool pays transaction fees to the miner, else we set $\txfee{\poolid,\cryptoid} = 0$. We modify Equation \ref{formula2} to include (non-negligible) transaction fees as follows (changes denoted in color, the equations for other cases can be modified in a similar way): 
\ifllncs
\else
	\begin{flalign}
	\label{formula3}
	\sum_{\cryptoid \in \cryptonum}&\Bigg(\sum_{\poolid=1}^{\poolsnum}\frac{(\minerpoolhashrate{\poolid,\cryptoid} + \poolhashrate{\poolid,\cryptoid})}{\blocktime{\cryptoid}\totalhashrate{\cryptoid}}(1 - e^{-\cara(\blockreward{\cryptoid}{\color{blue}+\txfee{\cryptoid}\txfee{\poolid,\cryptoid}})(1-\fee{\poolid,\cryptoid})\frac{\minerpoolhashrate{\poolid,\cryptoid}}{\minerpoolhashrate{\poolid,\cryptoid} + \poolhashrate{\poolid,\cryptoid}}}) &\nonumber \\ 
	&+ \frac{\minerpoolhashrate{0,\cryptoid}}{\blocktime{\cryptoid}\totalhashrate{\cryptoid}} (1-e^{-\cara(\blockreward{\cryptoid}{\color{blue}+\txfee{\cryptoid}\txfee{\poolid,\cryptoid}})})\Bigg) .&
	\end{flalign}
\fi

%% file: evaluation.tex
\section{Implementation of Our Model}
\label{sec:evaluation}

In this Section we present an implementation of our mining resources allocation mechanism.  
We developed a Python tool that automates the decision for an active miner, who owns either commercial off-the-shelf (COTS) hardware (e.g. CPUs/GPUs) or application-specific integrated circuit hardware (ASICs) \footnote{\label{github}Our implementation can be accessed at \url{http://smart-miner.cs.gmu.edu/}}. Our tool covers all the cases discussed in Section \ref{sec:model}. 
\ifllncs
\else
\subsubsection*{Choosing the Right Optimization Method} 
\fi
 In order to find the best possible allocation of the miner's hash power we tried a few optimization methods. First, we applied the sequential least squares programming algorithm (SLSQP), which uses the Han-Powell quasi-newton method with a BFGS update of the B-matrix and an L1-test function for the steplength algorithm \cite{kraft1988software}. Second, we implemented a modification of Newton's method that solves the Lagrange system of equations for the active constraints. To our surprise both mentioned gradient based optimization methods experienced difficulties in obtaining accurate solutions to the optimization problem for very small values of $\cara$, about $10^{-5}$. A possible explanation to that phenomenon is that the instances with a wide range of hashes/sec from a few to quintillion $10^{18}$, make the gradients calculated with significant computational errors. The presence of exponential functions sensitive to the scale of their argument is a contributing factor for the loss of the accuracy for the obtained gradients under the finite precision computer arithmetic. Even though those methods could be used for the cases with larger values of $\cara$, we abandoned them. 
 
 The most successful algorithm for the optimization problem was the constrained optimization by linear approximation (COBYLA) \cite{Powell1994}. COBYLA was developed for solving nonlinear constrained optimization problems via a sequence of linear programming subproblems, each solved on an updated simplex. COBYLA is a good fit for our optimization problem for the following reasons. First, our feasible set is a simplex, so the vertices of the feasible set form a good initial linear approximation. Second, our problem is low dimensional (no more than a dozen of variables). The low dimensionality of the problem results in a relatively small number of simplexes that need to be constructed before the solutions is found. Finally, COBYLA is a gradient free algorithm. Therefore to update iterates, COBYLA does not rely on the gradient obtained locally in one point, which may be not accurate for this problem. Instead, in its search COBYLA relies on the slope of a linear n-dimensional approximation calculated out of readily available n+1 feasible points. We believe that all these factors together contribute to the efficiency of the algorithm for finding the best possible allocation of the miner's hash power. Therefore, our tool utilizes COBYLA optimization solving method. 

\ifllncs
\else
\subsubsection*{Description and Instantiation Assumptions}
\fi
\label{sec:assumptions}
The basic single - cryptocurrency version of our tool, given as input the miner's hashing power $\minertotalhashrate$, coin's exchange rate $\exchange{}$, chosen pool data $[\poolhashrate{i},\fee{i}]_{i=1}^{M}$ (pool total hashrate and fee respectively) and risk aversion $\cara$, outputs the optimal distribution of his hashpower over these pools plus a ``solo-mining'' remainder. Some instantiations of our tool are outlined in Section~\ref{sec:SingleCurrencyEval} as examples for a typical value range of $\cara$. Our tool can also provide the optimal distribution for the ``multi-cryptocurrency, single PoW algorithm'' (Section \ref{multi-crypto-single-pow}) and ``multi-cryptocurrency, multi-PoW algorithm''  (Section \ref{multi-crypto-multi-pow}) extensions, and we also outline an extended instantiation in Section \ref{sec:MultiCurrencyEval}.

The results of our tool can be easily applied for mining in large-scale, where a miner can allocate a portion of his hardware to mine on a specific pool. The process of applying the results on a single mining hardware piece is not trivial, as to our knowledge, no ASIC or GPU miner application exists that enables the user to allocate his mining power over many pools by a specific percentage, even if theoretically it's technically feasible. The majority of ASICs utilize a fork of the \texttt{cgminer} tool, which initially offered a ``multipool strategy'' option for the miner, but was later deprecated as it was no longer compatible with the modern stratum mining protocol~\cite{cgminer}. Multi-pool mining in a round-robin fashion is not efficient for the miner as well, as this would result to a decrease in his overall reward, given the nature of reward schemes that prevent pool ``hopping''. We encourage ASIC manufacturers and mining application developers to enable user-specified multi-pool mining in future releases, for the benefit of the miners and the whole community.

We assume that the miner possess an average wealth of $\mathcal{W}$ = \$100k, while having typical values for the constant relative risk aversion CRRA metric between 1 and 10 \cite{RePEc:oxp:obooks:9780195102680}. Given that CARA = CRRA$/ \mathcal{W}$, we take as typical values for CARA $\cara$ between $10^{-5} - 10^{-4}$, which we mostly assume throughout the rest of this paper. Note that changing our assumption for our miner's wealth is equivalent to changing the typical value range for $\cara$ accordingly (we include additional evaluation analysis for a broader range of $\cara$ values).

\subsection{Single Cryptocurrency}
\label{sec:SingleCurrencyEval}

\ifllncs
\else
	\subsubsection{Evaluation I}
\fi
We instantiate the first experiment of our tool by using the following parameters: a miner with total hashpower $\minertotalhashrate$ = 40 hashes/sec, wishing to mine on a single cryptocurrency with block reward $\blockreward{}$ = \$50000, having picked 4 mining pools with parameters shown in Table \ref{single-crypto-ex}:
These values do not correspond to ``real'' mining pools (or an existing cryptocurrency), but are representatives for different classes of pools in terms of relative size, as larger real-world pools charge higher fees (but have less fluctuations on the miner's income) while smaller pools have lower (or even zero) fees to attract new miners to them. The results depicted on Figure \ref{single-crypto} show that our model produces the expected choices for rational miners.  For smaller values of $\rho$ the miner is willing to ``risk'' more, and would dedicate much of his hashpower to the small Pool 4, but for larger values of $\rho$ the miner would diversify among larger pools for a steadier income. Another important observation is that for $\cara > 6\cdot10^{-5}$ the miner would allocate some of his power at both Pools 1 and 2 to diversify his risk (which are the ``largest'' pools, having the same 2\% fee), although he would show a strong preference for Pool 1 which is 10 times larger than Pool 2. Note that for simplicity, we do not take any transaction fees kept by pools into account, however using the methodology discussed in Section \ref{pools-tx-fees} our evaluation would produce equivalent results.
\ifllncs
\else
	\begin{table}[t]			
	\caption{Evaluation I Pool parameters.}
	\label{single-crypto-ex}
	\centering		
	\begin{tabular}{|l|l|l|}
		\hline
		Pool 1&$\poolhashrate{1}$&$10^{6}$ hashes/sec\\
		\cline{2-3}
		&$ \fee{1}$ & 2\% \\
		\hline
		Pool 2&$\poolhashrate{2}$&$10^{5}$ hashes/sec\\
		\cline{2-3}
		&$ \fee{2}$ & 2\% \\
		\hline
		Pool 3&$\poolhashrate{3}$&$10^{4}$ hashes/sec\\
		\cline{2-3}
		&$ \fee{3}$ & 1\% \\
		\hline
		Pool 4&$\poolhashrate{4}$&$10^{3}$ hashes/sec\\
		\cline{2-3}
		&$ \fee{4}$ & 0\% \\
		\hline
	\end{tabular}
	\end{table}

	\begin{figure}[!t]
			\centering
		\resizebox{\figurewidth\textwidth}{!}{
			\input{single_crypto.pgf}
		}
		\caption{Single cryptocurrency diversification.} 
		\label{single-crypto}
\end{figure}
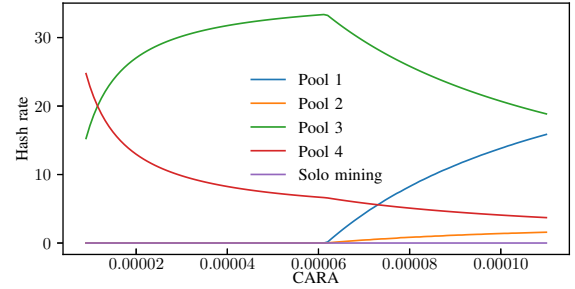
\fi

\ifllncs
\else
\subsubsection{Evaluation II}
\fi
We then pick some actual Bitcoin pools: Slush pool, ViaBTC and KanoPool. These pools, as indicated by their parameters shown in Table \ref{single-crypto-bitcoin} (values as of February 2019) are representatives of the options available to a miner, as they cover a wide range of pool hash-power $\poolhashrate{\poolid}$ and pool fee $\fee{\poolid}$. The above pools use either PPLNS or Score (variant of proportional) reward methods. We do not include a PPS pool in this example,  although taken into account in Equation \ref{SingleCryptPPS}, as the results turned out to be identical for the typical value range of $\cara$. Using our parameters, a PPS pool would participate in the diversification only for large values of $\cara$ that are not within the typical range (we show such an example later in this section).  For the other parameters, we consider a large-scale miner who owns total mining power of $\minertotalhashrate$ = 3000 TH/sec (roughly about 100 units of Antminer S15 ASICs), and the Bitcoin total block reward $\blockreward{}$ = \$45441 \footnote{\label{parameters-footnote}Parameters as of Jan 25 2019, retrieved from \url{https://btc.com/} and \url{https://bitinfocharts.com/}.}. The pool parameters are shown in Table \ref{single-crypto-bitcoin} and the resulting diversification graph in Figure \ref{single-crypto2}, where we observe a similar pattern to the previous ``representative'' pools example (i.e. a miner's preference for larger pools and steadier income as $\cara$ increases). 
\ifllncs
\else
	\begin{table}		
	\caption{Evaluation II Pool parameters}
	\label{single-crypto-bitcoin}
	\centering		
	\begin{tabular}{|l|l|l|}
		\hline
		Slush pool&$\poolhashrate{1}$&4040 PH/sec\\
		\cline{2-3}
		&$ \fee{1}$ & 2\% \\
		\hline
		ViaBTC&$\poolhashrate{2}$&3090 PH/sec\\
		\cline{2-3}
		&$ \fee{2}$ & 2\% \\
		\hline
		KanoPool&$\poolhashrate{3}$&48 PH/sec\\
		\cline{2-3}
		&$ \fee{3}$ & 0.9\% \\
		\hline
	\end{tabular}
\end{table}

	\begin{figure}
	\centering
	\resizebox{\figurewidth\textwidth}{!}{
		\input{single_crypto2.pgf}
	}
	\caption{Large-scale miner on Bitcoin pools.} \label{single-crypto2}	
	
\end{figure}
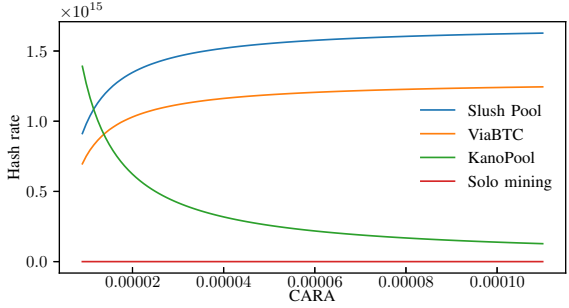
\begin{figure}
	\centering
	\resizebox{\figurewidth\textwidth}{!}{
		\input{single_crypto_small.pgf}
	}
	\caption{Small-scale miner on Bitcoin pools, parameters as in Table \ref{single-crypto-bitcoin}.}
	\label{single-crypto-small}	
\end{figure}
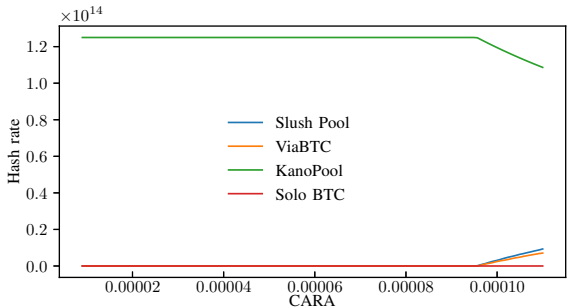
\fi

\subsubsection{Diversifying on a PPS Pool}
In the previous evaluation we showed that a PPS pool would not participate in the diversification using parameters for actual pools shown in Table \ref{single-crypto-bitcoin} and typical values for $\cara$. In Figure \ref{single-crypto-ppsfig} we show how a PPS pool would affect a miner's diversification for non-typical large values of $\cara$, using the parameters in Table \ref{single-crypto-pps}. This would be applicable only for a miner who is very risk-averse, as he would show a stronger preference to the steady income a PPS pool provides, as the value of $\cara$ increases.

\subsubsection{Small-scale miners}
An interesting observation is in regards of smaller scale miners and pools with higher fees. For instance, an active miner with 10 ASICs instead of 100, when following our method,  would allocate all  his hash power to the smaller pool (KanoPool) with the lowest fee.
In Figure \ref{single-crypto-small} we show a small-scale (or ``home'') Bitcoin miner as an example ($\minertotalhashrate$ = 125 TH/sec). Given his relatively small hashpower, he would only choose the lowest-fee pool to mine (KanoPool), without allocating any resources to larger pools with higher fees, except for the upper values of $\cara$. Essentially, it is shown that risk-aversion has less effect on small-scale miners.

\ifllncs
\else
\begin{table}[t]
	\caption{Bitcoin pool parameters including PPS pool}
	\label{single-crypto-pps}
	\centering		
	\begin{tabular}{|l|l|l|}
		\hline
		Slush pool&$\poolhashrate{1}$&4040 PH/sec\\
		\cline{2-3}
		&$ \fee{1}$ & 2\% \\
		\hline
		ViaBTC&$\poolhashrate{2}$&3090 PH/sec\\
		\cline{2-3}
		&$ \fee{2}$ & 2\% \\
		\hline
		KanoPool&$\poolhashrate{3}$&48 PH/sec\\
		\cline{2-3}
		&$ \fee{3}$ & 0.9\% \\
		\hline
		PPS Pool&$ \fee{4}$ & 4\% \\
		\hline
	\end{tabular}
\end{table}
\begin{figure}
	\centering
	\resizebox{\figurewidth\textwidth}{!}{
		\input{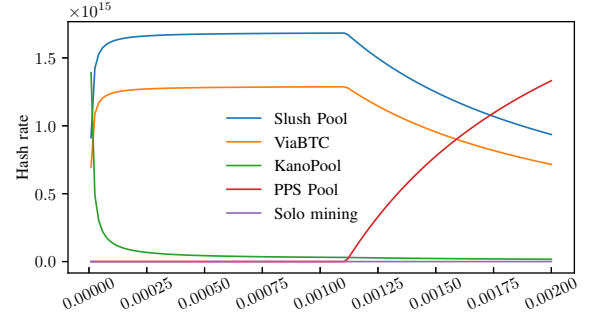}
	}
	\caption{Single currency with PPS pool and large values of $\cara$.} \label{single-crypto-ppsfig}	
\end{figure}
\fi

\subsection{Multiple  Cryptocurrencies}
\label{sec:MultiCurrencyEval}
We now consider a miner who diversifies over different cryptocurrencies. For simplicity, we just switch\footnote{Many pools as shown in the Appendix host pool mining services for multiple cryptocurrencies.} the currency in the 2nd pool (ViaBTC) from Bitcoin to Bitcoin Cash. The pool parameters are shown in Table \ref{multi-crypto-table}, $\totalhashrate{\textsf{BTC}}$ = 42.33 EH/sec, $\totalhashrate{\textsf{BCH}}$ = 1.43 EH/sec and $\blockreward{\textsf{BCH}}$ = \$1547 \textsuperscript{\ref{parameters-footnote}}. The resulting graph in Figure \ref{multi-crypto} show the diversification of his computational power for various values of $\cara$. We observe that in this instance, for small values of $\cara$ his optimal strategy would be to keep most of his resources for zero-fee Bitcoin solo mining. However, for increasing values of $\cara$ he would diversify his power to larger pools, and he would also choose to allocate some of his power to the Bitcoin Cash pool, even though it has the same fee and the pool might not be as profitable as the Bitcoin pool.  

\ifllncs
\else
\begin{table}[t]		
	\caption{Multi-currency pool parameters}
	\label{multi-crypto-table}
	\centering		
	\begin{tabular}{|l|l|l|}
		\hline
		Slush pool&$\poolhashrate{1}$&4040 PH/sec\\
		\cline{2-3}
		&$ \fee{1}$ & 2\% \\
		\hline
		ViaBTC&$\poolhashrate{2}$&135 PH/sec\\
		\cline{2-3}
		(Bitcoin Cash)&$ \fee{2}$ & 2\% \\
		\hline
		KanoPool&$\poolhashrate{3}$&48 PH/sec\\
		\cline{2-3}
		&$ \fee{3}$ & 0.9\% \\
		\hline
	\end{tabular}
\end{table}
\begin{figure}[t]
	\centering
	\resizebox{\figurewidth\textwidth}{!}{
		\input{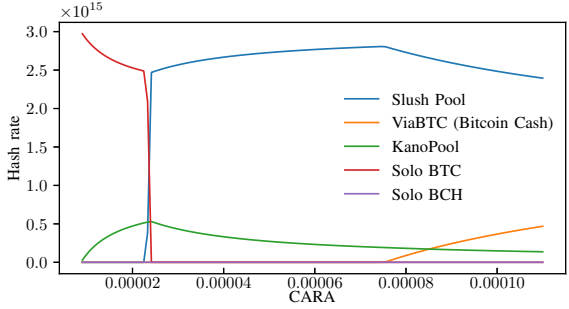}
	}
	\caption{Large-scale miner on SHA-256 pools.} \label{multi-crypto}	
\end{figure}
\fi

\subsubsection*{Impact of Exchange Rates} We noted that our results are highly sensitive even to very small changes to any of the parameters, such as the exchange rates. For instance, the same miner having chosen the same pools, based on the historical data, would allocate all of his power to the Bitcoin Cash pool the previous day, while after a few days he would transfer all of his power to the Bitcoin pools. In Figures \ref{multi-crypto-before-after}(a) and \ref{multi-crypto-before-after}(b) we show how small daily fluctuations in the exchange rate between two same-PoW cryptocurrencies can affect the miner's diversification for these cryptocurrencies. While in our previous instantiation the exchange rate was BTC/BCH = 0.034, a small increase in favor of Bitcoin Cash's value totally eliminates the presence of Bitcoin pools from the diversification, leaving only the Bitcoin Cash pool and Bitcoin Cash solo mining for the miner as his options. On the other hand, a small increase in favor of Bitcoin's value eliminates the presence of the Bitcoin Cash pool, and the miner would only diversify among the Bitcoin pools.

\ifllncs
\else

\begin{figure}
	\subfigure[BTC/BCH = 0.033.]{
		\resizebox{\figurewidth\textwidth}{!}{%
			\input{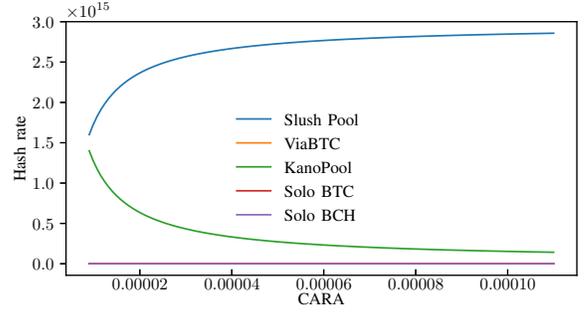}
	}}
	\quad
	\subfigure[BTC/BCH = 0.035.]{
		\resizebox{\figurewidth\textwidth}{!}{%
			\input{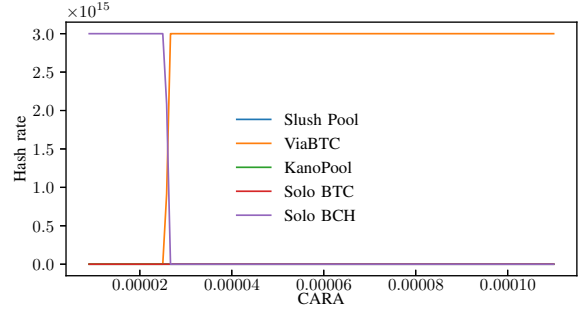}
		}
	}
	\caption{Large-scale miner on SHA-256 pools, parameters in Table \ref{multi-crypto-table} } \label{multi-crypto-before-after}
\end{figure}

\fi

%% file: single_crypto.pgf
\begingroup%
\makeatletter%
\begin{pgfpicture}%
\pgfpathrectangle{\pgfpointorigin}{\pgfqpoint{5.400000in}{2.700000in}}%
\pgfusepath{use as bounding box, clip}%
\begin{pgfscope}%
\pgfsetbuttcap%
\pgfsetmiterjoin%
\definecolor{currentfill}{rgb}{1.000000,1.000000,1.000000}%
\pgfsetfillcolor{currentfill}%
\pgfsetlinewidth{0.000000pt}%
\definecolor{currentstroke}{rgb}{1.000000,1.000000,1.000000}%
\pgfsetstrokecolor{currentstroke}%
\pgfsetdash{}{0pt}%
\pgfpathmoveto{\pgfqpoint{0.000000in}{0.000000in}}%
\pgfpathlineto{\pgfqpoint{5.400000in}{0.000000in}}%
\pgfpathlineto{\pgfqpoint{5.400000in}{2.700000in}}%
\pgfpathlineto{\pgfqpoint{0.000000in}{2.700000in}}%
\pgfpathclose%
\pgfusepath{fill}%
\end{pgfscope}%
\begin{pgfscope}%
\pgfsetbuttcap%
\pgfsetmiterjoin%
\definecolor{currentfill}{rgb}{1.000000,1.000000,1.000000}%
\pgfsetfillcolor{currentfill}%
\pgfsetlinewidth{0.000000pt}%
\definecolor{currentstroke}{rgb}{0.000000,0.000000,0.000000}%
\pgfsetstrokecolor{currentstroke}%
\pgfsetstrokeopacity{0.000000}%
\pgfsetdash{}{0pt}%
\pgfpathmoveto{\pgfqpoint{0.675000in}{0.297000in}}%
\pgfpathlineto{\pgfqpoint{4.860000in}{0.297000in}}%
\pgfpathlineto{\pgfqpoint{4.860000in}{2.376000in}}%
\pgfpathlineto{\pgfqpoint{0.675000in}{2.376000in}}%
\pgfpathclose%
\pgfusepath{fill}%
\end{pgfscope}%
\begin{pgfscope}%
\pgfsetbuttcap%
\pgfsetroundjoin%
\definecolor{currentfill}{rgb}{0.000000,0.000000,0.000000}%
\pgfsetfillcolor{currentfill}%
\pgfsetlinewidth{0.803000pt}%
\definecolor{currentstroke}{rgb}{0.000000,0.000000,0.000000}%
\pgfsetstrokecolor{currentstroke}%
\pgfsetdash{}{0pt}%
\pgfsys@defobject{currentmarker}{\pgfqpoint{0.000000in}{0.000000in}}{\pgfqpoint{0.000000in}{0.048611in}}{%
\pgfpathmoveto{\pgfqpoint{0.000000in}{0.000000in}}%
\pgfpathlineto{\pgfqpoint{0.000000in}{0.048611in}}%
\pgfusepath{stroke,fill}%
}%
\begin{pgfscope}%
\pgfsys@transformshift{1.279584in}{0.297000in}%
\pgfsys@useobject{currentmarker}{}%
\end{pgfscope}%
\end{pgfscope}%
\begin{pgfscope}%
\pgftext[x=1.279584in,y=0.269222in,,top]{\rmfamily\fontsize{10.000000}{12.000000}\selectfont \(\displaystyle 0.00002\)}%
\end{pgfscope}%
\begin{pgfscope}%
\pgfsetbuttcap%
\pgfsetroundjoin%
\definecolor{currentfill}{rgb}{0.000000,0.000000,0.000000}%
\pgfsetfillcolor{currentfill}%
\pgfsetlinewidth{0.803000pt}%
\definecolor{currentstroke}{rgb}{0.000000,0.000000,0.000000}%
\pgfsetstrokecolor{currentstroke}%
\pgfsetdash{}{0pt}%
\pgfsys@defobject{currentmarker}{\pgfqpoint{0.000000in}{0.000000in}}{\pgfqpoint{0.000000in}{0.048611in}}{%
\pgfpathmoveto{\pgfqpoint{0.000000in}{0.000000in}}%
\pgfpathlineto{\pgfqpoint{0.000000in}{0.048611in}}%
\pgfusepath{stroke,fill}%
}%
\begin{pgfscope}%
\pgfsys@transformshift{2.032959in}{0.297000in}%
\pgfsys@useobject{currentmarker}{}%
\end{pgfscope}%
\end{pgfscope}%
\begin{pgfscope}%
\pgftext[x=2.032959in,y=0.269222in,,top]{\rmfamily\fontsize{10.000000}{12.000000}\selectfont \(\displaystyle 0.00004\)}%
\end{pgfscope}%
\begin{pgfscope}%
\pgfsetbuttcap%
\pgfsetroundjoin%
\definecolor{currentfill}{rgb}{0.000000,0.000000,0.000000}%
\pgfsetfillcolor{currentfill}%
\pgfsetlinewidth{0.803000pt}%
\definecolor{currentstroke}{rgb}{0.000000,0.000000,0.000000}%
\pgfsetstrokecolor{currentstroke}%
\pgfsetdash{}{0pt}%
\pgfsys@defobject{currentmarker}{\pgfqpoint{0.000000in}{0.000000in}}{\pgfqpoint{0.000000in}{0.048611in}}{%
\pgfpathmoveto{\pgfqpoint{0.000000in}{0.000000in}}%
\pgfpathlineto{\pgfqpoint{0.000000in}{0.048611in}}%
\pgfusepath{stroke,fill}%
}%
\begin{pgfscope}%
\pgfsys@transformshift{2.786334in}{0.297000in}%
\pgfsys@useobject{currentmarker}{}%
\end{pgfscope}%
\end{pgfscope}%
\begin{pgfscope}%
\pgftext[x=2.786334in,y=0.269222in,,top]{\rmfamily\fontsize{10.000000}{12.000000}\selectfont \(\displaystyle 0.00006\)}%
\end{pgfscope}%
\begin{pgfscope}%
\pgfsetbuttcap%
\pgfsetroundjoin%
\definecolor{currentfill}{rgb}{0.000000,0.000000,0.000000}%
\pgfsetfillcolor{currentfill}%
\pgfsetlinewidth{0.803000pt}%
\definecolor{currentstroke}{rgb}{0.000000,0.000000,0.000000}%
\pgfsetstrokecolor{currentstroke}%
\pgfsetdash{}{0pt}%
\pgfsys@defobject{currentmarker}{\pgfqpoint{0.000000in}{0.000000in}}{\pgfqpoint{0.000000in}{0.048611in}}{%
\pgfpathmoveto{\pgfqpoint{0.000000in}{0.000000in}}%
\pgfpathlineto{\pgfqpoint{0.000000in}{0.048611in}}%
\pgfusepath{stroke,fill}%
}%
\begin{pgfscope}%
\pgfsys@transformshift{3.539710in}{0.297000in}%
\pgfsys@useobject{currentmarker}{}%
\end{pgfscope}%
\end{pgfscope}%
\begin{pgfscope}%
\pgftext[x=3.539710in,y=0.269222in,,top]{\rmfamily\fontsize{10.000000}{12.000000}\selectfont \(\displaystyle 0.00008\)}%
\end{pgfscope}%
\begin{pgfscope}%
\pgfsetbuttcap%
\pgfsetroundjoin%
\definecolor{currentfill}{rgb}{0.000000,0.000000,0.000000}%
\pgfsetfillcolor{currentfill}%
\pgfsetlinewidth{0.803000pt}%
\definecolor{currentstroke}{rgb}{0.000000,0.000000,0.000000}%
\pgfsetstrokecolor{currentstroke}%
\pgfsetdash{}{0pt}%
\pgfsys@defobject{currentmarker}{\pgfqpoint{0.000000in}{0.000000in}}{\pgfqpoint{0.000000in}{0.048611in}}{%
\pgfpathmoveto{\pgfqpoint{0.000000in}{0.000000in}}%
\pgfpathlineto{\pgfqpoint{0.000000in}{0.048611in}}%
\pgfusepath{stroke,fill}%
}%
\begin{pgfscope}%
\pgfsys@transformshift{4.293085in}{0.297000in}%
\pgfsys@useobject{currentmarker}{}%
\end{pgfscope}%
\end{pgfscope}%
\begin{pgfscope}%
\pgftext[x=4.293085in,y=0.269222in,,top]{\rmfamily\fontsize{10.000000}{12.000000}\selectfont \(\displaystyle 0.00010\)}%
\end{pgfscope}%
\begin{pgfscope}%
\pgftext[x=2.767500in,y=0.156972in,,top]{\rmfamily\fontsize{10.000000}{12.000000}\selectfont CARA}%
\end{pgfscope}%
\begin{pgfscope}%
\pgfsetbuttcap%
\pgfsetroundjoin%
\definecolor{currentfill}{rgb}{0.000000,0.000000,0.000000}%
\pgfsetfillcolor{currentfill}%
\pgfsetlinewidth{0.803000pt}%
\definecolor{currentstroke}{rgb}{0.000000,0.000000,0.000000}%
\pgfsetstrokecolor{currentstroke}%
\pgfsetdash{}{0pt}%
\pgfsys@defobject{currentmarker}{\pgfqpoint{-0.048611in}{0.000000in}}{\pgfqpoint{0.000000in}{0.000000in}}{%
\pgfpathmoveto{\pgfqpoint{0.000000in}{0.000000in}}%
\pgfpathlineto{\pgfqpoint{-0.048611in}{0.000000in}}%
\pgfusepath{stroke,fill}%
}%
\begin{pgfscope}%
\pgfsys@transformshift{0.675000in}{0.391500in}%
\pgfsys@useobject{currentmarker}{}%
\end{pgfscope}%
\end{pgfscope}%
\begin{pgfscope}%
\pgftext[x=0.508333in,y=0.343282in,left,base]{\rmfamily\fontsize{10.000000}{12.000000}\selectfont \(\displaystyle 0\)}%
\end{pgfscope}%
\begin{pgfscope}%
\pgfsetbuttcap%
\pgfsetroundjoin%
\definecolor{currentfill}{rgb}{0.000000,0.000000,0.000000}%
\pgfsetfillcolor{currentfill}%
\pgfsetlinewidth{0.803000pt}%
\definecolor{currentstroke}{rgb}{0.000000,0.000000,0.000000}%
\pgfsetstrokecolor{currentstroke}%
\pgfsetdash{}{0pt}%
\pgfsys@defobject{currentmarker}{\pgfqpoint{-0.048611in}{0.000000in}}{\pgfqpoint{0.000000in}{0.000000in}}{%
\pgfpathmoveto{\pgfqpoint{0.000000in}{0.000000in}}%
\pgfpathlineto{\pgfqpoint{-0.048611in}{0.000000in}}%
\pgfusepath{stroke,fill}%
}%
\begin{pgfscope}%
\pgfsys@transformshift{0.675000in}{0.958003in}%
\pgfsys@useobject{currentmarker}{}%
\end{pgfscope}%
\end{pgfscope}%
\begin{pgfscope}%
\pgftext[x=0.438888in,y=0.909785in,left,base]{\rmfamily\fontsize{10.000000}{12.000000}\selectfont \(\displaystyle 10\)}%
\end{pgfscope}%
\begin{pgfscope}%
\pgfsetbuttcap%
\pgfsetroundjoin%
\definecolor{currentfill}{rgb}{0.000000,0.000000,0.000000}%
\pgfsetfillcolor{currentfill}%
\pgfsetlinewidth{0.803000pt}%
\definecolor{currentstroke}{rgb}{0.000000,0.000000,0.000000}%
\pgfsetstrokecolor{currentstroke}%
\pgfsetdash{}{0pt}%
\pgfsys@defobject{currentmarker}{\pgfqpoint{-0.048611in}{0.000000in}}{\pgfqpoint{0.000000in}{0.000000in}}{%
\pgfpathmoveto{\pgfqpoint{0.000000in}{0.000000in}}%
\pgfpathlineto{\pgfqpoint{-0.048611in}{0.000000in}}%
\pgfusepath{stroke,fill}%
}%
\begin{pgfscope}%
\pgfsys@transformshift{0.675000in}{1.524506in}%
\pgfsys@useobject{currentmarker}{}%
\end{pgfscope}%
\end{pgfscope}%
\begin{pgfscope}%
\pgftext[x=0.438888in,y=1.476288in,left,base]{\rmfamily\fontsize{10.000000}{12.000000}\selectfont \(\displaystyle 20\)}%
\end{pgfscope}%
\begin{pgfscope}%
\pgfsetbuttcap%
\pgfsetroundjoin%
\definecolor{currentfill}{rgb}{0.000000,0.000000,0.000000}%
\pgfsetfillcolor{currentfill}%
\pgfsetlinewidth{0.803000pt}%
\definecolor{currentstroke}{rgb}{0.000000,0.000000,0.000000}%
\pgfsetstrokecolor{currentstroke}%
\pgfsetdash{}{0pt}%
\pgfsys@defobject{currentmarker}{\pgfqpoint{-0.048611in}{0.000000in}}{\pgfqpoint{0.000000in}{0.000000in}}{%
\pgfpathmoveto{\pgfqpoint{0.000000in}{0.000000in}}%
\pgfpathlineto{\pgfqpoint{-0.048611in}{0.000000in}}%
\pgfusepath{stroke,fill}%
}%
\begin{pgfscope}%
\pgfsys@transformshift{0.675000in}{2.091008in}%
\pgfsys@useobject{currentmarker}{}%
\end{pgfscope}%
\end{pgfscope}%
\begin{pgfscope}%
\pgftext[x=0.438888in,y=2.042791in,left,base]{\rmfamily\fontsize{10.000000}{12.000000}\selectfont \(\displaystyle 30\)}%
\end{pgfscope}%
\begin{pgfscope}%
\pgftext[x=0.383333in,y=1.336500in,,bottom,rotate=90.000000]{\rmfamily\fontsize{10.000000}{12.000000}\selectfont Hash rate}%
\end{pgfscope}%
\begin{pgfscope}%
\pgfpathrectangle{\pgfqpoint{0.675000in}{0.297000in}}{\pgfqpoint{4.185000in}{2.079000in}}%
\pgfusepath{clip}%
\pgfsetrectcap%
\pgfsetroundjoin%
\pgfsetlinewidth{1.003750pt}%
\definecolor{currentstroke}{rgb}{0.121569,0.466667,0.705882}%
\pgfsetstrokecolor{currentstroke}%
\pgfsetdash{}{0pt}%
\pgfpathmoveto{\pgfqpoint{0.865227in}{0.391500in}}%
\pgfpathlineto{\pgfqpoint{0.896932in}{0.391500in}}%
\pgfpathlineto{\pgfqpoint{0.928636in}{0.391500in}}%
\pgfpathlineto{\pgfqpoint{0.960341in}{0.391500in}}%
\pgfpathlineto{\pgfqpoint{0.992045in}{0.391500in}}%
\pgfpathlineto{\pgfqpoint{1.023750in}{0.391500in}}%
\pgfpathlineto{\pgfqpoint{1.055455in}{0.391500in}}%
\pgfpathlineto{\pgfqpoint{1.087159in}{0.391500in}}%
\pgfpathlineto{\pgfqpoint{1.118864in}{0.391500in}}%
\pgfpathlineto{\pgfqpoint{1.150568in}{0.391500in}}%
\pgfpathlineto{\pgfqpoint{1.182273in}{0.391500in}}%
\pgfpathlineto{\pgfqpoint{1.213977in}{0.391500in}}%
\pgfpathlineto{\pgfqpoint{1.245682in}{0.391500in}}%
\pgfpathlineto{\pgfqpoint{1.277386in}{0.391500in}}%
\pgfpathlineto{\pgfqpoint{1.309091in}{0.391500in}}%
\pgfpathlineto{\pgfqpoint{1.340795in}{0.391500in}}%
\pgfpathlineto{\pgfqpoint{1.372500in}{0.391500in}}%
\pgfpathlineto{\pgfqpoint{1.404205in}{0.391500in}}%
\pgfpathlineto{\pgfqpoint{1.435909in}{0.391500in}}%
\pgfpathlineto{\pgfqpoint{1.467614in}{0.391500in}}%
\pgfpathlineto{\pgfqpoint{1.499318in}{0.391500in}}%
\pgfpathlineto{\pgfqpoint{1.531023in}{0.391500in}}%
\pgfpathlineto{\pgfqpoint{1.562727in}{0.391500in}}%
\pgfpathlineto{\pgfqpoint{1.594432in}{0.391500in}}%
\pgfpathlineto{\pgfqpoint{1.626136in}{0.391500in}}%
\pgfpathlineto{\pgfqpoint{1.657841in}{0.391500in}}%
\pgfpathlineto{\pgfqpoint{1.689545in}{0.391500in}}%
\pgfpathlineto{\pgfqpoint{1.721250in}{0.391500in}}%
\pgfpathlineto{\pgfqpoint{1.752955in}{0.391500in}}%
\pgfpathlineto{\pgfqpoint{1.784659in}{0.391500in}}%
\pgfpathlineto{\pgfqpoint{1.816364in}{0.391500in}}%
\pgfpathlineto{\pgfqpoint{1.848068in}{0.391500in}}%
\pgfpathlineto{\pgfqpoint{1.879773in}{0.391500in}}%
\pgfpathlineto{\pgfqpoint{1.911477in}{0.391500in}}%
\pgfpathlineto{\pgfqpoint{1.943182in}{0.391500in}}%
\pgfpathlineto{\pgfqpoint{1.974886in}{0.391500in}}%
\pgfpathlineto{\pgfqpoint{2.006591in}{0.391500in}}%
\pgfpathlineto{\pgfqpoint{2.038295in}{0.391500in}}%
\pgfpathlineto{\pgfqpoint{2.070000in}{0.391500in}}%
\pgfpathlineto{\pgfqpoint{2.101705in}{0.391500in}}%
\pgfpathlineto{\pgfqpoint{2.133409in}{0.391500in}}%
\pgfpathlineto{\pgfqpoint{2.165114in}{0.391500in}}%
\pgfpathlineto{\pgfqpoint{2.196818in}{0.391500in}}%
\pgfpathlineto{\pgfqpoint{2.228523in}{0.391500in}}%
\pgfpathlineto{\pgfqpoint{2.260227in}{0.391500in}}%
\pgfpathlineto{\pgfqpoint{2.291932in}{0.391500in}}%
\pgfpathlineto{\pgfqpoint{2.323636in}{0.391500in}}%
\pgfpathlineto{\pgfqpoint{2.355341in}{0.391500in}}%
\pgfpathlineto{\pgfqpoint{2.387045in}{0.391500in}}%
\pgfpathlineto{\pgfqpoint{2.418750in}{0.391500in}}%
\pgfpathlineto{\pgfqpoint{2.450455in}{0.391500in}}%
\pgfpathlineto{\pgfqpoint{2.482159in}{0.391500in}}%
\pgfpathlineto{\pgfqpoint{2.513864in}{0.391500in}}%
\pgfpathlineto{\pgfqpoint{2.545568in}{0.391500in}}%
\pgfpathlineto{\pgfqpoint{2.577273in}{0.391500in}}%
\pgfpathlineto{\pgfqpoint{2.608977in}{0.391500in}}%
\pgfpathlineto{\pgfqpoint{2.640682in}{0.391500in}}%
\pgfpathlineto{\pgfqpoint{2.672386in}{0.391500in}}%
\pgfpathlineto{\pgfqpoint{2.704091in}{0.391500in}}%
\pgfpathlineto{\pgfqpoint{2.735795in}{0.391500in}}%
\pgfpathlineto{\pgfqpoint{2.767500in}{0.391500in}}%
\pgfpathlineto{\pgfqpoint{2.799205in}{0.391500in}}%
\pgfpathlineto{\pgfqpoint{2.830909in}{0.391500in}}%
\pgfpathlineto{\pgfqpoint{2.862614in}{0.400306in}}%
\pgfpathlineto{\pgfqpoint{2.894318in}{0.427929in}}%
\pgfpathlineto{\pgfqpoint{2.926023in}{0.454214in}}%
\pgfpathlineto{\pgfqpoint{2.957727in}{0.480212in}}%
\pgfpathlineto{\pgfqpoint{2.989432in}{0.505703in}}%
\pgfpathlineto{\pgfqpoint{3.021136in}{0.530212in}}%
\pgfpathlineto{\pgfqpoint{3.052841in}{0.554205in}}%
\pgfpathlineto{\pgfqpoint{3.084545in}{0.577623in}}%
\pgfpathlineto{\pgfqpoint{3.116250in}{0.600146in}}%
\pgfpathlineto{\pgfqpoint{3.147955in}{0.622253in}}%
\pgfpathlineto{\pgfqpoint{3.179659in}{0.644483in}}%
\pgfpathlineto{\pgfqpoint{3.211364in}{0.666723in}}%
\pgfpathlineto{\pgfqpoint{3.243068in}{0.686032in}}%
\pgfpathlineto{\pgfqpoint{3.274773in}{0.706535in}}%
\pgfpathlineto{\pgfqpoint{3.306477in}{0.726099in}}%
\pgfpathlineto{\pgfqpoint{3.338182in}{0.746702in}}%
\pgfpathlineto{\pgfqpoint{3.369886in}{0.765517in}}%
\pgfpathlineto{\pgfqpoint{3.401591in}{0.783926in}}%
\pgfpathlineto{\pgfqpoint{3.433295in}{0.802074in}}%
\pgfpathlineto{\pgfqpoint{3.465000in}{0.817567in}}%
\pgfpathlineto{\pgfqpoint{3.496705in}{0.835776in}}%
\pgfpathlineto{\pgfqpoint{3.528409in}{0.852622in}}%
\pgfpathlineto{\pgfqpoint{3.560114in}{0.869445in}}%
\pgfpathlineto{\pgfqpoint{3.591818in}{0.885567in}}%
\pgfpathlineto{\pgfqpoint{3.623523in}{0.902905in}}%
\pgfpathlineto{\pgfqpoint{3.655227in}{0.917144in}}%
\pgfpathlineto{\pgfqpoint{3.686932in}{0.932444in}}%
\pgfpathlineto{\pgfqpoint{3.718636in}{0.948351in}}%
\pgfpathlineto{\pgfqpoint{3.750341in}{0.961902in}}%
\pgfpathlineto{\pgfqpoint{3.782045in}{0.977306in}}%
\pgfpathlineto{\pgfqpoint{3.813750in}{0.991432in}}%
\pgfpathlineto{\pgfqpoint{3.845455in}{1.004051in}}%
\pgfpathlineto{\pgfqpoint{3.877159in}{1.017397in}}%
\pgfpathlineto{\pgfqpoint{3.908864in}{1.031072in}}%
\pgfpathlineto{\pgfqpoint{3.940568in}{1.044002in}}%
\pgfpathlineto{\pgfqpoint{3.972273in}{1.057862in}}%
\pgfpathlineto{\pgfqpoint{4.003977in}{1.069536in}}%
\pgfpathlineto{\pgfqpoint{4.035682in}{1.081796in}}%
\pgfpathlineto{\pgfqpoint{4.067386in}{1.094030in}}%
\pgfpathlineto{\pgfqpoint{4.099091in}{1.105918in}}%
\pgfpathlineto{\pgfqpoint{4.130795in}{1.118459in}}%
\pgfpathlineto{\pgfqpoint{4.162500in}{1.129108in}}%
\pgfpathlineto{\pgfqpoint{4.194205in}{1.140339in}}%
\pgfpathlineto{\pgfqpoint{4.225909in}{1.151485in}}%
\pgfpathlineto{\pgfqpoint{4.257614in}{1.162329in}}%
\pgfpathlineto{\pgfqpoint{4.289318in}{1.173185in}}%
\pgfpathlineto{\pgfqpoint{4.321023in}{1.183619in}}%
\pgfpathlineto{\pgfqpoint{4.352727in}{1.195018in}}%
\pgfpathlineto{\pgfqpoint{4.384432in}{1.204265in}}%
\pgfpathlineto{\pgfqpoint{4.416136in}{1.215157in}}%
\pgfpathlineto{\pgfqpoint{4.447841in}{1.225034in}}%
\pgfpathlineto{\pgfqpoint{4.479545in}{1.234696in}}%
\pgfpathlineto{\pgfqpoint{4.511250in}{1.244281in}}%
\pgfpathlineto{\pgfqpoint{4.542955in}{1.253663in}}%
\pgfpathlineto{\pgfqpoint{4.574659in}{1.262844in}}%
\pgfpathlineto{\pgfqpoint{4.606364in}{1.272011in}}%
\pgfpathlineto{\pgfqpoint{4.638068in}{1.280910in}}%
\pgfpathlineto{\pgfqpoint{4.669773in}{1.289831in}}%
\pgfusepath{stroke}%
\end{pgfscope}%
\begin{pgfscope}%
\pgfpathrectangle{\pgfqpoint{0.675000in}{0.297000in}}{\pgfqpoint{4.185000in}{2.079000in}}%
\pgfusepath{clip}%
\pgfsetrectcap%
\pgfsetroundjoin%
\pgfsetlinewidth{1.003750pt}%
\definecolor{currentstroke}{rgb}{1.000000,0.498039,0.054902}%
\pgfsetstrokecolor{currentstroke}%
\pgfsetdash{}{0pt}%
\pgfpathmoveto{\pgfqpoint{0.865227in}{0.391500in}}%
\pgfpathlineto{\pgfqpoint{0.896932in}{0.391500in}}%
\pgfpathlineto{\pgfqpoint{0.928636in}{0.391500in}}%
\pgfpathlineto{\pgfqpoint{0.960341in}{0.391500in}}%
\pgfpathlineto{\pgfqpoint{0.992045in}{0.391500in}}%
\pgfpathlineto{\pgfqpoint{1.023750in}{0.391500in}}%
\pgfpathlineto{\pgfqpoint{1.055455in}{0.391500in}}%
\pgfpathlineto{\pgfqpoint{1.087159in}{0.391500in}}%
\pgfpathlineto{\pgfqpoint{1.118864in}{0.391500in}}%
\pgfpathlineto{\pgfqpoint{1.150568in}{0.391500in}}%
\pgfpathlineto{\pgfqpoint{1.182273in}{0.391500in}}%
\pgfpathlineto{\pgfqpoint{1.213977in}{0.391500in}}%
\pgfpathlineto{\pgfqpoint{1.245682in}{0.391500in}}%
\pgfpathlineto{\pgfqpoint{1.277386in}{0.391500in}}%
\pgfpathlineto{\pgfqpoint{1.309091in}{0.391500in}}%
\pgfpathlineto{\pgfqpoint{1.340795in}{0.391500in}}%
\pgfpathlineto{\pgfqpoint{1.372500in}{0.391500in}}%
\pgfpathlineto{\pgfqpoint{1.404205in}{0.391500in}}%
\pgfpathlineto{\pgfqpoint{1.435909in}{0.391500in}}%
\pgfpathlineto{\pgfqpoint{1.467614in}{0.391500in}}%
\pgfpathlineto{\pgfqpoint{1.499318in}{0.391500in}}%
\pgfpathlineto{\pgfqpoint{1.531023in}{0.391500in}}%
\pgfpathlineto{\pgfqpoint{1.562727in}{0.391500in}}%
\pgfpathlineto{\pgfqpoint{1.594432in}{0.391500in}}%
\pgfpathlineto{\pgfqpoint{1.626136in}{0.391500in}}%
\pgfpathlineto{\pgfqpoint{1.657841in}{0.391500in}}%
\pgfpathlineto{\pgfqpoint{1.689545in}{0.391500in}}%
\pgfpathlineto{\pgfqpoint{1.721250in}{0.391500in}}%
\pgfpathlineto{\pgfqpoint{1.752955in}{0.391500in}}%
\pgfpathlineto{\pgfqpoint{1.784659in}{0.391500in}}%
\pgfpathlineto{\pgfqpoint{1.816364in}{0.391500in}}%
\pgfpathlineto{\pgfqpoint{1.848068in}{0.391500in}}%
\pgfpathlineto{\pgfqpoint{1.879773in}{0.391500in}}%
\pgfpathlineto{\pgfqpoint{1.911477in}{0.391500in}}%
\pgfpathlineto{\pgfqpoint{1.943182in}{0.391500in}}%
\pgfpathlineto{\pgfqpoint{1.974886in}{0.391500in}}%
\pgfpathlineto{\pgfqpoint{2.006591in}{0.391500in}}%
\pgfpathlineto{\pgfqpoint{2.038295in}{0.391500in}}%
\pgfpathlineto{\pgfqpoint{2.070000in}{0.391500in}}%
\pgfpathlineto{\pgfqpoint{2.101705in}{0.391500in}}%
\pgfpathlineto{\pgfqpoint{2.133409in}{0.391500in}}%
\pgfpathlineto{\pgfqpoint{2.165114in}{0.391500in}}%
\pgfpathlineto{\pgfqpoint{2.196818in}{0.391500in}}%
\pgfpathlineto{\pgfqpoint{2.228523in}{0.391500in}}%
\pgfpathlineto{\pgfqpoint{2.260227in}{0.391500in}}%
\pgfpathlineto{\pgfqpoint{2.291932in}{0.391500in}}%
\pgfpathlineto{\pgfqpoint{2.323636in}{0.391500in}}%
\pgfpathlineto{\pgfqpoint{2.355341in}{0.391500in}}%
\pgfpathlineto{\pgfqpoint{2.387045in}{0.391500in}}%
\pgfpathlineto{\pgfqpoint{2.418750in}{0.391500in}}%
\pgfpathlineto{\pgfqpoint{2.450455in}{0.391500in}}%
\pgfpathlineto{\pgfqpoint{2.482159in}{0.391500in}}%
\pgfpathlineto{\pgfqpoint{2.513864in}{0.391500in}}%
\pgfpathlineto{\pgfqpoint{2.545568in}{0.391500in}}%
\pgfpathlineto{\pgfqpoint{2.577273in}{0.391500in}}%
\pgfpathlineto{\pgfqpoint{2.608977in}{0.391500in}}%
\pgfpathlineto{\pgfqpoint{2.640682in}{0.391500in}}%
\pgfpathlineto{\pgfqpoint{2.672386in}{0.391500in}}%
\pgfpathlineto{\pgfqpoint{2.704091in}{0.391500in}}%
\pgfpathlineto{\pgfqpoint{2.735795in}{0.391500in}}%
\pgfpathlineto{\pgfqpoint{2.767500in}{0.391500in}}%
\pgfpathlineto{\pgfqpoint{2.799205in}{0.391500in}}%
\pgfpathlineto{\pgfqpoint{2.830909in}{0.391500in}}%
\pgfpathlineto{\pgfqpoint{2.862614in}{0.393265in}}%
\pgfpathlineto{\pgfqpoint{2.894318in}{0.395715in}}%
\pgfpathlineto{\pgfqpoint{2.926023in}{0.398641in}}%
\pgfpathlineto{\pgfqpoint{2.957727in}{0.401202in}}%
\pgfpathlineto{\pgfqpoint{2.989432in}{0.403486in}}%
\pgfpathlineto{\pgfqpoint{3.021136in}{0.405998in}}%
\pgfpathlineto{\pgfqpoint{3.052841in}{0.408395in}}%
\pgfpathlineto{\pgfqpoint{3.084545in}{0.410696in}}%
\pgfpathlineto{\pgfqpoint{3.116250in}{0.413188in}}%
\pgfpathlineto{\pgfqpoint{3.147955in}{0.415594in}}%
\pgfpathlineto{\pgfqpoint{3.179659in}{0.417277in}}%
\pgfpathlineto{\pgfqpoint{3.211364in}{0.418397in}}%
\pgfpathlineto{\pgfqpoint{3.243068in}{0.421785in}}%
\pgfpathlineto{\pgfqpoint{3.274773in}{0.423533in}}%
\pgfpathlineto{\pgfqpoint{3.306477in}{0.425689in}}%
\pgfpathlineto{\pgfqpoint{3.338182in}{0.426389in}}%
\pgfpathlineto{\pgfqpoint{3.369886in}{0.428356in}}%
\pgfpathlineto{\pgfqpoint{3.401591in}{0.430224in}}%
\pgfpathlineto{\pgfqpoint{3.433295in}{0.431964in}}%
\pgfpathlineto{\pgfqpoint{3.465000in}{0.435803in}}%
\pgfpathlineto{\pgfqpoint{3.496705in}{0.436621in}}%
\pgfpathlineto{\pgfqpoint{3.528409in}{0.438410in}}%
\pgfpathlineto{\pgfqpoint{3.560114in}{0.439809in}}%
\pgfpathlineto{\pgfqpoint{3.591818in}{0.441516in}}%
\pgfpathlineto{\pgfqpoint{3.623523in}{0.441726in}}%
\pgfpathlineto{\pgfqpoint{3.655227in}{0.444527in}}%
\pgfpathlineto{\pgfqpoint{3.686932in}{0.446022in}}%
\pgfpathlineto{\pgfqpoint{3.718636in}{0.446591in}}%
\pgfpathlineto{\pgfqpoint{3.750341in}{0.449099in}}%
\pgfpathlineto{\pgfqpoint{3.782045in}{0.449533in}}%
\pgfpathlineto{\pgfqpoint{3.813750in}{0.450910in}}%
\pgfpathlineto{\pgfqpoint{3.845455in}{0.453439in}}%
\pgfpathlineto{\pgfqpoint{3.877159in}{0.454967in}}%
\pgfpathlineto{\pgfqpoint{3.908864in}{0.455956in}}%
\pgfpathlineto{\pgfqpoint{3.940568in}{0.457375in}}%
\pgfpathlineto{\pgfqpoint{3.972273in}{0.457656in}}%
\pgfpathlineto{\pgfqpoint{4.003977in}{0.459743in}}%
\pgfpathlineto{\pgfqpoint{4.035682in}{0.461091in}}%
\pgfpathlineto{\pgfqpoint{4.067386in}{0.462181in}}%
\pgfpathlineto{\pgfqpoint{4.099091in}{0.463377in}}%
\pgfpathlineto{\pgfqpoint{4.130795in}{0.463770in}}%
\pgfpathlineto{\pgfqpoint{4.162500in}{0.465715in}}%
\pgfpathlineto{\pgfqpoint{4.194205in}{0.466899in}}%
\pgfpathlineto{\pgfqpoint{4.225909in}{0.467966in}}%
\pgfpathlineto{\pgfqpoint{4.257614in}{0.469110in}}%
\pgfpathlineto{\pgfqpoint{4.289318in}{0.470072in}}%
\pgfpathlineto{\pgfqpoint{4.321023in}{0.471234in}}%
\pgfpathlineto{\pgfqpoint{4.352727in}{0.471268in}}%
\pgfpathlineto{\pgfqpoint{4.384432in}{0.473205in}}%
\pgfpathlineto{\pgfqpoint{4.416136in}{0.473403in}}%
\pgfpathlineto{\pgfqpoint{4.447841in}{0.474386in}}%
\pgfpathlineto{\pgfqpoint{4.479545in}{0.475385in}}%
\pgfpathlineto{\pgfqpoint{4.511250in}{0.476329in}}%
\pgfpathlineto{\pgfqpoint{4.542955in}{0.477264in}}%
\pgfpathlineto{\pgfqpoint{4.574659in}{0.478277in}}%
\pgfpathlineto{\pgfqpoint{4.606364in}{0.479145in}}%
\pgfpathlineto{\pgfqpoint{4.638068in}{0.480090in}}%
\pgfpathlineto{\pgfqpoint{4.669773in}{0.480901in}}%
\pgfusepath{stroke}%
\end{pgfscope}%
\begin{pgfscope}%
\pgfpathrectangle{\pgfqpoint{0.675000in}{0.297000in}}{\pgfqpoint{4.185000in}{2.079000in}}%
\pgfusepath{clip}%
\pgfsetrectcap%
\pgfsetroundjoin%
\pgfsetlinewidth{1.003750pt}%
\definecolor{currentstroke}{rgb}{0.172549,0.627451,0.172549}%
\pgfsetstrokecolor{currentstroke}%
\pgfsetdash{}{0pt}%
\pgfpathmoveto{\pgfqpoint{0.865227in}{1.256472in}}%
\pgfpathlineto{\pgfqpoint{0.896932in}{1.361531in}}%
\pgfpathlineto{\pgfqpoint{0.928636in}{1.449538in}}%
\pgfpathlineto{\pgfqpoint{0.960341in}{1.524502in}}%
\pgfpathlineto{\pgfqpoint{0.992045in}{1.589045in}}%
\pgfpathlineto{\pgfqpoint{1.023750in}{1.645139in}}%
\pgfpathlineto{\pgfqpoint{1.055455in}{1.694449in}}%
\pgfpathlineto{\pgfqpoint{1.087159in}{1.738091in}}%
\pgfpathlineto{\pgfqpoint{1.118864in}{1.776943in}}%
\pgfpathlineto{\pgfqpoint{1.150568in}{1.811831in}}%
\pgfpathlineto{\pgfqpoint{1.182273in}{1.843296in}}%
\pgfpathlineto{\pgfqpoint{1.213977in}{1.871786in}}%
\pgfpathlineto{\pgfqpoint{1.245682in}{1.897766in}}%
\pgfpathlineto{\pgfqpoint{1.277386in}{1.921522in}}%
\pgfpathlineto{\pgfqpoint{1.309091in}{1.943306in}}%
\pgfpathlineto{\pgfqpoint{1.340795in}{1.963399in}}%
\pgfpathlineto{\pgfqpoint{1.372500in}{1.981967in}}%
\pgfpathlineto{\pgfqpoint{1.404205in}{1.999162in}}%
\pgfpathlineto{\pgfqpoint{1.435909in}{2.015164in}}%
\pgfpathlineto{\pgfqpoint{1.467614in}{2.030076in}}%
\pgfpathlineto{\pgfqpoint{1.499318in}{2.043992in}}%
\pgfpathlineto{\pgfqpoint{1.531023in}{2.057037in}}%
\pgfpathlineto{\pgfqpoint{1.562727in}{2.069277in}}%
\pgfpathlineto{\pgfqpoint{1.594432in}{2.080768in}}%
\pgfpathlineto{\pgfqpoint{1.626136in}{2.091605in}}%
\pgfpathlineto{\pgfqpoint{1.657841in}{2.101830in}}%
\pgfpathlineto{\pgfqpoint{1.689545in}{2.111481in}}%
\pgfpathlineto{\pgfqpoint{1.721250in}{2.120628in}}%
\pgfpathlineto{\pgfqpoint{1.752955in}{2.129298in}}%
\pgfpathlineto{\pgfqpoint{1.784659in}{2.137517in}}%
\pgfpathlineto{\pgfqpoint{1.816364in}{2.145341in}}%
\pgfpathlineto{\pgfqpoint{1.848068in}{2.152784in}}%
\pgfpathlineto{\pgfqpoint{1.879773in}{2.159869in}}%
\pgfpathlineto{\pgfqpoint{1.911477in}{2.166635in}}%
\pgfpathlineto{\pgfqpoint{1.943182in}{2.173096in}}%
\pgfpathlineto{\pgfqpoint{1.974886in}{2.179266in}}%
\pgfpathlineto{\pgfqpoint{2.006591in}{2.185178in}}%
\pgfpathlineto{\pgfqpoint{2.038295in}{2.190839in}}%
\pgfpathlineto{\pgfqpoint{2.070000in}{2.196259in}}%
\pgfpathlineto{\pgfqpoint{2.101705in}{2.201466in}}%
\pgfpathlineto{\pgfqpoint{2.133409in}{2.206466in}}%
\pgfpathlineto{\pgfqpoint{2.165114in}{2.211268in}}%
\pgfpathlineto{\pgfqpoint{2.196818in}{2.215890in}}%
\pgfpathlineto{\pgfqpoint{2.228523in}{2.220339in}}%
\pgfpathlineto{\pgfqpoint{2.260227in}{2.224620in}}%
\pgfpathlineto{\pgfqpoint{2.291932in}{2.228752in}}%
\pgfpathlineto{\pgfqpoint{2.323636in}{2.232737in}}%
\pgfpathlineto{\pgfqpoint{2.355341in}{2.236579in}}%
\pgfpathlineto{\pgfqpoint{2.387045in}{2.240292in}}%
\pgfpathlineto{\pgfqpoint{2.418750in}{2.243882in}}%
\pgfpathlineto{\pgfqpoint{2.450455in}{2.247347in}}%
\pgfpathlineto{\pgfqpoint{2.482159in}{2.250705in}}%
\pgfpathlineto{\pgfqpoint{2.513864in}{2.253954in}}%
\pgfpathlineto{\pgfqpoint{2.545568in}{2.257098in}}%
\pgfpathlineto{\pgfqpoint{2.577273in}{2.260147in}}%
\pgfpathlineto{\pgfqpoint{2.608977in}{2.263103in}}%
\pgfpathlineto{\pgfqpoint{2.640682in}{2.265967in}}%
\pgfpathlineto{\pgfqpoint{2.672386in}{2.268749in}}%
\pgfpathlineto{\pgfqpoint{2.704091in}{2.271450in}}%
\pgfpathlineto{\pgfqpoint{2.735795in}{2.274070in}}%
\pgfpathlineto{\pgfqpoint{2.767500in}{2.276618in}}%
\pgfpathlineto{\pgfqpoint{2.799205in}{2.279095in}}%
\pgfpathlineto{\pgfqpoint{2.830909in}{2.281500in}}%
\pgfpathlineto{\pgfqpoint{2.862614in}{2.274229in}}%
\pgfpathlineto{\pgfqpoint{2.894318in}{2.249172in}}%
\pgfpathlineto{\pgfqpoint{2.926023in}{2.224837in}}%
\pgfpathlineto{\pgfqpoint{2.957727in}{2.201026in}}%
\pgfpathlineto{\pgfqpoint{2.989432in}{2.177879in}}%
\pgfpathlineto{\pgfqpoint{3.021136in}{2.155368in}}%
\pgfpathlineto{\pgfqpoint{3.052841in}{2.133369in}}%
\pgfpathlineto{\pgfqpoint{3.084545in}{2.111934in}}%
\pgfpathlineto{\pgfqpoint{3.116250in}{2.091088in}}%
\pgfpathlineto{\pgfqpoint{3.147955in}{2.070657in}}%
\pgfpathlineto{\pgfqpoint{3.179659in}{2.050719in}}%
\pgfpathlineto{\pgfqpoint{3.211364in}{2.031247in}}%
\pgfpathlineto{\pgfqpoint{3.243068in}{2.012338in}}%
\pgfpathlineto{\pgfqpoint{3.274773in}{1.993790in}}%
\pgfpathlineto{\pgfqpoint{3.306477in}{1.975682in}}%
\pgfpathlineto{\pgfqpoint{3.338182in}{1.957923in}}%
\pgfpathlineto{\pgfqpoint{3.369886in}{1.940598in}}%
\pgfpathlineto{\pgfqpoint{3.401591in}{1.923700in}}%
\pgfpathlineto{\pgfqpoint{3.433295in}{1.907115in}}%
\pgfpathlineto{\pgfqpoint{3.465000in}{1.891013in}}%
\pgfpathlineto{\pgfqpoint{3.496705in}{1.875146in}}%
\pgfpathlineto{\pgfqpoint{3.528409in}{1.859603in}}%
\pgfpathlineto{\pgfqpoint{3.560114in}{1.844421in}}%
\pgfpathlineto{\pgfqpoint{3.591818in}{1.829561in}}%
\pgfpathlineto{\pgfqpoint{3.623523in}{1.814924in}}%
\pgfpathlineto{\pgfqpoint{3.655227in}{1.800725in}}%
\pgfpathlineto{\pgfqpoint{3.686932in}{1.786720in}}%
\pgfpathlineto{\pgfqpoint{3.718636in}{1.772981in}}%
\pgfpathlineto{\pgfqpoint{3.750341in}{1.759599in}}%
\pgfpathlineto{\pgfqpoint{3.782045in}{1.746390in}}%
\pgfpathlineto{\pgfqpoint{3.813750in}{1.733460in}}%
\pgfpathlineto{\pgfqpoint{3.845455in}{1.720836in}}%
\pgfpathlineto{\pgfqpoint{3.877159in}{1.708440in}}%
\pgfpathlineto{\pgfqpoint{3.908864in}{1.696211in}}%
\pgfpathlineto{\pgfqpoint{3.940568in}{1.684249in}}%
\pgfpathlineto{\pgfqpoint{3.972273in}{1.672455in}}%
\pgfpathlineto{\pgfqpoint{4.003977in}{1.660985in}}%
\pgfpathlineto{\pgfqpoint{4.035682in}{1.649634in}}%
\pgfpathlineto{\pgfqpoint{4.067386in}{1.638529in}}%
\pgfpathlineto{\pgfqpoint{4.099091in}{1.627624in}}%
\pgfpathlineto{\pgfqpoint{4.130795in}{1.616832in}}%
\pgfpathlineto{\pgfqpoint{4.162500in}{1.606332in}}%
\pgfpathlineto{\pgfqpoint{4.194205in}{1.595982in}}%
\pgfpathlineto{\pgfqpoint{4.225909in}{1.585800in}}%
\pgfpathlineto{\pgfqpoint{4.257614in}{1.575803in}}%
\pgfpathlineto{\pgfqpoint{4.289318in}{1.565950in}}%
\pgfpathlineto{\pgfqpoint{4.321023in}{1.556280in}}%
\pgfpathlineto{\pgfqpoint{4.352727in}{1.546745in}}%
\pgfpathlineto{\pgfqpoint{4.384432in}{1.537420in}}%
\pgfpathlineto{\pgfqpoint{4.416136in}{1.528166in}}%
\pgfpathlineto{\pgfqpoint{4.447841in}{1.519109in}}%
\pgfpathlineto{\pgfqpoint{4.479545in}{1.510224in}}%
\pgfpathlineto{\pgfqpoint{4.511250in}{1.501443in}}%
\pgfpathlineto{\pgfqpoint{4.542955in}{1.492842in}}%
\pgfpathlineto{\pgfqpoint{4.574659in}{1.484341in}}%
\pgfpathlineto{\pgfqpoint{4.606364in}{1.475970in}}%
\pgfpathlineto{\pgfqpoint{4.638068in}{1.467763in}}%
\pgfpathlineto{\pgfqpoint{4.669773in}{1.459645in}}%
\pgfusepath{stroke}%
\end{pgfscope}%
\begin{pgfscope}%
\pgfpathrectangle{\pgfqpoint{0.675000in}{0.297000in}}{\pgfqpoint{4.185000in}{2.079000in}}%
\pgfusepath{clip}%
\pgfsetrectcap%
\pgfsetroundjoin%
\pgfsetlinewidth{1.003750pt}%
\definecolor{currentstroke}{rgb}{0.839216,0.152941,0.156863}%
\pgfsetstrokecolor{currentstroke}%
\pgfsetdash{}{0pt}%
\pgfpathmoveto{\pgfqpoint{0.865227in}{1.792540in}}%
\pgfpathlineto{\pgfqpoint{0.896932in}{1.687480in}}%
\pgfpathlineto{\pgfqpoint{0.928636in}{1.599473in}}%
\pgfpathlineto{\pgfqpoint{0.960341in}{1.524509in}}%
\pgfpathlineto{\pgfqpoint{0.992045in}{1.459967in}}%
\pgfpathlineto{\pgfqpoint{1.023750in}{1.403872in}}%
\pgfpathlineto{\pgfqpoint{1.055455in}{1.354562in}}%
\pgfpathlineto{\pgfqpoint{1.087159in}{1.310920in}}%
\pgfpathlineto{\pgfqpoint{1.118864in}{1.272068in}}%
\pgfpathlineto{\pgfqpoint{1.150568in}{1.237180in}}%
\pgfpathlineto{\pgfqpoint{1.182273in}{1.205715in}}%
\pgfpathlineto{\pgfqpoint{1.213977in}{1.177225in}}%
\pgfpathlineto{\pgfqpoint{1.245682in}{1.151246in}}%
\pgfpathlineto{\pgfqpoint{1.277386in}{1.127489in}}%
\pgfpathlineto{\pgfqpoint{1.309091in}{1.105705in}}%
\pgfpathlineto{\pgfqpoint{1.340795in}{1.085612in}}%
\pgfpathlineto{\pgfqpoint{1.372500in}{1.067044in}}%
\pgfpathlineto{\pgfqpoint{1.404205in}{1.049850in}}%
\pgfpathlineto{\pgfqpoint{1.435909in}{1.033847in}}%
\pgfpathlineto{\pgfqpoint{1.467614in}{1.018935in}}%
\pgfpathlineto{\pgfqpoint{1.499318in}{1.005019in}}%
\pgfpathlineto{\pgfqpoint{1.531023in}{0.991974in}}%
\pgfpathlineto{\pgfqpoint{1.562727in}{0.979734in}}%
\pgfpathlineto{\pgfqpoint{1.594432in}{0.968243in}}%
\pgfpathlineto{\pgfqpoint{1.626136in}{0.957406in}}%
\pgfpathlineto{\pgfqpoint{1.657841in}{0.947181in}}%
\pgfpathlineto{\pgfqpoint{1.689545in}{0.937530in}}%
\pgfpathlineto{\pgfqpoint{1.721250in}{0.928383in}}%
\pgfpathlineto{\pgfqpoint{1.752955in}{0.919713in}}%
\pgfpathlineto{\pgfqpoint{1.784659in}{0.911494in}}%
\pgfpathlineto{\pgfqpoint{1.816364in}{0.903670in}}%
\pgfpathlineto{\pgfqpoint{1.848068in}{0.896227in}}%
\pgfpathlineto{\pgfqpoint{1.879773in}{0.889142in}}%
\pgfpathlineto{\pgfqpoint{1.911477in}{0.882376in}}%
\pgfpathlineto{\pgfqpoint{1.943182in}{0.875915in}}%
\pgfpathlineto{\pgfqpoint{1.974886in}{0.869745in}}%
\pgfpathlineto{\pgfqpoint{2.006591in}{0.863834in}}%
\pgfpathlineto{\pgfqpoint{2.038295in}{0.858173in}}%
\pgfpathlineto{\pgfqpoint{2.070000in}{0.852752in}}%
\pgfpathlineto{\pgfqpoint{2.101705in}{0.847545in}}%
\pgfpathlineto{\pgfqpoint{2.133409in}{0.842545in}}%
\pgfpathlineto{\pgfqpoint{2.165114in}{0.837744in}}%
\pgfpathlineto{\pgfqpoint{2.196818in}{0.833121in}}%
\pgfpathlineto{\pgfqpoint{2.228523in}{0.828672in}}%
\pgfpathlineto{\pgfqpoint{2.260227in}{0.824391in}}%
\pgfpathlineto{\pgfqpoint{2.291932in}{0.820259in}}%
\pgfpathlineto{\pgfqpoint{2.323636in}{0.816275in}}%
\pgfpathlineto{\pgfqpoint{2.355341in}{0.812432in}}%
\pgfpathlineto{\pgfqpoint{2.387045in}{0.808719in}}%
\pgfpathlineto{\pgfqpoint{2.418750in}{0.805129in}}%
\pgfpathlineto{\pgfqpoint{2.450455in}{0.801664in}}%
\pgfpathlineto{\pgfqpoint{2.482159in}{0.798306in}}%
\pgfpathlineto{\pgfqpoint{2.513864in}{0.795057in}}%
\pgfpathlineto{\pgfqpoint{2.545568in}{0.791913in}}%
\pgfpathlineto{\pgfqpoint{2.577273in}{0.788864in}}%
\pgfpathlineto{\pgfqpoint{2.608977in}{0.785908in}}%
\pgfpathlineto{\pgfqpoint{2.640682in}{0.783044in}}%
\pgfpathlineto{\pgfqpoint{2.672386in}{0.780262in}}%
\pgfpathlineto{\pgfqpoint{2.704091in}{0.777561in}}%
\pgfpathlineto{\pgfqpoint{2.735795in}{0.774941in}}%
\pgfpathlineto{\pgfqpoint{2.767500in}{0.772393in}}%
\pgfpathlineto{\pgfqpoint{2.799205in}{0.769917in}}%
\pgfpathlineto{\pgfqpoint{2.830909in}{0.767511in}}%
\pgfpathlineto{\pgfqpoint{2.862614in}{0.764211in}}%
\pgfpathlineto{\pgfqpoint{2.894318in}{0.759195in}}%
\pgfpathlineto{\pgfqpoint{2.926023in}{0.754319in}}%
\pgfpathlineto{\pgfqpoint{2.957727in}{0.749571in}}%
\pgfpathlineto{\pgfqpoint{2.989432in}{0.744943in}}%
\pgfpathlineto{\pgfqpoint{3.021136in}{0.740434in}}%
\pgfpathlineto{\pgfqpoint{3.052841in}{0.736043in}}%
\pgfpathlineto{\pgfqpoint{3.084545in}{0.731758in}}%
\pgfpathlineto{\pgfqpoint{3.116250in}{0.727588in}}%
\pgfpathlineto{\pgfqpoint{3.147955in}{0.723508in}}%
\pgfpathlineto{\pgfqpoint{3.179659in}{0.719532in}}%
\pgfpathlineto{\pgfqpoint{3.211364in}{0.715644in}}%
\pgfpathlineto{\pgfqpoint{3.243068in}{0.711856in}}%
\pgfpathlineto{\pgfqpoint{3.274773in}{0.708153in}}%
\pgfpathlineto{\pgfqpoint{3.306477in}{0.704541in}}%
\pgfpathlineto{\pgfqpoint{3.338182in}{0.700997in}}%
\pgfpathlineto{\pgfqpoint{3.369886in}{0.697539in}}%
\pgfpathlineto{\pgfqpoint{3.401591in}{0.694161in}}%
\pgfpathlineto{\pgfqpoint{3.433295in}{0.690858in}}%
\pgfpathlineto{\pgfqpoint{3.465000in}{0.687629in}}%
\pgfpathlineto{\pgfqpoint{3.496705in}{0.684467in}}%
\pgfpathlineto{\pgfqpoint{3.528409in}{0.681376in}}%
\pgfpathlineto{\pgfqpoint{3.560114in}{0.678335in}}%
\pgfpathlineto{\pgfqpoint{3.591818in}{0.675368in}}%
\pgfpathlineto{\pgfqpoint{3.623523in}{0.672457in}}%
\pgfpathlineto{\pgfqpoint{3.655227in}{0.669616in}}%
\pgfpathlineto{\pgfqpoint{3.686932in}{0.666826in}}%
\pgfpathlineto{\pgfqpoint{3.718636in}{0.664089in}}%
\pgfpathlineto{\pgfqpoint{3.750341in}{0.661411in}}%
\pgfpathlineto{\pgfqpoint{3.782045in}{0.658782in}}%
\pgfpathlineto{\pgfqpoint{3.813750in}{0.656208in}}%
\pgfpathlineto{\pgfqpoint{3.845455in}{0.653685in}}%
\pgfpathlineto{\pgfqpoint{3.877159in}{0.651208in}}%
\pgfpathlineto{\pgfqpoint{3.908864in}{0.648771in}}%
\pgfpathlineto{\pgfqpoint{3.940568in}{0.646386in}}%
\pgfpathlineto{\pgfqpoint{3.972273in}{0.644038in}}%
\pgfpathlineto{\pgfqpoint{4.003977in}{0.641747in}}%
\pgfpathlineto{\pgfqpoint{4.035682in}{0.639490in}}%
\pgfpathlineto{\pgfqpoint{4.067386in}{0.637271in}}%
\pgfpathlineto{\pgfqpoint{4.099091in}{0.635092in}}%
\pgfpathlineto{\pgfqpoint{4.130795in}{0.632950in}}%
\pgfpathlineto{\pgfqpoint{4.162500in}{0.630855in}}%
\pgfpathlineto{\pgfqpoint{4.194205in}{0.628790in}}%
\pgfpathlineto{\pgfqpoint{4.225909in}{0.626761in}}%
\pgfpathlineto{\pgfqpoint{4.257614in}{0.624769in}}%
\pgfpathlineto{\pgfqpoint{4.289318in}{0.622804in}}%
\pgfpathlineto{\pgfqpoint{4.321023in}{0.620878in}}%
\pgfpathlineto{\pgfqpoint{4.352727in}{0.618980in}}%
\pgfpathlineto{\pgfqpoint{4.384432in}{0.617121in}}%
\pgfpathlineto{\pgfqpoint{4.416136in}{0.615284in}}%
\pgfpathlineto{\pgfqpoint{4.447841in}{0.613482in}}%
\pgfpathlineto{\pgfqpoint{4.479545in}{0.611706in}}%
\pgfpathlineto{\pgfqpoint{4.511250in}{0.609958in}}%
\pgfpathlineto{\pgfqpoint{4.542955in}{0.608242in}}%
\pgfpathlineto{\pgfqpoint{4.574659in}{0.606550in}}%
\pgfpathlineto{\pgfqpoint{4.606364in}{0.604886in}}%
\pgfpathlineto{\pgfqpoint{4.638068in}{0.603248in}}%
\pgfpathlineto{\pgfqpoint{4.669773in}{0.601633in}}%
\pgfusepath{stroke}%
\end{pgfscope}%
\begin{pgfscope}%
\pgfpathrectangle{\pgfqpoint{0.675000in}{0.297000in}}{\pgfqpoint{4.185000in}{2.079000in}}%
\pgfusepath{clip}%
\pgfsetrectcap%
\pgfsetroundjoin%
\pgfsetlinewidth{1.003750pt}%
\definecolor{currentstroke}{rgb}{0.580392,0.403922,0.741176}%
\pgfsetstrokecolor{currentstroke}%
\pgfsetdash{}{0pt}%
\pgfpathmoveto{\pgfqpoint{0.865227in}{0.391500in}}%
\pgfpathlineto{\pgfqpoint{0.896932in}{0.391500in}}%
\pgfpathlineto{\pgfqpoint{0.928636in}{0.391500in}}%
\pgfpathlineto{\pgfqpoint{0.960341in}{0.391500in}}%
\pgfpathlineto{\pgfqpoint{0.992045in}{0.391500in}}%
\pgfpathlineto{\pgfqpoint{1.023750in}{0.391500in}}%
\pgfpathlineto{\pgfqpoint{1.055455in}{0.391500in}}%
\pgfpathlineto{\pgfqpoint{1.087159in}{0.391500in}}%
\pgfpathlineto{\pgfqpoint{1.118864in}{0.391500in}}%
\pgfpathlineto{\pgfqpoint{1.150568in}{0.391500in}}%
\pgfpathlineto{\pgfqpoint{1.182273in}{0.391500in}}%
\pgfpathlineto{\pgfqpoint{1.213977in}{0.391500in}}%
\pgfpathlineto{\pgfqpoint{1.245682in}{0.391500in}}%
\pgfpathlineto{\pgfqpoint{1.277386in}{0.391500in}}%
\pgfpathlineto{\pgfqpoint{1.309091in}{0.391500in}}%
\pgfpathlineto{\pgfqpoint{1.340795in}{0.391500in}}%
\pgfpathlineto{\pgfqpoint{1.372500in}{0.391500in}}%
\pgfpathlineto{\pgfqpoint{1.404205in}{0.391500in}}%
\pgfpathlineto{\pgfqpoint{1.435909in}{0.391500in}}%
\pgfpathlineto{\pgfqpoint{1.467614in}{0.391500in}}%
\pgfpathlineto{\pgfqpoint{1.499318in}{0.391500in}}%
\pgfpathlineto{\pgfqpoint{1.531023in}{0.391500in}}%
\pgfpathlineto{\pgfqpoint{1.562727in}{0.391500in}}%
\pgfpathlineto{\pgfqpoint{1.594432in}{0.391500in}}%
\pgfpathlineto{\pgfqpoint{1.626136in}{0.391500in}}%
\pgfpathlineto{\pgfqpoint{1.657841in}{0.391500in}}%
\pgfpathlineto{\pgfqpoint{1.689545in}{0.391500in}}%
\pgfpathlineto{\pgfqpoint{1.721250in}{0.391500in}}%
\pgfpathlineto{\pgfqpoint{1.752955in}{0.391500in}}%
\pgfpathlineto{\pgfqpoint{1.784659in}{0.391500in}}%
\pgfpathlineto{\pgfqpoint{1.816364in}{0.391500in}}%
\pgfpathlineto{\pgfqpoint{1.848068in}{0.391500in}}%
\pgfpathlineto{\pgfqpoint{1.879773in}{0.391500in}}%
\pgfpathlineto{\pgfqpoint{1.911477in}{0.391500in}}%
\pgfpathlineto{\pgfqpoint{1.943182in}{0.391500in}}%
\pgfpathlineto{\pgfqpoint{1.974886in}{0.391500in}}%
\pgfpathlineto{\pgfqpoint{2.006591in}{0.391500in}}%
\pgfpathlineto{\pgfqpoint{2.038295in}{0.391500in}}%
\pgfpathlineto{\pgfqpoint{2.070000in}{0.391500in}}%
\pgfpathlineto{\pgfqpoint{2.101705in}{0.391500in}}%
\pgfpathlineto{\pgfqpoint{2.133409in}{0.391500in}}%
\pgfpathlineto{\pgfqpoint{2.165114in}{0.391500in}}%
\pgfpathlineto{\pgfqpoint{2.196818in}{0.391500in}}%
\pgfpathlineto{\pgfqpoint{2.228523in}{0.391500in}}%
\pgfpathlineto{\pgfqpoint{2.260227in}{0.391500in}}%
\pgfpathlineto{\pgfqpoint{2.291932in}{0.391500in}}%
\pgfpathlineto{\pgfqpoint{2.323636in}{0.391500in}}%
\pgfpathlineto{\pgfqpoint{2.355341in}{0.391500in}}%
\pgfpathlineto{\pgfqpoint{2.387045in}{0.391500in}}%
\pgfpathlineto{\pgfqpoint{2.418750in}{0.391500in}}%
\pgfpathlineto{\pgfqpoint{2.450455in}{0.391500in}}%
\pgfpathlineto{\pgfqpoint{2.482159in}{0.391500in}}%
\pgfpathlineto{\pgfqpoint{2.513864in}{0.391500in}}%
\pgfpathlineto{\pgfqpoint{2.545568in}{0.391500in}}%
\pgfpathlineto{\pgfqpoint{2.577273in}{0.391500in}}%
\pgfpathlineto{\pgfqpoint{2.608977in}{0.391500in}}%
\pgfpathlineto{\pgfqpoint{2.640682in}{0.391500in}}%
\pgfpathlineto{\pgfqpoint{2.672386in}{0.391500in}}%
\pgfpathlineto{\pgfqpoint{2.704091in}{0.391500in}}%
\pgfpathlineto{\pgfqpoint{2.735795in}{0.391500in}}%
\pgfpathlineto{\pgfqpoint{2.767500in}{0.391500in}}%
\pgfpathlineto{\pgfqpoint{2.799205in}{0.391500in}}%
\pgfpathlineto{\pgfqpoint{2.830909in}{0.391500in}}%
\pgfpathlineto{\pgfqpoint{2.862614in}{0.391500in}}%
\pgfpathlineto{\pgfqpoint{2.894318in}{0.391500in}}%
\pgfpathlineto{\pgfqpoint{2.926023in}{0.391500in}}%
\pgfpathlineto{\pgfqpoint{2.957727in}{0.391500in}}%
\pgfpathlineto{\pgfqpoint{2.989432in}{0.391500in}}%
\pgfpathlineto{\pgfqpoint{3.021136in}{0.391500in}}%
\pgfpathlineto{\pgfqpoint{3.052841in}{0.391500in}}%
\pgfpathlineto{\pgfqpoint{3.084545in}{0.391500in}}%
\pgfpathlineto{\pgfqpoint{3.116250in}{0.391500in}}%
\pgfpathlineto{\pgfqpoint{3.147955in}{0.391500in}}%
\pgfpathlineto{\pgfqpoint{3.179659in}{0.391500in}}%
\pgfpathlineto{\pgfqpoint{3.211364in}{0.391500in}}%
\pgfpathlineto{\pgfqpoint{3.243068in}{0.391500in}}%
\pgfpathlineto{\pgfqpoint{3.274773in}{0.391500in}}%
\pgfpathlineto{\pgfqpoint{3.306477in}{0.391500in}}%
\pgfpathlineto{\pgfqpoint{3.338182in}{0.391500in}}%
\pgfpathlineto{\pgfqpoint{3.369886in}{0.391500in}}%
\pgfpathlineto{\pgfqpoint{3.401591in}{0.391500in}}%
\pgfpathlineto{\pgfqpoint{3.433295in}{0.391500in}}%
\pgfpathlineto{\pgfqpoint{3.465000in}{0.391500in}}%
\pgfpathlineto{\pgfqpoint{3.496705in}{0.391500in}}%
\pgfpathlineto{\pgfqpoint{3.528409in}{0.391500in}}%
\pgfpathlineto{\pgfqpoint{3.560114in}{0.391500in}}%
\pgfpathlineto{\pgfqpoint{3.591818in}{0.391500in}}%
\pgfpathlineto{\pgfqpoint{3.623523in}{0.391500in}}%
\pgfpathlineto{\pgfqpoint{3.655227in}{0.391500in}}%
\pgfpathlineto{\pgfqpoint{3.686932in}{0.391500in}}%
\pgfpathlineto{\pgfqpoint{3.718636in}{0.391500in}}%
\pgfpathlineto{\pgfqpoint{3.750341in}{0.391500in}}%
\pgfpathlineto{\pgfqpoint{3.782045in}{0.391500in}}%
\pgfpathlineto{\pgfqpoint{3.813750in}{0.391500in}}%
\pgfpathlineto{\pgfqpoint{3.845455in}{0.391500in}}%
\pgfpathlineto{\pgfqpoint{3.877159in}{0.391500in}}%
\pgfpathlineto{\pgfqpoint{3.908864in}{0.391500in}}%
\pgfpathlineto{\pgfqpoint{3.940568in}{0.391500in}}%
\pgfpathlineto{\pgfqpoint{3.972273in}{0.391500in}}%
\pgfpathlineto{\pgfqpoint{4.003977in}{0.391500in}}%
\pgfpathlineto{\pgfqpoint{4.035682in}{0.391500in}}%
\pgfpathlineto{\pgfqpoint{4.067386in}{0.391500in}}%
\pgfpathlineto{\pgfqpoint{4.099091in}{0.391500in}}%
\pgfpathlineto{\pgfqpoint{4.130795in}{0.391500in}}%
\pgfpathlineto{\pgfqpoint{4.162500in}{0.391500in}}%
\pgfpathlineto{\pgfqpoint{4.194205in}{0.391500in}}%
\pgfpathlineto{\pgfqpoint{4.225909in}{0.391500in}}%
\pgfpathlineto{\pgfqpoint{4.257614in}{0.391500in}}%
\pgfpathlineto{\pgfqpoint{4.289318in}{0.391500in}}%
\pgfpathlineto{\pgfqpoint{4.321023in}{0.391500in}}%
\pgfpathlineto{\pgfqpoint{4.352727in}{0.391500in}}%
\pgfpathlineto{\pgfqpoint{4.384432in}{0.391500in}}%
\pgfpathlineto{\pgfqpoint{4.416136in}{0.391500in}}%
\pgfpathlineto{\pgfqpoint{4.447841in}{0.391500in}}%
\pgfpathlineto{\pgfqpoint{4.479545in}{0.391500in}}%
\pgfpathlineto{\pgfqpoint{4.511250in}{0.391500in}}%
\pgfpathlineto{\pgfqpoint{4.542955in}{0.391500in}}%
\pgfpathlineto{\pgfqpoint{4.574659in}{0.391500in}}%
\pgfpathlineto{\pgfqpoint{4.606364in}{0.391500in}}%
\pgfpathlineto{\pgfqpoint{4.638068in}{0.391500in}}%
\pgfpathlineto{\pgfqpoint{4.669773in}{0.391500in}}%
\pgfusepath{stroke}%
\end{pgfscope}%
\begin{pgfscope}%
\pgfsetrectcap%
\pgfsetmiterjoin%
\pgfsetlinewidth{0.803000pt}%
\definecolor{currentstroke}{rgb}{0.000000,0.000000,0.000000}%
\pgfsetstrokecolor{currentstroke}%
\pgfsetdash{}{0pt}%
\pgfpathmoveto{\pgfqpoint{0.675000in}{0.297000in}}%
\pgfpathlineto{\pgfqpoint{0.675000in}{2.376000in}}%
\pgfusepath{stroke}%
\end{pgfscope}%
\begin{pgfscope}%
\pgfsetrectcap%
\pgfsetmiterjoin%
\pgfsetlinewidth{0.803000pt}%
\definecolor{currentstroke}{rgb}{0.000000,0.000000,0.000000}%
\pgfsetstrokecolor{currentstroke}%
\pgfsetdash{}{0pt}%
\pgfpathmoveto{\pgfqpoint{4.860000in}{0.297000in}}%
\pgfpathlineto{\pgfqpoint{4.860000in}{2.376000in}}%
\pgfusepath{stroke}%
\end{pgfscope}%
\begin{pgfscope}%
\pgfsetrectcap%
\pgfsetmiterjoin%
\pgfsetlinewidth{0.803000pt}%
\definecolor{currentstroke}{rgb}{0.000000,0.000000,0.000000}%
\pgfsetstrokecolor{currentstroke}%
\pgfsetdash{}{0pt}%
\pgfpathmoveto{\pgfqpoint{0.675000in}{0.297000in}}%
\pgfpathlineto{\pgfqpoint{4.860000in}{0.297000in}}%
\pgfusepath{stroke}%
\end{pgfscope}%
\begin{pgfscope}%
\pgfsetrectcap%
\pgfsetmiterjoin%
\pgfsetlinewidth{0.803000pt}%
\definecolor{currentstroke}{rgb}{0.000000,0.000000,0.000000}%
\pgfsetstrokecolor{currentstroke}%
\pgfsetdash{}{0pt}%
\pgfpathmoveto{\pgfqpoint{0.675000in}{2.376000in}}%
\pgfpathlineto{\pgfqpoint{4.860000in}{2.376000in}}%
\pgfusepath{stroke}%
\end{pgfscope}%
\begin{pgfscope}%
\pgfsetbuttcap%
\pgfsetmiterjoin%
\pgfsetlinewidth{0.000000pt}%
\definecolor{currentstroke}{rgb}{0.800000,0.800000,0.800000}%
\pgfsetstrokecolor{currentstroke}%
\pgfsetstrokeopacity{0.000000}%
\pgfsetdash{}{0pt}%
\pgfpathmoveto{\pgfqpoint{2.203786in}{0.824605in}}%
\pgfpathlineto{\pgfqpoint{3.331214in}{0.824605in}}%
\pgfpathquadraticcurveto{\pgfqpoint{3.358992in}{0.824605in}}{\pgfqpoint{3.358992in}{0.852383in}}%
\pgfpathlineto{\pgfqpoint{3.358992in}{1.820617in}}%
\pgfpathquadraticcurveto{\pgfqpoint{3.358992in}{1.848395in}}{\pgfqpoint{3.331214in}{1.848395in}}%
\pgfpathlineto{\pgfqpoint{2.203786in}{1.848395in}}%
\pgfpathquadraticcurveto{\pgfqpoint{2.176008in}{1.848395in}}{\pgfqpoint{2.176008in}{1.820617in}}%
\pgfpathlineto{\pgfqpoint{2.176008in}{0.852383in}}%
\pgfpathquadraticcurveto{\pgfqpoint{2.176008in}{0.824605in}}{\pgfqpoint{2.203786in}{0.824605in}}%
\pgfpathclose%
\pgfusepath{}%
\end{pgfscope}%
\begin{pgfscope}%
\pgfsetrectcap%
\pgfsetroundjoin%
\pgfsetlinewidth{1.003750pt}%
\definecolor{currentstroke}{rgb}{0.121569,0.466667,0.705882}%
\pgfsetstrokecolor{currentstroke}%
\pgfsetdash{}{0pt}%
\pgfpathmoveto{\pgfqpoint{2.231564in}{1.744228in}}%
\pgfpathlineto{\pgfqpoint{2.509342in}{1.744228in}}%
\pgfusepath{stroke}%
\end{pgfscope}%
\begin{pgfscope}%
\pgftext[x=2.620453in,y=1.695617in,left,base]{\rmfamily\fontsize{10.000000}{12.000000}\selectfont Pool 1}%
\end{pgfscope}%
\begin{pgfscope}%
\pgfsetrectcap%
\pgfsetroundjoin%
\pgfsetlinewidth{1.003750pt}%
\definecolor{currentstroke}{rgb}{1.000000,0.498039,0.054902}%
\pgfsetstrokecolor{currentstroke}%
\pgfsetdash{}{0pt}%
\pgfpathmoveto{\pgfqpoint{2.231564in}{1.547858in}}%
\pgfpathlineto{\pgfqpoint{2.509342in}{1.547858in}}%
\pgfusepath{stroke}%
\end{pgfscope}%
\begin{pgfscope}%
\pgftext[x=2.620453in,y=1.499247in,left,base]{\rmfamily\fontsize{10.000000}{12.000000}\selectfont Pool 2}%
\end{pgfscope}%
\begin{pgfscope}%
\pgfsetrectcap%
\pgfsetroundjoin%
\pgfsetlinewidth{1.003750pt}%
\definecolor{currentstroke}{rgb}{0.172549,0.627451,0.172549}%
\pgfsetstrokecolor{currentstroke}%
\pgfsetdash{}{0pt}%
\pgfpathmoveto{\pgfqpoint{2.231564in}{1.351487in}}%
\pgfpathlineto{\pgfqpoint{2.509342in}{1.351487in}}%
\pgfusepath{stroke}%
\end{pgfscope}%
\begin{pgfscope}%
\pgftext[x=2.620453in,y=1.302876in,left,base]{\rmfamily\fontsize{10.000000}{12.000000}\selectfont Pool 3}%
\end{pgfscope}%
\begin{pgfscope}%
\pgfsetrectcap%
\pgfsetroundjoin%
\pgfsetlinewidth{1.003750pt}%
\definecolor{currentstroke}{rgb}{0.839216,0.152941,0.156863}%
\pgfsetstrokecolor{currentstroke}%
\pgfsetdash{}{0pt}%
\pgfpathmoveto{\pgfqpoint{2.231564in}{1.155117in}}%
\pgfpathlineto{\pgfqpoint{2.509342in}{1.155117in}}%
\pgfusepath{stroke}%
\end{pgfscope}%
\begin{pgfscope}%
\pgftext[x=2.620453in,y=1.106506in,left,base]{\rmfamily\fontsize{10.000000}{12.000000}\selectfont Pool 4}%
\end{pgfscope}%
\begin{pgfscope}%
\pgfsetrectcap%
\pgfsetroundjoin%
\pgfsetlinewidth{1.003750pt}%
\definecolor{currentstroke}{rgb}{0.580392,0.403922,0.741176}%
\pgfsetstrokecolor{currentstroke}%
\pgfsetdash{}{0pt}%
\pgfpathmoveto{\pgfqpoint{2.231564in}{0.958747in}}%
\pgfpathlineto{\pgfqpoint{2.509342in}{0.958747in}}%
\pgfusepath{stroke}%
\end{pgfscope}%
\begin{pgfscope}%
\pgftext[x=2.620453in,y=0.910136in,left,base]{\rmfamily\fontsize{10.000000}{12.000000}\selectfont Solo mining}%
\end{pgfscope}%
\end{pgfpicture}%
\makeatother%
\endgroup%

%% file: single_crypto2.pgf
\begingroup%
\makeatletter%
\begin{pgfpicture}%
\pgfpathrectangle{\pgfpointorigin}{\pgfqpoint{5.400000in}{2.700000in}}%
\pgfusepath{use as bounding box, clip}%
\begin{pgfscope}%
\pgfsetbuttcap%
\pgfsetmiterjoin%
\definecolor{currentfill}{rgb}{1.000000,1.000000,1.000000}%
\pgfsetfillcolor{currentfill}%
\pgfsetlinewidth{0.000000pt}%
\definecolor{currentstroke}{rgb}{1.000000,1.000000,1.000000}%
\pgfsetstrokecolor{currentstroke}%
\pgfsetdash{}{0pt}%
\pgfpathmoveto{\pgfqpoint{0.000000in}{0.000000in}}%
\pgfpathlineto{\pgfqpoint{5.400000in}{0.000000in}}%
\pgfpathlineto{\pgfqpoint{5.400000in}{2.700000in}}%
\pgfpathlineto{\pgfqpoint{0.000000in}{2.700000in}}%
\pgfpathclose%
\pgfusepath{fill}%
\end{pgfscope}%
\begin{pgfscope}%
\pgfsetbuttcap%
\pgfsetmiterjoin%
\definecolor{currentfill}{rgb}{1.000000,1.000000,1.000000}%
\pgfsetfillcolor{currentfill}%
\pgfsetlinewidth{0.000000pt}%
\definecolor{currentstroke}{rgb}{0.000000,0.000000,0.000000}%
\pgfsetstrokecolor{currentstroke}%
\pgfsetstrokeopacity{0.000000}%
\pgfsetdash{}{0pt}%
\pgfpathmoveto{\pgfqpoint{0.675000in}{0.297000in}}%
\pgfpathlineto{\pgfqpoint{4.860000in}{0.297000in}}%
\pgfpathlineto{\pgfqpoint{4.860000in}{2.376000in}}%
\pgfpathlineto{\pgfqpoint{0.675000in}{2.376000in}}%
\pgfpathclose%
\pgfusepath{fill}%
\end{pgfscope}%
\begin{pgfscope}%
\pgfsetbuttcap%
\pgfsetroundjoin%
\definecolor{currentfill}{rgb}{0.000000,0.000000,0.000000}%
\pgfsetfillcolor{currentfill}%
\pgfsetlinewidth{0.803000pt}%
\definecolor{currentstroke}{rgb}{0.000000,0.000000,0.000000}%
\pgfsetstrokecolor{currentstroke}%
\pgfsetdash{}{0pt}%
\pgfsys@defobject{currentmarker}{\pgfqpoint{0.000000in}{0.000000in}}{\pgfqpoint{0.000000in}{0.048611in}}{%
\pgfpathmoveto{\pgfqpoint{0.000000in}{0.000000in}}%
\pgfpathlineto{\pgfqpoint{0.000000in}{0.048611in}}%
\pgfusepath{stroke,fill}%
}%
\begin{pgfscope}%
\pgfsys@transformshift{1.279584in}{0.297000in}%
\pgfsys@useobject{currentmarker}{}%
\end{pgfscope}%
\end{pgfscope}%
\begin{pgfscope}%
\pgftext[x=1.279584in,y=0.269222in,,top]{\rmfamily\fontsize{10.000000}{12.000000}\selectfont \(\displaystyle 0.00002\)}%
\end{pgfscope}%
\begin{pgfscope}%
\pgfsetbuttcap%
\pgfsetroundjoin%
\definecolor{currentfill}{rgb}{0.000000,0.000000,0.000000}%
\pgfsetfillcolor{currentfill}%
\pgfsetlinewidth{0.803000pt}%
\definecolor{currentstroke}{rgb}{0.000000,0.000000,0.000000}%
\pgfsetstrokecolor{currentstroke}%
\pgfsetdash{}{0pt}%
\pgfsys@defobject{currentmarker}{\pgfqpoint{0.000000in}{0.000000in}}{\pgfqpoint{0.000000in}{0.048611in}}{%
\pgfpathmoveto{\pgfqpoint{0.000000in}{0.000000in}}%
\pgfpathlineto{\pgfqpoint{0.000000in}{0.048611in}}%
\pgfusepath{stroke,fill}%
}%
\begin{pgfscope}%
\pgfsys@transformshift{2.032959in}{0.297000in}%
\pgfsys@useobject{currentmarker}{}%
\end{pgfscope}%
\end{pgfscope}%
\begin{pgfscope}%
\pgftext[x=2.032959in,y=0.269222in,,top]{\rmfamily\fontsize{10.000000}{12.000000}\selectfont \(\displaystyle 0.00004\)}%
\end{pgfscope}%
\begin{pgfscope}%
\pgfsetbuttcap%
\pgfsetroundjoin%
\definecolor{currentfill}{rgb}{0.000000,0.000000,0.000000}%
\pgfsetfillcolor{currentfill}%
\pgfsetlinewidth{0.803000pt}%
\definecolor{currentstroke}{rgb}{0.000000,0.000000,0.000000}%
\pgfsetstrokecolor{currentstroke}%
\pgfsetdash{}{0pt}%
\pgfsys@defobject{currentmarker}{\pgfqpoint{0.000000in}{0.000000in}}{\pgfqpoint{0.000000in}{0.048611in}}{%
\pgfpathmoveto{\pgfqpoint{0.000000in}{0.000000in}}%
\pgfpathlineto{\pgfqpoint{0.000000in}{0.048611in}}%
\pgfusepath{stroke,fill}%
}%
\begin{pgfscope}%
\pgfsys@transformshift{2.786334in}{0.297000in}%
\pgfsys@useobject{currentmarker}{}%
\end{pgfscope}%
\end{pgfscope}%
\begin{pgfscope}%
\pgftext[x=2.786334in,y=0.269222in,,top]{\rmfamily\fontsize{10.000000}{12.000000}\selectfont \(\displaystyle 0.00006\)}%
\end{pgfscope}%
\begin{pgfscope}%
\pgfsetbuttcap%
\pgfsetroundjoin%
\definecolor{currentfill}{rgb}{0.000000,0.000000,0.000000}%
\pgfsetfillcolor{currentfill}%
\pgfsetlinewidth{0.803000pt}%
\definecolor{currentstroke}{rgb}{0.000000,0.000000,0.000000}%
\pgfsetstrokecolor{currentstroke}%
\pgfsetdash{}{0pt}%
\pgfsys@defobject{currentmarker}{\pgfqpoint{0.000000in}{0.000000in}}{\pgfqpoint{0.000000in}{0.048611in}}{%
\pgfpathmoveto{\pgfqpoint{0.000000in}{0.000000in}}%
\pgfpathlineto{\pgfqpoint{0.000000in}{0.048611in}}%
\pgfusepath{stroke,fill}%
}%
\begin{pgfscope}%
\pgfsys@transformshift{3.539710in}{0.297000in}%
\pgfsys@useobject{currentmarker}{}%
\end{pgfscope}%
\end{pgfscope}%
\begin{pgfscope}%
\pgftext[x=3.539710in,y=0.269222in,,top]{\rmfamily\fontsize{10.000000}{12.000000}\selectfont \(\displaystyle 0.00008\)}%
\end{pgfscope}%
\begin{pgfscope}%
\pgfsetbuttcap%
\pgfsetroundjoin%
\definecolor{currentfill}{rgb}{0.000000,0.000000,0.000000}%
\pgfsetfillcolor{currentfill}%
\pgfsetlinewidth{0.803000pt}%
\definecolor{currentstroke}{rgb}{0.000000,0.000000,0.000000}%
\pgfsetstrokecolor{currentstroke}%
\pgfsetdash{}{0pt}%
\pgfsys@defobject{currentmarker}{\pgfqpoint{0.000000in}{0.000000in}}{\pgfqpoint{0.000000in}{0.048611in}}{%
\pgfpathmoveto{\pgfqpoint{0.000000in}{0.000000in}}%
\pgfpathlineto{\pgfqpoint{0.000000in}{0.048611in}}%
\pgfusepath{stroke,fill}%
}%
\begin{pgfscope}%
\pgfsys@transformshift{4.293085in}{0.297000in}%
\pgfsys@useobject{currentmarker}{}%
\end{pgfscope}%
\end{pgfscope}%
\begin{pgfscope}%
\pgftext[x=4.293085in,y=0.269222in,,top]{\rmfamily\fontsize{10.000000}{12.000000}\selectfont \(\displaystyle 0.00010\)}%
\end{pgfscope}%
\begin{pgfscope}%
\pgftext[x=2.767500in,y=0.156972in,,top]{\rmfamily\fontsize{10.000000}{12.000000}\selectfont CARA}%
\end{pgfscope}%
\begin{pgfscope}%
\pgfsetbuttcap%
\pgfsetroundjoin%
\definecolor{currentfill}{rgb}{0.000000,0.000000,0.000000}%
\pgfsetfillcolor{currentfill}%
\pgfsetlinewidth{0.803000pt}%
\definecolor{currentstroke}{rgb}{0.000000,0.000000,0.000000}%
\pgfsetstrokecolor{currentstroke}%
\pgfsetdash{}{0pt}%
\pgfsys@defobject{currentmarker}{\pgfqpoint{-0.048611in}{0.000000in}}{\pgfqpoint{0.000000in}{0.000000in}}{%
\pgfpathmoveto{\pgfqpoint{0.000000in}{0.000000in}}%
\pgfpathlineto{\pgfqpoint{-0.048611in}{0.000000in}}%
\pgfusepath{stroke,fill}%
}%
\begin{pgfscope}%
\pgfsys@transformshift{0.675000in}{0.391500in}%
\pgfsys@useobject{currentmarker}{}%
\end{pgfscope}%
\end{pgfscope}%
\begin{pgfscope}%
\pgftext[x=0.400308in,y=0.343282in,left,base]{\rmfamily\fontsize{10.000000}{12.000000}\selectfont \(\displaystyle 0.0\)}%
\end{pgfscope}%
\begin{pgfscope}%
\pgfsetbuttcap%
\pgfsetroundjoin%
\definecolor{currentfill}{rgb}{0.000000,0.000000,0.000000}%
\pgfsetfillcolor{currentfill}%
\pgfsetlinewidth{0.803000pt}%
\definecolor{currentstroke}{rgb}{0.000000,0.000000,0.000000}%
\pgfsetstrokecolor{currentstroke}%
\pgfsetdash{}{0pt}%
\pgfsys@defobject{currentmarker}{\pgfqpoint{-0.048611in}{0.000000in}}{\pgfqpoint{0.000000in}{0.000000in}}{%
\pgfpathmoveto{\pgfqpoint{0.000000in}{0.000000in}}%
\pgfpathlineto{\pgfqpoint{-0.048611in}{0.000000in}}%
\pgfusepath{stroke,fill}%
}%
\begin{pgfscope}%
\pgfsys@transformshift{0.675000in}{0.972170in}%
\pgfsys@useobject{currentmarker}{}%
\end{pgfscope}%
\end{pgfscope}%
\begin{pgfscope}%
\pgftext[x=0.400308in,y=0.923952in,left,base]{\rmfamily\fontsize{10.000000}{12.000000}\selectfont \(\displaystyle 0.5\)}%
\end{pgfscope}%
\begin{pgfscope}%
\pgfsetbuttcap%
\pgfsetroundjoin%
\definecolor{currentfill}{rgb}{0.000000,0.000000,0.000000}%
\pgfsetfillcolor{currentfill}%
\pgfsetlinewidth{0.803000pt}%
\definecolor{currentstroke}{rgb}{0.000000,0.000000,0.000000}%
\pgfsetstrokecolor{currentstroke}%
\pgfsetdash{}{0pt}%
\pgfsys@defobject{currentmarker}{\pgfqpoint{-0.048611in}{0.000000in}}{\pgfqpoint{0.000000in}{0.000000in}}{%
\pgfpathmoveto{\pgfqpoint{0.000000in}{0.000000in}}%
\pgfpathlineto{\pgfqpoint{-0.048611in}{0.000000in}}%
\pgfusepath{stroke,fill}%
}%
\begin{pgfscope}%
\pgfsys@transformshift{0.675000in}{1.552840in}%
\pgfsys@useobject{currentmarker}{}%
\end{pgfscope}%
\end{pgfscope}%
\begin{pgfscope}%
\pgftext[x=0.400308in,y=1.504622in,left,base]{\rmfamily\fontsize{10.000000}{12.000000}\selectfont \(\displaystyle 1.0\)}%
\end{pgfscope}%
\begin{pgfscope}%
\pgfsetbuttcap%
\pgfsetroundjoin%
\definecolor{currentfill}{rgb}{0.000000,0.000000,0.000000}%
\pgfsetfillcolor{currentfill}%
\pgfsetlinewidth{0.803000pt}%
\definecolor{currentstroke}{rgb}{0.000000,0.000000,0.000000}%
\pgfsetstrokecolor{currentstroke}%
\pgfsetdash{}{0pt}%
\pgfsys@defobject{currentmarker}{\pgfqpoint{-0.048611in}{0.000000in}}{\pgfqpoint{0.000000in}{0.000000in}}{%
\pgfpathmoveto{\pgfqpoint{0.000000in}{0.000000in}}%
\pgfpathlineto{\pgfqpoint{-0.048611in}{0.000000in}}%
\pgfusepath{stroke,fill}%
}%
\begin{pgfscope}%
\pgfsys@transformshift{0.675000in}{2.133510in}%
\pgfsys@useobject{currentmarker}{}%
\end{pgfscope}%
\end{pgfscope}%
\begin{pgfscope}%
\pgftext[x=0.400308in,y=2.085292in,left,base]{\rmfamily\fontsize{10.000000}{12.000000}\selectfont \(\displaystyle 1.5\)}%
\end{pgfscope}%
\begin{pgfscope}%
\pgftext[x=0.344753in,y=1.336500in,,bottom,rotate=90.000000]{\rmfamily\fontsize{10.000000}{12.000000}\selectfont Hash rate}%
\end{pgfscope}%
\begin{pgfscope}%
\pgftext[x=0.675000in,y=2.417667in,left,base]{\rmfamily\fontsize{10.000000}{12.000000}\selectfont \(\displaystyle \times10^{15}\)}%
\end{pgfscope}%
\begin{pgfscope}%
\pgfpathrectangle{\pgfqpoint{0.675000in}{0.297000in}}{\pgfqpoint{4.185000in}{2.079000in}}%
\pgfusepath{clip}%
\pgfsetrectcap%
\pgfsetroundjoin%
\pgfsetlinewidth{1.003750pt}%
\definecolor{currentstroke}{rgb}{0.121569,0.466667,0.705882}%
\pgfsetstrokecolor{currentstroke}%
\pgfsetdash{}{0pt}%
\pgfpathmoveto{\pgfqpoint{0.865227in}{1.451091in}}%
\pgfpathlineto{\pgfqpoint{0.896932in}{1.530211in}}%
\pgfpathlineto{\pgfqpoint{0.928636in}{1.597077in}}%
\pgfpathlineto{\pgfqpoint{0.960341in}{1.654990in}}%
\pgfpathlineto{\pgfqpoint{0.992045in}{1.703937in}}%
\pgfpathlineto{\pgfqpoint{1.023750in}{1.746362in}}%
\pgfpathlineto{\pgfqpoint{1.055455in}{1.783672in}}%
\pgfpathlineto{\pgfqpoint{1.087159in}{1.816086in}}%
\pgfpathlineto{\pgfqpoint{1.118864in}{1.846073in}}%
\pgfpathlineto{\pgfqpoint{1.150568in}{1.872349in}}%
\pgfpathlineto{\pgfqpoint{1.182273in}{1.895592in}}%
\pgfpathlineto{\pgfqpoint{1.213977in}{1.917555in}}%
\pgfpathlineto{\pgfqpoint{1.245682in}{1.937130in}}%
\pgfpathlineto{\pgfqpoint{1.277386in}{1.954983in}}%
\pgfpathlineto{\pgfqpoint{1.309091in}{1.971277in}}%
\pgfpathlineto{\pgfqpoint{1.340795in}{1.986367in}}%
\pgfpathlineto{\pgfqpoint{1.372500in}{2.000365in}}%
\pgfpathlineto{\pgfqpoint{1.404205in}{2.013330in}}%
\pgfpathlineto{\pgfqpoint{1.435909in}{2.025364in}}%
\pgfpathlineto{\pgfqpoint{1.467614in}{2.036193in}}%
\pgfpathlineto{\pgfqpoint{1.499318in}{2.046907in}}%
\pgfpathlineto{\pgfqpoint{1.531023in}{2.056675in}}%
\pgfpathlineto{\pgfqpoint{1.562727in}{2.065909in}}%
\pgfpathlineto{\pgfqpoint{1.594432in}{2.074537in}}%
\pgfpathlineto{\pgfqpoint{1.626136in}{2.082585in}}%
\pgfpathlineto{\pgfqpoint{1.657841in}{2.090320in}}%
\pgfpathlineto{\pgfqpoint{1.689545in}{2.097558in}}%
\pgfpathlineto{\pgfqpoint{1.721250in}{2.104428in}}%
\pgfpathlineto{\pgfqpoint{1.752955in}{2.110921in}}%
\pgfpathlineto{\pgfqpoint{1.784659in}{2.117055in}}%
\pgfpathlineto{\pgfqpoint{1.816364in}{2.122889in}}%
\pgfpathlineto{\pgfqpoint{1.848068in}{2.128433in}}%
\pgfpathlineto{\pgfqpoint{1.879773in}{2.133543in}}%
\pgfpathlineto{\pgfqpoint{1.911477in}{2.138554in}}%
\pgfpathlineto{\pgfqpoint{1.943182in}{2.143355in}}%
\pgfpathlineto{\pgfqpoint{1.974886in}{2.148213in}}%
\pgfpathlineto{\pgfqpoint{2.006591in}{2.152334in}}%
\pgfpathlineto{\pgfqpoint{2.038295in}{2.156569in}}%
\pgfpathlineto{\pgfqpoint{2.070000in}{2.160685in}}%
\pgfpathlineto{\pgfqpoint{2.101705in}{2.164614in}}%
\pgfpathlineto{\pgfqpoint{2.133409in}{2.168365in}}%
\pgfpathlineto{\pgfqpoint{2.165114in}{2.171979in}}%
\pgfpathlineto{\pgfqpoint{2.196818in}{2.175412in}}%
\pgfpathlineto{\pgfqpoint{2.228523in}{2.178707in}}%
\pgfpathlineto{\pgfqpoint{2.260227in}{2.181962in}}%
\pgfpathlineto{\pgfqpoint{2.291932in}{2.185039in}}%
\pgfpathlineto{\pgfqpoint{2.323636in}{2.188008in}}%
\pgfpathlineto{\pgfqpoint{2.355341in}{2.190841in}}%
\pgfpathlineto{\pgfqpoint{2.387045in}{2.193631in}}%
\pgfpathlineto{\pgfqpoint{2.418750in}{2.196306in}}%
\pgfpathlineto{\pgfqpoint{2.450455in}{2.198890in}}%
\pgfpathlineto{\pgfqpoint{2.482159in}{2.201466in}}%
\pgfpathlineto{\pgfqpoint{2.513864in}{2.203880in}}%
\pgfpathlineto{\pgfqpoint{2.545568in}{2.206225in}}%
\pgfpathlineto{\pgfqpoint{2.577273in}{2.208481in}}%
\pgfpathlineto{\pgfqpoint{2.608977in}{2.210700in}}%
\pgfpathlineto{\pgfqpoint{2.640682in}{2.212850in}}%
\pgfpathlineto{\pgfqpoint{2.672386in}{2.214883in}}%
\pgfpathlineto{\pgfqpoint{2.704091in}{2.216906in}}%
\pgfpathlineto{\pgfqpoint{2.735795in}{2.218866in}}%
\pgfpathlineto{\pgfqpoint{2.767500in}{2.220777in}}%
\pgfpathlineto{\pgfqpoint{2.799205in}{2.222636in}}%
\pgfpathlineto{\pgfqpoint{2.830909in}{2.224409in}}%
\pgfpathlineto{\pgfqpoint{2.862614in}{2.226186in}}%
\pgfpathlineto{\pgfqpoint{2.894318in}{2.227921in}}%
\pgfpathlineto{\pgfqpoint{2.926023in}{2.229551in}}%
\pgfpathlineto{\pgfqpoint{2.957727in}{2.231167in}}%
\pgfpathlineto{\pgfqpoint{2.989432in}{2.232729in}}%
\pgfpathlineto{\pgfqpoint{3.021136in}{2.234267in}}%
\pgfpathlineto{\pgfqpoint{3.052841in}{2.235752in}}%
\pgfpathlineto{\pgfqpoint{3.084545in}{2.237197in}}%
\pgfpathlineto{\pgfqpoint{3.116250in}{2.238621in}}%
\pgfpathlineto{\pgfqpoint{3.147955in}{2.239991in}}%
\pgfpathlineto{\pgfqpoint{3.179659in}{2.241382in}}%
\pgfpathlineto{\pgfqpoint{3.211364in}{2.242674in}}%
\pgfpathlineto{\pgfqpoint{3.243068in}{2.243979in}}%
\pgfpathlineto{\pgfqpoint{3.274773in}{2.245225in}}%
\pgfpathlineto{\pgfqpoint{3.306477in}{2.246474in}}%
\pgfpathlineto{\pgfqpoint{3.338182in}{2.247667in}}%
\pgfpathlineto{\pgfqpoint{3.369886in}{2.248808in}}%
\pgfpathlineto{\pgfqpoint{3.401591in}{2.249979in}}%
\pgfpathlineto{\pgfqpoint{3.433295in}{2.251111in}}%
\pgfpathlineto{\pgfqpoint{3.465000in}{2.252218in}}%
\pgfpathlineto{\pgfqpoint{3.496705in}{2.253262in}}%
\pgfpathlineto{\pgfqpoint{3.528409in}{2.254352in}}%
\pgfpathlineto{\pgfqpoint{3.560114in}{2.255487in}}%
\pgfpathlineto{\pgfqpoint{3.591818in}{2.256383in}}%
\pgfpathlineto{\pgfqpoint{3.623523in}{2.257366in}}%
\pgfpathlineto{\pgfqpoint{3.655227in}{2.258360in}}%
\pgfpathlineto{\pgfqpoint{3.686932in}{2.259307in}}%
\pgfpathlineto{\pgfqpoint{3.718636in}{2.260216in}}%
\pgfpathlineto{\pgfqpoint{3.750341in}{2.261144in}}%
\pgfpathlineto{\pgfqpoint{3.782045in}{2.262028in}}%
\pgfpathlineto{\pgfqpoint{3.813750in}{2.262888in}}%
\pgfpathlineto{\pgfqpoint{3.845455in}{2.263746in}}%
\pgfpathlineto{\pgfqpoint{3.877159in}{2.264615in}}%
\pgfpathlineto{\pgfqpoint{3.908864in}{2.265445in}}%
\pgfpathlineto{\pgfqpoint{3.940568in}{2.266240in}}%
\pgfpathlineto{\pgfqpoint{3.972273in}{2.267147in}}%
\pgfpathlineto{\pgfqpoint{4.003977in}{2.267799in}}%
\pgfpathlineto{\pgfqpoint{4.035682in}{2.268723in}}%
\pgfpathlineto{\pgfqpoint{4.067386in}{2.269351in}}%
\pgfpathlineto{\pgfqpoint{4.099091in}{2.270185in}}%
\pgfpathlineto{\pgfqpoint{4.130795in}{2.270797in}}%
\pgfpathlineto{\pgfqpoint{4.162500in}{2.271628in}}%
\pgfpathlineto{\pgfqpoint{4.194205in}{2.272221in}}%
\pgfpathlineto{\pgfqpoint{4.225909in}{2.272943in}}%
\pgfpathlineto{\pgfqpoint{4.257614in}{2.273711in}}%
\pgfpathlineto{\pgfqpoint{4.289318in}{2.274376in}}%
\pgfpathlineto{\pgfqpoint{4.321023in}{2.275043in}}%
\pgfpathlineto{\pgfqpoint{4.352727in}{2.275577in}}%
\pgfpathlineto{\pgfqpoint{4.384432in}{2.276218in}}%
\pgfpathlineto{\pgfqpoint{4.416136in}{2.276929in}}%
\pgfpathlineto{\pgfqpoint{4.447841in}{2.277531in}}%
\pgfpathlineto{\pgfqpoint{4.479545in}{2.278144in}}%
\pgfpathlineto{\pgfqpoint{4.511250in}{2.278756in}}%
\pgfpathlineto{\pgfqpoint{4.542955in}{2.279253in}}%
\pgfpathlineto{\pgfqpoint{4.574659in}{2.279811in}}%
\pgfpathlineto{\pgfqpoint{4.606364in}{2.280386in}}%
\pgfpathlineto{\pgfqpoint{4.638068in}{2.281019in}}%
\pgfpathlineto{\pgfqpoint{4.669773in}{2.281500in}}%
\pgfusepath{stroke}%
\end{pgfscope}%
\begin{pgfscope}%
\pgfpathrectangle{\pgfqpoint{0.675000in}{0.297000in}}{\pgfqpoint{4.185000in}{2.079000in}}%
\pgfusepath{clip}%
\pgfsetrectcap%
\pgfsetroundjoin%
\pgfsetlinewidth{1.003750pt}%
\definecolor{currentstroke}{rgb}{1.000000,0.498039,0.054902}%
\pgfsetstrokecolor{currentstroke}%
\pgfsetdash{}{0pt}%
\pgfpathmoveto{\pgfqpoint{0.865227in}{1.200575in}}%
\pgfpathlineto{\pgfqpoint{0.896932in}{1.263017in}}%
\pgfpathlineto{\pgfqpoint{0.928636in}{1.314441in}}%
\pgfpathlineto{\pgfqpoint{0.960341in}{1.357108in}}%
\pgfpathlineto{\pgfqpoint{0.992045in}{1.394640in}}%
\pgfpathlineto{\pgfqpoint{1.023750in}{1.427245in}}%
\pgfpathlineto{\pgfqpoint{1.055455in}{1.455794in}}%
\pgfpathlineto{\pgfqpoint{1.087159in}{1.481635in}}%
\pgfpathlineto{\pgfqpoint{1.118864in}{1.503450in}}%
\pgfpathlineto{\pgfqpoint{1.150568in}{1.523649in}}%
\pgfpathlineto{\pgfqpoint{1.182273in}{1.542284in}}%
\pgfpathlineto{\pgfqpoint{1.213977in}{1.558218in}}%
\pgfpathlineto{\pgfqpoint{1.245682in}{1.573173in}}%
\pgfpathlineto{\pgfqpoint{1.277386in}{1.586889in}}%
\pgfpathlineto{\pgfqpoint{1.309091in}{1.599512in}}%
\pgfpathlineto{\pgfqpoint{1.340795in}{1.611094in}}%
\pgfpathlineto{\pgfqpoint{1.372500in}{1.621727in}}%
\pgfpathlineto{\pgfqpoint{1.404205in}{1.631567in}}%
\pgfpathlineto{\pgfqpoint{1.435909in}{1.640742in}}%
\pgfpathlineto{\pgfqpoint{1.467614in}{1.649670in}}%
\pgfpathlineto{\pgfqpoint{1.499318in}{1.657390in}}%
\pgfpathlineto{\pgfqpoint{1.531023in}{1.664900in}}%
\pgfpathlineto{\pgfqpoint{1.562727in}{1.671865in}}%
\pgfpathlineto{\pgfqpoint{1.594432in}{1.678442in}}%
\pgfpathlineto{\pgfqpoint{1.626136in}{1.684736in}}%
\pgfpathlineto{\pgfqpoint{1.657841in}{1.690529in}}%
\pgfpathlineto{\pgfqpoint{1.689545in}{1.696054in}}%
\pgfpathlineto{\pgfqpoint{1.721250in}{1.701281in}}%
\pgfpathlineto{\pgfqpoint{1.752955in}{1.706247in}}%
\pgfpathlineto{\pgfqpoint{1.784659in}{1.710980in}}%
\pgfpathlineto{\pgfqpoint{1.816364in}{1.715478in}}%
\pgfpathlineto{\pgfqpoint{1.848068in}{1.719773in}}%
\pgfpathlineto{\pgfqpoint{1.879773in}{1.724018in}}%
\pgfpathlineto{\pgfqpoint{1.911477in}{1.727945in}}%
\pgfpathlineto{\pgfqpoint{1.943182in}{1.731677in}}%
\pgfpathlineto{\pgfqpoint{1.974886in}{1.734966in}}%
\pgfpathlineto{\pgfqpoint{2.006591in}{1.738646in}}%
\pgfpathlineto{\pgfqpoint{2.038295in}{1.741885in}}%
\pgfpathlineto{\pgfqpoint{2.070000in}{1.744924in}}%
\pgfpathlineto{\pgfqpoint{2.101705in}{1.747868in}}%
\pgfpathlineto{\pgfqpoint{2.133409in}{1.750716in}}%
\pgfpathlineto{\pgfqpoint{2.165114in}{1.753436in}}%
\pgfpathlineto{\pgfqpoint{2.196818in}{1.756101in}}%
\pgfpathlineto{\pgfqpoint{2.228523in}{1.758679in}}%
\pgfpathlineto{\pgfqpoint{2.260227in}{1.761067in}}%
\pgfpathlineto{\pgfqpoint{2.291932in}{1.763440in}}%
\pgfpathlineto{\pgfqpoint{2.323636in}{1.765726in}}%
\pgfpathlineto{\pgfqpoint{2.355341in}{1.767957in}}%
\pgfpathlineto{\pgfqpoint{2.387045in}{1.770064in}}%
\pgfpathlineto{\pgfqpoint{2.418750in}{1.772121in}}%
\pgfpathlineto{\pgfqpoint{2.450455in}{1.774107in}}%
\pgfpathlineto{\pgfqpoint{2.482159in}{1.775955in}}%
\pgfpathlineto{\pgfqpoint{2.513864in}{1.777825in}}%
\pgfpathlineto{\pgfqpoint{2.545568in}{1.779623in}}%
\pgfpathlineto{\pgfqpoint{2.577273in}{1.781386in}}%
\pgfpathlineto{\pgfqpoint{2.608977in}{1.783062in}}%
\pgfpathlineto{\pgfqpoint{2.640682in}{1.784686in}}%
\pgfpathlineto{\pgfqpoint{2.672386in}{1.786317in}}%
\pgfpathlineto{\pgfqpoint{2.704091in}{1.787853in}}%
\pgfpathlineto{\pgfqpoint{2.735795in}{1.789346in}}%
\pgfpathlineto{\pgfqpoint{2.767500in}{1.790791in}}%
\pgfpathlineto{\pgfqpoint{2.799205in}{1.792197in}}%
\pgfpathlineto{\pgfqpoint{2.830909in}{1.793590in}}%
\pgfpathlineto{\pgfqpoint{2.862614in}{1.794901in}}%
\pgfpathlineto{\pgfqpoint{2.894318in}{1.796168in}}%
\pgfpathlineto{\pgfqpoint{2.926023in}{1.797458in}}%
\pgfpathlineto{\pgfqpoint{2.957727in}{1.798687in}}%
\pgfpathlineto{\pgfqpoint{2.989432in}{1.799899in}}%
\pgfpathlineto{\pgfqpoint{3.021136in}{1.801060in}}%
\pgfpathlineto{\pgfqpoint{3.052841in}{1.802206in}}%
\pgfpathlineto{\pgfqpoint{3.084545in}{1.803331in}}%
\pgfpathlineto{\pgfqpoint{3.116250in}{1.804410in}}%
\pgfpathlineto{\pgfqpoint{3.147955in}{1.805485in}}%
\pgfpathlineto{\pgfqpoint{3.179659in}{1.806476in}}%
\pgfpathlineto{\pgfqpoint{3.211364in}{1.807510in}}%
\pgfpathlineto{\pgfqpoint{3.243068in}{1.808481in}}%
\pgfpathlineto{\pgfqpoint{3.274773in}{1.809456in}}%
\pgfpathlineto{\pgfqpoint{3.306477in}{1.810373in}}%
\pgfpathlineto{\pgfqpoint{3.338182in}{1.811302in}}%
\pgfpathlineto{\pgfqpoint{3.369886in}{1.812235in}}%
\pgfpathlineto{\pgfqpoint{3.401591in}{1.813089in}}%
\pgfpathlineto{\pgfqpoint{3.433295in}{1.813941in}}%
\pgfpathlineto{\pgfqpoint{3.465000in}{1.814774in}}%
\pgfpathlineto{\pgfqpoint{3.496705in}{1.815625in}}%
\pgfpathlineto{\pgfqpoint{3.528409in}{1.816394in}}%
\pgfpathlineto{\pgfqpoint{3.560114in}{1.817080in}}%
\pgfpathlineto{\pgfqpoint{3.591818in}{1.817962in}}%
\pgfpathlineto{\pgfqpoint{3.623523in}{1.818724in}}%
\pgfpathlineto{\pgfqpoint{3.655227in}{1.819438in}}%
\pgfpathlineto{\pgfqpoint{3.686932in}{1.820166in}}%
\pgfpathlineto{\pgfqpoint{3.718636in}{1.820898in}}%
\pgfpathlineto{\pgfqpoint{3.750341in}{1.821579in}}%
\pgfpathlineto{\pgfqpoint{3.782045in}{1.822269in}}%
\pgfpathlineto{\pgfqpoint{3.813750in}{1.822957in}}%
\pgfpathlineto{\pgfqpoint{3.845455in}{1.823617in}}%
\pgfpathlineto{\pgfqpoint{3.877159in}{1.824235in}}%
\pgfpathlineto{\pgfqpoint{3.908864in}{1.824866in}}%
\pgfpathlineto{\pgfqpoint{3.940568in}{1.825504in}}%
\pgfpathlineto{\pgfqpoint{3.972273in}{1.826002in}}%
\pgfpathlineto{\pgfqpoint{4.003977in}{1.826729in}}%
\pgfpathlineto{\pgfqpoint{4.035682in}{1.827164in}}%
\pgfpathlineto{\pgfqpoint{4.067386in}{1.827863in}}%
\pgfpathlineto{\pgfqpoint{4.099091in}{1.828339in}}%
\pgfpathlineto{\pgfqpoint{4.130795in}{1.829012in}}%
\pgfpathlineto{\pgfqpoint{4.162500in}{1.829441in}}%
\pgfpathlineto{\pgfqpoint{4.194205in}{1.830089in}}%
\pgfpathlineto{\pgfqpoint{4.225909in}{1.830586in}}%
\pgfpathlineto{\pgfqpoint{4.257614in}{1.831015in}}%
\pgfpathlineto{\pgfqpoint{4.289318in}{1.831528in}}%
\pgfpathlineto{\pgfqpoint{4.321023in}{1.832020in}}%
\pgfpathlineto{\pgfqpoint{4.352727in}{1.832622in}}%
\pgfpathlineto{\pgfqpoint{4.384432in}{1.833102in}}%
\pgfpathlineto{\pgfqpoint{4.416136in}{1.833493in}}%
\pgfpathlineto{\pgfqpoint{4.447841in}{1.833974in}}%
\pgfpathlineto{\pgfqpoint{4.479545in}{1.834427in}}%
\pgfpathlineto{\pgfqpoint{4.511250in}{1.834865in}}%
\pgfpathlineto{\pgfqpoint{4.542955in}{1.835400in}}%
\pgfpathlineto{\pgfqpoint{4.574659in}{1.835858in}}%
\pgfpathlineto{\pgfqpoint{4.606364in}{1.836285in}}%
\pgfpathlineto{\pgfqpoint{4.638068in}{1.836635in}}%
\pgfpathlineto{\pgfqpoint{4.669773in}{1.837124in}}%
\pgfusepath{stroke}%
\end{pgfscope}%
\begin{pgfscope}%
\pgfpathrectangle{\pgfqpoint{0.675000in}{0.297000in}}{\pgfqpoint{4.185000in}{2.079000in}}%
\pgfusepath{clip}%
\pgfsetrectcap%
\pgfsetroundjoin%
\pgfsetlinewidth{1.003750pt}%
\definecolor{currentstroke}{rgb}{0.172549,0.627451,0.172549}%
\pgfsetstrokecolor{currentstroke}%
\pgfsetdash{}{0pt}%
\pgfpathmoveto{\pgfqpoint{0.865227in}{2.006854in}}%
\pgfpathlineto{\pgfqpoint{0.896932in}{1.865292in}}%
\pgfpathlineto{\pgfqpoint{0.928636in}{1.747001in}}%
\pgfpathlineto{\pgfqpoint{0.960341in}{1.646422in}}%
\pgfpathlineto{\pgfqpoint{0.992045in}{1.559943in}}%
\pgfpathlineto{\pgfqpoint{1.023750in}{1.484913in}}%
\pgfpathlineto{\pgfqpoint{1.055455in}{1.419053in}}%
\pgfpathlineto{\pgfqpoint{1.087159in}{1.360799in}}%
\pgfpathlineto{\pgfqpoint{1.118864in}{1.308996in}}%
\pgfpathlineto{\pgfqpoint{1.150568in}{1.262522in}}%
\pgfpathlineto{\pgfqpoint{1.182273in}{1.220644in}}%
\pgfpathlineto{\pgfqpoint{1.213977in}{1.182746in}}%
\pgfpathlineto{\pgfqpoint{1.245682in}{1.148217in}}%
\pgfpathlineto{\pgfqpoint{1.277386in}{1.116647in}}%
\pgfpathlineto{\pgfqpoint{1.309091in}{1.087730in}}%
\pgfpathlineto{\pgfqpoint{1.340795in}{1.061058in}}%
\pgfpathlineto{\pgfqpoint{1.372500in}{1.036427in}}%
\pgfpathlineto{\pgfqpoint{1.404205in}{1.013623in}}%
\pgfpathlineto{\pgfqpoint{1.435909in}{0.992414in}}%
\pgfpathlineto{\pgfqpoint{1.467614in}{0.972656in}}%
\pgfpathlineto{\pgfqpoint{1.499318in}{0.954222in}}%
\pgfpathlineto{\pgfqpoint{1.531023in}{0.936944in}}%
\pgfpathlineto{\pgfqpoint{1.562727in}{0.920745in}}%
\pgfpathlineto{\pgfqpoint{1.594432in}{0.905541in}}%
\pgfpathlineto{\pgfqpoint{1.626136in}{0.891199in}}%
\pgfpathlineto{\pgfqpoint{1.657841in}{0.877671in}}%
\pgfpathlineto{\pgfqpoint{1.689545in}{0.864907in}}%
\pgfpathlineto{\pgfqpoint{1.721250in}{0.852811in}}%
\pgfpathlineto{\pgfqpoint{1.752955in}{0.841351in}}%
\pgfpathlineto{\pgfqpoint{1.784659in}{0.830485in}}%
\pgfpathlineto{\pgfqpoint{1.816364in}{0.820153in}}%
\pgfpathlineto{\pgfqpoint{1.848068in}{0.810313in}}%
\pgfpathlineto{\pgfqpoint{1.879773in}{0.800958in}}%
\pgfpathlineto{\pgfqpoint{1.911477in}{0.792021in}}%
\pgfpathlineto{\pgfqpoint{1.943182in}{0.783487in}}%
\pgfpathlineto{\pgfqpoint{1.974886in}{0.775340in}}%
\pgfpathlineto{\pgfqpoint{2.006591in}{0.767539in}}%
\pgfpathlineto{\pgfqpoint{2.038295in}{0.760064in}}%
\pgfpathlineto{\pgfqpoint{2.070000in}{0.752910in}}%
\pgfpathlineto{\pgfqpoint{2.101705in}{0.746037in}}%
\pgfpathlineto{\pgfqpoint{2.133409in}{0.739439in}}%
\pgfpathlineto{\pgfqpoint{2.165114in}{0.733105in}}%
\pgfpathlineto{\pgfqpoint{2.196818in}{0.727007in}}%
\pgfpathlineto{\pgfqpoint{2.228523in}{0.721134in}}%
\pgfpathlineto{\pgfqpoint{2.260227in}{0.715490in}}%
\pgfpathlineto{\pgfqpoint{2.291932in}{0.710040in}}%
\pgfpathlineto{\pgfqpoint{2.323636in}{0.704785in}}%
\pgfpathlineto{\pgfqpoint{2.355341in}{0.699722in}}%
\pgfpathlineto{\pgfqpoint{2.387045in}{0.694824in}}%
\pgfpathlineto{\pgfqpoint{2.418750in}{0.690093in}}%
\pgfpathlineto{\pgfqpoint{2.450455in}{0.685523in}}%
\pgfpathlineto{\pgfqpoint{2.482159in}{0.681098in}}%
\pgfpathlineto{\pgfqpoint{2.513864in}{0.676814in}}%
\pgfpathlineto{\pgfqpoint{2.545568in}{0.672671in}}%
\pgfpathlineto{\pgfqpoint{2.577273in}{0.668653in}}%
\pgfpathlineto{\pgfqpoint{2.608977in}{0.664757in}}%
\pgfpathlineto{\pgfqpoint{2.640682in}{0.660984in}}%
\pgfpathlineto{\pgfqpoint{2.672386in}{0.657319in}}%
\pgfpathlineto{\pgfqpoint{2.704091in}{0.653760in}}%
\pgfpathlineto{\pgfqpoint{2.735795in}{0.650307in}}%
\pgfpathlineto{\pgfqpoint{2.767500in}{0.646952in}}%
\pgfpathlineto{\pgfqpoint{2.799205in}{0.643687in}}%
\pgfpathlineto{\pgfqpoint{2.830909in}{0.640520in}}%
\pgfpathlineto{\pgfqpoint{2.862614in}{0.637433in}}%
\pgfpathlineto{\pgfqpoint{2.894318in}{0.634430in}}%
\pgfpathlineto{\pgfqpoint{2.926023in}{0.631511in}}%
\pgfpathlineto{\pgfqpoint{2.957727in}{0.628666in}}%
\pgfpathlineto{\pgfqpoint{2.989432in}{0.625892in}}%
\pgfpathlineto{\pgfqpoint{3.021136in}{0.623192in}}%
\pgfpathlineto{\pgfqpoint{3.052841in}{0.620561in}}%
\pgfpathlineto{\pgfqpoint{3.084545in}{0.617991in}}%
\pgfpathlineto{\pgfqpoint{3.116250in}{0.615489in}}%
\pgfpathlineto{\pgfqpoint{3.147955in}{0.613044in}}%
\pgfpathlineto{\pgfqpoint{3.179659in}{0.610661in}}%
\pgfpathlineto{\pgfqpoint{3.211364in}{0.608334in}}%
\pgfpathlineto{\pgfqpoint{3.243068in}{0.606059in}}%
\pgfpathlineto{\pgfqpoint{3.274773in}{0.603839in}}%
\pgfpathlineto{\pgfqpoint{3.306477in}{0.601672in}}%
\pgfpathlineto{\pgfqpoint{3.338182in}{0.599550in}}%
\pgfpathlineto{\pgfqpoint{3.369886in}{0.597476in}}%
\pgfpathlineto{\pgfqpoint{3.401591in}{0.595451in}}%
\pgfpathlineto{\pgfqpoint{3.433295in}{0.593467in}}%
\pgfpathlineto{\pgfqpoint{3.465000in}{0.591527in}}%
\pgfpathlineto{\pgfqpoint{3.496705in}{0.589632in}}%
\pgfpathlineto{\pgfqpoint{3.528409in}{0.587773in}}%
\pgfpathlineto{\pgfqpoint{3.560114in}{0.585953in}}%
\pgfpathlineto{\pgfqpoint{3.591818in}{0.584174in}}%
\pgfpathlineto{\pgfqpoint{3.623523in}{0.582429in}}%
\pgfpathlineto{\pgfqpoint{3.655227in}{0.580721in}}%
\pgfpathlineto{\pgfqpoint{3.686932in}{0.579046in}}%
\pgfpathlineto{\pgfqpoint{3.718636in}{0.577406in}}%
\pgfpathlineto{\pgfqpoint{3.750341in}{0.575797in}}%
\pgfpathlineto{\pgfqpoint{3.782045in}{0.574222in}}%
\pgfpathlineto{\pgfqpoint{3.813750in}{0.572674in}}%
\pgfpathlineto{\pgfqpoint{3.845455in}{0.571156in}}%
\pgfpathlineto{\pgfqpoint{3.877159in}{0.569669in}}%
\pgfpathlineto{\pgfqpoint{3.908864in}{0.568209in}}%
\pgfpathlineto{\pgfqpoint{3.940568in}{0.566776in}}%
\pgfpathlineto{\pgfqpoint{3.972273in}{0.565370in}}%
\pgfpathlineto{\pgfqpoint{4.003977in}{0.563990in}}%
\pgfpathlineto{\pgfqpoint{4.035682in}{0.562632in}}%
\pgfpathlineto{\pgfqpoint{4.067386in}{0.561305in}}%
\pgfpathlineto{\pgfqpoint{4.099091in}{0.559996in}}%
\pgfpathlineto{\pgfqpoint{4.130795in}{0.558710in}}%
\pgfpathlineto{\pgfqpoint{4.162500in}{0.557451in}}%
\pgfpathlineto{\pgfqpoint{4.194205in}{0.556209in}}%
\pgfpathlineto{\pgfqpoint{4.225909in}{0.554990in}}%
\pgfpathlineto{\pgfqpoint{4.257614in}{0.553793in}}%
\pgfpathlineto{\pgfqpoint{4.289318in}{0.552615in}}%
\pgfpathlineto{\pgfqpoint{4.321023in}{0.551457in}}%
\pgfpathlineto{\pgfqpoint{4.352727in}{0.550320in}}%
\pgfpathlineto{\pgfqpoint{4.384432in}{0.549200in}}%
\pgfpathlineto{\pgfqpoint{4.416136in}{0.548097in}}%
\pgfpathlineto{\pgfqpoint{4.447841in}{0.547015in}}%
\pgfpathlineto{\pgfqpoint{4.479545in}{0.545948in}}%
\pgfpathlineto{\pgfqpoint{4.511250in}{0.544898in}}%
\pgfpathlineto{\pgfqpoint{4.542955in}{0.543867in}}%
\pgfpathlineto{\pgfqpoint{4.574659in}{0.542851in}}%
\pgfpathlineto{\pgfqpoint{4.606364in}{0.541848in}}%
\pgfpathlineto{\pgfqpoint{4.638068in}{0.540865in}}%
\pgfpathlineto{\pgfqpoint{4.669773in}{0.539895in}}%
\pgfusepath{stroke}%
\end{pgfscope}%
\begin{pgfscope}%
\pgfpathrectangle{\pgfqpoint{0.675000in}{0.297000in}}{\pgfqpoint{4.185000in}{2.079000in}}%
\pgfusepath{clip}%
\pgfsetrectcap%
\pgfsetroundjoin%
\pgfsetlinewidth{1.003750pt}%
\definecolor{currentstroke}{rgb}{0.839216,0.152941,0.156863}%
\pgfsetstrokecolor{currentstroke}%
\pgfsetdash{}{0pt}%
\pgfpathmoveto{\pgfqpoint{0.865227in}{0.391500in}}%
\pgfpathlineto{\pgfqpoint{0.896932in}{0.391500in}}%
\pgfpathlineto{\pgfqpoint{0.928636in}{0.391500in}}%
\pgfpathlineto{\pgfqpoint{0.960341in}{0.391500in}}%
\pgfpathlineto{\pgfqpoint{0.992045in}{0.391500in}}%
\pgfpathlineto{\pgfqpoint{1.023750in}{0.391500in}}%
\pgfpathlineto{\pgfqpoint{1.055455in}{0.391500in}}%
\pgfpathlineto{\pgfqpoint{1.087159in}{0.391500in}}%
\pgfpathlineto{\pgfqpoint{1.118864in}{0.391500in}}%
\pgfpathlineto{\pgfqpoint{1.150568in}{0.391500in}}%
\pgfpathlineto{\pgfqpoint{1.182273in}{0.391500in}}%
\pgfpathlineto{\pgfqpoint{1.213977in}{0.391500in}}%
\pgfpathlineto{\pgfqpoint{1.245682in}{0.391500in}}%
\pgfpathlineto{\pgfqpoint{1.277386in}{0.391500in}}%
\pgfpathlineto{\pgfqpoint{1.309091in}{0.391500in}}%
\pgfpathlineto{\pgfqpoint{1.340795in}{0.391500in}}%
\pgfpathlineto{\pgfqpoint{1.372500in}{0.391500in}}%
\pgfpathlineto{\pgfqpoint{1.404205in}{0.391500in}}%
\pgfpathlineto{\pgfqpoint{1.435909in}{0.391500in}}%
\pgfpathlineto{\pgfqpoint{1.467614in}{0.391500in}}%
\pgfpathlineto{\pgfqpoint{1.499318in}{0.391500in}}%
\pgfpathlineto{\pgfqpoint{1.531023in}{0.391500in}}%
\pgfpathlineto{\pgfqpoint{1.562727in}{0.391500in}}%
\pgfpathlineto{\pgfqpoint{1.594432in}{0.391500in}}%
\pgfpathlineto{\pgfqpoint{1.626136in}{0.391500in}}%
\pgfpathlineto{\pgfqpoint{1.657841in}{0.391500in}}%
\pgfpathlineto{\pgfqpoint{1.689545in}{0.391500in}}%
\pgfpathlineto{\pgfqpoint{1.721250in}{0.391500in}}%
\pgfpathlineto{\pgfqpoint{1.752955in}{0.391500in}}%
\pgfpathlineto{\pgfqpoint{1.784659in}{0.391500in}}%
\pgfpathlineto{\pgfqpoint{1.816364in}{0.391500in}}%
\pgfpathlineto{\pgfqpoint{1.848068in}{0.391500in}}%
\pgfpathlineto{\pgfqpoint{1.879773in}{0.391500in}}%
\pgfpathlineto{\pgfqpoint{1.911477in}{0.391500in}}%
\pgfpathlineto{\pgfqpoint{1.943182in}{0.391500in}}%
\pgfpathlineto{\pgfqpoint{1.974886in}{0.391500in}}%
\pgfpathlineto{\pgfqpoint{2.006591in}{0.391500in}}%
\pgfpathlineto{\pgfqpoint{2.038295in}{0.391500in}}%
\pgfpathlineto{\pgfqpoint{2.070000in}{0.391500in}}%
\pgfpathlineto{\pgfqpoint{2.101705in}{0.391500in}}%
\pgfpathlineto{\pgfqpoint{2.133409in}{0.391500in}}%
\pgfpathlineto{\pgfqpoint{2.165114in}{0.391500in}}%
\pgfpathlineto{\pgfqpoint{2.196818in}{0.391500in}}%
\pgfpathlineto{\pgfqpoint{2.228523in}{0.391500in}}%
\pgfpathlineto{\pgfqpoint{2.260227in}{0.391500in}}%
\pgfpathlineto{\pgfqpoint{2.291932in}{0.391500in}}%
\pgfpathlineto{\pgfqpoint{2.323636in}{0.391500in}}%
\pgfpathlineto{\pgfqpoint{2.355341in}{0.391500in}}%
\pgfpathlineto{\pgfqpoint{2.387045in}{0.391500in}}%
\pgfpathlineto{\pgfqpoint{2.418750in}{0.391500in}}%
\pgfpathlineto{\pgfqpoint{2.450455in}{0.391500in}}%
\pgfpathlineto{\pgfqpoint{2.482159in}{0.391500in}}%
\pgfpathlineto{\pgfqpoint{2.513864in}{0.391500in}}%
\pgfpathlineto{\pgfqpoint{2.545568in}{0.391500in}}%
\pgfpathlineto{\pgfqpoint{2.577273in}{0.391500in}}%
\pgfpathlineto{\pgfqpoint{2.608977in}{0.391500in}}%
\pgfpathlineto{\pgfqpoint{2.640682in}{0.391500in}}%
\pgfpathlineto{\pgfqpoint{2.672386in}{0.391500in}}%
\pgfpathlineto{\pgfqpoint{2.704091in}{0.391500in}}%
\pgfpathlineto{\pgfqpoint{2.735795in}{0.391500in}}%
\pgfpathlineto{\pgfqpoint{2.767500in}{0.391500in}}%
\pgfpathlineto{\pgfqpoint{2.799205in}{0.391500in}}%
\pgfpathlineto{\pgfqpoint{2.830909in}{0.391500in}}%
\pgfpathlineto{\pgfqpoint{2.862614in}{0.391500in}}%
\pgfpathlineto{\pgfqpoint{2.894318in}{0.391500in}}%
\pgfpathlineto{\pgfqpoint{2.926023in}{0.391500in}}%
\pgfpathlineto{\pgfqpoint{2.957727in}{0.391500in}}%
\pgfpathlineto{\pgfqpoint{2.989432in}{0.391500in}}%
\pgfpathlineto{\pgfqpoint{3.021136in}{0.391500in}}%
\pgfpathlineto{\pgfqpoint{3.052841in}{0.391500in}}%
\pgfpathlineto{\pgfqpoint{3.084545in}{0.391500in}}%
\pgfpathlineto{\pgfqpoint{3.116250in}{0.391500in}}%
\pgfpathlineto{\pgfqpoint{3.147955in}{0.391500in}}%
\pgfpathlineto{\pgfqpoint{3.179659in}{0.391500in}}%
\pgfpathlineto{\pgfqpoint{3.211364in}{0.391500in}}%
\pgfpathlineto{\pgfqpoint{3.243068in}{0.391500in}}%
\pgfpathlineto{\pgfqpoint{3.274773in}{0.391500in}}%
\pgfpathlineto{\pgfqpoint{3.306477in}{0.391500in}}%
\pgfpathlineto{\pgfqpoint{3.338182in}{0.391500in}}%
\pgfpathlineto{\pgfqpoint{3.369886in}{0.391500in}}%
\pgfpathlineto{\pgfqpoint{3.401591in}{0.391500in}}%
\pgfpathlineto{\pgfqpoint{3.433295in}{0.391500in}}%
\pgfpathlineto{\pgfqpoint{3.465000in}{0.391500in}}%
\pgfpathlineto{\pgfqpoint{3.496705in}{0.391500in}}%
\pgfpathlineto{\pgfqpoint{3.528409in}{0.391500in}}%
\pgfpathlineto{\pgfqpoint{3.560114in}{0.391500in}}%
\pgfpathlineto{\pgfqpoint{3.591818in}{0.391500in}}%
\pgfpathlineto{\pgfqpoint{3.623523in}{0.391500in}}%
\pgfpathlineto{\pgfqpoint{3.655227in}{0.391500in}}%
\pgfpathlineto{\pgfqpoint{3.686932in}{0.391500in}}%
\pgfpathlineto{\pgfqpoint{3.718636in}{0.391500in}}%
\pgfpathlineto{\pgfqpoint{3.750341in}{0.391500in}}%
\pgfpathlineto{\pgfqpoint{3.782045in}{0.391500in}}%
\pgfpathlineto{\pgfqpoint{3.813750in}{0.391500in}}%
\pgfpathlineto{\pgfqpoint{3.845455in}{0.391500in}}%
\pgfpathlineto{\pgfqpoint{3.877159in}{0.391500in}}%
\pgfpathlineto{\pgfqpoint{3.908864in}{0.391500in}}%
\pgfpathlineto{\pgfqpoint{3.940568in}{0.391500in}}%
\pgfpathlineto{\pgfqpoint{3.972273in}{0.391500in}}%
\pgfpathlineto{\pgfqpoint{4.003977in}{0.391500in}}%
\pgfpathlineto{\pgfqpoint{4.035682in}{0.391500in}}%
\pgfpathlineto{\pgfqpoint{4.067386in}{0.391500in}}%
\pgfpathlineto{\pgfqpoint{4.099091in}{0.391500in}}%
\pgfpathlineto{\pgfqpoint{4.130795in}{0.391500in}}%
\pgfpathlineto{\pgfqpoint{4.162500in}{0.391500in}}%
\pgfpathlineto{\pgfqpoint{4.194205in}{0.391500in}}%
\pgfpathlineto{\pgfqpoint{4.225909in}{0.391500in}}%
\pgfpathlineto{\pgfqpoint{4.257614in}{0.391500in}}%
\pgfpathlineto{\pgfqpoint{4.289318in}{0.391500in}}%
\pgfpathlineto{\pgfqpoint{4.321023in}{0.391500in}}%
\pgfpathlineto{\pgfqpoint{4.352727in}{0.391500in}}%
\pgfpathlineto{\pgfqpoint{4.384432in}{0.391500in}}%
\pgfpathlineto{\pgfqpoint{4.416136in}{0.391500in}}%
\pgfpathlineto{\pgfqpoint{4.447841in}{0.391500in}}%
\pgfpathlineto{\pgfqpoint{4.479545in}{0.391500in}}%
\pgfpathlineto{\pgfqpoint{4.511250in}{0.391500in}}%
\pgfpathlineto{\pgfqpoint{4.542955in}{0.391500in}}%
\pgfpathlineto{\pgfqpoint{4.574659in}{0.391500in}}%
\pgfpathlineto{\pgfqpoint{4.606364in}{0.391500in}}%
\pgfpathlineto{\pgfqpoint{4.638068in}{0.391500in}}%
\pgfpathlineto{\pgfqpoint{4.669773in}{0.391500in}}%
\pgfusepath{stroke}%
\end{pgfscope}%
\begin{pgfscope}%
\pgfsetrectcap%
\pgfsetmiterjoin%
\pgfsetlinewidth{0.803000pt}%
\definecolor{currentstroke}{rgb}{0.000000,0.000000,0.000000}%
\pgfsetstrokecolor{currentstroke}%
\pgfsetdash{}{0pt}%
\pgfpathmoveto{\pgfqpoint{0.675000in}{0.297000in}}%
\pgfpathlineto{\pgfqpoint{0.675000in}{2.376000in}}%
\pgfusepath{stroke}%
\end{pgfscope}%
\begin{pgfscope}%
\pgfsetrectcap%
\pgfsetmiterjoin%
\pgfsetlinewidth{0.803000pt}%
\definecolor{currentstroke}{rgb}{0.000000,0.000000,0.000000}%
\pgfsetstrokecolor{currentstroke}%
\pgfsetdash{}{0pt}%
\pgfpathmoveto{\pgfqpoint{4.860000in}{0.297000in}}%
\pgfpathlineto{\pgfqpoint{4.860000in}{2.376000in}}%
\pgfusepath{stroke}%
\end{pgfscope}%
\begin{pgfscope}%
\pgfsetrectcap%
\pgfsetmiterjoin%
\pgfsetlinewidth{0.803000pt}%
\definecolor{currentstroke}{rgb}{0.000000,0.000000,0.000000}%
\pgfsetstrokecolor{currentstroke}%
\pgfsetdash{}{0pt}%
\pgfpathmoveto{\pgfqpoint{0.675000in}{0.297000in}}%
\pgfpathlineto{\pgfqpoint{4.860000in}{0.297000in}}%
\pgfusepath{stroke}%
\end{pgfscope}%
\begin{pgfscope}%
\pgfsetrectcap%
\pgfsetmiterjoin%
\pgfsetlinewidth{0.803000pt}%
\definecolor{currentstroke}{rgb}{0.000000,0.000000,0.000000}%
\pgfsetstrokecolor{currentstroke}%
\pgfsetdash{}{0pt}%
\pgfpathmoveto{\pgfqpoint{0.675000in}{2.376000in}}%
\pgfpathlineto{\pgfqpoint{4.860000in}{2.376000in}}%
\pgfusepath{stroke}%
\end{pgfscope}%
\begin{pgfscope}%
\pgfsetbuttcap%
\pgfsetmiterjoin%
\pgfsetlinewidth{0.000000pt}%
\definecolor{currentstroke}{rgb}{0.800000,0.800000,0.800000}%
\pgfsetstrokecolor{currentstroke}%
\pgfsetstrokeopacity{0.000000}%
\pgfsetdash{}{0pt}%
\pgfpathmoveto{\pgfqpoint{3.635350in}{0.922790in}}%
\pgfpathlineto{\pgfqpoint{4.762778in}{0.922790in}}%
\pgfpathquadraticcurveto{\pgfqpoint{4.790556in}{0.922790in}}{\pgfqpoint{4.790556in}{0.950568in}}%
\pgfpathlineto{\pgfqpoint{4.790556in}{1.722432in}}%
\pgfpathquadraticcurveto{\pgfqpoint{4.790556in}{1.750210in}}{\pgfqpoint{4.762778in}{1.750210in}}%
\pgfpathlineto{\pgfqpoint{3.635350in}{1.750210in}}%
\pgfpathquadraticcurveto{\pgfqpoint{3.607572in}{1.750210in}}{\pgfqpoint{3.607572in}{1.722432in}}%
\pgfpathlineto{\pgfqpoint{3.607572in}{0.950568in}}%
\pgfpathquadraticcurveto{\pgfqpoint{3.607572in}{0.922790in}}{\pgfqpoint{3.635350in}{0.922790in}}%
\pgfpathclose%
\pgfusepath{}%
\end{pgfscope}%
\begin{pgfscope}%
\pgfsetrectcap%
\pgfsetroundjoin%
\pgfsetlinewidth{1.003750pt}%
\definecolor{currentstroke}{rgb}{0.121569,0.466667,0.705882}%
\pgfsetstrokecolor{currentstroke}%
\pgfsetdash{}{0pt}%
\pgfpathmoveto{\pgfqpoint{3.663128in}{1.646043in}}%
\pgfpathlineto{\pgfqpoint{3.940906in}{1.646043in}}%
\pgfusepath{stroke}%
\end{pgfscope}%
\begin{pgfscope}%
\pgftext[x=4.052017in,y=1.597432in,left,base]{\rmfamily\fontsize{10.000000}{12.000000}\selectfont Slush Pool}%
\end{pgfscope}%
\begin{pgfscope}%
\pgfsetrectcap%
\pgfsetroundjoin%
\pgfsetlinewidth{1.003750pt}%
\definecolor{currentstroke}{rgb}{1.000000,0.498039,0.054902}%
\pgfsetstrokecolor{currentstroke}%
\pgfsetdash{}{0pt}%
\pgfpathmoveto{\pgfqpoint{3.663128in}{1.449673in}}%
\pgfpathlineto{\pgfqpoint{3.940906in}{1.449673in}}%
\pgfusepath{stroke}%
\end{pgfscope}%
\begin{pgfscope}%
\pgftext[x=4.052017in,y=1.401061in,left,base]{\rmfamily\fontsize{10.000000}{12.000000}\selectfont ViaBTC}%
\end{pgfscope}%
\begin{pgfscope}%
\pgfsetrectcap%
\pgfsetroundjoin%
\pgfsetlinewidth{1.003750pt}%
\definecolor{currentstroke}{rgb}{0.172549,0.627451,0.172549}%
\pgfsetstrokecolor{currentstroke}%
\pgfsetdash{}{0pt}%
\pgfpathmoveto{\pgfqpoint{3.663128in}{1.253302in}}%
\pgfpathlineto{\pgfqpoint{3.940906in}{1.253302in}}%
\pgfusepath{stroke}%
\end{pgfscope}%
\begin{pgfscope}%
\pgftext[x=4.052017in,y=1.204691in,left,base]{\rmfamily\fontsize{10.000000}{12.000000}\selectfont KanoPool}%
\end{pgfscope}%
\begin{pgfscope}%
\pgfsetrectcap%
\pgfsetroundjoin%
\pgfsetlinewidth{1.003750pt}%
\definecolor{currentstroke}{rgb}{0.839216,0.152941,0.156863}%
\pgfsetstrokecolor{currentstroke}%
\pgfsetdash{}{0pt}%
\pgfpathmoveto{\pgfqpoint{3.663128in}{1.056932in}}%
\pgfpathlineto{\pgfqpoint{3.940906in}{1.056932in}}%
\pgfusepath{stroke}%
\end{pgfscope}%
\begin{pgfscope}%
\pgftext[x=4.052017in,y=1.008321in,left,base]{\rmfamily\fontsize{10.000000}{12.000000}\selectfont Solo mining}%
\end{pgfscope}%
\end{pgfpicture}%
\makeatother%
\endgroup%

%% file: single_crypto_small.pgf
\begingroup%
\makeatletter%
\begin{pgfpicture}%
\pgfpathrectangle{\pgfpointorigin}{\pgfqpoint{5.400000in}{2.700000in}}%
\pgfusepath{use as bounding box, clip}%
\begin{pgfscope}%
\pgfsetbuttcap%
\pgfsetmiterjoin%
\definecolor{currentfill}{rgb}{1.000000,1.000000,1.000000}%
\pgfsetfillcolor{currentfill}%
\pgfsetlinewidth{0.000000pt}%
\definecolor{currentstroke}{rgb}{1.000000,1.000000,1.000000}%
\pgfsetstrokecolor{currentstroke}%
\pgfsetdash{}{0pt}%
\pgfpathmoveto{\pgfqpoint{0.000000in}{0.000000in}}%
\pgfpathlineto{\pgfqpoint{5.400000in}{0.000000in}}%
\pgfpathlineto{\pgfqpoint{5.400000in}{2.700000in}}%
\pgfpathlineto{\pgfqpoint{0.000000in}{2.700000in}}%
\pgfpathclose%
\pgfusepath{fill}%
\end{pgfscope}%
\begin{pgfscope}%
\pgfsetbuttcap%
\pgfsetmiterjoin%
\definecolor{currentfill}{rgb}{1.000000,1.000000,1.000000}%
\pgfsetfillcolor{currentfill}%
\pgfsetlinewidth{0.000000pt}%
\definecolor{currentstroke}{rgb}{0.000000,0.000000,0.000000}%
\pgfsetstrokecolor{currentstroke}%
\pgfsetstrokeopacity{0.000000}%
\pgfsetdash{}{0pt}%
\pgfpathmoveto{\pgfqpoint{0.675000in}{0.297000in}}%
\pgfpathlineto{\pgfqpoint{4.860000in}{0.297000in}}%
\pgfpathlineto{\pgfqpoint{4.860000in}{2.376000in}}%
\pgfpathlineto{\pgfqpoint{0.675000in}{2.376000in}}%
\pgfpathclose%
\pgfusepath{fill}%
\end{pgfscope}%
\begin{pgfscope}%
\pgfsetbuttcap%
\pgfsetroundjoin%
\definecolor{currentfill}{rgb}{0.000000,0.000000,0.000000}%
\pgfsetfillcolor{currentfill}%
\pgfsetlinewidth{0.803000pt}%
\definecolor{currentstroke}{rgb}{0.000000,0.000000,0.000000}%
\pgfsetstrokecolor{currentstroke}%
\pgfsetdash{}{0pt}%
\pgfsys@defobject{currentmarker}{\pgfqpoint{0.000000in}{0.000000in}}{\pgfqpoint{0.000000in}{0.048611in}}{%
\pgfpathmoveto{\pgfqpoint{0.000000in}{0.000000in}}%
\pgfpathlineto{\pgfqpoint{0.000000in}{0.048611in}}%
\pgfusepath{stroke,fill}%
}%
\begin{pgfscope}%
\pgfsys@transformshift{1.279584in}{0.297000in}%
\pgfsys@useobject{currentmarker}{}%
\end{pgfscope}%
\end{pgfscope}%
\begin{pgfscope}%
\pgftext[x=1.279584in,y=0.269222in,,top]{\rmfamily\fontsize{10.000000}{12.000000}\selectfont \(\displaystyle 0.00002\)}%
\end{pgfscope}%
\begin{pgfscope}%
\pgfsetbuttcap%
\pgfsetroundjoin%
\definecolor{currentfill}{rgb}{0.000000,0.000000,0.000000}%
\pgfsetfillcolor{currentfill}%
\pgfsetlinewidth{0.803000pt}%
\definecolor{currentstroke}{rgb}{0.000000,0.000000,0.000000}%
\pgfsetstrokecolor{currentstroke}%
\pgfsetdash{}{0pt}%
\pgfsys@defobject{currentmarker}{\pgfqpoint{0.000000in}{0.000000in}}{\pgfqpoint{0.000000in}{0.048611in}}{%
\pgfpathmoveto{\pgfqpoint{0.000000in}{0.000000in}}%
\pgfpathlineto{\pgfqpoint{0.000000in}{0.048611in}}%
\pgfusepath{stroke,fill}%
}%
\begin{pgfscope}%
\pgfsys@transformshift{2.032959in}{0.297000in}%
\pgfsys@useobject{currentmarker}{}%
\end{pgfscope}%
\end{pgfscope}%
\begin{pgfscope}%
\pgftext[x=2.032959in,y=0.269222in,,top]{\rmfamily\fontsize{10.000000}{12.000000}\selectfont \(\displaystyle 0.00004\)}%
\end{pgfscope}%
\begin{pgfscope}%
\pgfsetbuttcap%
\pgfsetroundjoin%
\definecolor{currentfill}{rgb}{0.000000,0.000000,0.000000}%
\pgfsetfillcolor{currentfill}%
\pgfsetlinewidth{0.803000pt}%
\definecolor{currentstroke}{rgb}{0.000000,0.000000,0.000000}%
\pgfsetstrokecolor{currentstroke}%
\pgfsetdash{}{0pt}%
\pgfsys@defobject{currentmarker}{\pgfqpoint{0.000000in}{0.000000in}}{\pgfqpoint{0.000000in}{0.048611in}}{%
\pgfpathmoveto{\pgfqpoint{0.000000in}{0.000000in}}%
\pgfpathlineto{\pgfqpoint{0.000000in}{0.048611in}}%
\pgfusepath{stroke,fill}%
}%
\begin{pgfscope}%
\pgfsys@transformshift{2.786334in}{0.297000in}%
\pgfsys@useobject{currentmarker}{}%
\end{pgfscope}%
\end{pgfscope}%
\begin{pgfscope}%
\pgftext[x=2.786334in,y=0.269222in,,top]{\rmfamily\fontsize{10.000000}{12.000000}\selectfont \(\displaystyle 0.00006\)}%
\end{pgfscope}%
\begin{pgfscope}%
\pgfsetbuttcap%
\pgfsetroundjoin%
\definecolor{currentfill}{rgb}{0.000000,0.000000,0.000000}%
\pgfsetfillcolor{currentfill}%
\pgfsetlinewidth{0.803000pt}%
\definecolor{currentstroke}{rgb}{0.000000,0.000000,0.000000}%
\pgfsetstrokecolor{currentstroke}%
\pgfsetdash{}{0pt}%
\pgfsys@defobject{currentmarker}{\pgfqpoint{0.000000in}{0.000000in}}{\pgfqpoint{0.000000in}{0.048611in}}{%
\pgfpathmoveto{\pgfqpoint{0.000000in}{0.000000in}}%
\pgfpathlineto{\pgfqpoint{0.000000in}{0.048611in}}%
\pgfusepath{stroke,fill}%
}%
\begin{pgfscope}%
\pgfsys@transformshift{3.539710in}{0.297000in}%
\pgfsys@useobject{currentmarker}{}%
\end{pgfscope}%
\end{pgfscope}%
\begin{pgfscope}%
\pgftext[x=3.539710in,y=0.269222in,,top]{\rmfamily\fontsize{10.000000}{12.000000}\selectfont \(\displaystyle 0.00008\)}%
\end{pgfscope}%
\begin{pgfscope}%
\pgfsetbuttcap%
\pgfsetroundjoin%
\definecolor{currentfill}{rgb}{0.000000,0.000000,0.000000}%
\pgfsetfillcolor{currentfill}%
\pgfsetlinewidth{0.803000pt}%
\definecolor{currentstroke}{rgb}{0.000000,0.000000,0.000000}%
\pgfsetstrokecolor{currentstroke}%
\pgfsetdash{}{0pt}%
\pgfsys@defobject{currentmarker}{\pgfqpoint{0.000000in}{0.000000in}}{\pgfqpoint{0.000000in}{0.048611in}}{%
\pgfpathmoveto{\pgfqpoint{0.000000in}{0.000000in}}%
\pgfpathlineto{\pgfqpoint{0.000000in}{0.048611in}}%
\pgfusepath{stroke,fill}%
}%
\begin{pgfscope}%
\pgfsys@transformshift{4.293085in}{0.297000in}%
\pgfsys@useobject{currentmarker}{}%
\end{pgfscope}%
\end{pgfscope}%
\begin{pgfscope}%
\pgftext[x=4.293085in,y=0.269222in,,top]{\rmfamily\fontsize{10.000000}{12.000000}\selectfont \(\displaystyle 0.00010\)}%
\end{pgfscope}%
\begin{pgfscope}%
\pgftext[x=2.767500in,y=0.156972in,,top]{\rmfamily\fontsize{10.000000}{12.000000}\selectfont CARA}%
\end{pgfscope}%
\begin{pgfscope}%
\pgfsetbuttcap%
\pgfsetroundjoin%
\definecolor{currentfill}{rgb}{0.000000,0.000000,0.000000}%
\pgfsetfillcolor{currentfill}%
\pgfsetlinewidth{0.803000pt}%
\definecolor{currentstroke}{rgb}{0.000000,0.000000,0.000000}%
\pgfsetstrokecolor{currentstroke}%
\pgfsetdash{}{0pt}%
\pgfsys@defobject{currentmarker}{\pgfqpoint{-0.048611in}{0.000000in}}{\pgfqpoint{0.000000in}{0.000000in}}{%
\pgfpathmoveto{\pgfqpoint{0.000000in}{0.000000in}}%
\pgfpathlineto{\pgfqpoint{-0.048611in}{0.000000in}}%
\pgfusepath{stroke,fill}%
}%
\begin{pgfscope}%
\pgfsys@transformshift{0.675000in}{0.391500in}%
\pgfsys@useobject{currentmarker}{}%
\end{pgfscope}%
\end{pgfscope}%
\begin{pgfscope}%
\pgftext[x=0.400308in,y=0.343282in,left,base]{\rmfamily\fontsize{10.000000}{12.000000}\selectfont \(\displaystyle 0.0\)}%
\end{pgfscope}%
\begin{pgfscope}%
\pgfsetbuttcap%
\pgfsetroundjoin%
\definecolor{currentfill}{rgb}{0.000000,0.000000,0.000000}%
\pgfsetfillcolor{currentfill}%
\pgfsetlinewidth{0.803000pt}%
\definecolor{currentstroke}{rgb}{0.000000,0.000000,0.000000}%
\pgfsetstrokecolor{currentstroke}%
\pgfsetdash{}{0pt}%
\pgfsys@defobject{currentmarker}{\pgfqpoint{-0.048611in}{0.000000in}}{\pgfqpoint{0.000000in}{0.000000in}}{%
\pgfpathmoveto{\pgfqpoint{0.000000in}{0.000000in}}%
\pgfpathlineto{\pgfqpoint{-0.048611in}{0.000000in}}%
\pgfusepath{stroke,fill}%
}%
\begin{pgfscope}%
\pgfsys@transformshift{0.675000in}{0.693900in}%
\pgfsys@useobject{currentmarker}{}%
\end{pgfscope}%
\end{pgfscope}%
\begin{pgfscope}%
\pgftext[x=0.400308in,y=0.645682in,left,base]{\rmfamily\fontsize{10.000000}{12.000000}\selectfont \(\displaystyle 0.2\)}%
\end{pgfscope}%
\begin{pgfscope}%
\pgfsetbuttcap%
\pgfsetroundjoin%
\definecolor{currentfill}{rgb}{0.000000,0.000000,0.000000}%
\pgfsetfillcolor{currentfill}%
\pgfsetlinewidth{0.803000pt}%
\definecolor{currentstroke}{rgb}{0.000000,0.000000,0.000000}%
\pgfsetstrokecolor{currentstroke}%
\pgfsetdash{}{0pt}%
\pgfsys@defobject{currentmarker}{\pgfqpoint{-0.048611in}{0.000000in}}{\pgfqpoint{0.000000in}{0.000000in}}{%
\pgfpathmoveto{\pgfqpoint{0.000000in}{0.000000in}}%
\pgfpathlineto{\pgfqpoint{-0.048611in}{0.000000in}}%
\pgfusepath{stroke,fill}%
}%
\begin{pgfscope}%
\pgfsys@transformshift{0.675000in}{0.996300in}%
\pgfsys@useobject{currentmarker}{}%
\end{pgfscope}%
\end{pgfscope}%
\begin{pgfscope}%
\pgftext[x=0.400308in,y=0.948082in,left,base]{\rmfamily\fontsize{10.000000}{12.000000}\selectfont \(\displaystyle 0.4\)}%
\end{pgfscope}%
\begin{pgfscope}%
\pgfsetbuttcap%
\pgfsetroundjoin%
\definecolor{currentfill}{rgb}{0.000000,0.000000,0.000000}%
\pgfsetfillcolor{currentfill}%
\pgfsetlinewidth{0.803000pt}%
\definecolor{currentstroke}{rgb}{0.000000,0.000000,0.000000}%
\pgfsetstrokecolor{currentstroke}%
\pgfsetdash{}{0pt}%
\pgfsys@defobject{currentmarker}{\pgfqpoint{-0.048611in}{0.000000in}}{\pgfqpoint{0.000000in}{0.000000in}}{%
\pgfpathmoveto{\pgfqpoint{0.000000in}{0.000000in}}%
\pgfpathlineto{\pgfqpoint{-0.048611in}{0.000000in}}%
\pgfusepath{stroke,fill}%
}%
\begin{pgfscope}%
\pgfsys@transformshift{0.675000in}{1.298700in}%
\pgfsys@useobject{currentmarker}{}%
\end{pgfscope}%
\end{pgfscope}%
\begin{pgfscope}%
\pgftext[x=0.400308in,y=1.250482in,left,base]{\rmfamily\fontsize{10.000000}{12.000000}\selectfont \(\displaystyle 0.6\)}%
\end{pgfscope}%
\begin{pgfscope}%
\pgfsetbuttcap%
\pgfsetroundjoin%
\definecolor{currentfill}{rgb}{0.000000,0.000000,0.000000}%
\pgfsetfillcolor{currentfill}%
\pgfsetlinewidth{0.803000pt}%
\definecolor{currentstroke}{rgb}{0.000000,0.000000,0.000000}%
\pgfsetstrokecolor{currentstroke}%
\pgfsetdash{}{0pt}%
\pgfsys@defobject{currentmarker}{\pgfqpoint{-0.048611in}{0.000000in}}{\pgfqpoint{0.000000in}{0.000000in}}{%
\pgfpathmoveto{\pgfqpoint{0.000000in}{0.000000in}}%
\pgfpathlineto{\pgfqpoint{-0.048611in}{0.000000in}}%
\pgfusepath{stroke,fill}%
}%
\begin{pgfscope}%
\pgfsys@transformshift{0.675000in}{1.601100in}%
\pgfsys@useobject{currentmarker}{}%
\end{pgfscope}%
\end{pgfscope}%
\begin{pgfscope}%
\pgftext[x=0.400308in,y=1.552882in,left,base]{\rmfamily\fontsize{10.000000}{12.000000}\selectfont \(\displaystyle 0.8\)}%
\end{pgfscope}%
\begin{pgfscope}%
\pgfsetbuttcap%
\pgfsetroundjoin%
\definecolor{currentfill}{rgb}{0.000000,0.000000,0.000000}%
\pgfsetfillcolor{currentfill}%
\pgfsetlinewidth{0.803000pt}%
\definecolor{currentstroke}{rgb}{0.000000,0.000000,0.000000}%
\pgfsetstrokecolor{currentstroke}%
\pgfsetdash{}{0pt}%
\pgfsys@defobject{currentmarker}{\pgfqpoint{-0.048611in}{0.000000in}}{\pgfqpoint{0.000000in}{0.000000in}}{%
\pgfpathmoveto{\pgfqpoint{0.000000in}{0.000000in}}%
\pgfpathlineto{\pgfqpoint{-0.048611in}{0.000000in}}%
\pgfusepath{stroke,fill}%
}%
\begin{pgfscope}%
\pgfsys@transformshift{0.675000in}{1.903500in}%
\pgfsys@useobject{currentmarker}{}%
\end{pgfscope}%
\end{pgfscope}%
\begin{pgfscope}%
\pgftext[x=0.400308in,y=1.855282in,left,base]{\rmfamily\fontsize{10.000000}{12.000000}\selectfont \(\displaystyle 1.0\)}%
\end{pgfscope}%
\begin{pgfscope}%
\pgfsetbuttcap%
\pgfsetroundjoin%
\definecolor{currentfill}{rgb}{0.000000,0.000000,0.000000}%
\pgfsetfillcolor{currentfill}%
\pgfsetlinewidth{0.803000pt}%
\definecolor{currentstroke}{rgb}{0.000000,0.000000,0.000000}%
\pgfsetstrokecolor{currentstroke}%
\pgfsetdash{}{0pt}%
\pgfsys@defobject{currentmarker}{\pgfqpoint{-0.048611in}{0.000000in}}{\pgfqpoint{0.000000in}{0.000000in}}{%
\pgfpathmoveto{\pgfqpoint{0.000000in}{0.000000in}}%
\pgfpathlineto{\pgfqpoint{-0.048611in}{0.000000in}}%
\pgfusepath{stroke,fill}%
}%
\begin{pgfscope}%
\pgfsys@transformshift{0.675000in}{2.205900in}%
\pgfsys@useobject{currentmarker}{}%
\end{pgfscope}%
\end{pgfscope}%
\begin{pgfscope}%
\pgftext[x=0.400308in,y=2.157682in,left,base]{\rmfamily\fontsize{10.000000}{12.000000}\selectfont \(\displaystyle 1.2\)}%
\end{pgfscope}%
\begin{pgfscope}%
\pgftext[x=0.344753in,y=1.336500in,,bottom,rotate=90.000000]{\rmfamily\fontsize{10.000000}{12.000000}\selectfont Hash rate}%
\end{pgfscope}%
\begin{pgfscope}%
\pgftext[x=0.675000in,y=2.417667in,left,base]{\rmfamily\fontsize{10.000000}{12.000000}\selectfont \(\displaystyle \times10^{14}\)}%
\end{pgfscope}%
\begin{pgfscope}%
\pgfpathrectangle{\pgfqpoint{0.675000in}{0.297000in}}{\pgfqpoint{4.185000in}{2.079000in}}%
\pgfusepath{clip}%
\pgfsetrectcap%
\pgfsetroundjoin%
\pgfsetlinewidth{1.003750pt}%
\definecolor{currentstroke}{rgb}{0.121569,0.466667,0.705882}%
\pgfsetstrokecolor{currentstroke}%
\pgfsetdash{}{0pt}%
\pgfpathmoveto{\pgfqpoint{0.865227in}{0.391500in}}%
\pgfpathlineto{\pgfqpoint{0.896932in}{0.391500in}}%
\pgfpathlineto{\pgfqpoint{0.928636in}{0.391500in}}%
\pgfpathlineto{\pgfqpoint{0.960341in}{0.391500in}}%
\pgfpathlineto{\pgfqpoint{0.992045in}{0.391500in}}%
\pgfpathlineto{\pgfqpoint{1.023750in}{0.391500in}}%
\pgfpathlineto{\pgfqpoint{1.055455in}{0.391500in}}%
\pgfpathlineto{\pgfqpoint{1.087159in}{0.391500in}}%
\pgfpathlineto{\pgfqpoint{1.118864in}{0.391500in}}%
\pgfpathlineto{\pgfqpoint{1.150568in}{0.391500in}}%
\pgfpathlineto{\pgfqpoint{1.182273in}{0.391500in}}%
\pgfpathlineto{\pgfqpoint{1.213977in}{0.391500in}}%
\pgfpathlineto{\pgfqpoint{1.245682in}{0.391500in}}%
\pgfpathlineto{\pgfqpoint{1.277386in}{0.391500in}}%
\pgfpathlineto{\pgfqpoint{1.309091in}{0.391500in}}%
\pgfpathlineto{\pgfqpoint{1.340795in}{0.391500in}}%
\pgfpathlineto{\pgfqpoint{1.372500in}{0.391500in}}%
\pgfpathlineto{\pgfqpoint{1.404205in}{0.391500in}}%
\pgfpathlineto{\pgfqpoint{1.435909in}{0.391500in}}%
\pgfpathlineto{\pgfqpoint{1.467614in}{0.391500in}}%
\pgfpathlineto{\pgfqpoint{1.499318in}{0.391500in}}%
\pgfpathlineto{\pgfqpoint{1.531023in}{0.391500in}}%
\pgfpathlineto{\pgfqpoint{1.562727in}{0.391500in}}%
\pgfpathlineto{\pgfqpoint{1.594432in}{0.391500in}}%
\pgfpathlineto{\pgfqpoint{1.626136in}{0.391500in}}%
\pgfpathlineto{\pgfqpoint{1.657841in}{0.391500in}}%
\pgfpathlineto{\pgfqpoint{1.689545in}{0.391500in}}%
\pgfpathlineto{\pgfqpoint{1.721250in}{0.391500in}}%
\pgfpathlineto{\pgfqpoint{1.752955in}{0.391500in}}%
\pgfpathlineto{\pgfqpoint{1.784659in}{0.391500in}}%
\pgfpathlineto{\pgfqpoint{1.816364in}{0.391500in}}%
\pgfpathlineto{\pgfqpoint{1.848068in}{0.391500in}}%
\pgfpathlineto{\pgfqpoint{1.879773in}{0.391500in}}%
\pgfpathlineto{\pgfqpoint{1.911477in}{0.391500in}}%
\pgfpathlineto{\pgfqpoint{1.943182in}{0.391500in}}%
\pgfpathlineto{\pgfqpoint{1.974886in}{0.391500in}}%
\pgfpathlineto{\pgfqpoint{2.006591in}{0.391500in}}%
\pgfpathlineto{\pgfqpoint{2.038295in}{0.391500in}}%
\pgfpathlineto{\pgfqpoint{2.070000in}{0.391500in}}%
\pgfpathlineto{\pgfqpoint{2.101705in}{0.391500in}}%
\pgfpathlineto{\pgfqpoint{2.133409in}{0.391500in}}%
\pgfpathlineto{\pgfqpoint{2.165114in}{0.391500in}}%
\pgfpathlineto{\pgfqpoint{2.196818in}{0.391500in}}%
\pgfpathlineto{\pgfqpoint{2.228523in}{0.391500in}}%
\pgfpathlineto{\pgfqpoint{2.260227in}{0.391500in}}%
\pgfpathlineto{\pgfqpoint{2.291932in}{0.391500in}}%
\pgfpathlineto{\pgfqpoint{2.323636in}{0.391500in}}%
\pgfpathlineto{\pgfqpoint{2.355341in}{0.391500in}}%
\pgfpathlineto{\pgfqpoint{2.387045in}{0.391500in}}%
\pgfpathlineto{\pgfqpoint{2.418750in}{0.391500in}}%
\pgfpathlineto{\pgfqpoint{2.450455in}{0.391500in}}%
\pgfpathlineto{\pgfqpoint{2.482159in}{0.391500in}}%
\pgfpathlineto{\pgfqpoint{2.513864in}{0.391500in}}%
\pgfpathlineto{\pgfqpoint{2.545568in}{0.391500in}}%
\pgfpathlineto{\pgfqpoint{2.577273in}{0.391500in}}%
\pgfpathlineto{\pgfqpoint{2.608977in}{0.391500in}}%
\pgfpathlineto{\pgfqpoint{2.640682in}{0.391500in}}%
\pgfpathlineto{\pgfqpoint{2.672386in}{0.391500in}}%
\pgfpathlineto{\pgfqpoint{2.704091in}{0.391500in}}%
\pgfpathlineto{\pgfqpoint{2.735795in}{0.391500in}}%
\pgfpathlineto{\pgfqpoint{2.767500in}{0.391500in}}%
\pgfpathlineto{\pgfqpoint{2.799205in}{0.391500in}}%
\pgfpathlineto{\pgfqpoint{2.830909in}{0.391500in}}%
\pgfpathlineto{\pgfqpoint{2.862614in}{0.391500in}}%
\pgfpathlineto{\pgfqpoint{2.894318in}{0.391500in}}%
\pgfpathlineto{\pgfqpoint{2.926023in}{0.391500in}}%
\pgfpathlineto{\pgfqpoint{2.957727in}{0.391500in}}%
\pgfpathlineto{\pgfqpoint{2.989432in}{0.391500in}}%
\pgfpathlineto{\pgfqpoint{3.021136in}{0.391500in}}%
\pgfpathlineto{\pgfqpoint{3.052841in}{0.391500in}}%
\pgfpathlineto{\pgfqpoint{3.084545in}{0.391500in}}%
\pgfpathlineto{\pgfqpoint{3.116250in}{0.391500in}}%
\pgfpathlineto{\pgfqpoint{3.147955in}{0.391500in}}%
\pgfpathlineto{\pgfqpoint{3.179659in}{0.391500in}}%
\pgfpathlineto{\pgfqpoint{3.211364in}{0.391500in}}%
\pgfpathlineto{\pgfqpoint{3.243068in}{0.391500in}}%
\pgfpathlineto{\pgfqpoint{3.274773in}{0.391500in}}%
\pgfpathlineto{\pgfqpoint{3.306477in}{0.391500in}}%
\pgfpathlineto{\pgfqpoint{3.338182in}{0.391500in}}%
\pgfpathlineto{\pgfqpoint{3.369886in}{0.391500in}}%
\pgfpathlineto{\pgfqpoint{3.401591in}{0.391500in}}%
\pgfpathlineto{\pgfqpoint{3.433295in}{0.391500in}}%
\pgfpathlineto{\pgfqpoint{3.465000in}{0.391500in}}%
\pgfpathlineto{\pgfqpoint{3.496705in}{0.391500in}}%
\pgfpathlineto{\pgfqpoint{3.528409in}{0.391500in}}%
\pgfpathlineto{\pgfqpoint{3.560114in}{0.391500in}}%
\pgfpathlineto{\pgfqpoint{3.591818in}{0.391500in}}%
\pgfpathlineto{\pgfqpoint{3.623523in}{0.391500in}}%
\pgfpathlineto{\pgfqpoint{3.655227in}{0.391500in}}%
\pgfpathlineto{\pgfqpoint{3.686932in}{0.391500in}}%
\pgfpathlineto{\pgfqpoint{3.718636in}{0.391500in}}%
\pgfpathlineto{\pgfqpoint{3.750341in}{0.391500in}}%
\pgfpathlineto{\pgfqpoint{3.782045in}{0.391500in}}%
\pgfpathlineto{\pgfqpoint{3.813750in}{0.391500in}}%
\pgfpathlineto{\pgfqpoint{3.845455in}{0.391500in}}%
\pgfpathlineto{\pgfqpoint{3.877159in}{0.391500in}}%
\pgfpathlineto{\pgfqpoint{3.908864in}{0.391500in}}%
\pgfpathlineto{\pgfqpoint{3.940568in}{0.391500in}}%
\pgfpathlineto{\pgfqpoint{3.972273in}{0.391500in}}%
\pgfpathlineto{\pgfqpoint{4.003977in}{0.391500in}}%
\pgfpathlineto{\pgfqpoint{4.035682in}{0.391500in}}%
\pgfpathlineto{\pgfqpoint{4.067386in}{0.391500in}}%
\pgfpathlineto{\pgfqpoint{4.099091in}{0.391500in}}%
\pgfpathlineto{\pgfqpoint{4.130795in}{0.393520in}}%
\pgfpathlineto{\pgfqpoint{4.162500in}{0.403266in}}%
\pgfpathlineto{\pgfqpoint{4.194205in}{0.410722in}}%
\pgfpathlineto{\pgfqpoint{4.225909in}{0.421086in}}%
\pgfpathlineto{\pgfqpoint{4.257614in}{0.429831in}}%
\pgfpathlineto{\pgfqpoint{4.289318in}{0.437005in}}%
\pgfpathlineto{\pgfqpoint{4.321023in}{0.445668in}}%
\pgfpathlineto{\pgfqpoint{4.352727in}{0.455476in}}%
\pgfpathlineto{\pgfqpoint{4.384432in}{0.463909in}}%
\pgfpathlineto{\pgfqpoint{4.416136in}{0.470850in}}%
\pgfpathlineto{\pgfqpoint{4.447841in}{0.478665in}}%
\pgfpathlineto{\pgfqpoint{4.479545in}{0.487550in}}%
\pgfpathlineto{\pgfqpoint{4.511250in}{0.494147in}}%
\pgfpathlineto{\pgfqpoint{4.542955in}{0.501871in}}%
\pgfpathlineto{\pgfqpoint{4.574659in}{0.509149in}}%
\pgfpathlineto{\pgfqpoint{4.606364in}{0.516792in}}%
\pgfpathlineto{\pgfqpoint{4.638068in}{0.523680in}}%
\pgfpathlineto{\pgfqpoint{4.669773in}{0.532150in}}%
\pgfusepath{stroke}%
\end{pgfscope}%
\begin{pgfscope}%
\pgfpathrectangle{\pgfqpoint{0.675000in}{0.297000in}}{\pgfqpoint{4.185000in}{2.079000in}}%
\pgfusepath{clip}%
\pgfsetrectcap%
\pgfsetroundjoin%
\pgfsetlinewidth{1.003750pt}%
\definecolor{currentstroke}{rgb}{1.000000,0.498039,0.054902}%
\pgfsetstrokecolor{currentstroke}%
\pgfsetdash{}{0pt}%
\pgfpathmoveto{\pgfqpoint{0.865227in}{0.391500in}}%
\pgfpathlineto{\pgfqpoint{0.896932in}{0.391500in}}%
\pgfpathlineto{\pgfqpoint{0.928636in}{0.391500in}}%
\pgfpathlineto{\pgfqpoint{0.960341in}{0.391500in}}%
\pgfpathlineto{\pgfqpoint{0.992045in}{0.391500in}}%
\pgfpathlineto{\pgfqpoint{1.023750in}{0.391500in}}%
\pgfpathlineto{\pgfqpoint{1.055455in}{0.391500in}}%
\pgfpathlineto{\pgfqpoint{1.087159in}{0.391500in}}%
\pgfpathlineto{\pgfqpoint{1.118864in}{0.391500in}}%
\pgfpathlineto{\pgfqpoint{1.150568in}{0.391500in}}%
\pgfpathlineto{\pgfqpoint{1.182273in}{0.391500in}}%
\pgfpathlineto{\pgfqpoint{1.213977in}{0.391500in}}%
\pgfpathlineto{\pgfqpoint{1.245682in}{0.391500in}}%
\pgfpathlineto{\pgfqpoint{1.277386in}{0.391500in}}%
\pgfpathlineto{\pgfqpoint{1.309091in}{0.391500in}}%
\pgfpathlineto{\pgfqpoint{1.340795in}{0.391500in}}%
\pgfpathlineto{\pgfqpoint{1.372500in}{0.391500in}}%
\pgfpathlineto{\pgfqpoint{1.404205in}{0.391500in}}%
\pgfpathlineto{\pgfqpoint{1.435909in}{0.391500in}}%
\pgfpathlineto{\pgfqpoint{1.467614in}{0.391500in}}%
\pgfpathlineto{\pgfqpoint{1.499318in}{0.391500in}}%
\pgfpathlineto{\pgfqpoint{1.531023in}{0.391500in}}%
\pgfpathlineto{\pgfqpoint{1.562727in}{0.391500in}}%
\pgfpathlineto{\pgfqpoint{1.594432in}{0.391500in}}%
\pgfpathlineto{\pgfqpoint{1.626136in}{0.391500in}}%
\pgfpathlineto{\pgfqpoint{1.657841in}{0.391500in}}%
\pgfpathlineto{\pgfqpoint{1.689545in}{0.391500in}}%
\pgfpathlineto{\pgfqpoint{1.721250in}{0.391500in}}%
\pgfpathlineto{\pgfqpoint{1.752955in}{0.391500in}}%
\pgfpathlineto{\pgfqpoint{1.784659in}{0.391500in}}%
\pgfpathlineto{\pgfqpoint{1.816364in}{0.391500in}}%
\pgfpathlineto{\pgfqpoint{1.848068in}{0.391500in}}%
\pgfpathlineto{\pgfqpoint{1.879773in}{0.391500in}}%
\pgfpathlineto{\pgfqpoint{1.911477in}{0.391500in}}%
\pgfpathlineto{\pgfqpoint{1.943182in}{0.391500in}}%
\pgfpathlineto{\pgfqpoint{1.974886in}{0.391500in}}%
\pgfpathlineto{\pgfqpoint{2.006591in}{0.391500in}}%
\pgfpathlineto{\pgfqpoint{2.038295in}{0.391500in}}%
\pgfpathlineto{\pgfqpoint{2.070000in}{0.391500in}}%
\pgfpathlineto{\pgfqpoint{2.101705in}{0.391500in}}%
\pgfpathlineto{\pgfqpoint{2.133409in}{0.391500in}}%
\pgfpathlineto{\pgfqpoint{2.165114in}{0.391500in}}%
\pgfpathlineto{\pgfqpoint{2.196818in}{0.391500in}}%
\pgfpathlineto{\pgfqpoint{2.228523in}{0.391500in}}%
\pgfpathlineto{\pgfqpoint{2.260227in}{0.391500in}}%
\pgfpathlineto{\pgfqpoint{2.291932in}{0.391500in}}%
\pgfpathlineto{\pgfqpoint{2.323636in}{0.391500in}}%
\pgfpathlineto{\pgfqpoint{2.355341in}{0.391500in}}%
\pgfpathlineto{\pgfqpoint{2.387045in}{0.391500in}}%
\pgfpathlineto{\pgfqpoint{2.418750in}{0.391500in}}%
\pgfpathlineto{\pgfqpoint{2.450455in}{0.391500in}}%
\pgfpathlineto{\pgfqpoint{2.482159in}{0.391500in}}%
\pgfpathlineto{\pgfqpoint{2.513864in}{0.391500in}}%
\pgfpathlineto{\pgfqpoint{2.545568in}{0.391500in}}%
\pgfpathlineto{\pgfqpoint{2.577273in}{0.391500in}}%
\pgfpathlineto{\pgfqpoint{2.608977in}{0.391500in}}%
\pgfpathlineto{\pgfqpoint{2.640682in}{0.391500in}}%
\pgfpathlineto{\pgfqpoint{2.672386in}{0.391500in}}%
\pgfpathlineto{\pgfqpoint{2.704091in}{0.391500in}}%
\pgfpathlineto{\pgfqpoint{2.735795in}{0.391500in}}%
\pgfpathlineto{\pgfqpoint{2.767500in}{0.391500in}}%
\pgfpathlineto{\pgfqpoint{2.799205in}{0.391500in}}%
\pgfpathlineto{\pgfqpoint{2.830909in}{0.391500in}}%
\pgfpathlineto{\pgfqpoint{2.862614in}{0.391500in}}%
\pgfpathlineto{\pgfqpoint{2.894318in}{0.391500in}}%
\pgfpathlineto{\pgfqpoint{2.926023in}{0.391500in}}%
\pgfpathlineto{\pgfqpoint{2.957727in}{0.391500in}}%
\pgfpathlineto{\pgfqpoint{2.989432in}{0.391500in}}%
\pgfpathlineto{\pgfqpoint{3.021136in}{0.391500in}}%
\pgfpathlineto{\pgfqpoint{3.052841in}{0.391500in}}%
\pgfpathlineto{\pgfqpoint{3.084545in}{0.391500in}}%
\pgfpathlineto{\pgfqpoint{3.116250in}{0.391500in}}%
\pgfpathlineto{\pgfqpoint{3.147955in}{0.391500in}}%
\pgfpathlineto{\pgfqpoint{3.179659in}{0.391500in}}%
\pgfpathlineto{\pgfqpoint{3.211364in}{0.391500in}}%
\pgfpathlineto{\pgfqpoint{3.243068in}{0.391500in}}%
\pgfpathlineto{\pgfqpoint{3.274773in}{0.391500in}}%
\pgfpathlineto{\pgfqpoint{3.306477in}{0.391500in}}%
\pgfpathlineto{\pgfqpoint{3.338182in}{0.391500in}}%
\pgfpathlineto{\pgfqpoint{3.369886in}{0.391500in}}%
\pgfpathlineto{\pgfqpoint{3.401591in}{0.391500in}}%
\pgfpathlineto{\pgfqpoint{3.433295in}{0.391500in}}%
\pgfpathlineto{\pgfqpoint{3.465000in}{0.391500in}}%
\pgfpathlineto{\pgfqpoint{3.496705in}{0.391500in}}%
\pgfpathlineto{\pgfqpoint{3.528409in}{0.391500in}}%
\pgfpathlineto{\pgfqpoint{3.560114in}{0.391500in}}%
\pgfpathlineto{\pgfqpoint{3.591818in}{0.391500in}}%
\pgfpathlineto{\pgfqpoint{3.623523in}{0.391500in}}%
\pgfpathlineto{\pgfqpoint{3.655227in}{0.391500in}}%
\pgfpathlineto{\pgfqpoint{3.686932in}{0.391500in}}%
\pgfpathlineto{\pgfqpoint{3.718636in}{0.391500in}}%
\pgfpathlineto{\pgfqpoint{3.750341in}{0.391500in}}%
\pgfpathlineto{\pgfqpoint{3.782045in}{0.391500in}}%
\pgfpathlineto{\pgfqpoint{3.813750in}{0.391500in}}%
\pgfpathlineto{\pgfqpoint{3.845455in}{0.391500in}}%
\pgfpathlineto{\pgfqpoint{3.877159in}{0.391500in}}%
\pgfpathlineto{\pgfqpoint{3.908864in}{0.391500in}}%
\pgfpathlineto{\pgfqpoint{3.940568in}{0.391500in}}%
\pgfpathlineto{\pgfqpoint{3.972273in}{0.391500in}}%
\pgfpathlineto{\pgfqpoint{4.003977in}{0.391500in}}%
\pgfpathlineto{\pgfqpoint{4.035682in}{0.391500in}}%
\pgfpathlineto{\pgfqpoint{4.067386in}{0.391500in}}%
\pgfpathlineto{\pgfqpoint{4.099091in}{0.391500in}}%
\pgfpathlineto{\pgfqpoint{4.130795in}{0.392202in}}%
\pgfpathlineto{\pgfqpoint{4.162500in}{0.398871in}}%
\pgfpathlineto{\pgfqpoint{4.194205in}{0.407527in}}%
\pgfpathlineto{\pgfqpoint{4.225909in}{0.413024in}}%
\pgfpathlineto{\pgfqpoint{4.257614in}{0.419837in}}%
\pgfpathlineto{\pgfqpoint{4.289318in}{0.428002in}}%
\pgfpathlineto{\pgfqpoint{4.321023in}{0.434380in}}%
\pgfpathlineto{\pgfqpoint{4.352727in}{0.439373in}}%
\pgfpathlineto{\pgfqpoint{4.384432in}{0.445505in}}%
\pgfpathlineto{\pgfqpoint{4.416136in}{0.452882in}}%
\pgfpathlineto{\pgfqpoint{4.447841in}{0.459160in}}%
\pgfpathlineto{\pgfqpoint{4.479545in}{0.464135in}}%
\pgfpathlineto{\pgfqpoint{4.511250in}{0.471179in}}%
\pgfpathlineto{\pgfqpoint{4.542955in}{0.476883in}}%
\pgfpathlineto{\pgfqpoint{4.574659in}{0.482826in}}%
\pgfpathlineto{\pgfqpoint{4.606364in}{0.488193in}}%
\pgfpathlineto{\pgfqpoint{4.638068in}{0.494097in}}%
\pgfpathlineto{\pgfqpoint{4.669773in}{0.498263in}}%
\pgfusepath{stroke}%
\end{pgfscope}%
\begin{pgfscope}%
\pgfpathrectangle{\pgfqpoint{0.675000in}{0.297000in}}{\pgfqpoint{4.185000in}{2.079000in}}%
\pgfusepath{clip}%
\pgfsetrectcap%
\pgfsetroundjoin%
\pgfsetlinewidth{1.003750pt}%
\definecolor{currentstroke}{rgb}{0.172549,0.627451,0.172549}%
\pgfsetstrokecolor{currentstroke}%
\pgfsetdash{}{0pt}%
\pgfpathmoveto{\pgfqpoint{0.865227in}{2.281500in}}%
\pgfpathlineto{\pgfqpoint{0.896932in}{2.281500in}}%
\pgfpathlineto{\pgfqpoint{0.928636in}{2.281500in}}%
\pgfpathlineto{\pgfqpoint{0.960341in}{2.281500in}}%
\pgfpathlineto{\pgfqpoint{0.992045in}{2.281500in}}%
\pgfpathlineto{\pgfqpoint{1.023750in}{2.281500in}}%
\pgfpathlineto{\pgfqpoint{1.055455in}{2.281500in}}%
\pgfpathlineto{\pgfqpoint{1.087159in}{2.281500in}}%
\pgfpathlineto{\pgfqpoint{1.118864in}{2.281500in}}%
\pgfpathlineto{\pgfqpoint{1.150568in}{2.281500in}}%
\pgfpathlineto{\pgfqpoint{1.182273in}{2.281500in}}%
\pgfpathlineto{\pgfqpoint{1.213977in}{2.281500in}}%
\pgfpathlineto{\pgfqpoint{1.245682in}{2.281500in}}%
\pgfpathlineto{\pgfqpoint{1.277386in}{2.281500in}}%
\pgfpathlineto{\pgfqpoint{1.309091in}{2.281500in}}%
\pgfpathlineto{\pgfqpoint{1.340795in}{2.281500in}}%
\pgfpathlineto{\pgfqpoint{1.372500in}{2.281500in}}%
\pgfpathlineto{\pgfqpoint{1.404205in}{2.281500in}}%
\pgfpathlineto{\pgfqpoint{1.435909in}{2.281500in}}%
\pgfpathlineto{\pgfqpoint{1.467614in}{2.281500in}}%
\pgfpathlineto{\pgfqpoint{1.499318in}{2.281500in}}%
\pgfpathlineto{\pgfqpoint{1.531023in}{2.281500in}}%
\pgfpathlineto{\pgfqpoint{1.562727in}{2.281500in}}%
\pgfpathlineto{\pgfqpoint{1.594432in}{2.281500in}}%
\pgfpathlineto{\pgfqpoint{1.626136in}{2.281500in}}%
\pgfpathlineto{\pgfqpoint{1.657841in}{2.281500in}}%
\pgfpathlineto{\pgfqpoint{1.689545in}{2.281500in}}%
\pgfpathlineto{\pgfqpoint{1.721250in}{2.281500in}}%
\pgfpathlineto{\pgfqpoint{1.752955in}{2.281500in}}%
\pgfpathlineto{\pgfqpoint{1.784659in}{2.281500in}}%
\pgfpathlineto{\pgfqpoint{1.816364in}{2.281500in}}%
\pgfpathlineto{\pgfqpoint{1.848068in}{2.281500in}}%
\pgfpathlineto{\pgfqpoint{1.879773in}{2.281500in}}%
\pgfpathlineto{\pgfqpoint{1.911477in}{2.281500in}}%
\pgfpathlineto{\pgfqpoint{1.943182in}{2.281500in}}%
\pgfpathlineto{\pgfqpoint{1.974886in}{2.281500in}}%
\pgfpathlineto{\pgfqpoint{2.006591in}{2.281500in}}%
\pgfpathlineto{\pgfqpoint{2.038295in}{2.281500in}}%
\pgfpathlineto{\pgfqpoint{2.070000in}{2.281500in}}%
\pgfpathlineto{\pgfqpoint{2.101705in}{2.281500in}}%
\pgfpathlineto{\pgfqpoint{2.133409in}{2.281500in}}%
\pgfpathlineto{\pgfqpoint{2.165114in}{2.281500in}}%
\pgfpathlineto{\pgfqpoint{2.196818in}{2.281500in}}%
\pgfpathlineto{\pgfqpoint{2.228523in}{2.281500in}}%
\pgfpathlineto{\pgfqpoint{2.260227in}{2.281500in}}%
\pgfpathlineto{\pgfqpoint{2.291932in}{2.281500in}}%
\pgfpathlineto{\pgfqpoint{2.323636in}{2.281500in}}%
\pgfpathlineto{\pgfqpoint{2.355341in}{2.281500in}}%
\pgfpathlineto{\pgfqpoint{2.387045in}{2.281500in}}%
\pgfpathlineto{\pgfqpoint{2.418750in}{2.281500in}}%
\pgfpathlineto{\pgfqpoint{2.450455in}{2.281500in}}%
\pgfpathlineto{\pgfqpoint{2.482159in}{2.281500in}}%
\pgfpathlineto{\pgfqpoint{2.513864in}{2.281500in}}%
\pgfpathlineto{\pgfqpoint{2.545568in}{2.281500in}}%
\pgfpathlineto{\pgfqpoint{2.577273in}{2.281500in}}%
\pgfpathlineto{\pgfqpoint{2.608977in}{2.281500in}}%
\pgfpathlineto{\pgfqpoint{2.640682in}{2.281500in}}%
\pgfpathlineto{\pgfqpoint{2.672386in}{2.281500in}}%
\pgfpathlineto{\pgfqpoint{2.704091in}{2.281500in}}%
\pgfpathlineto{\pgfqpoint{2.735795in}{2.281500in}}%
\pgfpathlineto{\pgfqpoint{2.767500in}{2.281500in}}%
\pgfpathlineto{\pgfqpoint{2.799205in}{2.281500in}}%
\pgfpathlineto{\pgfqpoint{2.830909in}{2.281500in}}%
\pgfpathlineto{\pgfqpoint{2.862614in}{2.281500in}}%
\pgfpathlineto{\pgfqpoint{2.894318in}{2.281500in}}%
\pgfpathlineto{\pgfqpoint{2.926023in}{2.281500in}}%
\pgfpathlineto{\pgfqpoint{2.957727in}{2.281500in}}%
\pgfpathlineto{\pgfqpoint{2.989432in}{2.281500in}}%
\pgfpathlineto{\pgfqpoint{3.021136in}{2.281500in}}%
\pgfpathlineto{\pgfqpoint{3.052841in}{2.281500in}}%
\pgfpathlineto{\pgfqpoint{3.084545in}{2.281500in}}%
\pgfpathlineto{\pgfqpoint{3.116250in}{2.281500in}}%
\pgfpathlineto{\pgfqpoint{3.147955in}{2.281500in}}%
\pgfpathlineto{\pgfqpoint{3.179659in}{2.281500in}}%
\pgfpathlineto{\pgfqpoint{3.211364in}{2.281500in}}%
\pgfpathlineto{\pgfqpoint{3.243068in}{2.281500in}}%
\pgfpathlineto{\pgfqpoint{3.274773in}{2.281500in}}%
\pgfpathlineto{\pgfqpoint{3.306477in}{2.281500in}}%
\pgfpathlineto{\pgfqpoint{3.338182in}{2.281500in}}%
\pgfpathlineto{\pgfqpoint{3.369886in}{2.281500in}}%
\pgfpathlineto{\pgfqpoint{3.401591in}{2.281500in}}%
\pgfpathlineto{\pgfqpoint{3.433295in}{2.281500in}}%
\pgfpathlineto{\pgfqpoint{3.465000in}{2.281500in}}%
\pgfpathlineto{\pgfqpoint{3.496705in}{2.281500in}}%
\pgfpathlineto{\pgfqpoint{3.528409in}{2.281500in}}%
\pgfpathlineto{\pgfqpoint{3.560114in}{2.281500in}}%
\pgfpathlineto{\pgfqpoint{3.591818in}{2.281500in}}%
\pgfpathlineto{\pgfqpoint{3.623523in}{2.281500in}}%
\pgfpathlineto{\pgfqpoint{3.655227in}{2.281500in}}%
\pgfpathlineto{\pgfqpoint{3.686932in}{2.281500in}}%
\pgfpathlineto{\pgfqpoint{3.718636in}{2.281500in}}%
\pgfpathlineto{\pgfqpoint{3.750341in}{2.281500in}}%
\pgfpathlineto{\pgfqpoint{3.782045in}{2.281500in}}%
\pgfpathlineto{\pgfqpoint{3.813750in}{2.281500in}}%
\pgfpathlineto{\pgfqpoint{3.845455in}{2.281500in}}%
\pgfpathlineto{\pgfqpoint{3.877159in}{2.281500in}}%
\pgfpathlineto{\pgfqpoint{3.908864in}{2.281500in}}%
\pgfpathlineto{\pgfqpoint{3.940568in}{2.281500in}}%
\pgfpathlineto{\pgfqpoint{3.972273in}{2.281500in}}%
\pgfpathlineto{\pgfqpoint{4.003977in}{2.281500in}}%
\pgfpathlineto{\pgfqpoint{4.035682in}{2.281500in}}%
\pgfpathlineto{\pgfqpoint{4.067386in}{2.281500in}}%
\pgfpathlineto{\pgfqpoint{4.099091in}{2.281500in}}%
\pgfpathlineto{\pgfqpoint{4.130795in}{2.278777in}}%
\pgfpathlineto{\pgfqpoint{4.162500in}{2.262363in}}%
\pgfpathlineto{\pgfqpoint{4.194205in}{2.246251in}}%
\pgfpathlineto{\pgfqpoint{4.225909in}{2.230390in}}%
\pgfpathlineto{\pgfqpoint{4.257614in}{2.214833in}}%
\pgfpathlineto{\pgfqpoint{4.289318in}{2.199493in}}%
\pgfpathlineto{\pgfqpoint{4.321023in}{2.184452in}}%
\pgfpathlineto{\pgfqpoint{4.352727in}{2.169651in}}%
\pgfpathlineto{\pgfqpoint{4.384432in}{2.155086in}}%
\pgfpathlineto{\pgfqpoint{4.416136in}{2.140768in}}%
\pgfpathlineto{\pgfqpoint{4.447841in}{2.126675in}}%
\pgfpathlineto{\pgfqpoint{4.479545in}{2.112815in}}%
\pgfpathlineto{\pgfqpoint{4.511250in}{2.099174in}}%
\pgfpathlineto{\pgfqpoint{4.542955in}{2.085746in}}%
\pgfpathlineto{\pgfqpoint{4.574659in}{2.072525in}}%
\pgfpathlineto{\pgfqpoint{4.606364in}{2.059515in}}%
\pgfpathlineto{\pgfqpoint{4.638068in}{2.046723in}}%
\pgfpathlineto{\pgfqpoint{4.669773in}{2.034087in}}%
\pgfusepath{stroke}%
\end{pgfscope}%
\begin{pgfscope}%
\pgfpathrectangle{\pgfqpoint{0.675000in}{0.297000in}}{\pgfqpoint{4.185000in}{2.079000in}}%
\pgfusepath{clip}%
\pgfsetrectcap%
\pgfsetroundjoin%
\pgfsetlinewidth{1.003750pt}%
\definecolor{currentstroke}{rgb}{0.839216,0.152941,0.156863}%
\pgfsetstrokecolor{currentstroke}%
\pgfsetdash{}{0pt}%
\pgfpathmoveto{\pgfqpoint{0.865227in}{0.391500in}}%
\pgfpathlineto{\pgfqpoint{0.896932in}{0.391500in}}%
\pgfpathlineto{\pgfqpoint{0.928636in}{0.391500in}}%
\pgfpathlineto{\pgfqpoint{0.960341in}{0.391500in}}%
\pgfpathlineto{\pgfqpoint{0.992045in}{0.391500in}}%
\pgfpathlineto{\pgfqpoint{1.023750in}{0.391500in}}%
\pgfpathlineto{\pgfqpoint{1.055455in}{0.391500in}}%
\pgfpathlineto{\pgfqpoint{1.087159in}{0.391500in}}%
\pgfpathlineto{\pgfqpoint{1.118864in}{0.391500in}}%
\pgfpathlineto{\pgfqpoint{1.150568in}{0.391500in}}%
\pgfpathlineto{\pgfqpoint{1.182273in}{0.391500in}}%
\pgfpathlineto{\pgfqpoint{1.213977in}{0.391500in}}%
\pgfpathlineto{\pgfqpoint{1.245682in}{0.391500in}}%
\pgfpathlineto{\pgfqpoint{1.277386in}{0.391500in}}%
\pgfpathlineto{\pgfqpoint{1.309091in}{0.391500in}}%
\pgfpathlineto{\pgfqpoint{1.340795in}{0.391500in}}%
\pgfpathlineto{\pgfqpoint{1.372500in}{0.391500in}}%
\pgfpathlineto{\pgfqpoint{1.404205in}{0.391500in}}%
\pgfpathlineto{\pgfqpoint{1.435909in}{0.391500in}}%
\pgfpathlineto{\pgfqpoint{1.467614in}{0.391500in}}%
\pgfpathlineto{\pgfqpoint{1.499318in}{0.391500in}}%
\pgfpathlineto{\pgfqpoint{1.531023in}{0.391500in}}%
\pgfpathlineto{\pgfqpoint{1.562727in}{0.391500in}}%
\pgfpathlineto{\pgfqpoint{1.594432in}{0.391500in}}%
\pgfpathlineto{\pgfqpoint{1.626136in}{0.391500in}}%
\pgfpathlineto{\pgfqpoint{1.657841in}{0.391500in}}%
\pgfpathlineto{\pgfqpoint{1.689545in}{0.391500in}}%
\pgfpathlineto{\pgfqpoint{1.721250in}{0.391500in}}%
\pgfpathlineto{\pgfqpoint{1.752955in}{0.391500in}}%
\pgfpathlineto{\pgfqpoint{1.784659in}{0.391500in}}%
\pgfpathlineto{\pgfqpoint{1.816364in}{0.391500in}}%
\pgfpathlineto{\pgfqpoint{1.848068in}{0.391500in}}%
\pgfpathlineto{\pgfqpoint{1.879773in}{0.391500in}}%
\pgfpathlineto{\pgfqpoint{1.911477in}{0.391500in}}%
\pgfpathlineto{\pgfqpoint{1.943182in}{0.391500in}}%
\pgfpathlineto{\pgfqpoint{1.974886in}{0.391500in}}%
\pgfpathlineto{\pgfqpoint{2.006591in}{0.391500in}}%
\pgfpathlineto{\pgfqpoint{2.038295in}{0.391500in}}%
\pgfpathlineto{\pgfqpoint{2.070000in}{0.391500in}}%
\pgfpathlineto{\pgfqpoint{2.101705in}{0.391500in}}%
\pgfpathlineto{\pgfqpoint{2.133409in}{0.391500in}}%
\pgfpathlineto{\pgfqpoint{2.165114in}{0.391500in}}%
\pgfpathlineto{\pgfqpoint{2.196818in}{0.391500in}}%
\pgfpathlineto{\pgfqpoint{2.228523in}{0.391500in}}%
\pgfpathlineto{\pgfqpoint{2.260227in}{0.391500in}}%
\pgfpathlineto{\pgfqpoint{2.291932in}{0.391500in}}%
\pgfpathlineto{\pgfqpoint{2.323636in}{0.391500in}}%
\pgfpathlineto{\pgfqpoint{2.355341in}{0.391500in}}%
\pgfpathlineto{\pgfqpoint{2.387045in}{0.391500in}}%
\pgfpathlineto{\pgfqpoint{2.418750in}{0.391500in}}%
\pgfpathlineto{\pgfqpoint{2.450455in}{0.391500in}}%
\pgfpathlineto{\pgfqpoint{2.482159in}{0.391500in}}%
\pgfpathlineto{\pgfqpoint{2.513864in}{0.391500in}}%
\pgfpathlineto{\pgfqpoint{2.545568in}{0.391500in}}%
\pgfpathlineto{\pgfqpoint{2.577273in}{0.391500in}}%
\pgfpathlineto{\pgfqpoint{2.608977in}{0.391500in}}%
\pgfpathlineto{\pgfqpoint{2.640682in}{0.391500in}}%
\pgfpathlineto{\pgfqpoint{2.672386in}{0.391500in}}%
\pgfpathlineto{\pgfqpoint{2.704091in}{0.391500in}}%
\pgfpathlineto{\pgfqpoint{2.735795in}{0.391500in}}%
\pgfpathlineto{\pgfqpoint{2.767500in}{0.391500in}}%
\pgfpathlineto{\pgfqpoint{2.799205in}{0.391500in}}%
\pgfpathlineto{\pgfqpoint{2.830909in}{0.391500in}}%
\pgfpathlineto{\pgfqpoint{2.862614in}{0.391500in}}%
\pgfpathlineto{\pgfqpoint{2.894318in}{0.391500in}}%
\pgfpathlineto{\pgfqpoint{2.926023in}{0.391500in}}%
\pgfpathlineto{\pgfqpoint{2.957727in}{0.391500in}}%
\pgfpathlineto{\pgfqpoint{2.989432in}{0.391500in}}%
\pgfpathlineto{\pgfqpoint{3.021136in}{0.391500in}}%
\pgfpathlineto{\pgfqpoint{3.052841in}{0.391500in}}%
\pgfpathlineto{\pgfqpoint{3.084545in}{0.391500in}}%
\pgfpathlineto{\pgfqpoint{3.116250in}{0.391500in}}%
\pgfpathlineto{\pgfqpoint{3.147955in}{0.391500in}}%
\pgfpathlineto{\pgfqpoint{3.179659in}{0.391500in}}%
\pgfpathlineto{\pgfqpoint{3.211364in}{0.391500in}}%
\pgfpathlineto{\pgfqpoint{3.243068in}{0.391500in}}%
\pgfpathlineto{\pgfqpoint{3.274773in}{0.391500in}}%
\pgfpathlineto{\pgfqpoint{3.306477in}{0.391500in}}%
\pgfpathlineto{\pgfqpoint{3.338182in}{0.391500in}}%
\pgfpathlineto{\pgfqpoint{3.369886in}{0.391500in}}%
\pgfpathlineto{\pgfqpoint{3.401591in}{0.391500in}}%
\pgfpathlineto{\pgfqpoint{3.433295in}{0.391500in}}%
\pgfpathlineto{\pgfqpoint{3.465000in}{0.391500in}}%
\pgfpathlineto{\pgfqpoint{3.496705in}{0.391500in}}%
\pgfpathlineto{\pgfqpoint{3.528409in}{0.391500in}}%
\pgfpathlineto{\pgfqpoint{3.560114in}{0.391500in}}%
\pgfpathlineto{\pgfqpoint{3.591818in}{0.391500in}}%
\pgfpathlineto{\pgfqpoint{3.623523in}{0.391500in}}%
\pgfpathlineto{\pgfqpoint{3.655227in}{0.391500in}}%
\pgfpathlineto{\pgfqpoint{3.686932in}{0.391500in}}%
\pgfpathlineto{\pgfqpoint{3.718636in}{0.391500in}}%
\pgfpathlineto{\pgfqpoint{3.750341in}{0.391500in}}%
\pgfpathlineto{\pgfqpoint{3.782045in}{0.391500in}}%
\pgfpathlineto{\pgfqpoint{3.813750in}{0.391500in}}%
\pgfpathlineto{\pgfqpoint{3.845455in}{0.391500in}}%
\pgfpathlineto{\pgfqpoint{3.877159in}{0.391500in}}%
\pgfpathlineto{\pgfqpoint{3.908864in}{0.391500in}}%
\pgfpathlineto{\pgfqpoint{3.940568in}{0.391500in}}%
\pgfpathlineto{\pgfqpoint{3.972273in}{0.391500in}}%
\pgfpathlineto{\pgfqpoint{4.003977in}{0.391500in}}%
\pgfpathlineto{\pgfqpoint{4.035682in}{0.391500in}}%
\pgfpathlineto{\pgfqpoint{4.067386in}{0.391500in}}%
\pgfpathlineto{\pgfqpoint{4.099091in}{0.391500in}}%
\pgfpathlineto{\pgfqpoint{4.130795in}{0.391500in}}%
\pgfpathlineto{\pgfqpoint{4.162500in}{0.391500in}}%
\pgfpathlineto{\pgfqpoint{4.194205in}{0.391500in}}%
\pgfpathlineto{\pgfqpoint{4.225909in}{0.391500in}}%
\pgfpathlineto{\pgfqpoint{4.257614in}{0.391500in}}%
\pgfpathlineto{\pgfqpoint{4.289318in}{0.391500in}}%
\pgfpathlineto{\pgfqpoint{4.321023in}{0.391500in}}%
\pgfpathlineto{\pgfqpoint{4.352727in}{0.391500in}}%
\pgfpathlineto{\pgfqpoint{4.384432in}{0.391500in}}%
\pgfpathlineto{\pgfqpoint{4.416136in}{0.391500in}}%
\pgfpathlineto{\pgfqpoint{4.447841in}{0.391500in}}%
\pgfpathlineto{\pgfqpoint{4.479545in}{0.391500in}}%
\pgfpathlineto{\pgfqpoint{4.511250in}{0.391500in}}%
\pgfpathlineto{\pgfqpoint{4.542955in}{0.391500in}}%
\pgfpathlineto{\pgfqpoint{4.574659in}{0.391500in}}%
\pgfpathlineto{\pgfqpoint{4.606364in}{0.391500in}}%
\pgfpathlineto{\pgfqpoint{4.638068in}{0.391500in}}%
\pgfpathlineto{\pgfqpoint{4.669773in}{0.391500in}}%
\pgfusepath{stroke}%
\end{pgfscope}%
\begin{pgfscope}%
\pgfsetrectcap%
\pgfsetmiterjoin%
\pgfsetlinewidth{0.803000pt}%
\definecolor{currentstroke}{rgb}{0.000000,0.000000,0.000000}%
\pgfsetstrokecolor{currentstroke}%
\pgfsetdash{}{0pt}%
\pgfpathmoveto{\pgfqpoint{0.675000in}{0.297000in}}%
\pgfpathlineto{\pgfqpoint{0.675000in}{2.376000in}}%
\pgfusepath{stroke}%
\end{pgfscope}%
\begin{pgfscope}%
\pgfsetrectcap%
\pgfsetmiterjoin%
\pgfsetlinewidth{0.803000pt}%
\definecolor{currentstroke}{rgb}{0.000000,0.000000,0.000000}%
\pgfsetstrokecolor{currentstroke}%
\pgfsetdash{}{0pt}%
\pgfpathmoveto{\pgfqpoint{4.860000in}{0.297000in}}%
\pgfpathlineto{\pgfqpoint{4.860000in}{2.376000in}}%
\pgfusepath{stroke}%
\end{pgfscope}%
\begin{pgfscope}%
\pgfsetrectcap%
\pgfsetmiterjoin%
\pgfsetlinewidth{0.803000pt}%
\definecolor{currentstroke}{rgb}{0.000000,0.000000,0.000000}%
\pgfsetstrokecolor{currentstroke}%
\pgfsetdash{}{0pt}%
\pgfpathmoveto{\pgfqpoint{0.675000in}{0.297000in}}%
\pgfpathlineto{\pgfqpoint{4.860000in}{0.297000in}}%
\pgfusepath{stroke}%
\end{pgfscope}%
\begin{pgfscope}%
\pgfsetrectcap%
\pgfsetmiterjoin%
\pgfsetlinewidth{0.803000pt}%
\definecolor{currentstroke}{rgb}{0.000000,0.000000,0.000000}%
\pgfsetstrokecolor{currentstroke}%
\pgfsetdash{}{0pt}%
\pgfpathmoveto{\pgfqpoint{0.675000in}{2.376000in}}%
\pgfpathlineto{\pgfqpoint{4.860000in}{2.376000in}}%
\pgfusepath{stroke}%
\end{pgfscope}%
\begin{pgfscope}%
\pgfsetbuttcap%
\pgfsetmiterjoin%
\pgfsetlinewidth{0.000000pt}%
\definecolor{currentstroke}{rgb}{0.800000,0.800000,0.800000}%
\pgfsetstrokecolor{currentstroke}%
\pgfsetstrokeopacity{0.000000}%
\pgfsetdash{}{0pt}%
\pgfpathmoveto{\pgfqpoint{2.046121in}{0.855708in}}%
\pgfpathlineto{\pgfqpoint{3.088778in}{0.855708in}}%
\pgfpathquadraticcurveto{\pgfqpoint{3.116556in}{0.855708in}}{\pgfqpoint{3.116556in}{0.883486in}}%
\pgfpathlineto{\pgfqpoint{3.116556in}{1.655078in}}%
\pgfpathquadraticcurveto{\pgfqpoint{3.116556in}{1.682856in}}{\pgfqpoint{3.088778in}{1.682856in}}%
\pgfpathlineto{\pgfqpoint{2.046121in}{1.682856in}}%
\pgfpathquadraticcurveto{\pgfqpoint{2.018343in}{1.682856in}}{\pgfqpoint{2.018343in}{1.655078in}}%
\pgfpathlineto{\pgfqpoint{2.018343in}{0.883486in}}%
\pgfpathquadraticcurveto{\pgfqpoint{2.018343in}{0.855708in}}{\pgfqpoint{2.046121in}{0.855708in}}%
\pgfpathclose%
\pgfusepath{}%
\end{pgfscope}%
\begin{pgfscope}%
\pgfsetrectcap%
\pgfsetroundjoin%
\pgfsetlinewidth{1.003750pt}%
\definecolor{currentstroke}{rgb}{0.121569,0.466667,0.705882}%
\pgfsetstrokecolor{currentstroke}%
\pgfsetdash{}{0pt}%
\pgfpathmoveto{\pgfqpoint{2.073899in}{1.578689in}}%
\pgfpathlineto{\pgfqpoint{2.351677in}{1.578689in}}%
\pgfusepath{stroke}%
\end{pgfscope}%
\begin{pgfscope}%
\pgftext[x=2.462788in,y=1.530078in,left,base]{\rmfamily\fontsize{10.000000}{12.000000}\selectfont Slush Pool}%
\end{pgfscope}%
\begin{pgfscope}%
\pgfsetrectcap%
\pgfsetroundjoin%
\pgfsetlinewidth{1.003750pt}%
\definecolor{currentstroke}{rgb}{1.000000,0.498039,0.054902}%
\pgfsetstrokecolor{currentstroke}%
\pgfsetdash{}{0pt}%
\pgfpathmoveto{\pgfqpoint{2.073899in}{1.382319in}}%
\pgfpathlineto{\pgfqpoint{2.351677in}{1.382319in}}%
\pgfusepath{stroke}%
\end{pgfscope}%
\begin{pgfscope}%
\pgftext[x=2.462788in,y=1.333707in,left,base]{\rmfamily\fontsize{10.000000}{12.000000}\selectfont ViaBTC}%
\end{pgfscope}%
\begin{pgfscope}%
\pgfsetrectcap%
\pgfsetroundjoin%
\pgfsetlinewidth{1.003750pt}%
\definecolor{currentstroke}{rgb}{0.172549,0.627451,0.172549}%
\pgfsetstrokecolor{currentstroke}%
\pgfsetdash{}{0pt}%
\pgfpathmoveto{\pgfqpoint{2.073899in}{1.185948in}}%
\pgfpathlineto{\pgfqpoint{2.351677in}{1.185948in}}%
\pgfusepath{stroke}%
\end{pgfscope}%
\begin{pgfscope}%
\pgftext[x=2.462788in,y=1.137337in,left,base]{\rmfamily\fontsize{10.000000}{12.000000}\selectfont KanoPool}%
\end{pgfscope}%
\begin{pgfscope}%
\pgfsetrectcap%
\pgfsetroundjoin%
\pgfsetlinewidth{1.003750pt}%
\definecolor{currentstroke}{rgb}{0.839216,0.152941,0.156863}%
\pgfsetstrokecolor{currentstroke}%
\pgfsetdash{}{0pt}%
\pgfpathmoveto{\pgfqpoint{2.073899in}{0.989578in}}%
\pgfpathlineto{\pgfqpoint{2.351677in}{0.989578in}}%
\pgfusepath{stroke}%
\end{pgfscope}%
\begin{pgfscope}%
\pgftext[x=2.462788in,y=0.940967in,left,base]{\rmfamily\fontsize{10.000000}{12.000000}\selectfont Solo BTC}%
\end{pgfscope}%
\end{pgfpicture}%
\makeatother%
\endgroup%

%% file: experiment.tex
\section{Evaluation and Simulated Results}
\label{simulated-results}
To showcase the advantage (and risks) of diversifying over multiple pools, we  present an evaluation of our model  using Bitcoin  data extracted from Smartbit Block Explorer API \cite{smartbit}, as shown in Table \ref{block-data}. First, we consider a Bitcoin miner owning some hashing power $\minertotalhashrate$ = 1200TH/sec, who would start mining passively on a single chosen pool on February 1st 2018 for $\Delta$ = 4 months. Then, we consider a Bitcoin miner having the same hashing power $\minertotalhashrate$, who using our tool over the same time period, would ``actively'' diversify every $\interval=3$ days over $\poolsnum = 3$ Bitcoin pools of his choosing and reallocate his hashpower accordingly. In Table \ref{simulation-params} we outline the mining pools chosen for our evaluation along with their respective fees and other simulation parameters. We choose these pools as representatives of different pool sizes and fees, and to show how such pools influence the active miner's diversification over time. As discussed in Section \ref{sec:assumptions}, we pick the mean value for $\cara$ = 0.00005. Note that as we mentioned in the introductory section, all of the above parameters constitute several degrees of freedom in our experiments. In the subsequent sections we show how each one of them can affect the end result derived from our main evaluation.

We retroactively compute the earnings on a daily basis for both miners. To take both overall reward and miner's income variance into account, we utilize the Sharpe ratio $\sharpe = \frac{\totalreward - \totalreward_{\textsf{PPS}}}{\sigma_{\Delta}}$ as our main comparison metric, where $\totalreward$ is the total accumulated reward of the miner during the time period $\Delta$, $\totalreward_{\textsf{PPS}}$ is the estimated miner's reward during time period $\Delta$ using a PPS pool\footnote{We approximate $\totalreward_{\textsf{PPS}}$ using a large Bitcoin pool which offers a relatively steady income, and subtract $\fee{\textsf{PPS}}$ from those rewards.} and ${\sigma_{\Delta}}$ stands for the Standard Deviation of miner's reward over the time period $\Delta$. 

\smallskip
\noindent \textbf{Remark.} Here we should note that while the CARA utility function is used to capture preferences over different wealth levels, one could argue that the miner should be amortizing his costs and revenue over a time period, since the \emph{average} payoff per block would have little variance and a diversification over multiple pools is not necessary. However the sum (or the discount rate weighted sum) of many Poisson variables linearly scales both the mean and variance of each single Poisson variable, so our analysis, which apparently looks like ``to evaluate the reward from a single block'', is indeed equivalent to capturing preferences over different wealth levels (this trap is a common ``fallacy of large numbers''~\cite{ross1999adding,samuelson1963risk}). 

\ifllncs
\else
\subsection{Evaluation Assumptions}
\fi 
We assume that both miners choose pools that do not use the PPS reward scheme and that all pool's fees and reward schemes remain constant over time. In addition, the miners are assumed to have constant hashpower (i.e. do not use any of their rewards to buy more mining hardware), and convert their rewards into USD on a daily basis. For our analysis, we derive the daily network hashrate $\totalhashrate{\blockday}$ from the daily network difficulty $\difficulty{\blockday}$ using the approximation $\totalhashrate{\blockday} = \frac{2^{32}}{600}\difficulty{\blockday}$. We also approximate the daily pool hashpower by $\poolhashrate{m,\blockday} = \frac{\blocksnum{\poolid,\blockday}}{\blocksnum{\blockday}}$ where  $\blocksnum{\poolid,\blockday}$ the number of blocks found by pool $\poolid$ and $\blocksnum{\blockday}$ the total actual number of blocks found on each day $\blockday$. The smaller the mining pool however, the less precise this approximation becomes (i.e. a small pool might be ``unlucky'' and would not find a block for several consecutive days, while on some day it might become ``lucky'' and find several blocks in a single day), so we employ averaging techniques over a time window of 14 days (which we believe is a reasonable period) to improve our approximation accuracy for computing $\poolhashrate{m,\blockday}$. 
However a miner in the ``real world'' would better use the self-reporting pool hashrates (based on submitted shares) for a more precise result (to our knowledge, such historical data is not available on any block explorer). Also as noted before, we do not take any transaction fees kept by pools into account for simplicity purposes. However, a miner can use the methodology in Section \ref{pools-tx-fees} to compute the average transaction fees over a recent period of time from online blockchain explorers and make a projection for fees in the future, then add $\txfee{\poolid,\cryptoid}$ and $\txfee{\cryptoid}$ to Equation \ref{formula3} accordingly. Finally, we  show the earnings in USD instead of cryptocurrency (Bitcoins) in order to take the exchange rate into account, which is an important parameter for the active miner (used to calculate the block reward $\blockreward{}$).

\ifllncs
\else
\begin{table}[!t]		
	\caption{Data from Bitcoin blockchain.} \label{block-data}
	\centering		
	\begin{tabular}{|l|l|}
		\hline
		Days $\in \Delta$ & $\blockday$ \\
		Exchange rate& $\exchange{\blockday}$ \\
		Network difficulty&$\difficulty{\blockday}$\\
		Participating pools& $\poolid_{\blockday}$\\
		Total number of blocks & $\blocksnum{\blockday}$\\
		Number of blocks found by pool $\poolid$ & $\blocksnum{\poolid,\blockday}$\\
		\hline
	\end{tabular}
\end{table}

\begin{table}[!t]		
	\caption{Simulation parameters}
	\label{simulation-params}
	\centering		
	\begin{tabular}{|l|l|}
		\hline
		Miners' hashpower $\minertotalhashrate$&1200TH/sec\\
		\hline
		Diversification interval $\interval$&3 days\\
		\hline
		Fee $\fee{\textsf{SlushPool}}$ & 2\%\\
		\hline
		Fee $\fee{\textsf{ViaBTC}}$ & 2\%\\
		\hline
		Fee $\fee{\textsf{DPOOL}}$ & 1\%\\
		\hline
		CARA $\rho$ & 0.00005 \\
		\hline
	\end{tabular}
\end{table}

\begin{figure}[t]
	\subfigure[Passive miner on Slush pool.]{
		\resizebox{\figurewidth\textwidth}{!}{%
			\input{passive_slush.pgf}
	}}
	\quad
	\subfigure[Active miner on 3 pools.]{
		\resizebox{\figurewidth\textwidth}{!}{%
			\input{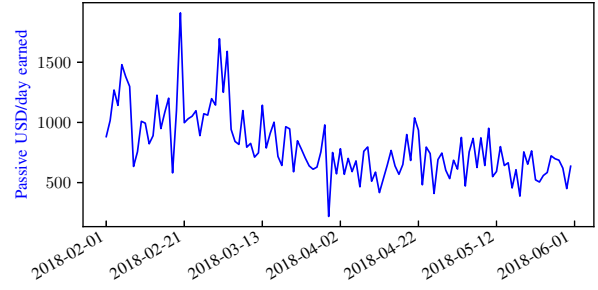}
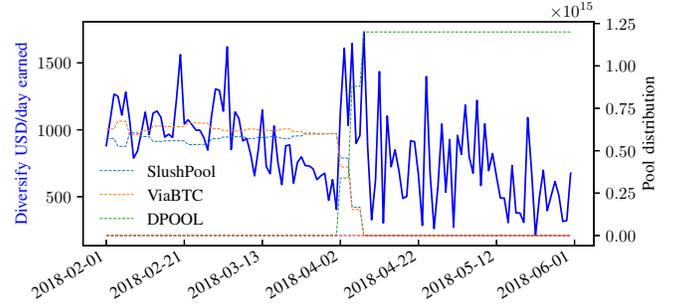
			\label{simulation-4mo-active}
		}
	}
	\caption{Hash power distribution over 4 month data} \label{simulation-4mo}
\end{figure}

\fi

\ifllncs
\else
\subsection{Main Evaluation Results}
\fi
\label{main-eval}
In Figure \ref{simulation-4mo} we show the earnings over time for both of the miners for comparison (in Figure \ref{simulation-4mo-active} we also show how the diversification changes over time in light colors). We observe a slightly increased variance for the active miner (blue line spikes), since he chose to include a third pool (DPOOL) with smaller total hashpower $\poolhashrate{}$. However his total accumulated reward would be $\totalreward_{\textsf{A}}$ = \$101221 for the active miner, compared to $\totalreward_{\textsf{P}}$ = \$97101, which eventually leads to a Sharpe ratio $\sharpe_{\mathsf{A}}$ = 0.156 compared to $\sharpe_{\mathsf{P}}$ = 0.060. Note that since we assumed both miners' hashpower $\minertotalhashrate$ remains constant throughout this period, we observe a general decline pattern in their daily reward, because of the increasing difficulty $\difficulty{}$ directly affecting $\frac{\minertotalhashrate}{\totalhashrate{}}$.  We also include an equivalent analysis using the same pools over an one year period in the next evaluation, where such a decline can be observed more clearly.  Lastly, an important parameter to consider is the block time $\blocktime{}$  \cite{DBLP:journals/corr/abs-1801-03998}. We used Bitcoin for our simulation, where $\blocktime{}$ is relatively high (roughly 10 minutes) - this resulted in high reward variance, especially when using smaller pools as shown in Figures \ref{simulation-4mo} and \ref{simulation-1yr}. If we used a cryptocurrency with more frequent blocks for our evaluation (e.g. Ethereum), then the result would be more predictable with lower observed overall variance values.

\ifllncs
\else
\subsection{Analysis for a Large $\Delta$}
\fi
In Figure \ref{simulation-1yr} we repeat the simulation discussed in Section \ref{main-eval} over the period of $\Delta$ = 1 year (January 1 2018 - December 31 2018). The decline of miner rewards due to the increasing difficulty can be observed more clearly. In such a case, a miner would most likely reinvest his earnings on mining hardware to keep $\frac{\minertotalhashrate}{\totalhashrate{}}$ as steady as possible. We also observe high variance around January 2018, generated by the high volatility in the Bitcoin/USD exchange rate $\exchange{\mathsf{BTC}}$. The results for this evaluation are $\totalreward_{\textsf{A}}$ = \$233293 and $\sharpe_{\mathsf{A}}$ = 0.064 vs. $\totalreward_{\textsf{P}}$ = \$224293 and $\sharpe_{\mathsf{P}}$ = 0.023, which are consistent with the results derived from the 4-month simulation.
\ifllncs
\else
\begin{figure}[t]
	\subfigure[Passive miner on Slush pool.]{
		\resizebox{\figurewidth\textwidth}{!}{%
			\input{passive_slush_1yr.pgf}
	}}
	\quad
	\subfigure[Active miner on 3 pools.]{
		\resizebox{\figurewidth\textwidth}{!}{%
			\input{active_miner_1yr.pgf}
		}
	}
	\caption{Hash power distribution over 1 year data} \label{simulation-1yr}
\end{figure}
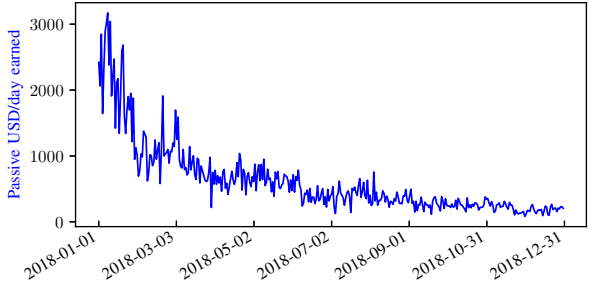
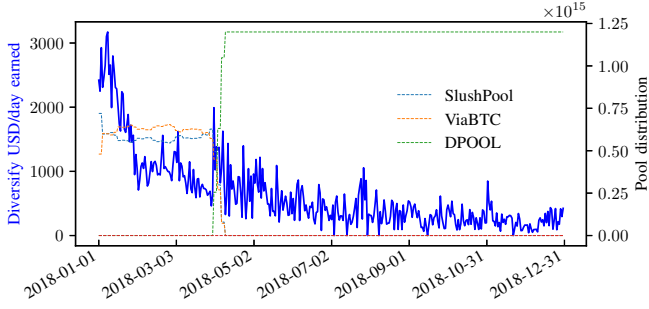
\fi

\ifllncs
\else
\subsection{Analysis for a Different $\Delta$ and  Different Pools}
\fi
We now replace the third small pool (DPOOL) from our default set of pools with a larger one (AntPool) and set our period from (January 1 2018 - June 30 2018) for $\Delta$ = 6 months, thus diversifying over the three largest PPLNS pools during that period. Again we observe an improvement in our metrics, $\totalreward_{\textsf{A}}$ = \$172719 and $\sharpe_{\mathsf{A}}$ = 0.041 vs. $\totalreward_{\textsf{P}}$ = \$172092 and $\sharpe_{\mathsf{P}}$ = 0.032. Detailed analysis shown in Figure \ref{simulation-6mo}.
\ifllncs
\else
\begin{figure}
	\subfigure[Passive miner on Slush pool.]{
		\resizebox{\figurewidth\textwidth}{!}{%
			\input{passive_slush_6mo.pgf}
	}}
	\quad
	\subfigure[Active miner on 3 pools.]{
		\resizebox{\figurewidth\textwidth}{!}{%
			\input{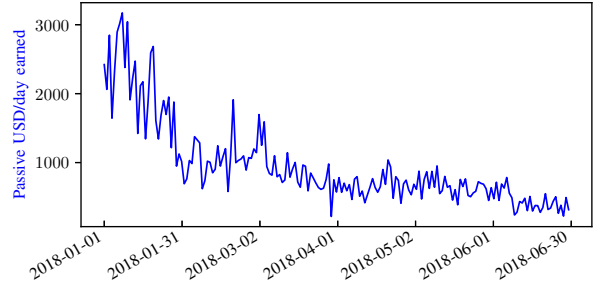}
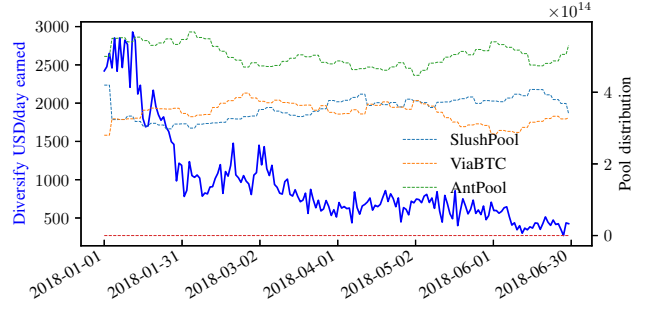
		}
	}
	\caption{Hash power distribution over 6 months data and different pool set.} \label{simulation-6mo}
\end{figure}

\begin{figure}
	\centering
	\resizebox{\figurewidth\textwidth}{!}{
		\input{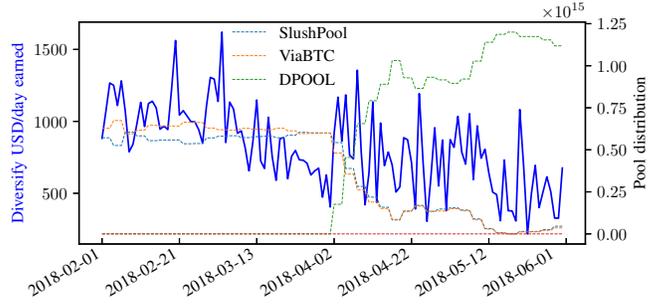}
	}
	\caption{Active miner diversifying over Bitcoin pools with increased $\cara$ = 0.0001.}
	\label{active-miner-large-rho}	
\end{figure}

\fi

\ifllncs
\else
\subsection{Analysis for Different Values of $\rho$}
\fi
We now repeat our retroactive analysis discussed in Section \ref{main-eval} by setting $\rho$ = 0.0001, which is at the upper bound of our considered typical values. By comparing Figure \ref{active-miner-large-rho} with Figure \ref{simulation-4mo-active}, we observe that the miner chose to diversify on the smaller pool (DPOOL) less frequently, since he is more ``sensitive'' to risk. As expected, this translates to a lower-variance graph and a more steady income. However his overall reward decreases, which offsets the previous benefit. Having kept the rest of the analysis parameters to our original default values, the miner's total accumulated reward would be now $\totalreward_{\textsf{A}}$ = \$99082, and the Sharpe ratio would be $\sharpe_{\mathsf{A}}$ = 0.104. 

By further experimenting with a broader value range for $\cara$, we derive Figure \ref{sharpe-rewards-vs-rho}, where we observe a decreasing trend for the Sharpe ratio as $\cara$ increases. This may be counterintuitive at first sight, as traditional portfolio theory would otherwise predict a flat relationship between a strategy’s Sharpe ratio and an investor’s risk aversion. The reason for this difference is that our model captures any potential impact of a miner's decisions to the whole equilibrium. If the miner is relatively small, his effect on the equilibrium is negligible through first-order Taylor expansion. In this case however, the miner is so large that his power is comparable to the third pool (DPOOL), and his effect on the whole equilibrium is no longer negligible, leading to the declining graph. If we replace DPOOL with a larger pool (Antpool) as shown in Figure \ref{sharpe-rewards-vs-rho-largepools}, we observe that the chosen $\cara$ has eventually no effect on the Sharpe ratio.
\ifllncs
\else
\begin{figure*}
	\begin{minipage}{0.47\textwidth}

		\resizebox{\textwidth}{!}{
			\input{sharpe-rewards-vs-rho.pgf}
		}
		\caption{Sharpe \& rewards vs $\cara$, SlushPool - ViaBTC - DPOOL.}\label{sharpe-rewards-vs-rho}
	\end{minipage}
	\begin{minipage}{0.47\textwidth}

		\resizebox{\textwidth}{!}{
			\input{sharpe-rewards-vs-rho-largepools.pgf}
		}
		\caption{Sharpe \& rewards vs $\cara$, SlushPool - ViaBTC - AntPool.}\label{sharpe-rewards-vs-rho-largepools}
	\end{minipage}

	\begin{minipage}{0.47\textwidth}

		\resizebox{\textwidth}{!}{
			\input{sharpe-rewards-vs-miningpower-dpool.pgf}
		}
		\caption{Sharpe \& rewards vs mining power, SlushPool - ViaBTC - DPOOL.}\label{sharpe-rewards-vs-miningpower-dpool}
	\end{minipage}
	\begin{minipage}{0.47\textwidth}

		\resizebox{\textwidth}{!}{
			\input{sharpe-rewards-vs-miningpower-large.pgf}
		}
		\caption{Active/Passive Sharpe ratios vs. mining power, SlushPool - ViaBTC - AntPool.}\label{sharpe-rewards-vs-miningpower-large}
	\end{minipage}

	\begin{minipage}{0.47\textwidth}

		\resizebox{\textwidth}{!}{
			\input{sharpe-rewards-vs-interval.pgf}
		}
		\caption{Sharpe \& rewards vs diversification interval $t$.}\label{sharpe-rewards-vs-interval}
	\end{minipage}
	\begin{minipage}{0.47\textwidth}

		\resizebox{\textwidth}{!}{
			\input{sharpe-rewards-vs-pools1.pgf}
		}
		\caption{Sharpe \& rewards vs type and order pools.}\label{sharpe-rewards-vs-pools1}
	\end{minipage}
\end{figure*}
\fi

\ifllncs
\else
\subsection{Analysis for Different Values of Miner Power}
\fi
By repeating our main evaluation for several different values of miner's computational power, we derive Figure \ref{sharpe-rewards-vs-miningpower-dpool} where we observe a negative correlation between Sharpe ratio and a miner's hash rates. As in the previous case, if the miner is large enough compared to the pools, our model captures his effect on the whole equilibrium which is negligible in typical portfolio analyses. When we replace the small pool (DPOOL) with a larger one (Antpool) and repeat our experiment, the traditional insight from portfolio theory reemerges: As shown in Figure \ref{sharpe-rewards-vs-miningpower-large}, for a relatively small miner, we now observe no significant correlation with miner hash rate and the resulting Sharpe ratio.

\ifllncs
\else
\subsection{Analysis for Different Diversification Intervals}
\fi
While our main evaluation assumed the active miner runs our tool every 3 days, in Figure \ref{sharpe-rewards-vs-interval} we show how different diversification intervals affect the Sharpe ratio. The general observation is that small intervals (less than 1 week) help slightly to improve the results, while large intervals (more than 1 month) are generally not recommended. After all, if the time period of  hashpower reallocation becomes very large, the miner is not very ``active'' and his behavior matches more that of a passive miner.

\ifllncs
\else
\subsection{Analysis for Different Number of Available Pools}
\fi
In Figure \ref{sharpe-rewards-vs-pools1} we examine how our main evaluation metrics are affected both by the total number and the specific pools available to the miner. We observe that if the miner only picks one pool (e.g. ViaBTC), effectively he can only diversify between that pool and solo mining, which usually  matches a passive miner for a typical value of $\cara$. However, as the miner includes additional pools into his consideration, his Sharpe ratio tends to increase, which indicates that a rational miner should consider as many pools as possible. However, given that we performed a retroactive analysis, adding ``bad luck'' pools into the miner's available pool set does not improve his Sharpe ratio any further. 

\ifllncs
\else
\subsection{Analysis for Different Cryptocurrencies}
\fi

As discussed in Section \ref{multi-crypto-single-pow}, our model also considers miners who diversify over multiple cryptocurrencies which use the same PoW algorithm. Extending our evaluation results to such a case is relatively straightforward (assuming the equivalent data shown in Table \ref{block-data} are available for \emph{all} considered cryptocurrencies) and we expect a similar derivation of results, as the only additional parameter in the problem is the reward over time ratio $\frac{\blockreward{\cryptoid}}{\blocktime{\cryptoid}}$, normalized to each cryptocurrency's total hashrate $\totalhashrate{\cryptoid}$, and a ``passive'' miner would just choose the most beneficial pool/cryptocurrency in the beginning of the experiment, without however taking any future changes of the above parameters into account.

We also note that the case of diversifying over different cryptocurrencies which \emph{also} employ different PoW algorithms as discussed in Section \ref{multi-crypto-multi-pow} is hard to execute in practice, since as discussed it excludes ASIC hardware, while it requires fine-tuning process priorities in CPUs and GPUs.

%% file: passive_slush.pgf
\begingroup%
\makeatletter%
\begin{pgfpicture}%
\pgfpathrectangle{\pgfpointorigin}{\pgfqpoint{5.400000in}{2.700000in}}%
\pgfusepath{use as bounding box, clip}%
\begin{pgfscope}%
\pgfsetbuttcap%
\pgfsetmiterjoin%
\definecolor{currentfill}{rgb}{1.000000,1.000000,1.000000}%
\pgfsetfillcolor{currentfill}%
\pgfsetlinewidth{0.000000pt}%
\definecolor{currentstroke}{rgb}{1.000000,1.000000,1.000000}%
\pgfsetstrokecolor{currentstroke}%
\pgfsetdash{}{0pt}%
\pgfpathmoveto{\pgfqpoint{0.000000in}{0.000000in}}%
\pgfpathlineto{\pgfqpoint{5.400000in}{0.000000in}}%
\pgfpathlineto{\pgfqpoint{5.400000in}{2.700000in}}%
\pgfpathlineto{\pgfqpoint{0.000000in}{2.700000in}}%
\pgfpathclose%
\pgfusepath{fill}%
\end{pgfscope}%
\begin{pgfscope}%
\pgfsetbuttcap%
\pgfsetmiterjoin%
\definecolor{currentfill}{rgb}{1.000000,1.000000,1.000000}%
\pgfsetfillcolor{currentfill}%
\pgfsetlinewidth{0.000000pt}%
\definecolor{currentstroke}{rgb}{0.000000,0.000000,0.000000}%
\pgfsetstrokecolor{currentstroke}%
\pgfsetstrokeopacity{0.000000}%
\pgfsetdash{}{0pt}%
\pgfpathmoveto{\pgfqpoint{0.675000in}{0.540000in}}%
\pgfpathlineto{\pgfqpoint{4.860000in}{0.540000in}}%
\pgfpathlineto{\pgfqpoint{4.860000in}{2.376000in}}%
\pgfpathlineto{\pgfqpoint{0.675000in}{2.376000in}}%
\pgfpathclose%
\pgfusepath{fill}%
\end{pgfscope}%
\begin{pgfscope}%
\pgfsetbuttcap%
\pgfsetroundjoin%
\definecolor{currentfill}{rgb}{0.000000,0.000000,0.000000}%
\pgfsetfillcolor{currentfill}%
\pgfsetlinewidth{0.803000pt}%
\definecolor{currentstroke}{rgb}{0.000000,0.000000,0.000000}%
\pgfsetstrokecolor{currentstroke}%
\pgfsetdash{}{0pt}%
\pgfsys@defobject{currentmarker}{\pgfqpoint{0.000000in}{0.000000in}}{\pgfqpoint{0.000000in}{0.048611in}}{%
\pgfpathmoveto{\pgfqpoint{0.000000in}{0.000000in}}%
\pgfpathlineto{\pgfqpoint{0.000000in}{0.048611in}}%
\pgfusepath{stroke,fill}%
}%
\begin{pgfscope}%
\pgfsys@transformshift{0.865227in}{0.540000in}%
\pgfsys@useobject{currentmarker}{}%
\end{pgfscope}%
\end{pgfscope}%
\begin{pgfscope}%
\pgftext[x=0.289141in,y=0.104678in,left,base,rotate=30.000000]{\rmfamily\fontsize{10.000000}{12.000000}\selectfont 2018-02-01}%
\end{pgfscope}%
\begin{pgfscope}%
\pgfsetbuttcap%
\pgfsetroundjoin%
\definecolor{currentfill}{rgb}{0.000000,0.000000,0.000000}%
\pgfsetfillcolor{currentfill}%
\pgfsetlinewidth{0.803000pt}%
\definecolor{currentstroke}{rgb}{0.000000,0.000000,0.000000}%
\pgfsetstrokecolor{currentstroke}%
\pgfsetdash{}{0pt}%
\pgfsys@defobject{currentmarker}{\pgfqpoint{0.000000in}{0.000000in}}{\pgfqpoint{0.000000in}{0.048611in}}{%
\pgfpathmoveto{\pgfqpoint{0.000000in}{0.000000in}}%
\pgfpathlineto{\pgfqpoint{0.000000in}{0.048611in}}%
\pgfusepath{stroke,fill}%
}%
\begin{pgfscope}%
\pgfsys@transformshift{1.504647in}{0.540000in}%
\pgfsys@useobject{currentmarker}{}%
\end{pgfscope}%
\end{pgfscope}%
\begin{pgfscope}%
\pgftext[x=0.928560in,y=0.104678in,left,base,rotate=30.000000]{\rmfamily\fontsize{10.000000}{12.000000}\selectfont 2018-02-21}%
\end{pgfscope}%
\begin{pgfscope}%
\pgfsetbuttcap%
\pgfsetroundjoin%
\definecolor{currentfill}{rgb}{0.000000,0.000000,0.000000}%
\pgfsetfillcolor{currentfill}%
\pgfsetlinewidth{0.803000pt}%
\definecolor{currentstroke}{rgb}{0.000000,0.000000,0.000000}%
\pgfsetstrokecolor{currentstroke}%
\pgfsetdash{}{0pt}%
\pgfsys@defobject{currentmarker}{\pgfqpoint{0.000000in}{0.000000in}}{\pgfqpoint{0.000000in}{0.048611in}}{%
\pgfpathmoveto{\pgfqpoint{0.000000in}{0.000000in}}%
\pgfpathlineto{\pgfqpoint{0.000000in}{0.048611in}}%
\pgfusepath{stroke,fill}%
}%
\begin{pgfscope}%
\pgfsys@transformshift{2.144066in}{0.540000in}%
\pgfsys@useobject{currentmarker}{}%
\end{pgfscope}%
\end{pgfscope}%
\begin{pgfscope}%
\pgftext[x=1.567980in,y=0.104678in,left,base,rotate=30.000000]{\rmfamily\fontsize{10.000000}{12.000000}\selectfont 2018-03-13}%
\end{pgfscope}%
\begin{pgfscope}%
\pgfsetbuttcap%
\pgfsetroundjoin%
\definecolor{currentfill}{rgb}{0.000000,0.000000,0.000000}%
\pgfsetfillcolor{currentfill}%
\pgfsetlinewidth{0.803000pt}%
\definecolor{currentstroke}{rgb}{0.000000,0.000000,0.000000}%
\pgfsetstrokecolor{currentstroke}%
\pgfsetdash{}{0pt}%
\pgfsys@defobject{currentmarker}{\pgfqpoint{0.000000in}{0.000000in}}{\pgfqpoint{0.000000in}{0.048611in}}{%
\pgfpathmoveto{\pgfqpoint{0.000000in}{0.000000in}}%
\pgfpathlineto{\pgfqpoint{0.000000in}{0.048611in}}%
\pgfusepath{stroke,fill}%
}%
\begin{pgfscope}%
\pgfsys@transformshift{2.783485in}{0.540000in}%
\pgfsys@useobject{currentmarker}{}%
\end{pgfscope}%
\end{pgfscope}%
\begin{pgfscope}%
\pgftext[x=2.207399in,y=0.104678in,left,base,rotate=30.000000]{\rmfamily\fontsize{10.000000}{12.000000}\selectfont 2018-04-02}%
\end{pgfscope}%
\begin{pgfscope}%
\pgfsetbuttcap%
\pgfsetroundjoin%
\definecolor{currentfill}{rgb}{0.000000,0.000000,0.000000}%
\pgfsetfillcolor{currentfill}%
\pgfsetlinewidth{0.803000pt}%
\definecolor{currentstroke}{rgb}{0.000000,0.000000,0.000000}%
\pgfsetstrokecolor{currentstroke}%
\pgfsetdash{}{0pt}%
\pgfsys@defobject{currentmarker}{\pgfqpoint{0.000000in}{0.000000in}}{\pgfqpoint{0.000000in}{0.048611in}}{%
\pgfpathmoveto{\pgfqpoint{0.000000in}{0.000000in}}%
\pgfpathlineto{\pgfqpoint{0.000000in}{0.048611in}}%
\pgfusepath{stroke,fill}%
}%
\begin{pgfscope}%
\pgfsys@transformshift{3.422905in}{0.540000in}%
\pgfsys@useobject{currentmarker}{}%
\end{pgfscope}%
\end{pgfscope}%
\begin{pgfscope}%
\pgftext[x=2.846819in,y=0.104678in,left,base,rotate=30.000000]{\rmfamily\fontsize{10.000000}{12.000000}\selectfont 2018-04-22}%
\end{pgfscope}%
\begin{pgfscope}%
\pgfsetbuttcap%
\pgfsetroundjoin%
\definecolor{currentfill}{rgb}{0.000000,0.000000,0.000000}%
\pgfsetfillcolor{currentfill}%
\pgfsetlinewidth{0.803000pt}%
\definecolor{currentstroke}{rgb}{0.000000,0.000000,0.000000}%
\pgfsetstrokecolor{currentstroke}%
\pgfsetdash{}{0pt}%
\pgfsys@defobject{currentmarker}{\pgfqpoint{0.000000in}{0.000000in}}{\pgfqpoint{0.000000in}{0.048611in}}{%
\pgfpathmoveto{\pgfqpoint{0.000000in}{0.000000in}}%
\pgfpathlineto{\pgfqpoint{0.000000in}{0.048611in}}%
\pgfusepath{stroke,fill}%
}%
\begin{pgfscope}%
\pgfsys@transformshift{4.062324in}{0.540000in}%
\pgfsys@useobject{currentmarker}{}%
\end{pgfscope}%
\end{pgfscope}%
\begin{pgfscope}%
\pgftext[x=3.486238in,y=0.104678in,left,base,rotate=30.000000]{\rmfamily\fontsize{10.000000}{12.000000}\selectfont 2018-05-12}%
\pgftext[x=4.104238in,y=0.104678in,left,base,rotate=30.000000]{\rmfamily\fontsize{10.000000}{12.000000}\selectfont 2018-06-01}%
\end{pgfscope}%
\begin{pgfscope}%
\pgfsetbuttcap%
\pgfsetroundjoin%
\definecolor{currentfill}{rgb}{0.000000,0.000000,0.000000}%
\pgfsetfillcolor{currentfill}%
\pgfsetlinewidth{0.803000pt}%
\definecolor{currentstroke}{rgb}{0.000000,0.000000,0.000000}%
\pgfsetstrokecolor{currentstroke}%
\pgfsetdash{}{0pt}%
\pgfsys@defobject{currentmarker}{\pgfqpoint{0.000000in}{0.000000in}}{\pgfqpoint{0.000000in}{0.048611in}}{%
\pgfpathmoveto{\pgfqpoint{0.000000in}{0.000000in}}%
\pgfpathlineto{\pgfqpoint{0.000000in}{0.048611in}}%
\pgfusepath{stroke,fill}%
}%
\begin{pgfscope}%
\pgfsys@transformshift{4.701744in}{0.540000in}%
\pgfsys@useobject{currentmarker}{}%
\end{pgfscope}%
\end{pgfscope}%
\begin{pgfscope}%
\pgfsetbuttcap%
\pgfsetroundjoin%
\definecolor{currentfill}{rgb}{0.000000,0.000000,0.000000}%
\pgfsetfillcolor{currentfill}%
\pgfsetlinewidth{0.803000pt}%
\definecolor{currentstroke}{rgb}{0.000000,0.000000,0.000000}%
\pgfsetstrokecolor{currentstroke}%
\pgfsetdash{}{0pt}%
\pgfsys@defobject{currentmarker}{\pgfqpoint{-0.048611in}{0.000000in}}{\pgfqpoint{0.000000in}{0.000000in}}{%
\pgfpathmoveto{\pgfqpoint{0.000000in}{0.000000in}}%
\pgfpathlineto{\pgfqpoint{-0.048611in}{0.000000in}}%
\pgfusepath{stroke,fill}%
}%
\begin{pgfscope}%
\pgfsys@transformshift{0.675000in}{0.901167in}%
\pgfsys@useobject{currentmarker}{}%
\end{pgfscope}%
\end{pgfscope}%
\begin{pgfscope}%
\pgftext[x=0.369444in,y=0.852950in,left,base]{\rmfamily\fontsize{10.000000}{12.000000}\selectfont \(\displaystyle 500\)}%
\end{pgfscope}%
\begin{pgfscope}%
\pgfsetbuttcap%
\pgfsetroundjoin%
\definecolor{currentfill}{rgb}{0.000000,0.000000,0.000000}%
\pgfsetfillcolor{currentfill}%
\pgfsetlinewidth{0.803000pt}%
\definecolor{currentstroke}{rgb}{0.000000,0.000000,0.000000}%
\pgfsetstrokecolor{currentstroke}%
\pgfsetdash{}{0pt}%
\pgfsys@defobject{currentmarker}{\pgfqpoint{-0.048611in}{0.000000in}}{\pgfqpoint{0.000000in}{0.000000in}}{%
\pgfpathmoveto{\pgfqpoint{0.000000in}{0.000000in}}%
\pgfpathlineto{\pgfqpoint{-0.048611in}{0.000000in}}%
\pgfusepath{stroke,fill}%
}%
\begin{pgfscope}%
\pgfsys@transformshift{0.675000in}{1.394174in}%
\pgfsys@useobject{currentmarker}{}%
\end{pgfscope}%
\end{pgfscope}%
\begin{pgfscope}%
\pgftext[x=0.299999in,y=1.345956in,left,base]{\rmfamily\fontsize{10.000000}{12.000000}\selectfont \(\displaystyle 1000\)}%
\end{pgfscope}%
\begin{pgfscope}%
\pgfsetbuttcap%
\pgfsetroundjoin%
\definecolor{currentfill}{rgb}{0.000000,0.000000,0.000000}%
\pgfsetfillcolor{currentfill}%
\pgfsetlinewidth{0.803000pt}%
\definecolor{currentstroke}{rgb}{0.000000,0.000000,0.000000}%
\pgfsetstrokecolor{currentstroke}%
\pgfsetdash{}{0pt}%
\pgfsys@defobject{currentmarker}{\pgfqpoint{-0.048611in}{0.000000in}}{\pgfqpoint{0.000000in}{0.000000in}}{%
\pgfpathmoveto{\pgfqpoint{0.000000in}{0.000000in}}%
\pgfpathlineto{\pgfqpoint{-0.048611in}{0.000000in}}%
\pgfusepath{stroke,fill}%
}%
\begin{pgfscope}%
\pgfsys@transformshift{0.675000in}{1.887181in}%
\pgfsys@useobject{currentmarker}{}%
\end{pgfscope}%
\end{pgfscope}%
\begin{pgfscope}%
\pgftext[x=0.299999in,y=1.838963in,left,base]{\rmfamily\fontsize{10.000000}{12.000000}\selectfont \(\displaystyle 1500\)}%
\end{pgfscope}%
\begin{pgfscope}%
\definecolor{textcolor}{rgb}{0.000000,0.000000,1.000000}%
\pgfsetstrokecolor{textcolor}%
\pgfsetfillcolor{textcolor}%
\pgftext[x=0.244444in,y=1.458000in,,bottom,rotate=90.000000]{\color{textcolor}\rmfamily\fontsize{10.000000}{12.000000}\selectfont Passive USD/day earned}%
\end{pgfscope}%
\begin{pgfscope}%
\pgfpathrectangle{\pgfqpoint{0.675000in}{0.540000in}}{\pgfqpoint{4.185000in}{1.836000in}}%
\pgfusepath{clip}%
\pgfsetrectcap%
\pgfsetroundjoin%
\pgfsetlinewidth{1.003750pt}%
\definecolor{currentstroke}{rgb}{0.000000,0.000000,1.000000}%
\pgfsetstrokecolor{currentstroke}%
\pgfsetdash{}{0pt}%
\pgfpathmoveto{\pgfqpoint{0.865227in}{1.277829in}}%
\pgfpathlineto{\pgfqpoint{0.897198in}{1.409946in}}%
\pgfpathlineto{\pgfqpoint{0.929169in}{1.660034in}}%
\pgfpathlineto{\pgfqpoint{0.961140in}{1.534168in}}%
\pgfpathlineto{\pgfqpoint{0.993111in}{1.867932in}}%
\pgfpathlineto{\pgfqpoint{1.025082in}{1.768694in}}%
\pgfpathlineto{\pgfqpoint{1.057053in}{1.689274in}}%
\pgfpathlineto{\pgfqpoint{1.089024in}{1.033445in}}%
\pgfpathlineto{\pgfqpoint{1.120995in}{1.153367in}}%
\pgfpathlineto{\pgfqpoint{1.152966in}{1.403624in}}%
\pgfpathlineto{\pgfqpoint{1.184937in}{1.388553in}}%
\pgfpathlineto{\pgfqpoint{1.216908in}{1.219184in}}%
\pgfpathlineto{\pgfqpoint{1.248879in}{1.285600in}}%
\pgfpathlineto{\pgfqpoint{1.280850in}{1.617575in}}%
\pgfpathlineto{\pgfqpoint{1.312821in}{1.344067in}}%
\pgfpathlineto{\pgfqpoint{1.344792in}{1.476799in}}%
\pgfpathlineto{\pgfqpoint{1.376763in}{1.592965in}}%
\pgfpathlineto{\pgfqpoint{1.408734in}{0.980600in}}%
\pgfpathlineto{\pgfqpoint{1.440705in}{1.493115in}}%
\pgfpathlineto{\pgfqpoint{1.472676in}{2.292545in}}%
\pgfpathlineto{\pgfqpoint{1.504647in}{1.391991in}}%
\pgfpathlineto{\pgfqpoint{1.536618in}{1.424634in}}%
\pgfpathlineto{\pgfqpoint{1.568589in}{1.445050in}}%
\pgfpathlineto{\pgfqpoint{1.600560in}{1.490426in}}%
\pgfpathlineto{\pgfqpoint{1.632531in}{1.286125in}}%
\pgfpathlineto{\pgfqpoint{1.664502in}{1.465471in}}%
\pgfpathlineto{\pgfqpoint{1.696472in}{1.454211in}}%
\pgfpathlineto{\pgfqpoint{1.728443in}{1.589389in}}%
\pgfpathlineto{\pgfqpoint{1.760414in}{1.536595in}}%
\pgfpathlineto{\pgfqpoint{1.792385in}{2.080825in}}%
\pgfpathlineto{\pgfqpoint{1.824356in}{1.641254in}}%
\pgfpathlineto{\pgfqpoint{1.856327in}{1.976801in}}%
\pgfpathlineto{\pgfqpoint{1.888298in}{1.337200in}}%
\pgfpathlineto{\pgfqpoint{1.920269in}{1.238404in}}%
\pgfpathlineto{\pgfqpoint{1.952240in}{1.213973in}}%
\pgfpathlineto{\pgfqpoint{1.984211in}{1.491742in}}%
\pgfpathlineto{\pgfqpoint{2.016182in}{1.191022in}}%
\pgfpathlineto{\pgfqpoint{2.048153in}{1.222437in}}%
\pgfpathlineto{\pgfqpoint{2.080124in}{1.109588in}}%
\pgfpathlineto{\pgfqpoint{2.112095in}{1.144801in}}%
\pgfpathlineto{\pgfqpoint{2.144066in}{1.535294in}}%
\pgfpathlineto{\pgfqpoint{2.176037in}{1.184988in}}%
\pgfpathlineto{\pgfqpoint{2.208008in}{1.303691in}}%
\pgfpathlineto{\pgfqpoint{2.239979in}{1.396312in}}%
\pgfpathlineto{\pgfqpoint{2.271950in}{1.115707in}}%
\pgfpathlineto{\pgfqpoint{2.303921in}{1.041382in}}%
\pgfpathlineto{\pgfqpoint{2.335892in}{1.359098in}}%
\pgfpathlineto{\pgfqpoint{2.367863in}{1.341901in}}%
\pgfpathlineto{\pgfqpoint{2.399834in}{0.989674in}}%
\pgfpathlineto{\pgfqpoint{2.431805in}{1.244164in}}%
\pgfpathlineto{\pgfqpoint{2.463776in}{1.175634in}}%
\pgfpathlineto{\pgfqpoint{2.495747in}{1.102809in}}%
\pgfpathlineto{\pgfqpoint{2.527718in}{1.037937in}}%
\pgfpathlineto{\pgfqpoint{2.559689in}{1.010307in}}%
\pgfpathlineto{\pgfqpoint{2.591660in}{1.028171in}}%
\pgfpathlineto{\pgfqpoint{2.623631in}{1.152869in}}%
\pgfpathlineto{\pgfqpoint{2.655602in}{1.373991in}}%
\pgfpathlineto{\pgfqpoint{2.687573in}{0.623455in}}%
\pgfpathlineto{\pgfqpoint{2.719544in}{1.147649in}}%
\pgfpathlineto{\pgfqpoint{2.751515in}{0.972972in}}%
\pgfpathlineto{\pgfqpoint{2.783485in}{1.178672in}}%
\pgfpathlineto{\pgfqpoint{2.815456in}{0.968857in}}%
\pgfpathlineto{\pgfqpoint{2.847427in}{1.099474in}}%
\pgfpathlineto{\pgfqpoint{2.879398in}{0.991085in}}%
\pgfpathlineto{\pgfqpoint{2.911369in}{1.079111in}}%
\pgfpathlineto{\pgfqpoint{2.943340in}{0.866542in}}%
\pgfpathlineto{\pgfqpoint{2.975311in}{1.157806in}}%
\pgfpathlineto{\pgfqpoint{3.007282in}{1.193195in}}%
\pgfpathlineto{\pgfqpoint{3.039253in}{0.912096in}}%
\pgfpathlineto{\pgfqpoint{3.071224in}{0.986744in}}%
\pgfpathlineto{\pgfqpoint{3.103195in}{0.819063in}}%
\pgfpathlineto{\pgfqpoint{3.135166in}{0.931589in}}%
\pgfpathlineto{\pgfqpoint{3.167137in}{1.043148in}}%
\pgfpathlineto{\pgfqpoint{3.199108in}{1.164692in}}%
\pgfpathlineto{\pgfqpoint{3.231079in}{1.037434in}}%
\pgfpathlineto{\pgfqpoint{3.263050in}{0.968861in}}%
\pgfpathlineto{\pgfqpoint{3.295021in}{1.049249in}}%
\pgfpathlineto{\pgfqpoint{3.326992in}{1.295085in}}%
\pgfpathlineto{\pgfqpoint{3.358963in}{1.082337in}}%
\pgfpathlineto{\pgfqpoint{3.390934in}{1.431839in}}%
\pgfpathlineto{\pgfqpoint{3.422905in}{1.330739in}}%
\pgfpathlineto{\pgfqpoint{3.454876in}{0.883121in}}%
\pgfpathlineto{\pgfqpoint{3.486847in}{1.192198in}}%
\pgfpathlineto{\pgfqpoint{3.518818in}{1.139105in}}%
\pgfpathlineto{\pgfqpoint{3.550789in}{0.812071in}}%
\pgfpathlineto{\pgfqpoint{3.582760in}{1.089700in}}%
\pgfpathlineto{\pgfqpoint{3.614731in}{1.143238in}}%
\pgfpathlineto{\pgfqpoint{3.646702in}{1.002341in}}%
\pgfpathlineto{\pgfqpoint{3.678673in}{0.933821in}}%
\pgfpathlineto{\pgfqpoint{3.710644in}{1.084186in}}%
\pgfpathlineto{\pgfqpoint{3.742615in}{1.011574in}}%
\pgfpathlineto{\pgfqpoint{3.774586in}{1.271273in}}%
\pgfpathlineto{\pgfqpoint{3.806557in}{0.873002in}}%
\pgfpathlineto{\pgfqpoint{3.838528in}{1.152814in}}%
\pgfpathlineto{\pgfqpoint{3.870498in}{1.263661in}}%
\pgfpathlineto{\pgfqpoint{3.902469in}{1.024939in}}%
\pgfpathlineto{\pgfqpoint{3.934440in}{1.268009in}}%
\pgfpathlineto{\pgfqpoint{3.966411in}{1.041150in}}%
\pgfpathlineto{\pgfqpoint{3.998382in}{1.346485in}}%
\pgfpathlineto{\pgfqpoint{4.030353in}{0.950015in}}%
\pgfpathlineto{\pgfqpoint{4.062324in}{0.991841in}}%
\pgfpathlineto{\pgfqpoint{4.094295in}{1.196165in}}%
\pgfpathlineto{\pgfqpoint{4.126266in}{1.041169in}}%
\pgfpathlineto{\pgfqpoint{4.158237in}{1.063537in}}%
\pgfpathlineto{\pgfqpoint{4.190208in}{0.857242in}}%
\pgfpathlineto{\pgfqpoint{4.222179in}{1.006496in}}%
\pgfpathlineto{\pgfqpoint{4.254150in}{0.790087in}}%
\pgfpathlineto{\pgfqpoint{4.286121in}{1.153028in}}%
\pgfpathlineto{\pgfqpoint{4.318092in}{1.051427in}}%
\pgfpathlineto{\pgfqpoint{4.350063in}{1.161056in}}%
\pgfpathlineto{\pgfqpoint{4.382034in}{0.923379in}}%
\pgfpathlineto{\pgfqpoint{4.414005in}{0.905727in}}%
\pgfpathlineto{\pgfqpoint{4.445976in}{0.959069in}}%
\pgfpathlineto{\pgfqpoint{4.477947in}{0.983796in}}%
\pgfpathlineto{\pgfqpoint{4.509918in}{1.120452in}}%
\pgfpathlineto{\pgfqpoint{4.541889in}{1.096631in}}%
\pgfpathlineto{\pgfqpoint{4.573860in}{1.084897in}}%
\pgfpathlineto{\pgfqpoint{4.605831in}{1.019615in}}%
\pgfpathlineto{\pgfqpoint{4.637802in}{0.851826in}}%
\pgfpathlineto{\pgfqpoint{4.669773in}{1.035168in}}%
\pgfusepath{stroke}%
\end{pgfscope}%
\begin{pgfscope}%
\pgfsetrectcap%
\pgfsetmiterjoin%
\pgfsetlinewidth{0.803000pt}%
\definecolor{currentstroke}{rgb}{0.000000,0.000000,0.000000}%
\pgfsetstrokecolor{currentstroke}%
\pgfsetdash{}{0pt}%
\pgfpathmoveto{\pgfqpoint{0.675000in}{0.540000in}}%
\pgfpathlineto{\pgfqpoint{0.675000in}{2.376000in}}%
\pgfusepath{stroke}%
\end{pgfscope}%
\begin{pgfscope}%
\pgfsetrectcap%
\pgfsetmiterjoin%
\pgfsetlinewidth{0.803000pt}%
\definecolor{currentstroke}{rgb}{0.000000,0.000000,0.000000}%
\pgfsetstrokecolor{currentstroke}%
\pgfsetdash{}{0pt}%
\pgfpathmoveto{\pgfqpoint{4.860000in}{0.540000in}}%
\pgfpathlineto{\pgfqpoint{4.860000in}{2.376000in}}%
\pgfusepath{stroke}%
\end{pgfscope}%
\begin{pgfscope}%
\pgfsetrectcap%
\pgfsetmiterjoin%
\pgfsetlinewidth{0.803000pt}%
\definecolor{currentstroke}{rgb}{0.000000,0.000000,0.000000}%
\pgfsetstrokecolor{currentstroke}%
\pgfsetdash{}{0pt}%
\pgfpathmoveto{\pgfqpoint{0.675000in}{0.540000in}}%
\pgfpathlineto{\pgfqpoint{4.860000in}{0.540000in}}%
\pgfusepath{stroke}%
\end{pgfscope}%
\begin{pgfscope}%
\pgfsetrectcap%
\pgfsetmiterjoin%
\pgfsetlinewidth{0.803000pt}%
\definecolor{currentstroke}{rgb}{0.000000,0.000000,0.000000}%
\pgfsetstrokecolor{currentstroke}%
\pgfsetdash{}{0pt}%
\pgfpathmoveto{\pgfqpoint{0.675000in}{2.376000in}}%
\pgfpathlineto{\pgfqpoint{4.860000in}{2.376000in}}%
\pgfusepath{stroke}%
\end{pgfscope}%
\begin{pgfscope}%
\pgfsetbuttcap%
\pgfsetmiterjoin%
\pgfsetlinewidth{0.000000pt}%
\definecolor{currentstroke}{rgb}{0.800000,0.800000,0.800000}%
\pgfsetstrokecolor{currentstroke}%
\pgfsetstrokeopacity{0.000000}%
\pgfsetdash{}{0pt}%
\pgfpathmoveto{\pgfqpoint{3.283222in}{0.884844in}}%
\pgfpathlineto{\pgfqpoint{3.338778in}{0.884844in}}%
\pgfpathquadraticcurveto{\pgfqpoint{3.366556in}{0.884844in}}{\pgfqpoint{3.366556in}{0.912622in}}%
\pgfpathlineto{\pgfqpoint{3.366556in}{0.968178in}}%
\pgfpathquadraticcurveto{\pgfqpoint{3.366556in}{0.995956in}}{\pgfqpoint{3.338778in}{0.995956in}}%
\pgfpathlineto{\pgfqpoint{3.283222in}{0.995956in}}%
\pgfpathquadraticcurveto{\pgfqpoint{3.255444in}{0.995956in}}{\pgfqpoint{3.255444in}{0.968178in}}%
\pgfpathlineto{\pgfqpoint{3.255444in}{0.912622in}}%
\pgfpathquadraticcurveto{\pgfqpoint{3.255444in}{0.884844in}}{\pgfqpoint{3.283222in}{0.884844in}}%
\pgfpathclose%
\pgfusepath{}%
\end{pgfscope}%
\end{pgfpicture}%
\makeatother%
\endgroup%

%% file: passive_slush_1yr.pgf
\begingroup%
\makeatletter%
\begin{pgfpicture}%
\pgfpathrectangle{\pgfpointorigin}{\pgfqpoint{5.400000in}{2.700000in}}%
\pgfusepath{use as bounding box, clip}%
\begin{pgfscope}%
\pgfsetbuttcap%
\pgfsetmiterjoin%
\definecolor{currentfill}{rgb}{1.000000,1.000000,1.000000}%
\pgfsetfillcolor{currentfill}%
\pgfsetlinewidth{0.000000pt}%
\definecolor{currentstroke}{rgb}{1.000000,1.000000,1.000000}%
\pgfsetstrokecolor{currentstroke}%
\pgfsetdash{}{0pt}%
\pgfpathmoveto{\pgfqpoint{0.000000in}{0.000000in}}%
\pgfpathlineto{\pgfqpoint{5.400000in}{0.000000in}}%
\pgfpathlineto{\pgfqpoint{5.400000in}{2.700000in}}%
\pgfpathlineto{\pgfqpoint{0.000000in}{2.700000in}}%
\pgfpathclose%
\pgfusepath{fill}%
\end{pgfscope}%
\begin{pgfscope}%
\pgfsetbuttcap%
\pgfsetmiterjoin%
\definecolor{currentfill}{rgb}{1.000000,1.000000,1.000000}%
\pgfsetfillcolor{currentfill}%
\pgfsetlinewidth{0.000000pt}%
\definecolor{currentstroke}{rgb}{0.000000,0.000000,0.000000}%
\pgfsetstrokecolor{currentstroke}%
\pgfsetstrokeopacity{0.000000}%
\pgfsetdash{}{0pt}%
\pgfpathmoveto{\pgfqpoint{0.675000in}{0.540000in}}%
\pgfpathlineto{\pgfqpoint{4.860000in}{0.540000in}}%
\pgfpathlineto{\pgfqpoint{4.860000in}{2.376000in}}%
\pgfpathlineto{\pgfqpoint{0.675000in}{2.376000in}}%
\pgfpathclose%
\pgfusepath{fill}%
\end{pgfscope}%
\begin{pgfscope}%
\pgfsetbuttcap%
\pgfsetroundjoin%
\definecolor{currentfill}{rgb}{0.000000,0.000000,0.000000}%
\pgfsetfillcolor{currentfill}%
\pgfsetlinewidth{0.803000pt}%
\definecolor{currentstroke}{rgb}{0.000000,0.000000,0.000000}%
\pgfsetstrokecolor{currentstroke}%
\pgfsetdash{}{0pt}%
\pgfsys@defobject{currentmarker}{\pgfqpoint{0.000000in}{0.000000in}}{\pgfqpoint{0.000000in}{0.048611in}}{%
\pgfpathmoveto{\pgfqpoint{0.000000in}{0.000000in}}%
\pgfpathlineto{\pgfqpoint{0.000000in}{0.048611in}}%
\pgfusepath{stroke,fill}%
}%
\begin{pgfscope}%
\pgfsys@transformshift{0.865227in}{0.540000in}%
\pgfsys@useobject{currentmarker}{}%
\end{pgfscope}%
\end{pgfscope}%
\begin{pgfscope}%
\pgftext[x=0.289141in,y=0.104678in,left,base,rotate=30.000000]{\rmfamily\fontsize{10.000000}{12.000000}\selectfont 2018-01-01}%
\end{pgfscope}%
\begin{pgfscope}%
\pgfsetbuttcap%
\pgfsetroundjoin%
\definecolor{currentfill}{rgb}{0.000000,0.000000,0.000000}%
\pgfsetfillcolor{currentfill}%
\pgfsetlinewidth{0.803000pt}%
\definecolor{currentstroke}{rgb}{0.000000,0.000000,0.000000}%
\pgfsetstrokecolor{currentstroke}%
\pgfsetdash{}{0pt}%
\pgfsys@defobject{currentmarker}{\pgfqpoint{0.000000in}{0.000000in}}{\pgfqpoint{0.000000in}{0.048611in}}{%
\pgfpathmoveto{\pgfqpoint{0.000000in}{0.000000in}}%
\pgfpathlineto{\pgfqpoint{0.000000in}{0.048611in}}%
\pgfusepath{stroke,fill}%
}%
\begin{pgfscope}%
\pgfsys@transformshift{1.501065in}{0.540000in}%
\pgfsys@useobject{currentmarker}{}%
\end{pgfscope}%
\end{pgfscope}%
\begin{pgfscope}%
\pgftext[x=0.924979in,y=0.104678in,left,base,rotate=30.000000]{\rmfamily\fontsize{10.000000}{12.000000}\selectfont 2018-03-03}%
\end{pgfscope}%
\begin{pgfscope}%
\pgfsetbuttcap%
\pgfsetroundjoin%
\definecolor{currentfill}{rgb}{0.000000,0.000000,0.000000}%
\pgfsetfillcolor{currentfill}%
\pgfsetlinewidth{0.803000pt}%
\definecolor{currentstroke}{rgb}{0.000000,0.000000,0.000000}%
\pgfsetstrokecolor{currentstroke}%
\pgfsetdash{}{0pt}%
\pgfsys@defobject{currentmarker}{\pgfqpoint{0.000000in}{0.000000in}}{\pgfqpoint{0.000000in}{0.048611in}}{%
\pgfpathmoveto{\pgfqpoint{0.000000in}{0.000000in}}%
\pgfpathlineto{\pgfqpoint{0.000000in}{0.048611in}}%
\pgfusepath{stroke,fill}%
}%
\begin{pgfscope}%
\pgfsys@transformshift{2.136903in}{0.540000in}%
\pgfsys@useobject{currentmarker}{}%
\end{pgfscope}%
\end{pgfscope}%
\begin{pgfscope}%
\pgftext[x=1.560816in,y=0.104678in,left,base,rotate=30.000000]{\rmfamily\fontsize{10.000000}{12.000000}\selectfont 2018-05-02}%
\end{pgfscope}%
\begin{pgfscope}%
\pgfsetbuttcap%
\pgfsetroundjoin%
\definecolor{currentfill}{rgb}{0.000000,0.000000,0.000000}%
\pgfsetfillcolor{currentfill}%
\pgfsetlinewidth{0.803000pt}%
\definecolor{currentstroke}{rgb}{0.000000,0.000000,0.000000}%
\pgfsetstrokecolor{currentstroke}%
\pgfsetdash{}{0pt}%
\pgfsys@defobject{currentmarker}{\pgfqpoint{0.000000in}{0.000000in}}{\pgfqpoint{0.000000in}{0.048611in}}{%
\pgfpathmoveto{\pgfqpoint{0.000000in}{0.000000in}}%
\pgfpathlineto{\pgfqpoint{0.000000in}{0.048611in}}%
\pgfusepath{stroke,fill}%
}%
\begin{pgfscope}%
\pgfsys@transformshift{2.772740in}{0.540000in}%
\pgfsys@useobject{currentmarker}{}%
\end{pgfscope}%
\end{pgfscope}%
\begin{pgfscope}%
\pgftext[x=2.196654in,y=0.104678in,left,base,rotate=30.000000]{\rmfamily\fontsize{10.000000}{12.000000}\selectfont 2018-07-02}%
\end{pgfscope}%
\begin{pgfscope}%
\pgfsetbuttcap%
\pgfsetroundjoin%
\definecolor{currentfill}{rgb}{0.000000,0.000000,0.000000}%
\pgfsetfillcolor{currentfill}%
\pgfsetlinewidth{0.803000pt}%
\definecolor{currentstroke}{rgb}{0.000000,0.000000,0.000000}%
\pgfsetstrokecolor{currentstroke}%
\pgfsetdash{}{0pt}%
\pgfsys@defobject{currentmarker}{\pgfqpoint{0.000000in}{0.000000in}}{\pgfqpoint{0.000000in}{0.048611in}}{%
\pgfpathmoveto{\pgfqpoint{0.000000in}{0.000000in}}%
\pgfpathlineto{\pgfqpoint{0.000000in}{0.048611in}}%
\pgfusepath{stroke,fill}%
}%
\begin{pgfscope}%
\pgfsys@transformshift{3.408578in}{0.540000in}%
\pgfsys@useobject{currentmarker}{}%
\end{pgfscope}%
\end{pgfscope}%
\begin{pgfscope}%
\pgftext[x=2.832492in,y=0.104678in,left,base,rotate=30.000000]{\rmfamily\fontsize{10.000000}{12.000000}\selectfont 2018-09-01}%
\end{pgfscope}%
\begin{pgfscope}%
\pgfsetbuttcap%
\pgfsetroundjoin%
\definecolor{currentfill}{rgb}{0.000000,0.000000,0.000000}%
\pgfsetfillcolor{currentfill}%
\pgfsetlinewidth{0.803000pt}%
\definecolor{currentstroke}{rgb}{0.000000,0.000000,0.000000}%
\pgfsetstrokecolor{currentstroke}%
\pgfsetdash{}{0pt}%
\pgfsys@defobject{currentmarker}{\pgfqpoint{0.000000in}{0.000000in}}{\pgfqpoint{0.000000in}{0.048611in}}{%
\pgfpathmoveto{\pgfqpoint{0.000000in}{0.000000in}}%
\pgfpathlineto{\pgfqpoint{0.000000in}{0.048611in}}%
\pgfusepath{stroke,fill}%
}%
\begin{pgfscope}%
\pgfsys@transformshift{4.044416in}{0.540000in}%
\pgfsys@useobject{currentmarker}{}%
\end{pgfscope}%
\end{pgfscope}%
\begin{pgfscope}%
\pgftext[x=3.468330in,y=0.104678in,left,base,rotate=30.000000]{\rmfamily\fontsize{10.000000}{12.000000}\selectfont 2018-10-31}%
\pgftext[x=4.104330in,y=0.104678in,left,base,rotate=30.000000]{\rmfamily\fontsize{10.000000}{12.000000}\selectfont 2018-12-31}%
\end{pgfscope}%
\begin{pgfscope}%
\pgfsetbuttcap%
\pgfsetroundjoin%
\definecolor{currentfill}{rgb}{0.000000,0.000000,0.000000}%
\pgfsetfillcolor{currentfill}%
\pgfsetlinewidth{0.803000pt}%
\definecolor{currentstroke}{rgb}{0.000000,0.000000,0.000000}%
\pgfsetstrokecolor{currentstroke}%
\pgfsetdash{}{0pt}%
\pgfsys@defobject{currentmarker}{\pgfqpoint{0.000000in}{0.000000in}}{\pgfqpoint{0.000000in}{0.048611in}}{%
\pgfpathmoveto{\pgfqpoint{0.000000in}{0.000000in}}%
\pgfpathlineto{\pgfqpoint{0.000000in}{0.048611in}}%
\pgfusepath{stroke,fill}%
}%
\begin{pgfscope}%
\pgfsys@transformshift{4.680254in}{0.540000in}%
\pgfsys@useobject{currentmarker}{}%
\end{pgfscope}%
\end{pgfscope}%
\begin{pgfscope}%
\pgfsetbuttcap%
\pgfsetroundjoin%
\definecolor{currentfill}{rgb}{0.000000,0.000000,0.000000}%
\pgfsetfillcolor{currentfill}%
\pgfsetlinewidth{0.803000pt}%
\definecolor{currentstroke}{rgb}{0.000000,0.000000,0.000000}%
\pgfsetstrokecolor{currentstroke}%
\pgfsetdash{}{0pt}%
\pgfsys@defobject{currentmarker}{\pgfqpoint{-0.048611in}{0.000000in}}{\pgfqpoint{0.000000in}{0.000000in}}{%
\pgfpathmoveto{\pgfqpoint{0.000000in}{0.000000in}}%
\pgfpathlineto{\pgfqpoint{-0.048611in}{0.000000in}}%
\pgfusepath{stroke,fill}%
}%
\begin{pgfscope}%
\pgfsys@transformshift{0.675000in}{0.579286in}%
\pgfsys@useobject{currentmarker}{}%
\end{pgfscope}%
\end{pgfscope}%
\begin{pgfscope}%
\pgftext[x=0.508333in,y=0.531068in,left,base]{\rmfamily\fontsize{10.000000}{12.000000}\selectfont \(\displaystyle 0\)}%
\end{pgfscope}%
\begin{pgfscope}%
\pgfsetbuttcap%
\pgfsetroundjoin%
\definecolor{currentfill}{rgb}{0.000000,0.000000,0.000000}%
\pgfsetfillcolor{currentfill}%
\pgfsetlinewidth{0.803000pt}%
\definecolor{currentstroke}{rgb}{0.000000,0.000000,0.000000}%
\pgfsetstrokecolor{currentstroke}%
\pgfsetdash{}{0pt}%
\pgfsys@defobject{currentmarker}{\pgfqpoint{-0.048611in}{0.000000in}}{\pgfqpoint{0.000000in}{0.000000in}}{%
\pgfpathmoveto{\pgfqpoint{0.000000in}{0.000000in}}%
\pgfpathlineto{\pgfqpoint{-0.048611in}{0.000000in}}%
\pgfusepath{stroke,fill}%
}%
\begin{pgfscope}%
\pgfsys@transformshift{0.675000in}{1.119470in}%
\pgfsys@useobject{currentmarker}{}%
\end{pgfscope}%
\end{pgfscope}%
\begin{pgfscope}%
\pgftext[x=0.299999in,y=1.071252in,left,base]{\rmfamily\fontsize{10.000000}{12.000000}\selectfont \(\displaystyle 1000\)}%
\end{pgfscope}%
\begin{pgfscope}%
\pgfsetbuttcap%
\pgfsetroundjoin%
\definecolor{currentfill}{rgb}{0.000000,0.000000,0.000000}%
\pgfsetfillcolor{currentfill}%
\pgfsetlinewidth{0.803000pt}%
\definecolor{currentstroke}{rgb}{0.000000,0.000000,0.000000}%
\pgfsetstrokecolor{currentstroke}%
\pgfsetdash{}{0pt}%
\pgfsys@defobject{currentmarker}{\pgfqpoint{-0.048611in}{0.000000in}}{\pgfqpoint{0.000000in}{0.000000in}}{%
\pgfpathmoveto{\pgfqpoint{0.000000in}{0.000000in}}%
\pgfpathlineto{\pgfqpoint{-0.048611in}{0.000000in}}%
\pgfusepath{stroke,fill}%
}%
\begin{pgfscope}%
\pgfsys@transformshift{0.675000in}{1.659654in}%
\pgfsys@useobject{currentmarker}{}%
\end{pgfscope}%
\end{pgfscope}%
\begin{pgfscope}%
\pgftext[x=0.299999in,y=1.611436in,left,base]{\rmfamily\fontsize{10.000000}{12.000000}\selectfont \(\displaystyle 2000\)}%
\end{pgfscope}%
\begin{pgfscope}%
\pgfsetbuttcap%
\pgfsetroundjoin%
\definecolor{currentfill}{rgb}{0.000000,0.000000,0.000000}%
\pgfsetfillcolor{currentfill}%
\pgfsetlinewidth{0.803000pt}%
\definecolor{currentstroke}{rgb}{0.000000,0.000000,0.000000}%
\pgfsetstrokecolor{currentstroke}%
\pgfsetdash{}{0pt}%
\pgfsys@defobject{currentmarker}{\pgfqpoint{-0.048611in}{0.000000in}}{\pgfqpoint{0.000000in}{0.000000in}}{%
\pgfpathmoveto{\pgfqpoint{0.000000in}{0.000000in}}%
\pgfpathlineto{\pgfqpoint{-0.048611in}{0.000000in}}%
\pgfusepath{stroke,fill}%
}%
\begin{pgfscope}%
\pgfsys@transformshift{0.675000in}{2.199837in}%
\pgfsys@useobject{currentmarker}{}%
\end{pgfscope}%
\end{pgfscope}%
\begin{pgfscope}%
\pgftext[x=0.299999in,y=2.151619in,left,base]{\rmfamily\fontsize{10.000000}{12.000000}\selectfont \(\displaystyle 3000\)}%
\end{pgfscope}%
\begin{pgfscope}%
\definecolor{textcolor}{rgb}{0.000000,0.000000,1.000000}%
\pgfsetstrokecolor{textcolor}%
\pgfsetfillcolor{textcolor}%
\pgftext[x=0.244444in,y=1.458000in,,bottom,rotate=90.000000]{\color{textcolor}\rmfamily\fontsize{10.000000}{12.000000}\selectfont Passive USD/day earned}%
\end{pgfscope}%
\begin{pgfscope}%
\pgfpathrectangle{\pgfqpoint{0.675000in}{0.540000in}}{\pgfqpoint{4.185000in}{1.836000in}}%
\pgfusepath{clip}%
\pgfsetrectcap%
\pgfsetroundjoin%
\pgfsetlinewidth{1.003750pt}%
\definecolor{currentstroke}{rgb}{0.000000,0.000000,1.000000}%
\pgfsetstrokecolor{currentstroke}%
\pgfsetdash{}{0pt}%
\pgfpathmoveto{\pgfqpoint{0.865227in}{1.887233in}}%
\pgfpathlineto{\pgfqpoint{0.875708in}{1.692937in}}%
\pgfpathlineto{\pgfqpoint{0.886189in}{2.117838in}}%
\pgfpathlineto{\pgfqpoint{0.896670in}{1.467451in}}%
\pgfpathlineto{\pgfqpoint{0.907151in}{1.837077in}}%
\pgfpathlineto{\pgfqpoint{0.917631in}{2.142213in}}%
\pgfpathlineto{\pgfqpoint{0.928112in}{2.209681in}}%
\pgfpathlineto{\pgfqpoint{0.938593in}{2.292545in}}%
\pgfpathlineto{\pgfqpoint{0.949074in}{1.864469in}}%
\pgfpathlineto{\pgfqpoint{0.959555in}{2.222787in}}%
\pgfpathlineto{\pgfqpoint{0.970036in}{1.611106in}}%
\pgfpathlineto{\pgfqpoint{0.980517in}{1.777947in}}%
\pgfpathlineto{\pgfqpoint{0.990997in}{1.914501in}}%
\pgfpathlineto{\pgfqpoint{1.001478in}{1.347835in}}%
\pgfpathlineto{\pgfqpoint{1.011959in}{1.721632in}}%
\pgfpathlineto{\pgfqpoint{1.022440in}{1.752943in}}%
\pgfpathlineto{\pgfqpoint{1.032921in}{1.305437in}}%
\pgfpathlineto{\pgfqpoint{1.043402in}{1.598632in}}%
\pgfpathlineto{\pgfqpoint{1.053882in}{1.979617in}}%
\pgfpathlineto{\pgfqpoint{1.064363in}{2.029174in}}%
\pgfpathlineto{\pgfqpoint{1.074844in}{1.447974in}}%
\pgfpathlineto{\pgfqpoint{1.085325in}{1.304291in}}%
\pgfpathlineto{\pgfqpoint{1.095806in}{1.488767in}}%
\pgfpathlineto{\pgfqpoint{1.106287in}{1.605631in}}%
\pgfpathlineto{\pgfqpoint{1.116767in}{1.496449in}}%
\pgfpathlineto{\pgfqpoint{1.127248in}{1.632978in}}%
\pgfpathlineto{\pgfqpoint{1.137729in}{1.236118in}}%
\pgfpathlineto{\pgfqpoint{1.148210in}{1.593933in}}%
\pgfpathlineto{\pgfqpoint{1.158691in}{1.091607in}}%
\pgfpathlineto{\pgfqpoint{1.169172in}{1.186761in}}%
\pgfpathlineto{\pgfqpoint{1.179653in}{1.128239in}}%
\pgfpathlineto{\pgfqpoint{1.190133in}{0.953377in}}%
\pgfpathlineto{\pgfqpoint{1.200614in}{0.992036in}}%
\pgfpathlineto{\pgfqpoint{1.211095in}{1.135669in}}%
\pgfpathlineto{\pgfqpoint{1.221576in}{1.111599in}}%
\pgfpathlineto{\pgfqpoint{1.232057in}{1.322290in}}%
\pgfpathlineto{\pgfqpoint{1.253018in}{1.273782in}}%
\pgfpathlineto{\pgfqpoint{1.263499in}{0.914601in}}%
\pgfpathlineto{\pgfqpoint{1.273980in}{0.977056in}}%
\pgfpathlineto{\pgfqpoint{1.284461in}{1.130259in}}%
\pgfpathlineto{\pgfqpoint{1.294942in}{1.122048in}}%
\pgfpathlineto{\pgfqpoint{1.305423in}{1.039296in}}%
\pgfpathlineto{\pgfqpoint{1.315903in}{1.067014in}}%
\pgfpathlineto{\pgfqpoint{1.326384in}{1.250939in}}%
\pgfpathlineto{\pgfqpoint{1.336865in}{1.092019in}}%
\pgfpathlineto{\pgfqpoint{1.347346in}{1.164736in}}%
\pgfpathlineto{\pgfqpoint{1.357827in}{1.228376in}}%
\pgfpathlineto{\pgfqpoint{1.368308in}{0.892895in}}%
\pgfpathlineto{\pgfqpoint{1.378789in}{1.173674in}}%
\pgfpathlineto{\pgfqpoint{1.389269in}{1.611639in}}%
\pgfpathlineto{\pgfqpoint{1.399750in}{1.118274in}}%
\pgfpathlineto{\pgfqpoint{1.410231in}{1.136157in}}%
\pgfpathlineto{\pgfqpoint{1.420712in}{1.147342in}}%
\pgfpathlineto{\pgfqpoint{1.431193in}{1.172201in}}%
\pgfpathlineto{\pgfqpoint{1.441674in}{1.060276in}}%
\pgfpathlineto{\pgfqpoint{1.452154in}{1.158530in}}%
\pgfpathlineto{\pgfqpoint{1.462635in}{1.152361in}}%
\pgfpathlineto{\pgfqpoint{1.473116in}{1.226417in}}%
\pgfpathlineto{\pgfqpoint{1.483597in}{1.197495in}}%
\pgfpathlineto{\pgfqpoint{1.494078in}{1.495649in}}%
\pgfpathlineto{\pgfqpoint{1.504559in}{1.254832in}}%
\pgfpathlineto{\pgfqpoint{1.515039in}{1.438660in}}%
\pgfpathlineto{\pgfqpoint{1.525520in}{1.088257in}}%
\pgfpathlineto{\pgfqpoint{1.536001in}{1.034132in}}%
\pgfpathlineto{\pgfqpoint{1.546482in}{1.020747in}}%
\pgfpathlineto{\pgfqpoint{1.556963in}{1.172922in}}%
\pgfpathlineto{\pgfqpoint{1.567444in}{1.008174in}}%
\pgfpathlineto{\pgfqpoint{1.577924in}{1.025384in}}%
\pgfpathlineto{\pgfqpoint{1.588405in}{0.963560in}}%
\pgfpathlineto{\pgfqpoint{1.598886in}{0.982852in}}%
\pgfpathlineto{\pgfqpoint{1.609367in}{1.196782in}}%
\pgfpathlineto{\pgfqpoint{1.619848in}{1.004868in}}%
\pgfpathlineto{\pgfqpoint{1.630329in}{1.069899in}}%
\pgfpathlineto{\pgfqpoint{1.640810in}{1.120641in}}%
\pgfpathlineto{\pgfqpoint{1.651290in}{0.966912in}}%
\pgfpathlineto{\pgfqpoint{1.661771in}{0.926194in}}%
\pgfpathlineto{\pgfqpoint{1.672252in}{1.100254in}}%
\pgfpathlineto{\pgfqpoint{1.682733in}{1.090832in}}%
\pgfpathlineto{\pgfqpoint{1.693214in}{0.897866in}}%
\pgfpathlineto{\pgfqpoint{1.703695in}{1.037287in}}%
\pgfpathlineto{\pgfqpoint{1.735137in}{0.924307in}}%
\pgfpathlineto{\pgfqpoint{1.745618in}{0.909170in}}%
\pgfpathlineto{\pgfqpoint{1.756099in}{0.918956in}}%
\pgfpathlineto{\pgfqpoint{1.766580in}{0.987272in}}%
\pgfpathlineto{\pgfqpoint{1.777060in}{1.108413in}}%
\pgfpathlineto{\pgfqpoint{1.787541in}{0.697234in}}%
\pgfpathlineto{\pgfqpoint{1.798022in}{0.984412in}}%
\pgfpathlineto{\pgfqpoint{1.808503in}{0.888716in}}%
\pgfpathlineto{\pgfqpoint{1.818984in}{1.001408in}}%
\pgfpathlineto{\pgfqpoint{1.829465in}{0.886462in}}%
\pgfpathlineto{\pgfqpoint{1.839946in}{0.958020in}}%
\pgfpathlineto{\pgfqpoint{1.850426in}{0.898639in}}%
\pgfpathlineto{\pgfqpoint{1.860907in}{0.946864in}}%
\pgfpathlineto{\pgfqpoint{1.871388in}{0.830408in}}%
\pgfpathlineto{\pgfqpoint{1.881869in}{0.989976in}}%
\pgfpathlineto{\pgfqpoint{1.892350in}{1.009364in}}%
\pgfpathlineto{\pgfqpoint{1.902831in}{0.855365in}}%
\pgfpathlineto{\pgfqpoint{1.913311in}{0.896261in}}%
\pgfpathlineto{\pgfqpoint{1.923792in}{0.804398in}}%
\pgfpathlineto{\pgfqpoint{1.955235in}{0.993749in}}%
\pgfpathlineto{\pgfqpoint{1.965716in}{0.924031in}}%
\pgfpathlineto{\pgfqpoint{1.976196in}{0.886464in}}%
\pgfpathlineto{\pgfqpoint{1.986677in}{0.930504in}}%
\pgfpathlineto{\pgfqpoint{1.997158in}{1.065184in}}%
\pgfpathlineto{\pgfqpoint{2.007639in}{0.948631in}}%
\pgfpathlineto{\pgfqpoint{2.018120in}{1.140104in}}%
\pgfpathlineto{\pgfqpoint{2.028601in}{1.084717in}}%
\pgfpathlineto{\pgfqpoint{2.039082in}{0.839491in}}%
\pgfpathlineto{\pgfqpoint{2.049562in}{1.008818in}}%
\pgfpathlineto{\pgfqpoint{2.060043in}{0.979731in}}%
\pgfpathlineto{\pgfqpoint{2.070524in}{0.800567in}}%
\pgfpathlineto{\pgfqpoint{2.081005in}{0.952665in}}%
\pgfpathlineto{\pgfqpoint{2.091486in}{0.981996in}}%
\pgfpathlineto{\pgfqpoint{2.101967in}{0.904806in}}%
\pgfpathlineto{\pgfqpoint{2.112447in}{0.867267in}}%
\pgfpathlineto{\pgfqpoint{2.122928in}{0.949644in}}%
\pgfpathlineto{\pgfqpoint{2.133409in}{0.909864in}}%
\pgfpathlineto{\pgfqpoint{2.143890in}{1.052139in}}%
\pgfpathlineto{\pgfqpoint{2.154371in}{0.833948in}}%
\pgfpathlineto{\pgfqpoint{2.164852in}{0.987242in}}%
\pgfpathlineto{\pgfqpoint{2.175332in}{1.047969in}}%
\pgfpathlineto{\pgfqpoint{2.185813in}{0.917186in}}%
\pgfpathlineto{\pgfqpoint{2.196294in}{1.050351in}}%
\pgfpathlineto{\pgfqpoint{2.206775in}{0.926067in}}%
\pgfpathlineto{\pgfqpoint{2.217256in}{1.093344in}}%
\pgfpathlineto{\pgfqpoint{2.227737in}{0.876139in}}%
\pgfpathlineto{\pgfqpoint{2.238218in}{0.899053in}}%
\pgfpathlineto{\pgfqpoint{2.248698in}{1.010991in}}%
\pgfpathlineto{\pgfqpoint{2.259179in}{0.926077in}}%
\pgfpathlineto{\pgfqpoint{2.269660in}{0.938331in}}%
\pgfpathlineto{\pgfqpoint{2.280141in}{0.825314in}}%
\pgfpathlineto{\pgfqpoint{2.290622in}{0.907082in}}%
\pgfpathlineto{\pgfqpoint{2.301103in}{0.788523in}}%
\pgfpathlineto{\pgfqpoint{2.311583in}{0.987359in}}%
\pgfpathlineto{\pgfqpoint{2.322064in}{0.931697in}}%
\pgfpathlineto{\pgfqpoint{2.332545in}{0.991757in}}%
\pgfpathlineto{\pgfqpoint{2.343026in}{0.861547in}}%
\pgfpathlineto{\pgfqpoint{2.353507in}{0.851876in}}%
\pgfpathlineto{\pgfqpoint{2.363988in}{0.881099in}}%
\pgfpathlineto{\pgfqpoint{2.374468in}{0.894646in}}%
\pgfpathlineto{\pgfqpoint{2.384949in}{0.969512in}}%
\pgfpathlineto{\pgfqpoint{2.395430in}{0.956462in}}%
\pgfpathlineto{\pgfqpoint{2.405911in}{0.950034in}}%
\pgfpathlineto{\pgfqpoint{2.416392in}{0.914269in}}%
\pgfpathlineto{\pgfqpoint{2.426873in}{0.822347in}}%
\pgfpathlineto{\pgfqpoint{2.437353in}{0.922790in}}%
\pgfpathlineto{\pgfqpoint{2.447834in}{0.836447in}}%
\pgfpathlineto{\pgfqpoint{2.458315in}{0.965421in}}%
\pgfpathlineto{\pgfqpoint{2.468796in}{0.823090in}}%
\pgfpathlineto{\pgfqpoint{2.479277in}{0.951385in}}%
\pgfpathlineto{\pgfqpoint{2.489758in}{0.919765in}}%
\pgfpathlineto{\pgfqpoint{2.500239in}{1.002555in}}%
\pgfpathlineto{\pgfqpoint{2.510719in}{0.879820in}}%
\pgfpathlineto{\pgfqpoint{2.521200in}{0.843924in}}%
\pgfpathlineto{\pgfqpoint{2.531681in}{0.708621in}}%
\pgfpathlineto{\pgfqpoint{2.542162in}{0.731195in}}%
\pgfpathlineto{\pgfqpoint{2.552643in}{0.814092in}}%
\pgfpathlineto{\pgfqpoint{2.563124in}{0.802635in}}%
\pgfpathlineto{\pgfqpoint{2.573604in}{0.839866in}}%
\pgfpathlineto{\pgfqpoint{2.584085in}{0.742934in}}%
\pgfpathlineto{\pgfqpoint{2.594566in}{0.853783in}}%
\pgfpathlineto{\pgfqpoint{2.605047in}{0.739708in}}%
\pgfpathlineto{\pgfqpoint{2.615528in}{0.782975in}}%
\pgfpathlineto{\pgfqpoint{2.626009in}{0.782272in}}%
\pgfpathlineto{\pgfqpoint{2.636489in}{0.728868in}}%
\pgfpathlineto{\pgfqpoint{2.646970in}{0.767534in}}%
\pgfpathlineto{\pgfqpoint{2.657451in}{0.875594in}}%
\pgfpathlineto{\pgfqpoint{2.667932in}{0.751690in}}%
\pgfpathlineto{\pgfqpoint{2.678413in}{0.762894in}}%
\pgfpathlineto{\pgfqpoint{2.688894in}{0.815628in}}%
\pgfpathlineto{\pgfqpoint{2.699375in}{0.851869in}}%
\pgfpathlineto{\pgfqpoint{2.709855in}{0.721859in}}%
\pgfpathlineto{\pgfqpoint{2.720336in}{0.784848in}}%
\pgfpathlineto{\pgfqpoint{2.730817in}{0.700079in}}%
\pgfpathlineto{\pgfqpoint{2.741298in}{0.845578in}}%
\pgfpathlineto{\pgfqpoint{2.751779in}{0.750083in}}%
\pgfpathlineto{\pgfqpoint{2.762260in}{0.789223in}}%
\pgfpathlineto{\pgfqpoint{2.772740in}{0.804975in}}%
\pgfpathlineto{\pgfqpoint{2.783221in}{0.867521in}}%
\pgfpathlineto{\pgfqpoint{2.793702in}{0.697289in}}%
\pgfpathlineto{\pgfqpoint{2.804183in}{0.647230in}}%
\pgfpathlineto{\pgfqpoint{2.814664in}{0.786344in}}%
\pgfpathlineto{\pgfqpoint{2.825145in}{0.810663in}}%
\pgfpathlineto{\pgfqpoint{2.835625in}{0.912932in}}%
\pgfpathlineto{\pgfqpoint{2.846106in}{0.818986in}}%
\pgfpathlineto{\pgfqpoint{2.867068in}{0.776959in}}%
\pgfpathlineto{\pgfqpoint{2.877549in}{0.714628in}}%
\pgfpathlineto{\pgfqpoint{2.888030in}{0.771856in}}%
\pgfpathlineto{\pgfqpoint{2.898511in}{0.818971in}}%
\pgfpathlineto{\pgfqpoint{2.908991in}{0.832873in}}%
\pgfpathlineto{\pgfqpoint{2.919472in}{0.789739in}}%
\pgfpathlineto{\pgfqpoint{2.929953in}{0.655857in}}%
\pgfpathlineto{\pgfqpoint{2.940434in}{0.853328in}}%
\pgfpathlineto{\pgfqpoint{2.950915in}{0.835477in}}%
\pgfpathlineto{\pgfqpoint{2.961396in}{0.864906in}}%
\pgfpathlineto{\pgfqpoint{2.971876in}{0.820693in}}%
\pgfpathlineto{\pgfqpoint{2.982357in}{0.795567in}}%
\pgfpathlineto{\pgfqpoint{2.992838in}{0.879486in}}%
\pgfpathlineto{\pgfqpoint{3.003319in}{0.933524in}}%
\pgfpathlineto{\pgfqpoint{3.013800in}{0.780428in}}%
\pgfpathlineto{\pgfqpoint{3.024281in}{0.827677in}}%
\pgfpathlineto{\pgfqpoint{3.034761in}{0.903572in}}%
\pgfpathlineto{\pgfqpoint{3.045242in}{0.793361in}}%
\pgfpathlineto{\pgfqpoint{3.055723in}{0.769671in}}%
\pgfpathlineto{\pgfqpoint{3.066204in}{0.920418in}}%
\pgfpathlineto{\pgfqpoint{3.076685in}{0.733637in}}%
\pgfpathlineto{\pgfqpoint{3.087166in}{0.798272in}}%
\pgfpathlineto{\pgfqpoint{3.097647in}{0.712682in}}%
\pgfpathlineto{\pgfqpoint{3.108127in}{0.725137in}}%
\pgfpathlineto{\pgfqpoint{3.118608in}{0.988475in}}%
\pgfpathlineto{\pgfqpoint{3.129089in}{0.750752in}}%
\pgfpathlineto{\pgfqpoint{3.139570in}{0.822710in}}%
\pgfpathlineto{\pgfqpoint{3.150051in}{0.715796in}}%
\pgfpathlineto{\pgfqpoint{3.160532in}{0.756081in}}%
\pgfpathlineto{\pgfqpoint{3.171012in}{0.753084in}}%
\pgfpathlineto{\pgfqpoint{3.181493in}{0.773564in}}%
\pgfpathlineto{\pgfqpoint{3.191974in}{0.833840in}}%
\pgfpathlineto{\pgfqpoint{3.202455in}{0.810680in}}%
\pgfpathlineto{\pgfqpoint{3.212936in}{0.741466in}}%
\pgfpathlineto{\pgfqpoint{3.223417in}{0.790599in}}%
\pgfpathlineto{\pgfqpoint{3.233897in}{0.764740in}}%
\pgfpathlineto{\pgfqpoint{3.244378in}{0.698861in}}%
\pgfpathlineto{\pgfqpoint{3.254859in}{0.747664in}}%
\pgfpathlineto{\pgfqpoint{3.265340in}{0.743443in}}%
\pgfpathlineto{\pgfqpoint{3.275821in}{0.722601in}}%
\pgfpathlineto{\pgfqpoint{3.286302in}{0.769160in}}%
\pgfpathlineto{\pgfqpoint{3.296782in}{0.749215in}}%
\pgfpathlineto{\pgfqpoint{3.307263in}{0.823732in}}%
\pgfpathlineto{\pgfqpoint{3.317744in}{0.785173in}}%
\pgfpathlineto{\pgfqpoint{3.328225in}{0.730733in}}%
\pgfpathlineto{\pgfqpoint{3.338706in}{0.732174in}}%
\pgfpathlineto{\pgfqpoint{3.349187in}{0.727189in}}%
\pgfpathlineto{\pgfqpoint{3.359668in}{0.792850in}}%
\pgfpathlineto{\pgfqpoint{3.370148in}{0.789964in}}%
\pgfpathlineto{\pgfqpoint{3.380629in}{0.844474in}}%
\pgfpathlineto{\pgfqpoint{3.391110in}{0.756477in}}%
\pgfpathlineto{\pgfqpoint{3.401591in}{0.735887in}}%
\pgfpathlineto{\pgfqpoint{3.422553in}{0.847675in}}%
\pgfpathlineto{\pgfqpoint{3.433033in}{0.729501in}}%
\pgfpathlineto{\pgfqpoint{3.443514in}{0.748502in}}%
\pgfpathlineto{\pgfqpoint{3.453995in}{0.658991in}}%
\pgfpathlineto{\pgfqpoint{3.464476in}{0.747948in}}%
\pgfpathlineto{\pgfqpoint{3.474957in}{0.670831in}}%
\pgfpathlineto{\pgfqpoint{3.485438in}{0.721259in}}%
\pgfpathlineto{\pgfqpoint{3.495918in}{0.717941in}}%
\pgfpathlineto{\pgfqpoint{3.506399in}{0.782230in}}%
\pgfpathlineto{\pgfqpoint{3.527361in}{0.663112in}}%
\pgfpathlineto{\pgfqpoint{3.537842in}{0.741056in}}%
\pgfpathlineto{\pgfqpoint{3.548323in}{0.719624in}}%
\pgfpathlineto{\pgfqpoint{3.558804in}{0.718649in}}%
\pgfpathlineto{\pgfqpoint{3.569284in}{0.696528in}}%
\pgfpathlineto{\pgfqpoint{3.579765in}{0.717805in}}%
\pgfpathlineto{\pgfqpoint{3.590246in}{0.641243in}}%
\pgfpathlineto{\pgfqpoint{3.600727in}{0.748393in}}%
\pgfpathlineto{\pgfqpoint{3.611208in}{0.777433in}}%
\pgfpathlineto{\pgfqpoint{3.621689in}{0.785778in}}%
\pgfpathlineto{\pgfqpoint{3.632169in}{0.734761in}}%
\pgfpathlineto{\pgfqpoint{3.653131in}{0.701809in}}%
\pgfpathlineto{\pgfqpoint{3.663612in}{0.668870in}}%
\pgfpathlineto{\pgfqpoint{3.674093in}{0.786661in}}%
\pgfpathlineto{\pgfqpoint{3.684574in}{0.746128in}}%
\pgfpathlineto{\pgfqpoint{3.695054in}{0.688549in}}%
\pgfpathlineto{\pgfqpoint{3.705535in}{0.768512in}}%
\pgfpathlineto{\pgfqpoint{3.716016in}{0.716097in}}%
\pgfpathlineto{\pgfqpoint{3.726497in}{0.708454in}}%
\pgfpathlineto{\pgfqpoint{3.736978in}{0.710207in}}%
\pgfpathlineto{\pgfqpoint{3.747459in}{0.696619in}}%
\pgfpathlineto{\pgfqpoint{3.757940in}{0.725801in}}%
\pgfpathlineto{\pgfqpoint{3.768420in}{0.700231in}}%
\pgfpathlineto{\pgfqpoint{3.778901in}{0.702716in}}%
\pgfpathlineto{\pgfqpoint{3.789382in}{0.756457in}}%
\pgfpathlineto{\pgfqpoint{3.799863in}{0.683814in}}%
\pgfpathlineto{\pgfqpoint{3.810344in}{0.772047in}}%
\pgfpathlineto{\pgfqpoint{3.831305in}{0.717079in}}%
\pgfpathlineto{\pgfqpoint{3.841786in}{0.716849in}}%
\pgfpathlineto{\pgfqpoint{3.852267in}{0.698063in}}%
\pgfpathlineto{\pgfqpoint{3.862748in}{0.684845in}}%
\pgfpathlineto{\pgfqpoint{3.873229in}{0.660923in}}%
\pgfpathlineto{\pgfqpoint{3.883710in}{0.776011in}}%
\pgfpathlineto{\pgfqpoint{3.894190in}{0.695072in}}%
\pgfpathlineto{\pgfqpoint{3.904671in}{0.719594in}}%
\pgfpathlineto{\pgfqpoint{3.915152in}{0.694368in}}%
\pgfpathlineto{\pgfqpoint{3.925633in}{0.723253in}}%
\pgfpathlineto{\pgfqpoint{3.936114in}{0.705964in}}%
\pgfpathlineto{\pgfqpoint{3.946595in}{0.735702in}}%
\pgfpathlineto{\pgfqpoint{3.957076in}{0.732214in}}%
\pgfpathlineto{\pgfqpoint{3.967556in}{0.719465in}}%
\pgfpathlineto{\pgfqpoint{3.978037in}{0.676228in}}%
\pgfpathlineto{\pgfqpoint{3.988518in}{0.697058in}}%
\pgfpathlineto{\pgfqpoint{4.009480in}{0.699684in}}%
\pgfpathlineto{\pgfqpoint{4.019961in}{0.717512in}}%
\pgfpathlineto{\pgfqpoint{4.030441in}{0.784117in}}%
\pgfpathlineto{\pgfqpoint{4.040922in}{0.765059in}}%
\pgfpathlineto{\pgfqpoint{4.051403in}{0.770879in}}%
\pgfpathlineto{\pgfqpoint{4.061884in}{0.735869in}}%
\pgfpathlineto{\pgfqpoint{4.072365in}{0.708120in}}%
\pgfpathlineto{\pgfqpoint{4.082846in}{0.745022in}}%
\pgfpathlineto{\pgfqpoint{4.093326in}{0.715992in}}%
\pgfpathlineto{\pgfqpoint{4.103807in}{0.655532in}}%
\pgfpathlineto{\pgfqpoint{4.114288in}{0.667994in}}%
\pgfpathlineto{\pgfqpoint{4.124769in}{0.717015in}}%
\pgfpathlineto{\pgfqpoint{4.135250in}{0.722305in}}%
\pgfpathlineto{\pgfqpoint{4.145731in}{0.735762in}}%
\pgfpathlineto{\pgfqpoint{4.156211in}{0.697575in}}%
\pgfpathlineto{\pgfqpoint{4.166692in}{0.706131in}}%
\pgfpathlineto{\pgfqpoint{4.177173in}{0.706142in}}%
\pgfpathlineto{\pgfqpoint{4.187654in}{0.747242in}}%
\pgfpathlineto{\pgfqpoint{4.198135in}{0.728122in}}%
\pgfpathlineto{\pgfqpoint{4.208616in}{0.698362in}}%
\pgfpathlineto{\pgfqpoint{4.219097in}{0.682759in}}%
\pgfpathlineto{\pgfqpoint{4.229577in}{0.739177in}}%
\pgfpathlineto{\pgfqpoint{4.240058in}{0.716313in}}%
\pgfpathlineto{\pgfqpoint{4.250539in}{0.716174in}}%
\pgfpathlineto{\pgfqpoint{4.261020in}{0.681931in}}%
\pgfpathlineto{\pgfqpoint{4.271501in}{0.635940in}}%
\pgfpathlineto{\pgfqpoint{4.281982in}{0.676450in}}%
\pgfpathlineto{\pgfqpoint{4.292462in}{0.651369in}}%
\pgfpathlineto{\pgfqpoint{4.302943in}{0.643790in}}%
\pgfpathlineto{\pgfqpoint{4.313424in}{0.651887in}}%
\pgfpathlineto{\pgfqpoint{4.323905in}{0.650328in}}%
\pgfpathlineto{\pgfqpoint{4.344867in}{0.664113in}}%
\pgfpathlineto{\pgfqpoint{4.355347in}{0.623455in}}%
\pgfpathlineto{\pgfqpoint{4.365828in}{0.642132in}}%
\pgfpathlineto{\pgfqpoint{4.376309in}{0.673260in}}%
\pgfpathlineto{\pgfqpoint{4.386790in}{0.671668in}}%
\pgfpathlineto{\pgfqpoint{4.397271in}{0.664242in}}%
\pgfpathlineto{\pgfqpoint{4.407752in}{0.708145in}}%
\pgfpathlineto{\pgfqpoint{4.418233in}{0.721952in}}%
\pgfpathlineto{\pgfqpoint{4.428713in}{0.685046in}}%
\pgfpathlineto{\pgfqpoint{4.439194in}{0.657387in}}%
\pgfpathlineto{\pgfqpoint{4.449675in}{0.647183in}}%
\pgfpathlineto{\pgfqpoint{4.460156in}{0.671054in}}%
\pgfpathlineto{\pgfqpoint{4.470637in}{0.679352in}}%
\pgfpathlineto{\pgfqpoint{4.481118in}{0.672002in}}%
\pgfpathlineto{\pgfqpoint{4.491598in}{0.680967in}}%
\pgfpathlineto{\pgfqpoint{4.502079in}{0.630190in}}%
\pgfpathlineto{\pgfqpoint{4.512560in}{0.665144in}}%
\pgfpathlineto{\pgfqpoint{4.523041in}{0.708405in}}%
\pgfpathlineto{\pgfqpoint{4.533522in}{0.700554in}}%
\pgfpathlineto{\pgfqpoint{4.544003in}{0.639075in}}%
\pgfpathlineto{\pgfqpoint{4.554483in}{0.632606in}}%
\pgfpathlineto{\pgfqpoint{4.564964in}{0.697402in}}%
\pgfpathlineto{\pgfqpoint{4.575445in}{0.723028in}}%
\pgfpathlineto{\pgfqpoint{4.585926in}{0.680387in}}%
\pgfpathlineto{\pgfqpoint{4.596407in}{0.689427in}}%
\pgfpathlineto{\pgfqpoint{4.606888in}{0.693120in}}%
\pgfpathlineto{\pgfqpoint{4.617369in}{0.662798in}}%
\pgfpathlineto{\pgfqpoint{4.627849in}{0.691112in}}%
\pgfpathlineto{\pgfqpoint{4.638330in}{0.685234in}}%
\pgfpathlineto{\pgfqpoint{4.648811in}{0.703381in}}%
\pgfpathlineto{\pgfqpoint{4.659292in}{0.703725in}}%
\pgfpathlineto{\pgfqpoint{4.669773in}{0.690039in}}%
\pgfpathlineto{\pgfqpoint{4.669773in}{0.690039in}}%
\pgfusepath{stroke}%
\end{pgfscope}%
\begin{pgfscope}%
\pgfsetrectcap%
\pgfsetmiterjoin%
\pgfsetlinewidth{0.803000pt}%
\definecolor{currentstroke}{rgb}{0.000000,0.000000,0.000000}%
\pgfsetstrokecolor{currentstroke}%
\pgfsetdash{}{0pt}%
\pgfpathmoveto{\pgfqpoint{0.675000in}{0.540000in}}%
\pgfpathlineto{\pgfqpoint{0.675000in}{2.376000in}}%
\pgfusepath{stroke}%
\end{pgfscope}%
\begin{pgfscope}%
\pgfsetrectcap%
\pgfsetmiterjoin%
\pgfsetlinewidth{0.803000pt}%
\definecolor{currentstroke}{rgb}{0.000000,0.000000,0.000000}%
\pgfsetstrokecolor{currentstroke}%
\pgfsetdash{}{0pt}%
\pgfpathmoveto{\pgfqpoint{4.860000in}{0.540000in}}%
\pgfpathlineto{\pgfqpoint{4.860000in}{2.376000in}}%
\pgfusepath{stroke}%
\end{pgfscope}%
\begin{pgfscope}%
\pgfsetrectcap%
\pgfsetmiterjoin%
\pgfsetlinewidth{0.803000pt}%
\definecolor{currentstroke}{rgb}{0.000000,0.000000,0.000000}%
\pgfsetstrokecolor{currentstroke}%
\pgfsetdash{}{0pt}%
\pgfpathmoveto{\pgfqpoint{0.675000in}{0.540000in}}%
\pgfpathlineto{\pgfqpoint{4.860000in}{0.540000in}}%
\pgfusepath{stroke}%
\end{pgfscope}%
\begin{pgfscope}%
\pgfsetrectcap%
\pgfsetmiterjoin%
\pgfsetlinewidth{0.803000pt}%
\definecolor{currentstroke}{rgb}{0.000000,0.000000,0.000000}%
\pgfsetstrokecolor{currentstroke}%
\pgfsetdash{}{0pt}%
\pgfpathmoveto{\pgfqpoint{0.675000in}{2.376000in}}%
\pgfpathlineto{\pgfqpoint{4.860000in}{2.376000in}}%
\pgfusepath{stroke}%
\end{pgfscope}%
\begin{pgfscope}%
\pgfsetbuttcap%
\pgfsetmiterjoin%
\pgfsetlinewidth{0.000000pt}%
\definecolor{currentstroke}{rgb}{0.800000,0.800000,0.800000}%
\pgfsetstrokecolor{currentstroke}%
\pgfsetstrokeopacity{0.000000}%
\pgfsetdash{}{0pt}%
\pgfpathmoveto{\pgfqpoint{3.283222in}{0.884844in}}%
\pgfpathlineto{\pgfqpoint{3.338778in}{0.884844in}}%
\pgfpathquadraticcurveto{\pgfqpoint{3.366556in}{0.884844in}}{\pgfqpoint{3.366556in}{0.912622in}}%
\pgfpathlineto{\pgfqpoint{3.366556in}{0.968178in}}%
\pgfpathquadraticcurveto{\pgfqpoint{3.366556in}{0.995956in}}{\pgfqpoint{3.338778in}{0.995956in}}%
\pgfpathlineto{\pgfqpoint{3.283222in}{0.995956in}}%
\pgfpathquadraticcurveto{\pgfqpoint{3.255444in}{0.995956in}}{\pgfqpoint{3.255444in}{0.968178in}}%
\pgfpathlineto{\pgfqpoint{3.255444in}{0.912622in}}%
\pgfpathquadraticcurveto{\pgfqpoint{3.255444in}{0.884844in}}{\pgfqpoint{3.283222in}{0.884844in}}%
\pgfpathclose%
\pgfusepath{}%
\end{pgfscope}%
\end{pgfpicture}%
\makeatother%
\endgroup%

%% file: active_miner_1yr.pgf
\begingroup%
\makeatletter%
\begin{pgfpicture}%
\pgfpathrectangle{\pgfpointorigin}{\pgfqpoint{5.400000in}{2.700000in}}%
\pgfusepath{use as bounding box, clip}%
\begin{pgfscope}%
\pgfsetbuttcap%
\pgfsetmiterjoin%
\definecolor{currentfill}{rgb}{1.000000,1.000000,1.000000}%
\pgfsetfillcolor{currentfill}%
\pgfsetlinewidth{0.000000pt}%
\definecolor{currentstroke}{rgb}{1.000000,1.000000,1.000000}%
\pgfsetstrokecolor{currentstroke}%
\pgfsetdash{}{0pt}%
\pgfpathmoveto{\pgfqpoint{0.000000in}{0.000000in}}%
\pgfpathlineto{\pgfqpoint{5.400000in}{0.000000in}}%
\pgfpathlineto{\pgfqpoint{5.400000in}{2.700000in}}%
\pgfpathlineto{\pgfqpoint{0.000000in}{2.700000in}}%
\pgfpathclose%
\pgfusepath{fill}%
\end{pgfscope}%
\begin{pgfscope}%
\pgfsetbuttcap%
\pgfsetmiterjoin%
\definecolor{currentfill}{rgb}{1.000000,1.000000,1.000000}%
\pgfsetfillcolor{currentfill}%
\pgfsetlinewidth{0.000000pt}%
\definecolor{currentstroke}{rgb}{0.000000,0.000000,0.000000}%
\pgfsetstrokecolor{currentstroke}%
\pgfsetstrokeopacity{0.000000}%
\pgfsetdash{}{0pt}%
\pgfpathmoveto{\pgfqpoint{0.675000in}{0.540000in}}%
\pgfpathlineto{\pgfqpoint{4.860000in}{0.540000in}}%
\pgfpathlineto{\pgfqpoint{4.860000in}{2.376000in}}%
\pgfpathlineto{\pgfqpoint{0.675000in}{2.376000in}}%
\pgfpathclose%
\pgfusepath{fill}%
\end{pgfscope}%
\begin{pgfscope}%
\pgfsetbuttcap%
\pgfsetroundjoin%
\definecolor{currentfill}{rgb}{0.000000,0.000000,0.000000}%
\pgfsetfillcolor{currentfill}%
\pgfsetlinewidth{0.803000pt}%
\definecolor{currentstroke}{rgb}{0.000000,0.000000,0.000000}%
\pgfsetstrokecolor{currentstroke}%
\pgfsetdash{}{0pt}%
\pgfsys@defobject{currentmarker}{\pgfqpoint{0.000000in}{0.000000in}}{\pgfqpoint{0.000000in}{0.048611in}}{%
\pgfpathmoveto{\pgfqpoint{0.000000in}{0.000000in}}%
\pgfpathlineto{\pgfqpoint{0.000000in}{0.048611in}}%
\pgfusepath{stroke,fill}%
}%
\begin{pgfscope}%
\pgfsys@transformshift{0.865227in}{0.540000in}%
\pgfsys@useobject{currentmarker}{}%
\end{pgfscope}%
\end{pgfscope}%
\begin{pgfscope}%
\pgftext[x=0.289141in,y=0.104678in,left,base,rotate=30.000000]{\rmfamily\fontsize{10.000000}{12.000000}\selectfont 2018-01-01}%
\end{pgfscope}%
\begin{pgfscope}%
\pgfsetbuttcap%
\pgfsetroundjoin%
\definecolor{currentfill}{rgb}{0.000000,0.000000,0.000000}%
\pgfsetfillcolor{currentfill}%
\pgfsetlinewidth{0.803000pt}%
\definecolor{currentstroke}{rgb}{0.000000,0.000000,0.000000}%
\pgfsetstrokecolor{currentstroke}%
\pgfsetdash{}{0pt}%
\pgfsys@defobject{currentmarker}{\pgfqpoint{0.000000in}{0.000000in}}{\pgfqpoint{0.000000in}{0.048611in}}{%
\pgfpathmoveto{\pgfqpoint{0.000000in}{0.000000in}}%
\pgfpathlineto{\pgfqpoint{0.000000in}{0.048611in}}%
\pgfusepath{stroke,fill}%
}%
\begin{pgfscope}%
\pgfsys@transformshift{1.501065in}{0.540000in}%
\pgfsys@useobject{currentmarker}{}%
\end{pgfscope}%
\end{pgfscope}%
\begin{pgfscope}%
\pgftext[x=0.924979in,y=0.104678in,left,base,rotate=30.000000]{\rmfamily\fontsize{10.000000}{12.000000}\selectfont 2018-03-03}%
\end{pgfscope}%
\begin{pgfscope}%
\pgfsetbuttcap%
\pgfsetroundjoin%
\definecolor{currentfill}{rgb}{0.000000,0.000000,0.000000}%
\pgfsetfillcolor{currentfill}%
\pgfsetlinewidth{0.803000pt}%
\definecolor{currentstroke}{rgb}{0.000000,0.000000,0.000000}%
\pgfsetstrokecolor{currentstroke}%
\pgfsetdash{}{0pt}%
\pgfsys@defobject{currentmarker}{\pgfqpoint{0.000000in}{0.000000in}}{\pgfqpoint{0.000000in}{0.048611in}}{%
\pgfpathmoveto{\pgfqpoint{0.000000in}{0.000000in}}%
\pgfpathlineto{\pgfqpoint{0.000000in}{0.048611in}}%
\pgfusepath{stroke,fill}%
}%
\begin{pgfscope}%
\pgfsys@transformshift{2.136903in}{0.540000in}%
\pgfsys@useobject{currentmarker}{}%
\end{pgfscope}%
\end{pgfscope}%
\begin{pgfscope}%
\pgftext[x=1.560816in,y=0.104678in,left,base,rotate=30.000000]{\rmfamily\fontsize{10.000000}{12.000000}\selectfont 2018-05-02}%
\end{pgfscope}%
\begin{pgfscope}%
\pgfsetbuttcap%
\pgfsetroundjoin%
\definecolor{currentfill}{rgb}{0.000000,0.000000,0.000000}%
\pgfsetfillcolor{currentfill}%
\pgfsetlinewidth{0.803000pt}%
\definecolor{currentstroke}{rgb}{0.000000,0.000000,0.000000}%
\pgfsetstrokecolor{currentstroke}%
\pgfsetdash{}{0pt}%
\pgfsys@defobject{currentmarker}{\pgfqpoint{0.000000in}{0.000000in}}{\pgfqpoint{0.000000in}{0.048611in}}{%
\pgfpathmoveto{\pgfqpoint{0.000000in}{0.000000in}}%
\pgfpathlineto{\pgfqpoint{0.000000in}{0.048611in}}%
\pgfusepath{stroke,fill}%
}%
\begin{pgfscope}%
\pgfsys@transformshift{2.772740in}{0.540000in}%
\pgfsys@useobject{currentmarker}{}%
\end{pgfscope}%
\end{pgfscope}%
\begin{pgfscope}%
\pgftext[x=2.196654in,y=0.104678in,left,base,rotate=30.000000]{\rmfamily\fontsize{10.000000}{12.000000}\selectfont 2018-07-02}%
\end{pgfscope}%
\begin{pgfscope}%
\pgfsetbuttcap%
\pgfsetroundjoin%
\definecolor{currentfill}{rgb}{0.000000,0.000000,0.000000}%
\pgfsetfillcolor{currentfill}%
\pgfsetlinewidth{0.803000pt}%
\definecolor{currentstroke}{rgb}{0.000000,0.000000,0.000000}%
\pgfsetstrokecolor{currentstroke}%
\pgfsetdash{}{0pt}%
\pgfsys@defobject{currentmarker}{\pgfqpoint{0.000000in}{0.000000in}}{\pgfqpoint{0.000000in}{0.048611in}}{%
\pgfpathmoveto{\pgfqpoint{0.000000in}{0.000000in}}%
\pgfpathlineto{\pgfqpoint{0.000000in}{0.048611in}}%
\pgfusepath{stroke,fill}%
}%
\begin{pgfscope}%
\pgfsys@transformshift{3.408578in}{0.540000in}%
\pgfsys@useobject{currentmarker}{}%
\end{pgfscope}%
\end{pgfscope}%
\begin{pgfscope}%
\pgftext[x=2.832492in,y=0.104678in,left,base,rotate=30.000000]{\rmfamily\fontsize{10.000000}{12.000000}\selectfont 2018-09-01}%
\end{pgfscope}%
\begin{pgfscope}%
\pgfsetbuttcap%
\pgfsetroundjoin%
\definecolor{currentfill}{rgb}{0.000000,0.000000,0.000000}%
\pgfsetfillcolor{currentfill}%
\pgfsetlinewidth{0.803000pt}%
\definecolor{currentstroke}{rgb}{0.000000,0.000000,0.000000}%
\pgfsetstrokecolor{currentstroke}%
\pgfsetdash{}{0pt}%
\pgfsys@defobject{currentmarker}{\pgfqpoint{0.000000in}{0.000000in}}{\pgfqpoint{0.000000in}{0.048611in}}{%
\pgfpathmoveto{\pgfqpoint{0.000000in}{0.000000in}}%
\pgfpathlineto{\pgfqpoint{0.000000in}{0.048611in}}%
\pgfusepath{stroke,fill}%
}%
\begin{pgfscope}%
\pgfsys@transformshift{4.044416in}{0.540000in}%
\pgfsys@useobject{currentmarker}{}%
\end{pgfscope}%
\end{pgfscope}%
\begin{pgfscope}%
\pgftext[x=3.468330in,y=0.104678in,left,base,rotate=30.000000]{\rmfamily\fontsize{10.000000}{12.000000}\selectfont 2018-10-31}%
\pgftext[x=4.104330in,y=0.104678in,left,base,rotate=30.000000]{\rmfamily\fontsize{10.000000}{12.000000}\selectfont 2018-12-31}%
\end{pgfscope}%
\begin{pgfscope}%
\pgfsetbuttcap%
\pgfsetroundjoin%
\definecolor{currentfill}{rgb}{0.000000,0.000000,0.000000}%
\pgfsetfillcolor{currentfill}%
\pgfsetlinewidth{0.803000pt}%
\definecolor{currentstroke}{rgb}{0.000000,0.000000,0.000000}%
\pgfsetstrokecolor{currentstroke}%
\pgfsetdash{}{0pt}%
\pgfsys@defobject{currentmarker}{\pgfqpoint{0.000000in}{0.000000in}}{\pgfqpoint{0.000000in}{0.048611in}}{%
\pgfpathmoveto{\pgfqpoint{0.000000in}{0.000000in}}%
\pgfpathlineto{\pgfqpoint{0.000000in}{0.048611in}}%
\pgfusepath{stroke,fill}%
}%
\begin{pgfscope}%
\pgfsys@transformshift{4.680254in}{0.540000in}%
\pgfsys@useobject{currentmarker}{}%
\end{pgfscope}%
\end{pgfscope}%
\begin{pgfscope}%
\pgfsetbuttcap%
\pgfsetroundjoin%
\definecolor{currentfill}{rgb}{0.000000,0.000000,0.000000}%
\pgfsetfillcolor{currentfill}%
\pgfsetlinewidth{0.803000pt}%
\definecolor{currentstroke}{rgb}{0.000000,0.000000,0.000000}%
\pgfsetstrokecolor{currentstroke}%
\pgfsetdash{}{0pt}%
\pgfsys@defobject{currentmarker}{\pgfqpoint{-0.048611in}{0.000000in}}{\pgfqpoint{0.000000in}{0.000000in}}{%
\pgfpathmoveto{\pgfqpoint{0.000000in}{0.000000in}}%
\pgfpathlineto{\pgfqpoint{-0.048611in}{0.000000in}}%
\pgfusepath{stroke,fill}%
}%
\begin{pgfscope}%
\pgfsys@transformshift{0.675000in}{0.623455in}%
\pgfsys@useobject{currentmarker}{}%
\end{pgfscope}%
\end{pgfscope}%
\begin{pgfscope}%
\pgftext[x=0.508333in,y=0.575237in,left,base]{\rmfamily\fontsize{10.000000}{12.000000}\selectfont \(\displaystyle 0\)}%
\end{pgfscope}%
\begin{pgfscope}%
\pgfsetbuttcap%
\pgfsetroundjoin%
\definecolor{currentfill}{rgb}{0.000000,0.000000,0.000000}%
\pgfsetfillcolor{currentfill}%
\pgfsetlinewidth{0.803000pt}%
\definecolor{currentstroke}{rgb}{0.000000,0.000000,0.000000}%
\pgfsetstrokecolor{currentstroke}%
\pgfsetdash{}{0pt}%
\pgfsys@defobject{currentmarker}{\pgfqpoint{-0.048611in}{0.000000in}}{\pgfqpoint{0.000000in}{0.000000in}}{%
\pgfpathmoveto{\pgfqpoint{0.000000in}{0.000000in}}%
\pgfpathlineto{\pgfqpoint{-0.048611in}{0.000000in}}%
\pgfusepath{stroke,fill}%
}%
\begin{pgfscope}%
\pgfsys@transformshift{0.675000in}{1.149590in}%
\pgfsys@useobject{currentmarker}{}%
\end{pgfscope}%
\end{pgfscope}%
\begin{pgfscope}%
\pgftext[x=0.299999in,y=1.101372in,left,base]{\rmfamily\fontsize{10.000000}{12.000000}\selectfont \(\displaystyle 1000\)}%
\end{pgfscope}%
\begin{pgfscope}%
\pgfsetbuttcap%
\pgfsetroundjoin%
\definecolor{currentfill}{rgb}{0.000000,0.000000,0.000000}%
\pgfsetfillcolor{currentfill}%
\pgfsetlinewidth{0.803000pt}%
\definecolor{currentstroke}{rgb}{0.000000,0.000000,0.000000}%
\pgfsetstrokecolor{currentstroke}%
\pgfsetdash{}{0pt}%
\pgfsys@defobject{currentmarker}{\pgfqpoint{-0.048611in}{0.000000in}}{\pgfqpoint{0.000000in}{0.000000in}}{%
\pgfpathmoveto{\pgfqpoint{0.000000in}{0.000000in}}%
\pgfpathlineto{\pgfqpoint{-0.048611in}{0.000000in}}%
\pgfusepath{stroke,fill}%
}%
\begin{pgfscope}%
\pgfsys@transformshift{0.675000in}{1.675726in}%
\pgfsys@useobject{currentmarker}{}%
\end{pgfscope}%
\end{pgfscope}%
\begin{pgfscope}%
\pgftext[x=0.299999in,y=1.627508in,left,base]{\rmfamily\fontsize{10.000000}{12.000000}\selectfont \(\displaystyle 2000\)}%
\end{pgfscope}%
\begin{pgfscope}%
\pgfsetbuttcap%
\pgfsetroundjoin%
\definecolor{currentfill}{rgb}{0.000000,0.000000,0.000000}%
\pgfsetfillcolor{currentfill}%
\pgfsetlinewidth{0.803000pt}%
\definecolor{currentstroke}{rgb}{0.000000,0.000000,0.000000}%
\pgfsetstrokecolor{currentstroke}%
\pgfsetdash{}{0pt}%
\pgfsys@defobject{currentmarker}{\pgfqpoint{-0.048611in}{0.000000in}}{\pgfqpoint{0.000000in}{0.000000in}}{%
\pgfpathmoveto{\pgfqpoint{0.000000in}{0.000000in}}%
\pgfpathlineto{\pgfqpoint{-0.048611in}{0.000000in}}%
\pgfusepath{stroke,fill}%
}%
\begin{pgfscope}%
\pgfsys@transformshift{0.675000in}{2.201862in}%
\pgfsys@useobject{currentmarker}{}%
\end{pgfscope}%
\end{pgfscope}%
\begin{pgfscope}%
\pgftext[x=0.299999in,y=2.153644in,left,base]{\rmfamily\fontsize{10.000000}{12.000000}\selectfont \(\displaystyle 3000\)}%
\end{pgfscope}%
\begin{pgfscope}%
\definecolor{textcolor}{rgb}{0.000000,0.000000,1.000000}%
\pgfsetstrokecolor{textcolor}%
\pgfsetfillcolor{textcolor}%
\pgftext[x=0.244444in,y=1.458000in,,bottom,rotate=90.000000]{\color{textcolor}\rmfamily\fontsize{10.000000}{12.000000}\selectfont Diversify USD/day earned}%
\end{pgfscope}%
\begin{pgfscope}%
\pgfpathrectangle{\pgfqpoint{0.675000in}{0.540000in}}{\pgfqpoint{4.185000in}{1.836000in}}%
\pgfusepath{clip}%
\pgfsetrectcap%
\pgfsetroundjoin%
\pgfsetlinewidth{1.003750pt}%
\definecolor{currentstroke}{rgb}{0.000000,0.000000,1.000000}%
\pgfsetstrokecolor{currentstroke}%
\pgfsetdash{}{0pt}%
\pgfpathmoveto{\pgfqpoint{0.865227in}{1.897189in}}%
\pgfpathlineto{\pgfqpoint{0.875708in}{1.806911in}}%
\pgfpathlineto{\pgfqpoint{0.886189in}{2.163753in}}%
\pgfpathlineto{\pgfqpoint{0.896670in}{1.838546in}}%
\pgfpathlineto{\pgfqpoint{0.907151in}{1.916684in}}%
\pgfpathlineto{\pgfqpoint{0.917631in}{2.017481in}}%
\pgfpathlineto{\pgfqpoint{0.928112in}{2.256006in}}%
\pgfpathlineto{\pgfqpoint{0.938593in}{2.292545in}}%
\pgfpathlineto{\pgfqpoint{0.949074in}{1.945021in}}%
\pgfpathlineto{\pgfqpoint{0.959555in}{2.021848in}}%
\pgfpathlineto{\pgfqpoint{0.970036in}{1.672551in}}%
\pgfpathlineto{\pgfqpoint{0.980517in}{2.096069in}}%
\pgfpathlineto{\pgfqpoint{0.990997in}{1.980930in}}%
\pgfpathlineto{\pgfqpoint{1.001478in}{1.837624in}}%
\pgfpathlineto{\pgfqpoint{1.011959in}{1.826152in}}%
\pgfpathlineto{\pgfqpoint{1.022440in}{1.690100in}}%
\pgfpathlineto{\pgfqpoint{1.032921in}{1.540237in}}%
\pgfpathlineto{\pgfqpoint{1.043402in}{1.480360in}}%
\pgfpathlineto{\pgfqpoint{1.053882in}{1.802545in}}%
\pgfpathlineto{\pgfqpoint{1.064363in}{1.784957in}}%
\pgfpathlineto{\pgfqpoint{1.074844in}{1.495926in}}%
\pgfpathlineto{\pgfqpoint{1.085325in}{1.469201in}}%
\pgfpathlineto{\pgfqpoint{1.095806in}{1.532919in}}%
\pgfpathlineto{\pgfqpoint{1.106287in}{1.616033in}}%
\pgfpathlineto{\pgfqpoint{1.116767in}{1.392878in}}%
\pgfpathlineto{\pgfqpoint{1.127248in}{1.528641in}}%
\pgfpathlineto{\pgfqpoint{1.137729in}{1.389759in}}%
\pgfpathlineto{\pgfqpoint{1.148210in}{1.445997in}}%
\pgfpathlineto{\pgfqpoint{1.158691in}{1.122793in}}%
\pgfpathlineto{\pgfqpoint{1.169172in}{1.287953in}}%
\pgfpathlineto{\pgfqpoint{1.179653in}{1.170844in}}%
\pgfpathlineto{\pgfqpoint{1.190133in}{0.996445in}}%
\pgfpathlineto{\pgfqpoint{1.200614in}{1.075567in}}%
\pgfpathlineto{\pgfqpoint{1.211095in}{1.189538in}}%
\pgfpathlineto{\pgfqpoint{1.221576in}{1.218724in}}%
\pgfpathlineto{\pgfqpoint{1.232057in}{1.173444in}}%
\pgfpathlineto{\pgfqpoint{1.242538in}{1.270501in}}%
\pgfpathlineto{\pgfqpoint{1.253018in}{1.164178in}}%
\pgfpathlineto{\pgfqpoint{1.263499in}{1.029243in}}%
\pgfpathlineto{\pgfqpoint{1.273980in}{1.053181in}}%
\pgfpathlineto{\pgfqpoint{1.284461in}{1.137140in}}%
\pgfpathlineto{\pgfqpoint{1.294942in}{1.229865in}}%
\pgfpathlineto{\pgfqpoint{1.305423in}{1.147670in}}%
\pgfpathlineto{\pgfqpoint{1.315903in}{1.218770in}}%
\pgfpathlineto{\pgfqpoint{1.326384in}{1.232129in}}%
\pgfpathlineto{\pgfqpoint{1.336865in}{1.199203in}}%
\pgfpathlineto{\pgfqpoint{1.347346in}{1.120275in}}%
\pgfpathlineto{\pgfqpoint{1.357827in}{1.131358in}}%
\pgfpathlineto{\pgfqpoint{1.368308in}{1.119890in}}%
\pgfpathlineto{\pgfqpoint{1.378789in}{1.270277in}}%
\pgfpathlineto{\pgfqpoint{1.389269in}{1.444889in}}%
\pgfpathlineto{\pgfqpoint{1.399750in}{1.172979in}}%
\pgfpathlineto{\pgfqpoint{1.410231in}{1.188379in}}%
\pgfpathlineto{\pgfqpoint{1.431193in}{1.148951in}}%
\pgfpathlineto{\pgfqpoint{1.441674in}{1.147275in}}%
\pgfpathlineto{\pgfqpoint{1.452154in}{1.120051in}}%
\pgfpathlineto{\pgfqpoint{1.462635in}{1.073274in}}%
\pgfpathlineto{\pgfqpoint{1.473116in}{1.204443in}}%
\pgfpathlineto{\pgfqpoint{1.483597in}{1.312083in}}%
\pgfpathlineto{\pgfqpoint{1.494078in}{1.307811in}}%
\pgfpathlineto{\pgfqpoint{1.504559in}{1.222700in}}%
\pgfpathlineto{\pgfqpoint{1.515039in}{1.475994in}}%
\pgfpathlineto{\pgfqpoint{1.525520in}{1.072707in}}%
\pgfpathlineto{\pgfqpoint{1.536001in}{1.218845in}}%
\pgfpathlineto{\pgfqpoint{1.546482in}{1.192888in}}%
\pgfpathlineto{\pgfqpoint{1.556963in}{1.104476in}}%
\pgfpathlineto{\pgfqpoint{1.567444in}{1.114743in}}%
\pgfpathlineto{\pgfqpoint{1.577924in}{1.053495in}}%
\pgfpathlineto{\pgfqpoint{1.588405in}{0.968719in}}%
\pgfpathlineto{\pgfqpoint{1.598886in}{1.075382in}}%
\pgfpathlineto{\pgfqpoint{1.609367in}{1.227811in}}%
\pgfpathlineto{\pgfqpoint{1.619848in}{1.005094in}}%
\pgfpathlineto{\pgfqpoint{1.630329in}{0.974953in}}%
\pgfpathlineto{\pgfqpoint{1.640810in}{1.164817in}}%
\pgfpathlineto{\pgfqpoint{1.651290in}{1.020135in}}%
\pgfpathlineto{\pgfqpoint{1.661771in}{0.934021in}}%
\pgfpathlineto{\pgfqpoint{1.672252in}{1.086908in}}%
\pgfpathlineto{\pgfqpoint{1.682733in}{1.090259in}}%
\pgfpathlineto{\pgfqpoint{1.693214in}{0.939742in}}%
\pgfpathlineto{\pgfqpoint{1.703695in}{1.020729in}}%
\pgfpathlineto{\pgfqpoint{1.714175in}{1.043201in}}%
\pgfpathlineto{\pgfqpoint{1.724656in}{1.009076in}}%
\pgfpathlineto{\pgfqpoint{1.735137in}{1.008043in}}%
\pgfpathlineto{\pgfqpoint{1.745618in}{0.997487in}}%
\pgfpathlineto{\pgfqpoint{1.756099in}{0.953396in}}%
\pgfpathlineto{\pgfqpoint{1.766580in}{0.968168in}}%
\pgfpathlineto{\pgfqpoint{1.777060in}{0.986237in}}%
\pgfpathlineto{\pgfqpoint{1.787541in}{0.866359in}}%
\pgfpathlineto{\pgfqpoint{1.798022in}{0.956616in}}%
\pgfpathlineto{\pgfqpoint{1.808503in}{1.671904in}}%
\pgfpathlineto{\pgfqpoint{1.818984in}{1.161867in}}%
\pgfpathlineto{\pgfqpoint{1.829465in}{1.351497in}}%
\pgfpathlineto{\pgfqpoint{1.839946in}{1.326311in}}%
\pgfpathlineto{\pgfqpoint{1.850426in}{1.347037in}}%
\pgfpathlineto{\pgfqpoint{1.860907in}{1.055403in}}%
\pgfpathlineto{\pgfqpoint{1.871388in}{1.172640in}}%
\pgfpathlineto{\pgfqpoint{1.881869in}{1.479052in}}%
\pgfpathlineto{\pgfqpoint{1.892350in}{1.067344in}}%
\pgfpathlineto{\pgfqpoint{1.902831in}{0.795838in}}%
\pgfpathlineto{\pgfqpoint{1.913311in}{0.951648in}}%
\pgfpathlineto{\pgfqpoint{1.923792in}{1.378461in}}%
\pgfpathlineto{\pgfqpoint{1.934273in}{0.783306in}}%
\pgfpathlineto{\pgfqpoint{1.944754in}{1.203788in}}%
\pgfpathlineto{\pgfqpoint{1.955235in}{1.004265in}}%
\pgfpathlineto{\pgfqpoint{1.965716in}{1.070864in}}%
\pgfpathlineto{\pgfqpoint{1.976196in}{0.988876in}}%
\pgfpathlineto{\pgfqpoint{1.986677in}{0.879674in}}%
\pgfpathlineto{\pgfqpoint{1.997158in}{0.887439in}}%
\pgfpathlineto{\pgfqpoint{2.007639in}{1.106149in}}%
\pgfpathlineto{\pgfqpoint{2.018120in}{1.103145in}}%
\pgfpathlineto{\pgfqpoint{2.028601in}{0.972491in}}%
\pgfpathlineto{\pgfqpoint{2.039082in}{0.773454in}}%
\pgfpathlineto{\pgfqpoint{2.049562in}{1.359059in}}%
\pgfpathlineto{\pgfqpoint{2.060043in}{0.975852in}}%
\pgfpathlineto{\pgfqpoint{2.070524in}{0.761286in}}%
\pgfpathlineto{\pgfqpoint{2.081005in}{0.927984in}}%
\pgfpathlineto{\pgfqpoint{2.091486in}{1.172155in}}%
\pgfpathlineto{\pgfqpoint{2.101967in}{0.904399in}}%
\pgfpathlineto{\pgfqpoint{2.112447in}{1.111316in}}%
\pgfpathlineto{\pgfqpoint{2.122928in}{0.766209in}}%
\pgfpathlineto{\pgfqpoint{2.133409in}{1.128072in}}%
\pgfpathlineto{\pgfqpoint{2.143890in}{1.053292in}}%
\pgfpathlineto{\pgfqpoint{2.154371in}{1.247252in}}%
\pgfpathlineto{\pgfqpoint{2.164852in}{1.043614in}}%
\pgfpathlineto{\pgfqpoint{2.175332in}{0.979253in}}%
\pgfpathlineto{\pgfqpoint{2.185813in}{1.265235in}}%
\pgfpathlineto{\pgfqpoint{2.196294in}{0.930670in}}%
\pgfpathlineto{\pgfqpoint{2.206775in}{1.172845in}}%
\pgfpathlineto{\pgfqpoint{2.217256in}{0.989008in}}%
\pgfpathlineto{\pgfqpoint{2.227737in}{1.056486in}}%
\pgfpathlineto{\pgfqpoint{2.248698in}{0.881153in}}%
\pgfpathlineto{\pgfqpoint{2.259179in}{0.881025in}}%
\pgfpathlineto{\pgfqpoint{2.269660in}{0.784336in}}%
\pgfpathlineto{\pgfqpoint{2.280141in}{1.009204in}}%
\pgfpathlineto{\pgfqpoint{2.290622in}{0.822744in}}%
\pgfpathlineto{\pgfqpoint{2.301103in}{0.822391in}}%
\pgfpathlineto{\pgfqpoint{2.311583in}{0.785270in}}%
\pgfpathlineto{\pgfqpoint{2.322064in}{1.197727in}}%
\pgfpathlineto{\pgfqpoint{2.332545in}{0.983157in}}%
\pgfpathlineto{\pgfqpoint{2.343026in}{0.733814in}}%
\pgfpathlineto{\pgfqpoint{2.353507in}{0.884466in}}%
\pgfpathlineto{\pgfqpoint{2.363988in}{0.990437in}}%
\pgfpathlineto{\pgfqpoint{2.374468in}{0.831569in}}%
\pgfpathlineto{\pgfqpoint{2.395430in}{0.946597in}}%
\pgfpathlineto{\pgfqpoint{2.405911in}{0.890116in}}%
\pgfpathlineto{\pgfqpoint{2.416392in}{0.789071in}}%
\pgfpathlineto{\pgfqpoint{2.426873in}{0.792631in}}%
\pgfpathlineto{\pgfqpoint{2.437353in}{0.979192in}}%
\pgfpathlineto{\pgfqpoint{2.447834in}{0.802625in}}%
\pgfpathlineto{\pgfqpoint{2.458315in}{0.926893in}}%
\pgfpathlineto{\pgfqpoint{2.468796in}{0.748121in}}%
\pgfpathlineto{\pgfqpoint{2.489758in}{1.044234in}}%
\pgfpathlineto{\pgfqpoint{2.500239in}{1.078706in}}%
\pgfpathlineto{\pgfqpoint{2.510719in}{0.746595in}}%
\pgfpathlineto{\pgfqpoint{2.521200in}{0.990828in}}%
\pgfpathlineto{\pgfqpoint{2.531681in}{1.031972in}}%
\pgfpathlineto{\pgfqpoint{2.542162in}{0.940938in}}%
\pgfpathlineto{\pgfqpoint{2.552643in}{0.826561in}}%
\pgfpathlineto{\pgfqpoint{2.563124in}{0.874339in}}%
\pgfpathlineto{\pgfqpoint{2.573604in}{0.811299in}}%
\pgfpathlineto{\pgfqpoint{2.584085in}{0.765766in}}%
\pgfpathlineto{\pgfqpoint{2.594566in}{0.863400in}}%
\pgfpathlineto{\pgfqpoint{2.605047in}{0.852689in}}%
\pgfpathlineto{\pgfqpoint{2.615528in}{0.896949in}}%
\pgfpathlineto{\pgfqpoint{2.626009in}{0.758819in}}%
\pgfpathlineto{\pgfqpoint{2.636489in}{0.718840in}}%
\pgfpathlineto{\pgfqpoint{2.646970in}{0.812776in}}%
\pgfpathlineto{\pgfqpoint{2.657451in}{0.818330in}}%
\pgfpathlineto{\pgfqpoint{2.667932in}{1.041398in}}%
\pgfpathlineto{\pgfqpoint{2.678413in}{0.973394in}}%
\pgfpathlineto{\pgfqpoint{2.688894in}{0.665137in}}%
\pgfpathlineto{\pgfqpoint{2.699375in}{0.924260in}}%
\pgfpathlineto{\pgfqpoint{2.709855in}{0.916170in}}%
\pgfpathlineto{\pgfqpoint{2.720336in}{0.829774in}}%
\pgfpathlineto{\pgfqpoint{2.730817in}{0.859429in}}%
\pgfpathlineto{\pgfqpoint{2.741298in}{0.810558in}}%
\pgfpathlineto{\pgfqpoint{2.751779in}{0.747274in}}%
\pgfpathlineto{\pgfqpoint{2.762260in}{0.789598in}}%
\pgfpathlineto{\pgfqpoint{2.772740in}{0.754249in}}%
\pgfpathlineto{\pgfqpoint{2.783221in}{0.873659in}}%
\pgfpathlineto{\pgfqpoint{2.793702in}{0.623455in}}%
\pgfpathlineto{\pgfqpoint{2.804183in}{0.896768in}}%
\pgfpathlineto{\pgfqpoint{2.814664in}{0.947633in}}%
\pgfpathlineto{\pgfqpoint{2.825145in}{0.814065in}}%
\pgfpathlineto{\pgfqpoint{2.835625in}{0.750360in}}%
\pgfpathlineto{\pgfqpoint{2.846106in}{0.785448in}}%
\pgfpathlineto{\pgfqpoint{2.856587in}{0.783617in}}%
\pgfpathlineto{\pgfqpoint{2.867068in}{0.747406in}}%
\pgfpathlineto{\pgfqpoint{2.877549in}{0.963793in}}%
\pgfpathlineto{\pgfqpoint{2.888030in}{0.818566in}}%
\pgfpathlineto{\pgfqpoint{2.898511in}{0.623455in}}%
\pgfpathlineto{\pgfqpoint{2.908991in}{0.749487in}}%
\pgfpathlineto{\pgfqpoint{2.919472in}{0.796161in}}%
\pgfpathlineto{\pgfqpoint{2.929953in}{0.715245in}}%
\pgfpathlineto{\pgfqpoint{2.940434in}{0.732524in}}%
\pgfpathlineto{\pgfqpoint{2.950915in}{0.781091in}}%
\pgfpathlineto{\pgfqpoint{2.961396in}{0.849224in}}%
\pgfpathlineto{\pgfqpoint{2.971876in}{0.927160in}}%
\pgfpathlineto{\pgfqpoint{2.982357in}{0.689163in}}%
\pgfpathlineto{\pgfqpoint{2.992838in}{0.954430in}}%
\pgfpathlineto{\pgfqpoint{3.003319in}{1.028892in}}%
\pgfpathlineto{\pgfqpoint{3.013800in}{1.089110in}}%
\pgfpathlineto{\pgfqpoint{3.024281in}{0.758655in}}%
\pgfpathlineto{\pgfqpoint{3.034761in}{1.178975in}}%
\pgfpathlineto{\pgfqpoint{3.045242in}{0.915909in}}%
\pgfpathlineto{\pgfqpoint{3.055723in}{1.075168in}}%
\pgfpathlineto{\pgfqpoint{3.066204in}{0.623455in}}%
\pgfpathlineto{\pgfqpoint{3.076685in}{0.799347in}}%
\pgfpathlineto{\pgfqpoint{3.087166in}{0.789979in}}%
\pgfpathlineto{\pgfqpoint{3.097647in}{0.787138in}}%
\pgfpathlineto{\pgfqpoint{3.108127in}{0.730923in}}%
\pgfpathlineto{\pgfqpoint{3.118608in}{0.783795in}}%
\pgfpathlineto{\pgfqpoint{3.129089in}{0.884240in}}%
\pgfpathlineto{\pgfqpoint{3.139570in}{0.997050in}}%
\pgfpathlineto{\pgfqpoint{3.160532in}{0.623455in}}%
\pgfpathlineto{\pgfqpoint{3.171012in}{0.818762in}}%
\pgfpathlineto{\pgfqpoint{3.181493in}{0.720685in}}%
\pgfpathlineto{\pgfqpoint{3.191974in}{0.777881in}}%
\pgfpathlineto{\pgfqpoint{3.202455in}{0.915551in}}%
\pgfpathlineto{\pgfqpoint{3.212936in}{0.826863in}}%
\pgfpathlineto{\pgfqpoint{3.223417in}{0.716630in}}%
\pgfpathlineto{\pgfqpoint{3.233897in}{0.723778in}}%
\pgfpathlineto{\pgfqpoint{3.244378in}{0.725915in}}%
\pgfpathlineto{\pgfqpoint{3.254859in}{0.780481in}}%
\pgfpathlineto{\pgfqpoint{3.265340in}{0.912694in}}%
\pgfpathlineto{\pgfqpoint{3.275821in}{0.895098in}}%
\pgfpathlineto{\pgfqpoint{3.286302in}{0.671926in}}%
\pgfpathlineto{\pgfqpoint{3.296782in}{0.830171in}}%
\pgfpathlineto{\pgfqpoint{3.307263in}{0.886345in}}%
\pgfpathlineto{\pgfqpoint{3.317744in}{0.673884in}}%
\pgfpathlineto{\pgfqpoint{3.328225in}{0.828020in}}%
\pgfpathlineto{\pgfqpoint{3.338706in}{0.821027in}}%
\pgfpathlineto{\pgfqpoint{3.349187in}{0.672282in}}%
\pgfpathlineto{\pgfqpoint{3.359668in}{0.873986in}}%
\pgfpathlineto{\pgfqpoint{3.370148in}{0.732735in}}%
\pgfpathlineto{\pgfqpoint{3.380629in}{0.785553in}}%
\pgfpathlineto{\pgfqpoint{3.391110in}{0.826897in}}%
\pgfpathlineto{\pgfqpoint{3.401591in}{0.907963in}}%
\pgfpathlineto{\pgfqpoint{3.412072in}{0.897747in}}%
\pgfpathlineto{\pgfqpoint{3.422553in}{0.860593in}}%
\pgfpathlineto{\pgfqpoint{3.433033in}{0.772846in}}%
\pgfpathlineto{\pgfqpoint{3.443514in}{0.856287in}}%
\pgfpathlineto{\pgfqpoint{3.453995in}{0.881474in}}%
\pgfpathlineto{\pgfqpoint{3.464476in}{0.815542in}}%
\pgfpathlineto{\pgfqpoint{3.474957in}{0.770190in}}%
\pgfpathlineto{\pgfqpoint{3.485438in}{0.827209in}}%
\pgfpathlineto{\pgfqpoint{3.506399in}{0.689479in}}%
\pgfpathlineto{\pgfqpoint{3.516880in}{0.691668in}}%
\pgfpathlineto{\pgfqpoint{3.527361in}{0.657737in}}%
\pgfpathlineto{\pgfqpoint{3.537842in}{0.762868in}}%
\pgfpathlineto{\pgfqpoint{3.548323in}{0.696362in}}%
\pgfpathlineto{\pgfqpoint{3.558804in}{0.623455in}}%
\pgfpathlineto{\pgfqpoint{3.579765in}{0.794975in}}%
\pgfpathlineto{\pgfqpoint{3.590246in}{0.788036in}}%
\pgfpathlineto{\pgfqpoint{3.600727in}{0.711366in}}%
\pgfpathlineto{\pgfqpoint{3.611208in}{0.769122in}}%
\pgfpathlineto{\pgfqpoint{3.621689in}{0.784005in}}%
\pgfpathlineto{\pgfqpoint{3.632169in}{0.834667in}}%
\pgfpathlineto{\pgfqpoint{3.642650in}{0.739672in}}%
\pgfpathlineto{\pgfqpoint{3.653131in}{0.685680in}}%
\pgfpathlineto{\pgfqpoint{3.663612in}{0.947078in}}%
\pgfpathlineto{\pgfqpoint{3.674093in}{0.784096in}}%
\pgfpathlineto{\pgfqpoint{3.684574in}{0.728650in}}%
\pgfpathlineto{\pgfqpoint{3.695054in}{0.680255in}}%
\pgfpathlineto{\pgfqpoint{3.705535in}{0.681920in}}%
\pgfpathlineto{\pgfqpoint{3.716016in}{0.834571in}}%
\pgfpathlineto{\pgfqpoint{3.726497in}{0.873776in}}%
\pgfpathlineto{\pgfqpoint{3.747459in}{0.777556in}}%
\pgfpathlineto{\pgfqpoint{3.757940in}{0.905303in}}%
\pgfpathlineto{\pgfqpoint{3.768420in}{0.803125in}}%
\pgfpathlineto{\pgfqpoint{3.778901in}{0.795180in}}%
\pgfpathlineto{\pgfqpoint{3.789382in}{0.712369in}}%
\pgfpathlineto{\pgfqpoint{3.799863in}{0.668844in}}%
\pgfpathlineto{\pgfqpoint{3.810344in}{0.753939in}}%
\pgfpathlineto{\pgfqpoint{3.820825in}{0.671495in}}%
\pgfpathlineto{\pgfqpoint{3.831305in}{0.718179in}}%
\pgfpathlineto{\pgfqpoint{3.841786in}{0.880860in}}%
\pgfpathlineto{\pgfqpoint{3.852267in}{0.746187in}}%
\pgfpathlineto{\pgfqpoint{3.862748in}{0.703817in}}%
\pgfpathlineto{\pgfqpoint{3.873229in}{0.746215in}}%
\pgfpathlineto{\pgfqpoint{3.883710in}{0.752881in}}%
\pgfpathlineto{\pgfqpoint{3.894190in}{0.712620in}}%
\pgfpathlineto{\pgfqpoint{3.904671in}{0.836288in}}%
\pgfpathlineto{\pgfqpoint{3.915152in}{0.715965in}}%
\pgfpathlineto{\pgfqpoint{3.925633in}{0.813572in}}%
\pgfpathlineto{\pgfqpoint{3.936114in}{0.771326in}}%
\pgfpathlineto{\pgfqpoint{3.946595in}{0.674193in}}%
\pgfpathlineto{\pgfqpoint{3.957076in}{0.769025in}}%
\pgfpathlineto{\pgfqpoint{3.967556in}{0.723788in}}%
\pgfpathlineto{\pgfqpoint{3.978037in}{0.673444in}}%
\pgfpathlineto{\pgfqpoint{3.988518in}{0.769056in}}%
\pgfpathlineto{\pgfqpoint{3.998999in}{0.677392in}}%
\pgfpathlineto{\pgfqpoint{4.009480in}{0.834032in}}%
\pgfpathlineto{\pgfqpoint{4.019961in}{0.822519in}}%
\pgfpathlineto{\pgfqpoint{4.030441in}{0.726166in}}%
\pgfpathlineto{\pgfqpoint{4.040922in}{0.727572in}}%
\pgfpathlineto{\pgfqpoint{4.051403in}{1.070243in}}%
\pgfpathlineto{\pgfqpoint{4.061884in}{0.890850in}}%
\pgfpathlineto{\pgfqpoint{4.072365in}{0.833396in}}%
\pgfpathlineto{\pgfqpoint{4.082846in}{0.904941in}}%
\pgfpathlineto{\pgfqpoint{4.093326in}{0.743526in}}%
\pgfpathlineto{\pgfqpoint{4.103807in}{0.739491in}}%
\pgfpathlineto{\pgfqpoint{4.114288in}{0.702229in}}%
\pgfpathlineto{\pgfqpoint{4.124769in}{0.808746in}}%
\pgfpathlineto{\pgfqpoint{4.135250in}{0.695564in}}%
\pgfpathlineto{\pgfqpoint{4.145731in}{0.730510in}}%
\pgfpathlineto{\pgfqpoint{4.156211in}{0.659013in}}%
\pgfpathlineto{\pgfqpoint{4.166692in}{0.799271in}}%
\pgfpathlineto{\pgfqpoint{4.177173in}{0.730815in}}%
\pgfpathlineto{\pgfqpoint{4.187654in}{0.757323in}}%
\pgfpathlineto{\pgfqpoint{4.198135in}{0.715582in}}%
\pgfpathlineto{\pgfqpoint{4.208616in}{0.777639in}}%
\pgfpathlineto{\pgfqpoint{4.219097in}{0.777110in}}%
\pgfpathlineto{\pgfqpoint{4.229577in}{0.789405in}}%
\pgfpathlineto{\pgfqpoint{4.240058in}{0.786024in}}%
\pgfpathlineto{\pgfqpoint{4.250539in}{0.765953in}}%
\pgfpathlineto{\pgfqpoint{4.261020in}{0.623455in}}%
\pgfpathlineto{\pgfqpoint{4.281982in}{0.731196in}}%
\pgfpathlineto{\pgfqpoint{4.292462in}{0.700726in}}%
\pgfpathlineto{\pgfqpoint{4.302943in}{0.804067in}}%
\pgfpathlineto{\pgfqpoint{4.313424in}{0.759680in}}%
\pgfpathlineto{\pgfqpoint{4.323905in}{0.663317in}}%
\pgfpathlineto{\pgfqpoint{4.334386in}{0.664811in}}%
\pgfpathlineto{\pgfqpoint{4.344867in}{0.690490in}}%
\pgfpathlineto{\pgfqpoint{4.355347in}{0.709972in}}%
\pgfpathlineto{\pgfqpoint{4.365828in}{0.708602in}}%
\pgfpathlineto{\pgfqpoint{4.376309in}{0.710603in}}%
\pgfpathlineto{\pgfqpoint{4.386790in}{0.646119in}}%
\pgfpathlineto{\pgfqpoint{4.397271in}{0.719459in}}%
\pgfpathlineto{\pgfqpoint{4.407752in}{0.678545in}}%
\pgfpathlineto{\pgfqpoint{4.418233in}{0.649756in}}%
\pgfpathlineto{\pgfqpoint{4.428713in}{0.673290in}}%
\pgfpathlineto{\pgfqpoint{4.439194in}{0.672909in}}%
\pgfpathlineto{\pgfqpoint{4.449675in}{0.675045in}}%
\pgfpathlineto{\pgfqpoint{4.460156in}{0.653281in}}%
\pgfpathlineto{\pgfqpoint{4.470637in}{0.748836in}}%
\pgfpathlineto{\pgfqpoint{4.481118in}{0.768805in}}%
\pgfpathlineto{\pgfqpoint{4.491598in}{0.733828in}}%
\pgfpathlineto{\pgfqpoint{4.502079in}{0.753190in}}%
\pgfpathlineto{\pgfqpoint{4.512560in}{0.702834in}}%
\pgfpathlineto{\pgfqpoint{4.523041in}{0.804811in}}%
\pgfpathlineto{\pgfqpoint{4.533522in}{0.728891in}}%
\pgfpathlineto{\pgfqpoint{4.544003in}{0.820887in}}%
\pgfpathlineto{\pgfqpoint{4.554483in}{0.704606in}}%
\pgfpathlineto{\pgfqpoint{4.564964in}{0.852424in}}%
\pgfpathlineto{\pgfqpoint{4.575445in}{0.833723in}}%
\pgfpathlineto{\pgfqpoint{4.585926in}{0.675573in}}%
\pgfpathlineto{\pgfqpoint{4.596407in}{0.750941in}}%
\pgfpathlineto{\pgfqpoint{4.606888in}{0.844734in}}%
\pgfpathlineto{\pgfqpoint{4.617369in}{0.670915in}}%
\pgfpathlineto{\pgfqpoint{4.627849in}{0.737248in}}%
\pgfpathlineto{\pgfqpoint{4.638330in}{0.716217in}}%
\pgfpathlineto{\pgfqpoint{4.648811in}{0.842571in}}%
\pgfpathlineto{\pgfqpoint{4.659292in}{0.778860in}}%
\pgfpathlineto{\pgfqpoint{4.669773in}{0.846654in}}%
\pgfpathlineto{\pgfqpoint{4.669773in}{0.846654in}}%
\pgfusepath{stroke}%
\end{pgfscope}%
\begin{pgfscope}%
\pgfsetrectcap%
\pgfsetmiterjoin%
\pgfsetlinewidth{0.803000pt}%
\definecolor{currentstroke}{rgb}{0.000000,0.000000,0.000000}%
\pgfsetstrokecolor{currentstroke}%
\pgfsetdash{}{0pt}%
\pgfpathmoveto{\pgfqpoint{0.675000in}{0.540000in}}%
\pgfpathlineto{\pgfqpoint{0.675000in}{2.376000in}}%
\pgfusepath{stroke}%
\end{pgfscope}%
\begin{pgfscope}%
\pgfsetrectcap%
\pgfsetmiterjoin%
\pgfsetlinewidth{0.803000pt}%
\definecolor{currentstroke}{rgb}{0.000000,0.000000,0.000000}%
\pgfsetstrokecolor{currentstroke}%
\pgfsetdash{}{0pt}%
\pgfpathmoveto{\pgfqpoint{4.860000in}{0.540000in}}%
\pgfpathlineto{\pgfqpoint{4.860000in}{2.376000in}}%
\pgfusepath{stroke}%
\end{pgfscope}%
\begin{pgfscope}%
\pgfsetrectcap%
\pgfsetmiterjoin%
\pgfsetlinewidth{0.803000pt}%
\definecolor{currentstroke}{rgb}{0.000000,0.000000,0.000000}%
\pgfsetstrokecolor{currentstroke}%
\pgfsetdash{}{0pt}%
\pgfpathmoveto{\pgfqpoint{0.675000in}{0.540000in}}%
\pgfpathlineto{\pgfqpoint{4.860000in}{0.540000in}}%
\pgfusepath{stroke}%
\end{pgfscope}%
\begin{pgfscope}%
\pgfsetrectcap%
\pgfsetmiterjoin%
\pgfsetlinewidth{0.803000pt}%
\definecolor{currentstroke}{rgb}{0.000000,0.000000,0.000000}%
\pgfsetstrokecolor{currentstroke}%
\pgfsetdash{}{0pt}%
\pgfpathmoveto{\pgfqpoint{0.675000in}{2.376000in}}%
\pgfpathlineto{\pgfqpoint{4.860000in}{2.376000in}}%
\pgfusepath{stroke}%
\end{pgfscope}%
\begin{pgfscope}%
\pgfsetbuttcap%
\pgfsetroundjoin%
\definecolor{currentfill}{rgb}{0.000000,0.000000,0.000000}%
\pgfsetfillcolor{currentfill}%
\pgfsetlinewidth{0.803000pt}%
\definecolor{currentstroke}{rgb}{0.000000,0.000000,0.000000}%
\pgfsetstrokecolor{currentstroke}%
\pgfsetdash{}{0pt}%
\pgfsys@defobject{currentmarker}{\pgfqpoint{0.000000in}{0.000000in}}{\pgfqpoint{0.048611in}{0.000000in}}{%
\pgfpathmoveto{\pgfqpoint{0.000000in}{0.000000in}}%
\pgfpathlineto{\pgfqpoint{0.048611in}{0.000000in}}%
\pgfusepath{stroke,fill}%
}%
\begin{pgfscope}%
\pgfsys@transformshift{4.860000in}{0.623455in}%
\pgfsys@useobject{currentmarker}{}%
\end{pgfscope}%
\end{pgfscope}%
\begin{pgfscope}%
\pgftext[x=4.957222in,y=0.575237in,left,base]{\rmfamily\fontsize{10.000000}{12.000000}\selectfont \(\displaystyle 0.00\)}%
\end{pgfscope}%
\begin{pgfscope}%
\pgfsetbuttcap%
\pgfsetroundjoin%
\definecolor{currentfill}{rgb}{0.000000,0.000000,0.000000}%
\pgfsetfillcolor{currentfill}%
\pgfsetlinewidth{0.803000pt}%
\definecolor{currentstroke}{rgb}{0.000000,0.000000,0.000000}%
\pgfsetstrokecolor{currentstroke}%
\pgfsetdash{}{0pt}%
\pgfsys@defobject{currentmarker}{\pgfqpoint{0.000000in}{0.000000in}}{\pgfqpoint{0.048611in}{0.000000in}}{%
\pgfpathmoveto{\pgfqpoint{0.000000in}{0.000000in}}%
\pgfpathlineto{\pgfqpoint{0.048611in}{0.000000in}}%
\pgfusepath{stroke,fill}%
}%
\begin{pgfscope}%
\pgfsys@transformshift{4.860000in}{0.971182in}%
\pgfsys@useobject{currentmarker}{}%
\end{pgfscope}%
\end{pgfscope}%
\begin{pgfscope}%
\pgftext[x=4.957222in,y=0.922964in,left,base]{\rmfamily\fontsize{10.000000}{12.000000}\selectfont \(\displaystyle 0.25\)}%
\end{pgfscope}%
\begin{pgfscope}%
\pgfsetbuttcap%
\pgfsetroundjoin%
\definecolor{currentfill}{rgb}{0.000000,0.000000,0.000000}%
\pgfsetfillcolor{currentfill}%
\pgfsetlinewidth{0.803000pt}%
\definecolor{currentstroke}{rgb}{0.000000,0.000000,0.000000}%
\pgfsetstrokecolor{currentstroke}%
\pgfsetdash{}{0pt}%
\pgfsys@defobject{currentmarker}{\pgfqpoint{0.000000in}{0.000000in}}{\pgfqpoint{0.048611in}{0.000000in}}{%
\pgfpathmoveto{\pgfqpoint{0.000000in}{0.000000in}}%
\pgfpathlineto{\pgfqpoint{0.048611in}{0.000000in}}%
\pgfusepath{stroke,fill}%
}%
\begin{pgfscope}%
\pgfsys@transformshift{4.860000in}{1.318909in}%
\pgfsys@useobject{currentmarker}{}%
\end{pgfscope}%
\end{pgfscope}%
\begin{pgfscope}%
\pgftext[x=4.957222in,y=1.270691in,left,base]{\rmfamily\fontsize{10.000000}{12.000000}\selectfont \(\displaystyle 0.50\)}%
\end{pgfscope}%
\begin{pgfscope}%
\pgfsetbuttcap%
\pgfsetroundjoin%
\definecolor{currentfill}{rgb}{0.000000,0.000000,0.000000}%
\pgfsetfillcolor{currentfill}%
\pgfsetlinewidth{0.803000pt}%
\definecolor{currentstroke}{rgb}{0.000000,0.000000,0.000000}%
\pgfsetstrokecolor{currentstroke}%
\pgfsetdash{}{0pt}%
\pgfsys@defobject{currentmarker}{\pgfqpoint{0.000000in}{0.000000in}}{\pgfqpoint{0.048611in}{0.000000in}}{%
\pgfpathmoveto{\pgfqpoint{0.000000in}{0.000000in}}%
\pgfpathlineto{\pgfqpoint{0.048611in}{0.000000in}}%
\pgfusepath{stroke,fill}%
}%
\begin{pgfscope}%
\pgfsys@transformshift{4.860000in}{1.666636in}%
\pgfsys@useobject{currentmarker}{}%
\end{pgfscope}%
\end{pgfscope}%
\begin{pgfscope}%
\pgftext[x=4.957222in,y=1.618419in,left,base]{\rmfamily\fontsize{10.000000}{12.000000}\selectfont \(\displaystyle 0.75\)}%
\end{pgfscope}%
\begin{pgfscope}%
\pgfsetbuttcap%
\pgfsetroundjoin%
\definecolor{currentfill}{rgb}{0.000000,0.000000,0.000000}%
\pgfsetfillcolor{currentfill}%
\pgfsetlinewidth{0.803000pt}%
\definecolor{currentstroke}{rgb}{0.000000,0.000000,0.000000}%
\pgfsetstrokecolor{currentstroke}%
\pgfsetdash{}{0pt}%
\pgfsys@defobject{currentmarker}{\pgfqpoint{0.000000in}{0.000000in}}{\pgfqpoint{0.048611in}{0.000000in}}{%
\pgfpathmoveto{\pgfqpoint{0.000000in}{0.000000in}}%
\pgfpathlineto{\pgfqpoint{0.048611in}{0.000000in}}%
\pgfusepath{stroke,fill}%
}%
\begin{pgfscope}%
\pgfsys@transformshift{4.860000in}{2.014364in}%
\pgfsys@useobject{currentmarker}{}%
\end{pgfscope}%
\end{pgfscope}%
\begin{pgfscope}%
\pgftext[x=4.957222in,y=1.966146in,left,base]{\rmfamily\fontsize{10.000000}{12.000000}\selectfont \(\displaystyle 1.00\)}%
\end{pgfscope}%
\begin{pgfscope}%
\pgfsetbuttcap%
\pgfsetroundjoin%
\definecolor{currentfill}{rgb}{0.000000,0.000000,0.000000}%
\pgfsetfillcolor{currentfill}%
\pgfsetlinewidth{0.803000pt}%
\definecolor{currentstroke}{rgb}{0.000000,0.000000,0.000000}%
\pgfsetstrokecolor{currentstroke}%
\pgfsetdash{}{0pt}%
\pgfsys@defobject{currentmarker}{\pgfqpoint{0.000000in}{0.000000in}}{\pgfqpoint{0.048611in}{0.000000in}}{%
\pgfpathmoveto{\pgfqpoint{0.000000in}{0.000000in}}%
\pgfpathlineto{\pgfqpoint{0.048611in}{0.000000in}}%
\pgfusepath{stroke,fill}%
}%
\begin{pgfscope}%
\pgfsys@transformshift{4.860000in}{2.362091in}%
\pgfsys@useobject{currentmarker}{}%
\end{pgfscope}%
\end{pgfscope}%
\begin{pgfscope}%
\pgftext[x=4.957222in,y=2.313873in,left,base]{\rmfamily\fontsize{10.000000}{12.000000}\selectfont \(\displaystyle 1.25\)}%
\end{pgfscope}%
\begin{pgfscope}%
\pgftext[x=5.259692in,y=1.458000in,,top,rotate=90.000000]{\rmfamily\fontsize{10.000000}{12.000000}\selectfont Pool distribution}%
\end{pgfscope}%
\begin{pgfscope}%
\pgftext[x=4.860000in,y=2.417667in,right,base]{\rmfamily\fontsize{10.000000}{12.000000}\selectfont \(\displaystyle \times10^{15}\)}%
\end{pgfscope}%
\begin{pgfscope}%
\pgfpathrectangle{\pgfqpoint{0.675000in}{0.540000in}}{\pgfqpoint{4.185000in}{1.836000in}}%
\pgfusepath{clip}%
\pgfsetbuttcap%
\pgfsetroundjoin%
\pgfsetlinewidth{0.501875pt}%
\definecolor{currentstroke}{rgb}{0.121569,0.466667,0.705882}%
\pgfsetstrokecolor{currentstroke}%
\pgfsetdash{{1.850000pt}{0.800000pt}}{0.000000pt}%
\pgfpathmoveto{\pgfqpoint{0.865227in}{1.624754in}}%
\pgfpathlineto{\pgfqpoint{0.886189in}{1.624754in}}%
\pgfpathlineto{\pgfqpoint{0.896670in}{1.455354in}}%
\pgfpathlineto{\pgfqpoint{0.917631in}{1.455354in}}%
\pgfpathlineto{\pgfqpoint{0.928112in}{1.458225in}}%
\pgfpathlineto{\pgfqpoint{0.949074in}{1.458225in}}%
\pgfpathlineto{\pgfqpoint{0.959555in}{1.467610in}}%
\pgfpathlineto{\pgfqpoint{0.980517in}{1.467610in}}%
\pgfpathlineto{\pgfqpoint{0.990997in}{1.446172in}}%
\pgfpathlineto{\pgfqpoint{1.011959in}{1.446172in}}%
\pgfpathlineto{\pgfqpoint{1.022440in}{1.402158in}}%
\pgfpathlineto{\pgfqpoint{1.043402in}{1.402158in}}%
\pgfpathlineto{\pgfqpoint{1.053882in}{1.405033in}}%
\pgfpathlineto{\pgfqpoint{1.074844in}{1.405033in}}%
\pgfpathlineto{\pgfqpoint{1.085325in}{1.400863in}}%
\pgfpathlineto{\pgfqpoint{1.106287in}{1.400863in}}%
\pgfpathlineto{\pgfqpoint{1.116767in}{1.387188in}}%
\pgfpathlineto{\pgfqpoint{1.137729in}{1.387188in}}%
\pgfpathlineto{\pgfqpoint{1.148210in}{1.402767in}}%
\pgfpathlineto{\pgfqpoint{1.169172in}{1.402767in}}%
\pgfpathlineto{\pgfqpoint{1.179653in}{1.420885in}}%
\pgfpathlineto{\pgfqpoint{1.200614in}{1.420885in}}%
\pgfpathlineto{\pgfqpoint{1.211095in}{1.415969in}}%
\pgfpathlineto{\pgfqpoint{1.232057in}{1.415969in}}%
\pgfpathlineto{\pgfqpoint{1.242538in}{1.425128in}}%
\pgfpathlineto{\pgfqpoint{1.294942in}{1.425700in}}%
\pgfpathlineto{\pgfqpoint{1.305423in}{1.404589in}}%
\pgfpathlineto{\pgfqpoint{1.326384in}{1.404589in}}%
\pgfpathlineto{\pgfqpoint{1.336865in}{1.393163in}}%
\pgfpathlineto{\pgfqpoint{1.357827in}{1.393163in}}%
\pgfpathlineto{\pgfqpoint{1.368308in}{1.398506in}}%
\pgfpathlineto{\pgfqpoint{1.389269in}{1.398506in}}%
\pgfpathlineto{\pgfqpoint{1.399750in}{1.387831in}}%
\pgfpathlineto{\pgfqpoint{1.420712in}{1.387831in}}%
\pgfpathlineto{\pgfqpoint{1.431193in}{1.382854in}}%
\pgfpathlineto{\pgfqpoint{1.452154in}{1.382854in}}%
\pgfpathlineto{\pgfqpoint{1.462635in}{1.399468in}}%
\pgfpathlineto{\pgfqpoint{1.483597in}{1.399468in}}%
\pgfpathlineto{\pgfqpoint{1.494078in}{1.431746in}}%
\pgfpathlineto{\pgfqpoint{1.515039in}{1.431746in}}%
\pgfpathlineto{\pgfqpoint{1.525520in}{1.442777in}}%
\pgfpathlineto{\pgfqpoint{1.546482in}{1.442777in}}%
\pgfpathlineto{\pgfqpoint{1.556963in}{1.422936in}}%
\pgfpathlineto{\pgfqpoint{1.577924in}{1.422936in}}%
\pgfpathlineto{\pgfqpoint{1.588405in}{1.424099in}}%
\pgfpathlineto{\pgfqpoint{1.609367in}{1.424099in}}%
\pgfpathlineto{\pgfqpoint{1.619848in}{1.418309in}}%
\pgfpathlineto{\pgfqpoint{1.640810in}{1.418309in}}%
\pgfpathlineto{\pgfqpoint{1.651290in}{1.420579in}}%
\pgfpathlineto{\pgfqpoint{1.672252in}{1.420579in}}%
\pgfpathlineto{\pgfqpoint{1.682733in}{1.425394in}}%
\pgfpathlineto{\pgfqpoint{1.703695in}{1.425394in}}%
\pgfpathlineto{\pgfqpoint{1.714175in}{1.454397in}}%
\pgfpathlineto{\pgfqpoint{1.735137in}{1.454397in}}%
\pgfpathlineto{\pgfqpoint{1.745618in}{1.447752in}}%
\pgfpathlineto{\pgfqpoint{1.766580in}{1.447752in}}%
\pgfpathlineto{\pgfqpoint{1.777060in}{1.496256in}}%
\pgfpathlineto{\pgfqpoint{1.798022in}{1.496256in}}%
\pgfpathlineto{\pgfqpoint{1.808503in}{1.312042in}}%
\pgfpathlineto{\pgfqpoint{1.829465in}{1.312042in}}%
\pgfpathlineto{\pgfqpoint{1.839946in}{1.039280in}}%
\pgfpathlineto{\pgfqpoint{1.860907in}{1.039280in}}%
\pgfpathlineto{\pgfqpoint{1.871388in}{0.733997in}}%
\pgfpathlineto{\pgfqpoint{1.892350in}{0.733997in}}%
\pgfpathlineto{\pgfqpoint{1.902831in}{0.623455in}}%
\pgfpathlineto{\pgfqpoint{4.669773in}{0.623455in}}%
\pgfpathlineto{\pgfqpoint{4.669773in}{0.623455in}}%
\pgfusepath{stroke}%
\end{pgfscope}%
\begin{pgfscope}%
\pgfpathrectangle{\pgfqpoint{0.675000in}{0.540000in}}{\pgfqpoint{4.185000in}{1.836000in}}%
\pgfusepath{clip}%
\pgfsetbuttcap%
\pgfsetroundjoin%
\pgfsetlinewidth{0.501875pt}%
\definecolor{currentstroke}{rgb}{1.000000,0.498039,0.054902}%
\pgfsetstrokecolor{currentstroke}%
\pgfsetdash{{1.850000pt}{0.800000pt}}{0.000000pt}%
\pgfpathmoveto{\pgfqpoint{0.865227in}{1.291246in}}%
\pgfpathlineto{\pgfqpoint{0.886189in}{1.291246in}}%
\pgfpathlineto{\pgfqpoint{0.896670in}{1.460646in}}%
\pgfpathlineto{\pgfqpoint{0.917631in}{1.460646in}}%
\pgfpathlineto{\pgfqpoint{0.928112in}{1.457775in}}%
\pgfpathlineto{\pgfqpoint{0.949074in}{1.457775in}}%
\pgfpathlineto{\pgfqpoint{0.959555in}{1.448390in}}%
\pgfpathlineto{\pgfqpoint{0.980517in}{1.448390in}}%
\pgfpathlineto{\pgfqpoint{0.990997in}{1.469828in}}%
\pgfpathlineto{\pgfqpoint{1.011959in}{1.469828in}}%
\pgfpathlineto{\pgfqpoint{1.022440in}{1.513842in}}%
\pgfpathlineto{\pgfqpoint{1.043402in}{1.513842in}}%
\pgfpathlineto{\pgfqpoint{1.053882in}{1.510967in}}%
\pgfpathlineto{\pgfqpoint{1.074844in}{1.510967in}}%
\pgfpathlineto{\pgfqpoint{1.085325in}{1.515137in}}%
\pgfpathlineto{\pgfqpoint{1.106287in}{1.515137in}}%
\pgfpathlineto{\pgfqpoint{1.116767in}{1.528812in}}%
\pgfpathlineto{\pgfqpoint{1.137729in}{1.528812in}}%
\pgfpathlineto{\pgfqpoint{1.148210in}{1.513233in}}%
\pgfpathlineto{\pgfqpoint{1.169172in}{1.513233in}}%
\pgfpathlineto{\pgfqpoint{1.179653in}{1.495115in}}%
\pgfpathlineto{\pgfqpoint{1.200614in}{1.495115in}}%
\pgfpathlineto{\pgfqpoint{1.211095in}{1.500031in}}%
\pgfpathlineto{\pgfqpoint{1.232057in}{1.500031in}}%
\pgfpathlineto{\pgfqpoint{1.242538in}{1.490872in}}%
\pgfpathlineto{\pgfqpoint{1.294942in}{1.490300in}}%
\pgfpathlineto{\pgfqpoint{1.305423in}{1.511411in}}%
\pgfpathlineto{\pgfqpoint{1.326384in}{1.511411in}}%
\pgfpathlineto{\pgfqpoint{1.336865in}{1.522837in}}%
\pgfpathlineto{\pgfqpoint{1.357827in}{1.522837in}}%
\pgfpathlineto{\pgfqpoint{1.368308in}{1.517494in}}%
\pgfpathlineto{\pgfqpoint{1.389269in}{1.517494in}}%
\pgfpathlineto{\pgfqpoint{1.399750in}{1.528169in}}%
\pgfpathlineto{\pgfqpoint{1.420712in}{1.528169in}}%
\pgfpathlineto{\pgfqpoint{1.431193in}{1.533146in}}%
\pgfpathlineto{\pgfqpoint{1.452154in}{1.533146in}}%
\pgfpathlineto{\pgfqpoint{1.462635in}{1.516532in}}%
\pgfpathlineto{\pgfqpoint{1.483597in}{1.516532in}}%
\pgfpathlineto{\pgfqpoint{1.494078in}{1.484254in}}%
\pgfpathlineto{\pgfqpoint{1.515039in}{1.484254in}}%
\pgfpathlineto{\pgfqpoint{1.525520in}{1.473223in}}%
\pgfpathlineto{\pgfqpoint{1.546482in}{1.473223in}}%
\pgfpathlineto{\pgfqpoint{1.556963in}{1.493064in}}%
\pgfpathlineto{\pgfqpoint{1.577924in}{1.493064in}}%
\pgfpathlineto{\pgfqpoint{1.588405in}{1.491901in}}%
\pgfpathlineto{\pgfqpoint{1.609367in}{1.491901in}}%
\pgfpathlineto{\pgfqpoint{1.619848in}{1.497691in}}%
\pgfpathlineto{\pgfqpoint{1.640810in}{1.497691in}}%
\pgfpathlineto{\pgfqpoint{1.651290in}{1.495421in}}%
\pgfpathlineto{\pgfqpoint{1.672252in}{1.495421in}}%
\pgfpathlineto{\pgfqpoint{1.682733in}{1.490606in}}%
\pgfpathlineto{\pgfqpoint{1.703695in}{1.490606in}}%
\pgfpathlineto{\pgfqpoint{1.714175in}{1.461603in}}%
\pgfpathlineto{\pgfqpoint{1.735137in}{1.461603in}}%
\pgfpathlineto{\pgfqpoint{1.745618in}{1.468248in}}%
\pgfpathlineto{\pgfqpoint{1.766580in}{1.468248in}}%
\pgfpathlineto{\pgfqpoint{1.777060in}{1.419744in}}%
\pgfpathlineto{\pgfqpoint{1.798022in}{1.419744in}}%
\pgfpathlineto{\pgfqpoint{1.808503in}{1.248676in}}%
\pgfpathlineto{\pgfqpoint{1.829465in}{1.248676in}}%
\pgfpathlineto{\pgfqpoint{1.839946in}{0.998532in}}%
\pgfpathlineto{\pgfqpoint{1.860907in}{0.998532in}}%
\pgfpathlineto{\pgfqpoint{1.871388in}{0.724115in}}%
\pgfpathlineto{\pgfqpoint{1.892350in}{0.724115in}}%
\pgfpathlineto{\pgfqpoint{1.902831in}{0.623455in}}%
\pgfpathlineto{\pgfqpoint{4.669773in}{0.623455in}}%
\pgfpathlineto{\pgfqpoint{4.669773in}{0.623455in}}%
\pgfusepath{stroke}%
\end{pgfscope}%
\begin{pgfscope}%
\pgfpathrectangle{\pgfqpoint{0.675000in}{0.540000in}}{\pgfqpoint{4.185000in}{1.836000in}}%
\pgfusepath{clip}%
\pgfsetbuttcap%
\pgfsetroundjoin%
\pgfsetlinewidth{0.501875pt}%
\definecolor{currentstroke}{rgb}{0.172549,0.627451,0.172549}%
\pgfsetstrokecolor{currentstroke}%
\pgfsetdash{{1.850000pt}{0.800000pt}}{0.000000pt}%
\pgfpathmoveto{\pgfqpoint{0.865227in}{0.623455in}}%
\pgfpathlineto{\pgfqpoint{1.798022in}{0.623455in}}%
\pgfpathlineto{\pgfqpoint{1.808503in}{0.978737in}}%
\pgfpathlineto{\pgfqpoint{1.829465in}{0.978737in}}%
\pgfpathlineto{\pgfqpoint{1.839946in}{1.501643in}}%
\pgfpathlineto{\pgfqpoint{1.860907in}{1.501643in}}%
\pgfpathlineto{\pgfqpoint{1.871388in}{2.081343in}}%
\pgfpathlineto{\pgfqpoint{1.892350in}{2.081343in}}%
\pgfpathlineto{\pgfqpoint{1.902831in}{2.292545in}}%
\pgfpathlineto{\pgfqpoint{4.669773in}{2.292545in}}%
\pgfpathlineto{\pgfqpoint{4.669773in}{2.292545in}}%
\pgfusepath{stroke}%
\end{pgfscope}%
\begin{pgfscope}%
\pgfpathrectangle{\pgfqpoint{0.675000in}{0.540000in}}{\pgfqpoint{4.185000in}{1.836000in}}%
\pgfusepath{clip}%
\pgfsetbuttcap%
\pgfsetroundjoin%
\pgfsetlinewidth{0.501875pt}%
\definecolor{currentstroke}{rgb}{0.839216,0.152941,0.156863}%
\pgfsetstrokecolor{currentstroke}%
\pgfsetdash{{1.850000pt}{0.800000pt}}{0.000000pt}%
\pgfpathmoveto{\pgfqpoint{0.865227in}{0.623455in}}%
\pgfpathlineto{\pgfqpoint{4.669773in}{0.623455in}}%
\pgfpathlineto{\pgfqpoint{4.669773in}{0.623455in}}%
\pgfusepath{stroke}%
\end{pgfscope}%
\begin{pgfscope}%
\pgfsetrectcap%
\pgfsetmiterjoin%
\pgfsetlinewidth{0.803000pt}%
\definecolor{currentstroke}{rgb}{0.000000,0.000000,0.000000}%
\pgfsetstrokecolor{currentstroke}%
\pgfsetdash{}{0pt}%
\pgfpathmoveto{\pgfqpoint{0.675000in}{0.540000in}}%
\pgfpathlineto{\pgfqpoint{0.675000in}{2.376000in}}%
\pgfusepath{stroke}%
\end{pgfscope}%
\begin{pgfscope}%
\pgfsetrectcap%
\pgfsetmiterjoin%
\pgfsetlinewidth{0.803000pt}%
\definecolor{currentstroke}{rgb}{0.000000,0.000000,0.000000}%
\pgfsetstrokecolor{currentstroke}%
\pgfsetdash{}{0pt}%
\pgfpathmoveto{\pgfqpoint{4.860000in}{0.540000in}}%
\pgfpathlineto{\pgfqpoint{4.860000in}{2.376000in}}%
\pgfusepath{stroke}%
\end{pgfscope}%
\begin{pgfscope}%
\pgfsetrectcap%
\pgfsetmiterjoin%
\pgfsetlinewidth{0.803000pt}%
\definecolor{currentstroke}{rgb}{0.000000,0.000000,0.000000}%
\pgfsetstrokecolor{currentstroke}%
\pgfsetdash{}{0pt}%
\pgfpathmoveto{\pgfqpoint{0.675000in}{0.540000in}}%
\pgfpathlineto{\pgfqpoint{4.860000in}{0.540000in}}%
\pgfusepath{stroke}%
\end{pgfscope}%
\begin{pgfscope}%
\pgfsetrectcap%
\pgfsetmiterjoin%
\pgfsetlinewidth{0.803000pt}%
\definecolor{currentstroke}{rgb}{0.000000,0.000000,0.000000}%
\pgfsetstrokecolor{currentstroke}%
\pgfsetdash{}{0pt}%
\pgfpathmoveto{\pgfqpoint{0.675000in}{2.376000in}}%
\pgfpathlineto{\pgfqpoint{4.860000in}{2.376000in}}%
\pgfusepath{stroke}%
\end{pgfscope}%
\begin{pgfscope}%
\pgfsetbuttcap%
\pgfsetmiterjoin%
\pgfsetlinewidth{0.000000pt}%
\definecolor{currentstroke}{rgb}{0.800000,0.800000,0.800000}%
\pgfsetstrokecolor{currentstroke}%
\pgfsetstrokeopacity{0.000000}%
\pgfsetdash{}{0pt}%
\pgfpathmoveto{\pgfqpoint{3.283222in}{1.252044in}}%
\pgfpathlineto{\pgfqpoint{4.291157in}{1.252044in}}%
\pgfpathquadraticcurveto{\pgfqpoint{4.318935in}{1.252044in}}{\pgfqpoint{4.318935in}{1.279822in}}%
\pgfpathlineto{\pgfqpoint{4.318935in}{1.855044in}}%
\pgfpathquadraticcurveto{\pgfqpoint{4.318935in}{1.882822in}}{\pgfqpoint{4.291157in}{1.882822in}}%
\pgfpathlineto{\pgfqpoint{3.283222in}{1.882822in}}%
\pgfpathquadraticcurveto{\pgfqpoint{3.255444in}{1.882822in}}{\pgfqpoint{3.255444in}{1.855044in}}%
\pgfpathlineto{\pgfqpoint{3.255444in}{1.279822in}}%
\pgfpathquadraticcurveto{\pgfqpoint{3.255444in}{1.252044in}}{\pgfqpoint{3.283222in}{1.252044in}}%
\pgfpathclose%
\pgfusepath{}%
\end{pgfscope}%
\begin{pgfscope}%
\pgfsetbuttcap%
\pgfsetroundjoin%
\pgfsetlinewidth{0.501875pt}%
\definecolor{currentstroke}{rgb}{0.121569,0.466667,0.705882}%
\pgfsetstrokecolor{currentstroke}%
\pgfsetdash{{1.850000pt}{0.800000pt}}{0.000000pt}%
\pgfpathmoveto{\pgfqpoint{3.311000in}{1.778655in}}%
\pgfpathlineto{\pgfqpoint{3.588778in}{1.778655in}}%
\pgfusepath{stroke}%
\end{pgfscope}%
\begin{pgfscope}%
\pgftext[x=3.699889in,y=1.730044in,left,base]{\rmfamily\fontsize{10.000000}{12.000000}\selectfont SlushPool}%
\end{pgfscope}%
\begin{pgfscope}%
\pgfsetbuttcap%
\pgfsetroundjoin%
\pgfsetlinewidth{0.501875pt}%
\definecolor{currentstroke}{rgb}{1.000000,0.498039,0.054902}%
\pgfsetstrokecolor{currentstroke}%
\pgfsetdash{{1.850000pt}{0.800000pt}}{0.000000pt}%
\pgfpathmoveto{\pgfqpoint{3.311000in}{1.582285in}}%
\pgfpathlineto{\pgfqpoint{3.588778in}{1.582285in}}%
\pgfusepath{stroke}%
\end{pgfscope}%
\begin{pgfscope}%
\pgftext[x=3.699889in,y=1.533674in,left,base]{\rmfamily\fontsize{10.000000}{12.000000}\selectfont ViaBTC}%
\end{pgfscope}%
\begin{pgfscope}%
\pgfsetbuttcap%
\pgfsetroundjoin%
\pgfsetlinewidth{0.501875pt}%
\definecolor{currentstroke}{rgb}{0.172549,0.627451,0.172549}%
\pgfsetstrokecolor{currentstroke}%
\pgfsetdash{{1.850000pt}{0.800000pt}}{0.000000pt}%
\pgfpathmoveto{\pgfqpoint{3.311000in}{1.385915in}}%
\pgfpathlineto{\pgfqpoint{3.588778in}{1.385915in}}%
\pgfusepath{stroke}%
\end{pgfscope}%
\begin{pgfscope}%
\pgftext[x=3.699889in,y=1.337304in,left,base]{\rmfamily\fontsize{10.000000}{12.000000}\selectfont DPOOL}%
\end{pgfscope}%
\end{pgfpicture}%
\makeatother%
\endgroup%

%% file: passive_slush_6mo.pgf
\begingroup%
\makeatletter%
\begin{pgfpicture}%
\pgfpathrectangle{\pgfpointorigin}{\pgfqpoint{5.400000in}{2.700000in}}%
\pgfusepath{use as bounding box, clip}%
\begin{pgfscope}%
\pgfsetbuttcap%
\pgfsetmiterjoin%
\definecolor{currentfill}{rgb}{1.000000,1.000000,1.000000}%
\pgfsetfillcolor{currentfill}%
\pgfsetlinewidth{0.000000pt}%
\definecolor{currentstroke}{rgb}{1.000000,1.000000,1.000000}%
\pgfsetstrokecolor{currentstroke}%
\pgfsetdash{}{0pt}%
\pgfpathmoveto{\pgfqpoint{0.000000in}{0.000000in}}%
\pgfpathlineto{\pgfqpoint{5.400000in}{0.000000in}}%
\pgfpathlineto{\pgfqpoint{5.400000in}{2.700000in}}%
\pgfpathlineto{\pgfqpoint{0.000000in}{2.700000in}}%
\pgfpathclose%
\pgfusepath{fill}%
\end{pgfscope}%
\begin{pgfscope}%
\pgfsetbuttcap%
\pgfsetmiterjoin%
\definecolor{currentfill}{rgb}{1.000000,1.000000,1.000000}%
\pgfsetfillcolor{currentfill}%
\pgfsetlinewidth{0.000000pt}%
\definecolor{currentstroke}{rgb}{0.000000,0.000000,0.000000}%
\pgfsetstrokecolor{currentstroke}%
\pgfsetstrokeopacity{0.000000}%
\pgfsetdash{}{0pt}%
\pgfpathmoveto{\pgfqpoint{0.675000in}{0.540000in}}%
\pgfpathlineto{\pgfqpoint{4.860000in}{0.540000in}}%
\pgfpathlineto{\pgfqpoint{4.860000in}{2.376000in}}%
\pgfpathlineto{\pgfqpoint{0.675000in}{2.376000in}}%
\pgfpathclose%
\pgfusepath{fill}%
\end{pgfscope}%
\begin{pgfscope}%
\pgfsetbuttcap%
\pgfsetroundjoin%
\definecolor{currentfill}{rgb}{0.000000,0.000000,0.000000}%
\pgfsetfillcolor{currentfill}%
\pgfsetlinewidth{0.803000pt}%
\definecolor{currentstroke}{rgb}{0.000000,0.000000,0.000000}%
\pgfsetstrokecolor{currentstroke}%
\pgfsetdash{}{0pt}%
\pgfsys@defobject{currentmarker}{\pgfqpoint{0.000000in}{0.000000in}}{\pgfqpoint{0.000000in}{0.048611in}}{%
\pgfpathmoveto{\pgfqpoint{0.000000in}{0.000000in}}%
\pgfpathlineto{\pgfqpoint{0.000000in}{0.048611in}}%
\pgfusepath{stroke,fill}%
}%
\begin{pgfscope}%
\pgfsys@transformshift{0.865227in}{0.540000in}%
\pgfsys@useobject{currentmarker}{}%
\end{pgfscope}%
\end{pgfscope}%
\begin{pgfscope}%
\pgftext[x=0.289141in,y=0.104678in,left,base,rotate=30.000000]{\rmfamily\fontsize{10.000000}{12.000000}\selectfont 2018-01-01}%
\end{pgfscope}%
\begin{pgfscope}%
\pgfsetbuttcap%
\pgfsetroundjoin%
\definecolor{currentfill}{rgb}{0.000000,0.000000,0.000000}%
\pgfsetfillcolor{currentfill}%
\pgfsetlinewidth{0.803000pt}%
\definecolor{currentstroke}{rgb}{0.000000,0.000000,0.000000}%
\pgfsetstrokecolor{currentstroke}%
\pgfsetdash{}{0pt}%
\pgfsys@defobject{currentmarker}{\pgfqpoint{0.000000in}{0.000000in}}{\pgfqpoint{0.000000in}{0.048611in}}{%
\pgfpathmoveto{\pgfqpoint{0.000000in}{0.000000in}}%
\pgfpathlineto{\pgfqpoint{0.000000in}{0.048611in}}%
\pgfusepath{stroke,fill}%
}%
\begin{pgfscope}%
\pgfsys@transformshift{1.502841in}{0.540000in}%
\pgfsys@useobject{currentmarker}{}%
\end{pgfscope}%
\end{pgfscope}%
\begin{pgfscope}%
\pgftext[x=0.926755in,y=0.104678in,left,base,rotate=30.000000]{\rmfamily\fontsize{10.000000}{12.000000}\selectfont 2018-01-31}%
\end{pgfscope}%
\begin{pgfscope}%
\pgfsetbuttcap%
\pgfsetroundjoin%
\definecolor{currentfill}{rgb}{0.000000,0.000000,0.000000}%
\pgfsetfillcolor{currentfill}%
\pgfsetlinewidth{0.803000pt}%
\definecolor{currentstroke}{rgb}{0.000000,0.000000,0.000000}%
\pgfsetstrokecolor{currentstroke}%
\pgfsetdash{}{0pt}%
\pgfsys@defobject{currentmarker}{\pgfqpoint{0.000000in}{0.000000in}}{\pgfqpoint{0.000000in}{0.048611in}}{%
\pgfpathmoveto{\pgfqpoint{0.000000in}{0.000000in}}%
\pgfpathlineto{\pgfqpoint{0.000000in}{0.048611in}}%
\pgfusepath{stroke,fill}%
}%
\begin{pgfscope}%
\pgfsys@transformshift{2.140455in}{0.540000in}%
\pgfsys@useobject{currentmarker}{}%
\end{pgfscope}%
\end{pgfscope}%
\begin{pgfscope}%
\pgftext[x=1.564368in,y=0.104678in,left,base,rotate=30.000000]{\rmfamily\fontsize{10.000000}{12.000000}\selectfont 2018-03-02}%
\end{pgfscope}%
\begin{pgfscope}%
\pgfsetbuttcap%
\pgfsetroundjoin%
\definecolor{currentfill}{rgb}{0.000000,0.000000,0.000000}%
\pgfsetfillcolor{currentfill}%
\pgfsetlinewidth{0.803000pt}%
\definecolor{currentstroke}{rgb}{0.000000,0.000000,0.000000}%
\pgfsetstrokecolor{currentstroke}%
\pgfsetdash{}{0pt}%
\pgfsys@defobject{currentmarker}{\pgfqpoint{0.000000in}{0.000000in}}{\pgfqpoint{0.000000in}{0.048611in}}{%
\pgfpathmoveto{\pgfqpoint{0.000000in}{0.000000in}}%
\pgfpathlineto{\pgfqpoint{0.000000in}{0.048611in}}%
\pgfusepath{stroke,fill}%
}%
\begin{pgfscope}%
\pgfsys@transformshift{2.778068in}{0.540000in}%
\pgfsys@useobject{currentmarker}{}%
\end{pgfscope}%
\end{pgfscope}%
\begin{pgfscope}%
\pgftext[x=2.201982in,y=0.104678in,left,base,rotate=30.000000]{\rmfamily\fontsize{10.000000}{12.000000}\selectfont 2018-04-01}%
\end{pgfscope}%
\begin{pgfscope}%
\pgfsetbuttcap%
\pgfsetroundjoin%
\definecolor{currentfill}{rgb}{0.000000,0.000000,0.000000}%
\pgfsetfillcolor{currentfill}%
\pgfsetlinewidth{0.803000pt}%
\definecolor{currentstroke}{rgb}{0.000000,0.000000,0.000000}%
\pgfsetstrokecolor{currentstroke}%
\pgfsetdash{}{0pt}%
\pgfsys@defobject{currentmarker}{\pgfqpoint{0.000000in}{0.000000in}}{\pgfqpoint{0.000000in}{0.048611in}}{%
\pgfpathmoveto{\pgfqpoint{0.000000in}{0.000000in}}%
\pgfpathlineto{\pgfqpoint{0.000000in}{0.048611in}}%
\pgfusepath{stroke,fill}%
}%
\begin{pgfscope}%
\pgfsys@transformshift{3.415682in}{0.540000in}%
\pgfsys@useobject{currentmarker}{}%
\end{pgfscope}%
\end{pgfscope}%
\begin{pgfscope}%
\pgftext[x=2.839596in,y=0.104678in,left,base,rotate=30.000000]{\rmfamily\fontsize{10.000000}{12.000000}\selectfont 2018-05-02}%
\end{pgfscope}%
\begin{pgfscope}%
\pgfsetbuttcap%
\pgfsetroundjoin%
\definecolor{currentfill}{rgb}{0.000000,0.000000,0.000000}%
\pgfsetfillcolor{currentfill}%
\pgfsetlinewidth{0.803000pt}%
\definecolor{currentstroke}{rgb}{0.000000,0.000000,0.000000}%
\pgfsetstrokecolor{currentstroke}%
\pgfsetdash{}{0pt}%
\pgfsys@defobject{currentmarker}{\pgfqpoint{0.000000in}{0.000000in}}{\pgfqpoint{0.000000in}{0.048611in}}{%
\pgfpathmoveto{\pgfqpoint{0.000000in}{0.000000in}}%
\pgfpathlineto{\pgfqpoint{0.000000in}{0.048611in}}%
\pgfusepath{stroke,fill}%
}%
\begin{pgfscope}%
\pgfsys@transformshift{4.053295in}{0.540000in}%
\pgfsys@useobject{currentmarker}{}%
\end{pgfscope}%
\end{pgfscope}%
\begin{pgfscope}%
\pgftext[x=3.477209in,y=0.104678in,left,base,rotate=30.000000]{\rmfamily\fontsize{10.000000}{12.000000}\selectfont 2018-06-01}%
\pgftext[x=4.104209in,y=0.104678in,left,base,rotate=30.000000]{\rmfamily\fontsize{10.000000}{12.000000}\selectfont 2018-06-30}%
\end{pgfscope}%
\begin{pgfscope}%
\pgfsetbuttcap%
\pgfsetroundjoin%
\definecolor{currentfill}{rgb}{0.000000,0.000000,0.000000}%
\pgfsetfillcolor{currentfill}%
\pgfsetlinewidth{0.803000pt}%
\definecolor{currentstroke}{rgb}{0.000000,0.000000,0.000000}%
\pgfsetstrokecolor{currentstroke}%
\pgfsetdash{}{0pt}%
\pgfsys@defobject{currentmarker}{\pgfqpoint{0.000000in}{0.000000in}}{\pgfqpoint{0.000000in}{0.048611in}}{%
\pgfpathmoveto{\pgfqpoint{0.000000in}{0.000000in}}%
\pgfpathlineto{\pgfqpoint{0.000000in}{0.048611in}}%
\pgfusepath{stroke,fill}%
}%
\begin{pgfscope}%
\pgfsys@transformshift{4.690909in}{0.540000in}%
\pgfsys@useobject{currentmarker}{}%
\end{pgfscope}%
\end{pgfscope}%
\begin{pgfscope}%
\pgfsetbuttcap%
\pgfsetroundjoin%
\definecolor{currentfill}{rgb}{0.000000,0.000000,0.000000}%
\pgfsetfillcolor{currentfill}%
\pgfsetlinewidth{0.803000pt}%
\definecolor{currentstroke}{rgb}{0.000000,0.000000,0.000000}%
\pgfsetstrokecolor{currentstroke}%
\pgfsetdash{}{0pt}%
\pgfsys@defobject{currentmarker}{\pgfqpoint{-0.048611in}{0.000000in}}{\pgfqpoint{0.000000in}{0.000000in}}{%
\pgfpathmoveto{\pgfqpoint{0.000000in}{0.000000in}}%
\pgfpathlineto{\pgfqpoint{-0.048611in}{0.000000in}}%
\pgfusepath{stroke,fill}%
}%
\begin{pgfscope}%
\pgfsys@transformshift{0.675000in}{1.065218in}%
\pgfsys@useobject{currentmarker}{}%
\end{pgfscope}%
\end{pgfscope}%
\begin{pgfscope}%
\pgftext[x=0.299999in,y=1.017000in,left,base]{\rmfamily\fontsize{10.000000}{12.000000}\selectfont \(\displaystyle 1000\)}%
\end{pgfscope}%
\begin{pgfscope}%
\pgfsetbuttcap%
\pgfsetroundjoin%
\definecolor{currentfill}{rgb}{0.000000,0.000000,0.000000}%
\pgfsetfillcolor{currentfill}%
\pgfsetlinewidth{0.803000pt}%
\definecolor{currentstroke}{rgb}{0.000000,0.000000,0.000000}%
\pgfsetstrokecolor{currentstroke}%
\pgfsetdash{}{0pt}%
\pgfsys@defobject{currentmarker}{\pgfqpoint{-0.048611in}{0.000000in}}{\pgfqpoint{0.000000in}{0.000000in}}{%
\pgfpathmoveto{\pgfqpoint{0.000000in}{0.000000in}}%
\pgfpathlineto{\pgfqpoint{-0.048611in}{0.000000in}}%
\pgfusepath{stroke,fill}%
}%
\begin{pgfscope}%
\pgfsys@transformshift{0.675000in}{1.630384in}%
\pgfsys@useobject{currentmarker}{}%
\end{pgfscope}%
\end{pgfscope}%
\begin{pgfscope}%
\pgftext[x=0.299999in,y=1.582166in,left,base]{\rmfamily\fontsize{10.000000}{12.000000}\selectfont \(\displaystyle 2000\)}%
\end{pgfscope}%
\begin{pgfscope}%
\pgfsetbuttcap%
\pgfsetroundjoin%
\definecolor{currentfill}{rgb}{0.000000,0.000000,0.000000}%
\pgfsetfillcolor{currentfill}%
\pgfsetlinewidth{0.803000pt}%
\definecolor{currentstroke}{rgb}{0.000000,0.000000,0.000000}%
\pgfsetstrokecolor{currentstroke}%
\pgfsetdash{}{0pt}%
\pgfsys@defobject{currentmarker}{\pgfqpoint{-0.048611in}{0.000000in}}{\pgfqpoint{0.000000in}{0.000000in}}{%
\pgfpathmoveto{\pgfqpoint{0.000000in}{0.000000in}}%
\pgfpathlineto{\pgfqpoint{-0.048611in}{0.000000in}}%
\pgfusepath{stroke,fill}%
}%
\begin{pgfscope}%
\pgfsys@transformshift{0.675000in}{2.195550in}%
\pgfsys@useobject{currentmarker}{}%
\end{pgfscope}%
\end{pgfscope}%
\begin{pgfscope}%
\pgftext[x=0.299999in,y=2.147332in,left,base]{\rmfamily\fontsize{10.000000}{12.000000}\selectfont \(\displaystyle 3000\)}%
\end{pgfscope}%
\begin{pgfscope}%
\definecolor{textcolor}{rgb}{0.000000,0.000000,1.000000}%
\pgfsetstrokecolor{textcolor}%
\pgfsetfillcolor{textcolor}%
\pgftext[x=0.244444in,y=1.458000in,,bottom,rotate=90.000000]{\color{textcolor}\rmfamily\fontsize{10.000000}{12.000000}\selectfont Passive USD/day earned}%
\end{pgfscope}%
\begin{pgfscope}%
\pgfpathrectangle{\pgfqpoint{0.675000in}{0.540000in}}{\pgfqpoint{4.185000in}{1.836000in}}%
\pgfusepath{clip}%
\pgfsetrectcap%
\pgfsetroundjoin%
\pgfsetlinewidth{1.003750pt}%
\definecolor{currentstroke}{rgb}{0.000000,0.000000,1.000000}%
\pgfsetstrokecolor{currentstroke}%
\pgfsetdash{}{0pt}%
\pgfpathmoveto{\pgfqpoint{0.865227in}{1.868488in}}%
\pgfpathlineto{\pgfqpoint{0.886364in}{1.665206in}}%
\pgfpathlineto{\pgfqpoint{0.907500in}{2.109758in}}%
\pgfpathlineto{\pgfqpoint{0.928636in}{1.429292in}}%
\pgfpathlineto{\pgfqpoint{0.949773in}{1.816012in}}%
\pgfpathlineto{\pgfqpoint{0.970909in}{2.135261in}}%
\pgfpathlineto{\pgfqpoint{0.992045in}{2.205849in}}%
\pgfpathlineto{\pgfqpoint{1.013182in}{2.292545in}}%
\pgfpathlineto{\pgfqpoint{1.034318in}{1.844672in}}%
\pgfpathlineto{\pgfqpoint{1.055455in}{2.219561in}}%
\pgfpathlineto{\pgfqpoint{1.076591in}{1.579591in}}%
\pgfpathlineto{\pgfqpoint{1.097727in}{1.754148in}}%
\pgfpathlineto{\pgfqpoint{1.118864in}{1.897018in}}%
\pgfpathlineto{\pgfqpoint{1.140000in}{1.304144in}}%
\pgfpathlineto{\pgfqpoint{1.161136in}{1.695229in}}%
\pgfpathlineto{\pgfqpoint{1.182273in}{1.727988in}}%
\pgfpathlineto{\pgfqpoint{1.203409in}{1.259785in}}%
\pgfpathlineto{\pgfqpoint{1.224545in}{1.566540in}}%
\pgfpathlineto{\pgfqpoint{1.245682in}{1.965145in}}%
\pgfpathlineto{\pgfqpoint{1.266818in}{2.016994in}}%
\pgfpathlineto{\pgfqpoint{1.287955in}{1.408915in}}%
\pgfpathlineto{\pgfqpoint{1.309091in}{1.258587in}}%
\pgfpathlineto{\pgfqpoint{1.330227in}{1.451594in}}%
\pgfpathlineto{\pgfqpoint{1.351364in}{1.573862in}}%
\pgfpathlineto{\pgfqpoint{1.372500in}{1.459631in}}%
\pgfpathlineto{\pgfqpoint{1.393636in}{1.602474in}}%
\pgfpathlineto{\pgfqpoint{1.414773in}{1.187261in}}%
\pgfpathlineto{\pgfqpoint{1.435909in}{1.561624in}}%
\pgfpathlineto{\pgfqpoint{1.457045in}{1.036066in}}%
\pgfpathlineto{\pgfqpoint{1.478182in}{1.135620in}}%
\pgfpathlineto{\pgfqpoint{1.499318in}{1.074393in}}%
\pgfpathlineto{\pgfqpoint{1.520455in}{0.891443in}}%
\pgfpathlineto{\pgfqpoint{1.541591in}{0.931890in}}%
\pgfpathlineto{\pgfqpoint{1.562727in}{1.082166in}}%
\pgfpathlineto{\pgfqpoint{1.583864in}{1.056983in}}%
\pgfpathlineto{\pgfqpoint{1.605000in}{1.277418in}}%
\pgfpathlineto{\pgfqpoint{1.647273in}{1.226666in}}%
\pgfpathlineto{\pgfqpoint{1.668409in}{0.850874in}}%
\pgfpathlineto{\pgfqpoint{1.689545in}{0.916217in}}%
\pgfpathlineto{\pgfqpoint{1.710682in}{1.076506in}}%
\pgfpathlineto{\pgfqpoint{1.731818in}{1.067915in}}%
\pgfpathlineto{\pgfqpoint{1.752955in}{0.981336in}}%
\pgfpathlineto{\pgfqpoint{1.774091in}{1.010335in}}%
\pgfpathlineto{\pgfqpoint{1.795227in}{1.202767in}}%
\pgfpathlineto{\pgfqpoint{1.816364in}{1.036497in}}%
\pgfpathlineto{\pgfqpoint{1.837500in}{1.112577in}}%
\pgfpathlineto{\pgfqpoint{1.858636in}{1.179161in}}%
\pgfpathlineto{\pgfqpoint{1.879773in}{0.828164in}}%
\pgfpathlineto{\pgfqpoint{1.900909in}{1.121929in}}%
\pgfpathlineto{\pgfqpoint{1.922045in}{1.580149in}}%
\pgfpathlineto{\pgfqpoint{1.943182in}{1.063967in}}%
\pgfpathlineto{\pgfqpoint{1.964318in}{1.082677in}}%
\pgfpathlineto{\pgfqpoint{1.985455in}{1.094379in}}%
\pgfpathlineto{\pgfqpoint{2.006591in}{1.120388in}}%
\pgfpathlineto{\pgfqpoint{2.027727in}{1.003286in}}%
\pgfpathlineto{\pgfqpoint{2.048864in}{1.106084in}}%
\pgfpathlineto{\pgfqpoint{2.070000in}{1.099630in}}%
\pgfpathlineto{\pgfqpoint{2.091136in}{1.177111in}}%
\pgfpathlineto{\pgfqpoint{2.112273in}{1.146851in}}%
\pgfpathlineto{\pgfqpoint{2.133409in}{1.458794in}}%
\pgfpathlineto{\pgfqpoint{2.154545in}{1.206840in}}%
\pgfpathlineto{\pgfqpoint{2.175682in}{1.399169in}}%
\pgfpathlineto{\pgfqpoint{2.196818in}{1.032561in}}%
\pgfpathlineto{\pgfqpoint{2.217955in}{0.975933in}}%
\pgfpathlineto{\pgfqpoint{2.239091in}{0.961929in}}%
\pgfpathlineto{\pgfqpoint{2.260227in}{1.121142in}}%
\pgfpathlineto{\pgfqpoint{2.281364in}{0.948774in}}%
\pgfpathlineto{\pgfqpoint{2.302500in}{0.966781in}}%
\pgfpathlineto{\pgfqpoint{2.323636in}{0.902098in}}%
\pgfpathlineto{\pgfqpoint{2.344773in}{0.922282in}}%
\pgfpathlineto{\pgfqpoint{2.365909in}{1.146105in}}%
\pgfpathlineto{\pgfqpoint{2.387045in}{0.945316in}}%
\pgfpathlineto{\pgfqpoint{2.408182in}{1.013355in}}%
\pgfpathlineto{\pgfqpoint{2.429318in}{1.066443in}}%
\pgfpathlineto{\pgfqpoint{2.450455in}{0.905605in}}%
\pgfpathlineto{\pgfqpoint{2.471591in}{0.863003in}}%
\pgfpathlineto{\pgfqpoint{2.492727in}{1.045113in}}%
\pgfpathlineto{\pgfqpoint{2.513864in}{1.035256in}}%
\pgfpathlineto{\pgfqpoint{2.535000in}{0.833365in}}%
\pgfpathlineto{\pgfqpoint{2.556136in}{0.979234in}}%
\pgfpathlineto{\pgfqpoint{2.577273in}{0.939954in}}%
\pgfpathlineto{\pgfqpoint{2.598409in}{0.898212in}}%
\pgfpathlineto{\pgfqpoint{2.619545in}{0.861029in}}%
\pgfpathlineto{\pgfqpoint{2.640682in}{0.845192in}}%
\pgfpathlineto{\pgfqpoint{2.661818in}{0.855431in}}%
\pgfpathlineto{\pgfqpoint{2.682955in}{0.926906in}}%
\pgfpathlineto{\pgfqpoint{2.704091in}{1.053649in}}%
\pgfpathlineto{\pgfqpoint{2.725227in}{0.623455in}}%
\pgfpathlineto{\pgfqpoint{2.746364in}{0.923914in}}%
\pgfpathlineto{\pgfqpoint{2.767500in}{0.823792in}}%
\pgfpathlineto{\pgfqpoint{2.788636in}{0.941696in}}%
\pgfpathlineto{\pgfqpoint{2.809773in}{0.821433in}}%
\pgfpathlineto{\pgfqpoint{2.830909in}{0.896301in}}%
\pgfpathlineto{\pgfqpoint{2.852045in}{0.834174in}}%
\pgfpathlineto{\pgfqpoint{2.873182in}{0.884629in}}%
\pgfpathlineto{\pgfqpoint{2.894318in}{0.762788in}}%
\pgfpathlineto{\pgfqpoint{2.915455in}{0.929736in}}%
\pgfpathlineto{\pgfqpoint{2.936591in}{0.950020in}}%
\pgfpathlineto{\pgfqpoint{2.957727in}{0.788899in}}%
\pgfpathlineto{\pgfqpoint{2.978864in}{0.831686in}}%
\pgfpathlineto{\pgfqpoint{3.000000in}{0.735574in}}%
\pgfpathlineto{\pgfqpoint{3.042273in}{0.864016in}}%
\pgfpathlineto{\pgfqpoint{3.063409in}{0.933683in}}%
\pgfpathlineto{\pgfqpoint{3.084545in}{0.860741in}}%
\pgfpathlineto{\pgfqpoint{3.105682in}{0.821436in}}%
\pgfpathlineto{\pgfqpoint{3.126818in}{0.867512in}}%
\pgfpathlineto{\pgfqpoint{3.147955in}{1.008421in}}%
\pgfpathlineto{\pgfqpoint{3.169091in}{0.886478in}}%
\pgfpathlineto{\pgfqpoint{3.190227in}{1.086807in}}%
\pgfpathlineto{\pgfqpoint{3.211364in}{1.028858in}}%
\pgfpathlineto{\pgfqpoint{3.232500in}{0.772291in}}%
\pgfpathlineto{\pgfqpoint{3.253636in}{0.949449in}}%
\pgfpathlineto{\pgfqpoint{3.274773in}{0.919017in}}%
\pgfpathlineto{\pgfqpoint{3.295909in}{0.731566in}}%
\pgfpathlineto{\pgfqpoint{3.317045in}{0.890698in}}%
\pgfpathlineto{\pgfqpoint{3.338182in}{0.921386in}}%
\pgfpathlineto{\pgfqpoint{3.359318in}{0.840626in}}%
\pgfpathlineto{\pgfqpoint{3.380455in}{0.801351in}}%
\pgfpathlineto{\pgfqpoint{3.401591in}{0.887538in}}%
\pgfpathlineto{\pgfqpoint{3.422727in}{0.845918in}}%
\pgfpathlineto{\pgfqpoint{3.443864in}{0.994773in}}%
\pgfpathlineto{\pgfqpoint{3.465000in}{0.766491in}}%
\pgfpathlineto{\pgfqpoint{3.486136in}{0.926874in}}%
\pgfpathlineto{\pgfqpoint{3.507273in}{0.990410in}}%
\pgfpathlineto{\pgfqpoint{3.528409in}{0.853578in}}%
\pgfpathlineto{\pgfqpoint{3.549545in}{0.992902in}}%
\pgfpathlineto{\pgfqpoint{3.570682in}{0.862870in}}%
\pgfpathlineto{\pgfqpoint{3.591818in}{1.037883in}}%
\pgfpathlineto{\pgfqpoint{3.612955in}{0.810634in}}%
\pgfpathlineto{\pgfqpoint{3.634091in}{0.834608in}}%
\pgfpathlineto{\pgfqpoint{3.655227in}{0.951722in}}%
\pgfpathlineto{\pgfqpoint{3.676364in}{0.862881in}}%
\pgfpathlineto{\pgfqpoint{3.697500in}{0.875702in}}%
\pgfpathlineto{\pgfqpoint{3.718636in}{0.757457in}}%
\pgfpathlineto{\pgfqpoint{3.739773in}{0.843007in}}%
\pgfpathlineto{\pgfqpoint{3.760909in}{0.718965in}}%
\pgfpathlineto{\pgfqpoint{3.782045in}{0.926997in}}%
\pgfpathlineto{\pgfqpoint{3.803182in}{0.868761in}}%
\pgfpathlineto{\pgfqpoint{3.824318in}{0.931599in}}%
\pgfpathlineto{\pgfqpoint{3.845455in}{0.795366in}}%
\pgfpathlineto{\pgfqpoint{3.866591in}{0.785248in}}%
\pgfpathlineto{\pgfqpoint{3.887727in}{0.815823in}}%
\pgfpathlineto{\pgfqpoint{3.908864in}{0.829996in}}%
\pgfpathlineto{\pgfqpoint{3.930000in}{0.908325in}}%
\pgfpathlineto{\pgfqpoint{3.951136in}{0.894671in}}%
\pgfpathlineto{\pgfqpoint{3.972273in}{0.887946in}}%
\pgfpathlineto{\pgfqpoint{3.993409in}{0.850527in}}%
\pgfpathlineto{\pgfqpoint{4.014545in}{0.754353in}}%
\pgfpathlineto{\pgfqpoint{4.035682in}{0.859442in}}%
\pgfpathlineto{\pgfqpoint{4.056818in}{0.769105in}}%
\pgfpathlineto{\pgfqpoint{4.077955in}{0.904045in}}%
\pgfpathlineto{\pgfqpoint{4.099091in}{0.755131in}}%
\pgfpathlineto{\pgfqpoint{4.120227in}{0.889359in}}%
\pgfpathlineto{\pgfqpoint{4.141364in}{0.856277in}}%
\pgfpathlineto{\pgfqpoint{4.162500in}{0.942895in}}%
\pgfpathlineto{\pgfqpoint{4.183636in}{0.814485in}}%
\pgfpathlineto{\pgfqpoint{4.204773in}{0.776928in}}%
\pgfpathlineto{\pgfqpoint{4.225909in}{0.635368in}}%
\pgfpathlineto{\pgfqpoint{4.247045in}{0.658986in}}%
\pgfpathlineto{\pgfqpoint{4.268182in}{0.745717in}}%
\pgfpathlineto{\pgfqpoint{4.289318in}{0.733730in}}%
\pgfpathlineto{\pgfqpoint{4.310455in}{0.772683in}}%
\pgfpathlineto{\pgfqpoint{4.331591in}{0.671268in}}%
\pgfpathlineto{\pgfqpoint{4.352727in}{0.787244in}}%
\pgfpathlineto{\pgfqpoint{4.373864in}{0.667893in}}%
\pgfpathlineto{\pgfqpoint{4.395000in}{0.713161in}}%
\pgfpathlineto{\pgfqpoint{4.416136in}{0.712426in}}%
\pgfpathlineto{\pgfqpoint{4.437273in}{0.656552in}}%
\pgfpathlineto{\pgfqpoint{4.458409in}{0.697006in}}%
\pgfpathlineto{\pgfqpoint{4.479545in}{0.810063in}}%
\pgfpathlineto{\pgfqpoint{4.500682in}{0.680429in}}%
\pgfpathlineto{\pgfqpoint{4.521818in}{0.692151in}}%
\pgfpathlineto{\pgfqpoint{4.542955in}{0.747324in}}%
\pgfpathlineto{\pgfqpoint{4.564091in}{0.785241in}}%
\pgfpathlineto{\pgfqpoint{4.585227in}{0.649218in}}%
\pgfpathlineto{\pgfqpoint{4.606364in}{0.715120in}}%
\pgfpathlineto{\pgfqpoint{4.627500in}{0.626431in}}%
\pgfpathlineto{\pgfqpoint{4.648636in}{0.778659in}}%
\pgfpathlineto{\pgfqpoint{4.669773in}{0.678748in}}%
\pgfpathlineto{\pgfqpoint{4.669773in}{0.678748in}}%
\pgfusepath{stroke}%
\end{pgfscope}%
\begin{pgfscope}%
\pgfsetrectcap%
\pgfsetmiterjoin%
\pgfsetlinewidth{0.803000pt}%
\definecolor{currentstroke}{rgb}{0.000000,0.000000,0.000000}%
\pgfsetstrokecolor{currentstroke}%
\pgfsetdash{}{0pt}%
\pgfpathmoveto{\pgfqpoint{0.675000in}{0.540000in}}%
\pgfpathlineto{\pgfqpoint{0.675000in}{2.376000in}}%
\pgfusepath{stroke}%
\end{pgfscope}%
\begin{pgfscope}%
\pgfsetrectcap%
\pgfsetmiterjoin%
\pgfsetlinewidth{0.803000pt}%
\definecolor{currentstroke}{rgb}{0.000000,0.000000,0.000000}%
\pgfsetstrokecolor{currentstroke}%
\pgfsetdash{}{0pt}%
\pgfpathmoveto{\pgfqpoint{4.860000in}{0.540000in}}%
\pgfpathlineto{\pgfqpoint{4.860000in}{2.376000in}}%
\pgfusepath{stroke}%
\end{pgfscope}%
\begin{pgfscope}%
\pgfsetrectcap%
\pgfsetmiterjoin%
\pgfsetlinewidth{0.803000pt}%
\definecolor{currentstroke}{rgb}{0.000000,0.000000,0.000000}%
\pgfsetstrokecolor{currentstroke}%
\pgfsetdash{}{0pt}%
\pgfpathmoveto{\pgfqpoint{0.675000in}{0.540000in}}%
\pgfpathlineto{\pgfqpoint{4.860000in}{0.540000in}}%
\pgfusepath{stroke}%
\end{pgfscope}%
\begin{pgfscope}%
\pgfsetrectcap%
\pgfsetmiterjoin%
\pgfsetlinewidth{0.803000pt}%
\definecolor{currentstroke}{rgb}{0.000000,0.000000,0.000000}%
\pgfsetstrokecolor{currentstroke}%
\pgfsetdash{}{0pt}%
\pgfpathmoveto{\pgfqpoint{0.675000in}{2.376000in}}%
\pgfpathlineto{\pgfqpoint{4.860000in}{2.376000in}}%
\pgfusepath{stroke}%
\end{pgfscope}%
\begin{pgfscope}%
\pgfsetbuttcap%
\pgfsetmiterjoin%
\pgfsetlinewidth{0.000000pt}%
\definecolor{currentstroke}{rgb}{0.800000,0.800000,0.800000}%
\pgfsetstrokecolor{currentstroke}%
\pgfsetstrokeopacity{0.000000}%
\pgfsetdash{}{0pt}%
\pgfpathmoveto{\pgfqpoint{3.283222in}{0.884844in}}%
\pgfpathlineto{\pgfqpoint{3.338778in}{0.884844in}}%
\pgfpathquadraticcurveto{\pgfqpoint{3.366556in}{0.884844in}}{\pgfqpoint{3.366556in}{0.912622in}}%
\pgfpathlineto{\pgfqpoint{3.366556in}{0.968178in}}%
\pgfpathquadraticcurveto{\pgfqpoint{3.366556in}{0.995956in}}{\pgfqpoint{3.338778in}{0.995956in}}%
\pgfpathlineto{\pgfqpoint{3.283222in}{0.995956in}}%
\pgfpathquadraticcurveto{\pgfqpoint{3.255444in}{0.995956in}}{\pgfqpoint{3.255444in}{0.968178in}}%
\pgfpathlineto{\pgfqpoint{3.255444in}{0.912622in}}%
\pgfpathquadraticcurveto{\pgfqpoint{3.255444in}{0.884844in}}{\pgfqpoint{3.283222in}{0.884844in}}%
\pgfpathclose%
\pgfusepath{}%
\end{pgfscope}%
\end{pgfpicture}%
\makeatother%
\endgroup%

%% file: sharpe-rewards-vs-rho.pgf
\begingroup%
\makeatletter%
\begin{pgfpicture}%
\pgfpathrectangle{\pgfpointorigin}{\pgfqpoint{5.400000in}{2.700000in}}%
\pgfusepath{use as bounding box, clip}%
\begin{pgfscope}%
\pgfsetbuttcap%
\pgfsetmiterjoin%
\definecolor{currentfill}{rgb}{1.000000,1.000000,1.000000}%
\pgfsetfillcolor{currentfill}%
\pgfsetlinewidth{0.000000pt}%
\definecolor{currentstroke}{rgb}{1.000000,1.000000,1.000000}%
\pgfsetstrokecolor{currentstroke}%
\pgfsetdash{}{0pt}%
\pgfpathmoveto{\pgfqpoint{0.000000in}{0.000000in}}%
\pgfpathlineto{\pgfqpoint{5.400000in}{0.000000in}}%
\pgfpathlineto{\pgfqpoint{5.400000in}{2.700000in}}%
\pgfpathlineto{\pgfqpoint{0.000000in}{2.700000in}}%
\pgfpathclose%
\pgfusepath{fill}%
\end{pgfscope}%
\begin{pgfscope}%
\pgfsetbuttcap%
\pgfsetmiterjoin%
\definecolor{currentfill}{rgb}{1.000000,1.000000,1.000000}%
\pgfsetfillcolor{currentfill}%
\pgfsetlinewidth{0.000000pt}%
\definecolor{currentstroke}{rgb}{0.000000,0.000000,0.000000}%
\pgfsetstrokecolor{currentstroke}%
\pgfsetstrokeopacity{0.000000}%
\pgfsetdash{}{0pt}%
\pgfpathmoveto{\pgfqpoint{0.675831in}{0.555056in}}%
\pgfpathlineto{\pgfqpoint{5.215000in}{0.555056in}}%
\pgfpathlineto{\pgfqpoint{5.215000in}{2.515000in}}%
\pgfpathlineto{\pgfqpoint{0.675831in}{2.515000in}}%
\pgfpathclose%
\pgfusepath{fill}%
\end{pgfscope}%
\begin{pgfscope}%
\pgfsetbuttcap%
\pgfsetroundjoin%
\definecolor{currentfill}{rgb}{0.000000,0.000000,0.000000}%
\pgfsetfillcolor{currentfill}%
\pgfsetlinewidth{0.803000pt}%
\definecolor{currentstroke}{rgb}{0.000000,0.000000,0.000000}%
\pgfsetstrokecolor{currentstroke}%
\pgfsetdash{}{0pt}%
\pgfsys@defobject{currentmarker}{\pgfqpoint{0.000000in}{-0.048611in}}{\pgfqpoint{0.000000in}{0.000000in}}{%
\pgfpathmoveto{\pgfqpoint{0.000000in}{0.000000in}}%
\pgfpathlineto{\pgfqpoint{0.000000in}{-0.048611in}}%
\pgfusepath{stroke,fill}%
}%
\begin{pgfscope}%
\pgfsys@transformshift{0.744607in}{0.555056in}%
\pgfsys@useobject{currentmarker}{}%
\end{pgfscope}%
\end{pgfscope}%
\begin{pgfscope}%
\pgftext[x=0.744607in,y=0.457834in,,top]{\rmfamily\fontsize{10.000000}{12.000000}\selectfont \(\displaystyle 0.00000\)}%
\end{pgfscope}%
\begin{pgfscope}%
\pgfsetbuttcap%
\pgfsetroundjoin%
\definecolor{currentfill}{rgb}{0.000000,0.000000,0.000000}%
\pgfsetfillcolor{currentfill}%
\pgfsetlinewidth{0.803000pt}%
\definecolor{currentstroke}{rgb}{0.000000,0.000000,0.000000}%
\pgfsetstrokecolor{currentstroke}%
\pgfsetdash{}{0pt}%
\pgfsys@defobject{currentmarker}{\pgfqpoint{0.000000in}{-0.048611in}}{\pgfqpoint{0.000000in}{0.000000in}}{%
\pgfpathmoveto{\pgfqpoint{0.000000in}{0.000000in}}%
\pgfpathlineto{\pgfqpoint{0.000000in}{-0.048611in}}%
\pgfusepath{stroke,fill}%
}%
\begin{pgfscope}%
\pgfsys@transformshift{1.727111in}{0.555056in}%
\pgfsys@useobject{currentmarker}{}%
\end{pgfscope}%
\end{pgfscope}%
\begin{pgfscope}%
\pgftext[x=1.727111in,y=0.457834in,,top]{\rmfamily\fontsize{10.000000}{12.000000}\selectfont \(\displaystyle 0.00005\)}%
\end{pgfscope}%
\begin{pgfscope}%
\pgfsetbuttcap%
\pgfsetroundjoin%
\definecolor{currentfill}{rgb}{0.000000,0.000000,0.000000}%
\pgfsetfillcolor{currentfill}%
\pgfsetlinewidth{0.803000pt}%
\definecolor{currentstroke}{rgb}{0.000000,0.000000,0.000000}%
\pgfsetstrokecolor{currentstroke}%
\pgfsetdash{}{0pt}%
\pgfsys@defobject{currentmarker}{\pgfqpoint{0.000000in}{-0.048611in}}{\pgfqpoint{0.000000in}{0.000000in}}{%
\pgfpathmoveto{\pgfqpoint{0.000000in}{0.000000in}}%
\pgfpathlineto{\pgfqpoint{0.000000in}{-0.048611in}}%
\pgfusepath{stroke,fill}%
}%
\begin{pgfscope}%
\pgfsys@transformshift{2.709615in}{0.555056in}%
\pgfsys@useobject{currentmarker}{}%
\end{pgfscope}%
\end{pgfscope}%
\begin{pgfscope}%
\pgftext[x=2.709615in,y=0.457834in,,top]{\rmfamily\fontsize{10.000000}{12.000000}\selectfont \(\displaystyle 0.00010\)}%
\end{pgfscope}%
\begin{pgfscope}%
\pgfsetbuttcap%
\pgfsetroundjoin%
\definecolor{currentfill}{rgb}{0.000000,0.000000,0.000000}%
\pgfsetfillcolor{currentfill}%
\pgfsetlinewidth{0.803000pt}%
\definecolor{currentstroke}{rgb}{0.000000,0.000000,0.000000}%
\pgfsetstrokecolor{currentstroke}%
\pgfsetdash{}{0pt}%
\pgfsys@defobject{currentmarker}{\pgfqpoint{0.000000in}{-0.048611in}}{\pgfqpoint{0.000000in}{0.000000in}}{%
\pgfpathmoveto{\pgfqpoint{0.000000in}{0.000000in}}%
\pgfpathlineto{\pgfqpoint{0.000000in}{-0.048611in}}%
\pgfusepath{stroke,fill}%
}%
\begin{pgfscope}%
\pgfsys@transformshift{3.692119in}{0.555056in}%
\pgfsys@useobject{currentmarker}{}%
\end{pgfscope}%
\end{pgfscope}%
\begin{pgfscope}%
\pgftext[x=3.692119in,y=0.457834in,,top]{\rmfamily\fontsize{10.000000}{12.000000}\selectfont \(\displaystyle 0.00015\)}%
\end{pgfscope}%
\begin{pgfscope}%
\pgfsetbuttcap%
\pgfsetroundjoin%
\definecolor{currentfill}{rgb}{0.000000,0.000000,0.000000}%
\pgfsetfillcolor{currentfill}%
\pgfsetlinewidth{0.803000pt}%
\definecolor{currentstroke}{rgb}{0.000000,0.000000,0.000000}%
\pgfsetstrokecolor{currentstroke}%
\pgfsetdash{}{0pt}%
\pgfsys@defobject{currentmarker}{\pgfqpoint{0.000000in}{-0.048611in}}{\pgfqpoint{0.000000in}{0.000000in}}{%
\pgfpathmoveto{\pgfqpoint{0.000000in}{0.000000in}}%
\pgfpathlineto{\pgfqpoint{0.000000in}{-0.048611in}}%
\pgfusepath{stroke,fill}%
}%
\begin{pgfscope}%
\pgfsys@transformshift{4.674623in}{0.555056in}%
\pgfsys@useobject{currentmarker}{}%
\end{pgfscope}%
\end{pgfscope}%
\begin{pgfscope}%
\pgftext[x=4.674623in,y=0.457834in,,top]{\rmfamily\fontsize{10.000000}{12.000000}\selectfont \(\displaystyle 0.00020\)}%
\end{pgfscope}%
\begin{pgfscope}%
\pgftext[x=2.945416in,y=0.276139in,,top]{\rmfamily\fontsize{10.000000}{12.000000}\selectfont CARA \(\displaystyle \rho\)}%
\end{pgfscope}%
\begin{pgfscope}%
\pgfsetbuttcap%
\pgfsetroundjoin%
\definecolor{currentfill}{rgb}{0.000000,0.000000,0.000000}%
\pgfsetfillcolor{currentfill}%
\pgfsetlinewidth{0.803000pt}%
\definecolor{currentstroke}{rgb}{0.000000,0.000000,0.000000}%
\pgfsetstrokecolor{currentstroke}%
\pgfsetdash{}{0pt}%
\pgfsys@defobject{currentmarker}{\pgfqpoint{-0.048611in}{0.000000in}}{\pgfqpoint{0.000000in}{0.000000in}}{%
\pgfpathmoveto{\pgfqpoint{0.000000in}{0.000000in}}%
\pgfpathlineto{\pgfqpoint{-0.048611in}{0.000000in}}%
\pgfusepath{stroke,fill}%
}%
\begin{pgfscope}%
\pgfsys@transformshift{0.675831in}{0.686017in}%
\pgfsys@useobject{currentmarker}{}%
\end{pgfscope}%
\end{pgfscope}%
\begin{pgfscope}%
\pgftext[x=0.331695in,y=0.637800in,left,base]{\rmfamily\fontsize{10.000000}{12.000000}\selectfont \(\displaystyle 0.08\)}%
\end{pgfscope}%
\begin{pgfscope}%
\pgfsetbuttcap%
\pgfsetroundjoin%
\definecolor{currentfill}{rgb}{0.000000,0.000000,0.000000}%
\pgfsetfillcolor{currentfill}%
\pgfsetlinewidth{0.803000pt}%
\definecolor{currentstroke}{rgb}{0.000000,0.000000,0.000000}%
\pgfsetstrokecolor{currentstroke}%
\pgfsetdash{}{0pt}%
\pgfsys@defobject{currentmarker}{\pgfqpoint{-0.048611in}{0.000000in}}{\pgfqpoint{0.000000in}{0.000000in}}{%
\pgfpathmoveto{\pgfqpoint{0.000000in}{0.000000in}}%
\pgfpathlineto{\pgfqpoint{-0.048611in}{0.000000in}}%
\pgfusepath{stroke,fill}%
}%
\begin{pgfscope}%
\pgfsys@transformshift{0.675831in}{0.974797in}%
\pgfsys@useobject{currentmarker}{}%
\end{pgfscope}%
\end{pgfscope}%
\begin{pgfscope}%
\pgftext[x=0.331695in,y=0.926579in,left,base]{\rmfamily\fontsize{10.000000}{12.000000}\selectfont \(\displaystyle 0.10\)}%
\end{pgfscope}%
\begin{pgfscope}%
\pgfsetbuttcap%
\pgfsetroundjoin%
\definecolor{currentfill}{rgb}{0.000000,0.000000,0.000000}%
\pgfsetfillcolor{currentfill}%
\pgfsetlinewidth{0.803000pt}%
\definecolor{currentstroke}{rgb}{0.000000,0.000000,0.000000}%
\pgfsetstrokecolor{currentstroke}%
\pgfsetdash{}{0pt}%
\pgfsys@defobject{currentmarker}{\pgfqpoint{-0.048611in}{0.000000in}}{\pgfqpoint{0.000000in}{0.000000in}}{%
\pgfpathmoveto{\pgfqpoint{0.000000in}{0.000000in}}%
\pgfpathlineto{\pgfqpoint{-0.048611in}{0.000000in}}%
\pgfusepath{stroke,fill}%
}%
\begin{pgfscope}%
\pgfsys@transformshift{0.675831in}{1.263576in}%
\pgfsys@useobject{currentmarker}{}%
\end{pgfscope}%
\end{pgfscope}%
\begin{pgfscope}%
\pgftext[x=0.331695in,y=1.215358in,left,base]{\rmfamily\fontsize{10.000000}{12.000000}\selectfont \(\displaystyle 0.12\)}%
\end{pgfscope}%
\begin{pgfscope}%
\pgfsetbuttcap%
\pgfsetroundjoin%
\definecolor{currentfill}{rgb}{0.000000,0.000000,0.000000}%
\pgfsetfillcolor{currentfill}%
\pgfsetlinewidth{0.803000pt}%
\definecolor{currentstroke}{rgb}{0.000000,0.000000,0.000000}%
\pgfsetstrokecolor{currentstroke}%
\pgfsetdash{}{0pt}%
\pgfsys@defobject{currentmarker}{\pgfqpoint{-0.048611in}{0.000000in}}{\pgfqpoint{0.000000in}{0.000000in}}{%
\pgfpathmoveto{\pgfqpoint{0.000000in}{0.000000in}}%
\pgfpathlineto{\pgfqpoint{-0.048611in}{0.000000in}}%
\pgfusepath{stroke,fill}%
}%
\begin{pgfscope}%
\pgfsys@transformshift{0.675831in}{1.552355in}%
\pgfsys@useobject{currentmarker}{}%
\end{pgfscope}%
\end{pgfscope}%
\begin{pgfscope}%
\pgftext[x=0.331695in,y=1.504137in,left,base]{\rmfamily\fontsize{10.000000}{12.000000}\selectfont \(\displaystyle 0.14\)}%
\end{pgfscope}%
\begin{pgfscope}%
\pgfsetbuttcap%
\pgfsetroundjoin%
\definecolor{currentfill}{rgb}{0.000000,0.000000,0.000000}%
\pgfsetfillcolor{currentfill}%
\pgfsetlinewidth{0.803000pt}%
\definecolor{currentstroke}{rgb}{0.000000,0.000000,0.000000}%
\pgfsetstrokecolor{currentstroke}%
\pgfsetdash{}{0pt}%
\pgfsys@defobject{currentmarker}{\pgfqpoint{-0.048611in}{0.000000in}}{\pgfqpoint{0.000000in}{0.000000in}}{%
\pgfpathmoveto{\pgfqpoint{0.000000in}{0.000000in}}%
\pgfpathlineto{\pgfqpoint{-0.048611in}{0.000000in}}%
\pgfusepath{stroke,fill}%
}%
\begin{pgfscope}%
\pgfsys@transformshift{0.675831in}{1.841134in}%
\pgfsys@useobject{currentmarker}{}%
\end{pgfscope}%
\end{pgfscope}%
\begin{pgfscope}%
\pgftext[x=0.331695in,y=1.792916in,left,base]{\rmfamily\fontsize{10.000000}{12.000000}\selectfont \(\displaystyle 0.16\)}%
\end{pgfscope}%
\begin{pgfscope}%
\pgfsetbuttcap%
\pgfsetroundjoin%
\definecolor{currentfill}{rgb}{0.000000,0.000000,0.000000}%
\pgfsetfillcolor{currentfill}%
\pgfsetlinewidth{0.803000pt}%
\definecolor{currentstroke}{rgb}{0.000000,0.000000,0.000000}%
\pgfsetstrokecolor{currentstroke}%
\pgfsetdash{}{0pt}%
\pgfsys@defobject{currentmarker}{\pgfqpoint{-0.048611in}{0.000000in}}{\pgfqpoint{0.000000in}{0.000000in}}{%
\pgfpathmoveto{\pgfqpoint{0.000000in}{0.000000in}}%
\pgfpathlineto{\pgfqpoint{-0.048611in}{0.000000in}}%
\pgfusepath{stroke,fill}%
}%
\begin{pgfscope}%
\pgfsys@transformshift{0.675831in}{2.129913in}%
\pgfsys@useobject{currentmarker}{}%
\end{pgfscope}%
\end{pgfscope}%
\begin{pgfscope}%
\pgftext[x=0.331695in,y=2.081695in,left,base]{\rmfamily\fontsize{10.000000}{12.000000}\selectfont \(\displaystyle 0.18\)}%
\end{pgfscope}%
\begin{pgfscope}%
\pgfsetbuttcap%
\pgfsetroundjoin%
\definecolor{currentfill}{rgb}{0.000000,0.000000,0.000000}%
\pgfsetfillcolor{currentfill}%
\pgfsetlinewidth{0.803000pt}%
\definecolor{currentstroke}{rgb}{0.000000,0.000000,0.000000}%
\pgfsetstrokecolor{currentstroke}%
\pgfsetdash{}{0pt}%
\pgfsys@defobject{currentmarker}{\pgfqpoint{-0.048611in}{0.000000in}}{\pgfqpoint{0.000000in}{0.000000in}}{%
\pgfpathmoveto{\pgfqpoint{0.000000in}{0.000000in}}%
\pgfpathlineto{\pgfqpoint{-0.048611in}{0.000000in}}%
\pgfusepath{stroke,fill}%
}%
\begin{pgfscope}%
\pgfsys@transformshift{0.675831in}{2.418692in}%
\pgfsys@useobject{currentmarker}{}%
\end{pgfscope}%
\end{pgfscope}%
\begin{pgfscope}%
\pgftext[x=0.331695in,y=2.370474in,left,base]{\rmfamily\fontsize{10.000000}{12.000000}\selectfont \(\displaystyle 0.20\)}%
\end{pgfscope}%
\begin{pgfscope}%
\definecolor{textcolor}{rgb}{0.000000,0.000000,1.000000}%
\pgfsetstrokecolor{textcolor}%
\pgfsetfillcolor{textcolor}%
\pgftext[x=0.276139in,y=1.535028in,,bottom,rotate=90.000000]{\color{textcolor}\rmfamily\fontsize{10.000000}{12.000000}\selectfont Sharpe ratio}%
\end{pgfscope}%
\begin{pgfscope}%
\pgfpathrectangle{\pgfqpoint{0.675831in}{0.555056in}}{\pgfqpoint{4.539169in}{1.959944in}}%
\pgfusepath{clip}%
\pgfsetrectcap%
\pgfsetroundjoin%
\pgfsetlinewidth{1.003750pt}%
\definecolor{currentstroke}{rgb}{0.000000,0.000000,1.000000}%
\pgfsetstrokecolor{currentstroke}%
\pgfsetdash{}{0pt}%
\pgfpathmoveto{\pgfqpoint{0.882157in}{2.425912in}}%
\pgfpathlineto{\pgfqpoint{1.000058in}{2.425912in}}%
\pgfpathlineto{\pgfqpoint{1.117958in}{2.360936in}}%
\pgfpathlineto{\pgfqpoint{1.235859in}{2.256976in}}%
\pgfpathlineto{\pgfqpoint{1.353759in}{2.158791in}}%
\pgfpathlineto{\pgfqpoint{1.471660in}{2.060606in}}%
\pgfpathlineto{\pgfqpoint{1.589560in}{1.920548in}}%
\pgfpathlineto{\pgfqpoint{1.707461in}{1.803593in}}%
\pgfpathlineto{\pgfqpoint{1.825361in}{1.698188in}}%
\pgfpathlineto{\pgfqpoint{1.943261in}{1.600003in}}%
\pgfpathlineto{\pgfqpoint{2.061162in}{1.513370in}}%
\pgfpathlineto{\pgfqpoint{2.179062in}{1.433955in}}%
\pgfpathlineto{\pgfqpoint{2.296963in}{1.329995in}}%
\pgfpathlineto{\pgfqpoint{2.414863in}{1.226034in}}%
\pgfpathlineto{\pgfqpoint{2.532764in}{1.136513in}}%
\pgfpathlineto{\pgfqpoint{2.650664in}{1.059986in}}%
\pgfpathlineto{\pgfqpoint{2.768565in}{0.997899in}}%
\pgfpathlineto{\pgfqpoint{2.886465in}{0.957470in}}%
\pgfpathlineto{\pgfqpoint{3.004366in}{0.931480in}}%
\pgfpathlineto{\pgfqpoint{3.122266in}{0.906933in}}%
\pgfpathlineto{\pgfqpoint{3.240167in}{0.883831in}}%
\pgfpathlineto{\pgfqpoint{3.358067in}{0.862173in}}%
\pgfpathlineto{\pgfqpoint{3.475968in}{0.840514in}}%
\pgfpathlineto{\pgfqpoint{3.593868in}{0.820300in}}%
\pgfpathlineto{\pgfqpoint{3.711769in}{0.801529in}}%
\pgfpathlineto{\pgfqpoint{3.829669in}{0.784202in}}%
\pgfpathlineto{\pgfqpoint{3.947570in}{0.766876in}}%
\pgfpathlineto{\pgfqpoint{4.065470in}{0.749549in}}%
\pgfpathlineto{\pgfqpoint{4.183371in}{0.735110in}}%
\pgfpathlineto{\pgfqpoint{4.301271in}{0.719227in}}%
\pgfpathlineto{\pgfqpoint{4.419172in}{0.706232in}}%
\pgfpathlineto{\pgfqpoint{4.537072in}{0.691793in}}%
\pgfpathlineto{\pgfqpoint{4.654973in}{0.678798in}}%
\pgfpathlineto{\pgfqpoint{4.772873in}{0.667247in}}%
\pgfpathlineto{\pgfqpoint{4.890774in}{0.655696in}}%
\pgfpathlineto{\pgfqpoint{5.008674in}{0.644144in}}%
\pgfusepath{stroke}%
\end{pgfscope}%
\begin{pgfscope}%
\pgfsetrectcap%
\pgfsetmiterjoin%
\pgfsetlinewidth{0.803000pt}%
\definecolor{currentstroke}{rgb}{0.000000,0.000000,0.000000}%
\pgfsetstrokecolor{currentstroke}%
\pgfsetdash{}{0pt}%
\pgfpathmoveto{\pgfqpoint{0.675831in}{0.555056in}}%
\pgfpathlineto{\pgfqpoint{0.675831in}{2.515000in}}%
\pgfusepath{stroke}%
\end{pgfscope}%
\begin{pgfscope}%
\pgfsetrectcap%
\pgfsetmiterjoin%
\pgfsetlinewidth{0.803000pt}%
\definecolor{currentstroke}{rgb}{0.000000,0.000000,0.000000}%
\pgfsetstrokecolor{currentstroke}%
\pgfsetdash{}{0pt}%
\pgfpathmoveto{\pgfqpoint{5.215000in}{0.555056in}}%
\pgfpathlineto{\pgfqpoint{5.215000in}{2.515000in}}%
\pgfusepath{stroke}%
\end{pgfscope}%
\begin{pgfscope}%
\pgfsetrectcap%
\pgfsetmiterjoin%
\pgfsetlinewidth{0.803000pt}%
\definecolor{currentstroke}{rgb}{0.000000,0.000000,0.000000}%
\pgfsetstrokecolor{currentstroke}%
\pgfsetdash{}{0pt}%
\pgfpathmoveto{\pgfqpoint{0.675831in}{0.555056in}}%
\pgfpathlineto{\pgfqpoint{5.215000in}{0.555056in}}%
\pgfusepath{stroke}%
\end{pgfscope}%
\begin{pgfscope}%
\pgfsetrectcap%
\pgfsetmiterjoin%
\pgfsetlinewidth{0.803000pt}%
\definecolor{currentstroke}{rgb}{0.000000,0.000000,0.000000}%
\pgfsetstrokecolor{currentstroke}%
\pgfsetdash{}{0pt}%
\pgfpathmoveto{\pgfqpoint{0.675831in}{2.515000in}}%
\pgfpathlineto{\pgfqpoint{5.215000in}{2.515000in}}%
\pgfusepath{stroke}%
\end{pgfscope}%
\end{pgfpicture}%
\makeatother%
\endgroup%

%% file: sharpe-rewards-vs-rho-largepools.pgf
\begingroup%
\makeatletter%
\begin{pgfpicture}%
\pgfpathrectangle{\pgfpointorigin}{\pgfqpoint{5.400000in}{2.700000in}}%
\pgfusepath{use as bounding box, clip}%
\begin{pgfscope}%
\pgfsetbuttcap%
\pgfsetmiterjoin%
\definecolor{currentfill}{rgb}{1.000000,1.000000,1.000000}%
\pgfsetfillcolor{currentfill}%
\pgfsetlinewidth{0.000000pt}%
\definecolor{currentstroke}{rgb}{1.000000,1.000000,1.000000}%
\pgfsetstrokecolor{currentstroke}%
\pgfsetdash{}{0pt}%
\pgfpathmoveto{\pgfqpoint{0.000000in}{0.000000in}}%
\pgfpathlineto{\pgfqpoint{5.400000in}{0.000000in}}%
\pgfpathlineto{\pgfqpoint{5.400000in}{2.700000in}}%
\pgfpathlineto{\pgfqpoint{0.000000in}{2.700000in}}%
\pgfpathclose%
\pgfusepath{fill}%
\end{pgfscope}%
\begin{pgfscope}%
\pgfsetbuttcap%
\pgfsetmiterjoin%
\definecolor{currentfill}{rgb}{1.000000,1.000000,1.000000}%
\pgfsetfillcolor{currentfill}%
\pgfsetlinewidth{0.000000pt}%
\definecolor{currentstroke}{rgb}{0.000000,0.000000,0.000000}%
\pgfsetstrokecolor{currentstroke}%
\pgfsetstrokeopacity{0.000000}%
\pgfsetdash{}{0pt}%
\pgfpathmoveto{\pgfqpoint{0.814721in}{0.555056in}}%
\pgfpathlineto{\pgfqpoint{5.215000in}{0.555056in}}%
\pgfpathlineto{\pgfqpoint{5.215000in}{2.515000in}}%
\pgfpathlineto{\pgfqpoint{0.814721in}{2.515000in}}%
\pgfpathclose%
\pgfusepath{fill}%
\end{pgfscope}%
\begin{pgfscope}%
\pgfsetbuttcap%
\pgfsetroundjoin%
\definecolor{currentfill}{rgb}{0.000000,0.000000,0.000000}%
\pgfsetfillcolor{currentfill}%
\pgfsetlinewidth{0.803000pt}%
\definecolor{currentstroke}{rgb}{0.000000,0.000000,0.000000}%
\pgfsetstrokecolor{currentstroke}%
\pgfsetdash{}{0pt}%
\pgfsys@defobject{currentmarker}{\pgfqpoint{0.000000in}{-0.048611in}}{\pgfqpoint{0.000000in}{0.000000in}}{%
\pgfpathmoveto{\pgfqpoint{0.000000in}{0.000000in}}%
\pgfpathlineto{\pgfqpoint{0.000000in}{-0.048611in}}%
\pgfusepath{stroke,fill}%
}%
\begin{pgfscope}%
\pgfsys@transformshift{0.881391in}{0.555056in}%
\pgfsys@useobject{currentmarker}{}%
\end{pgfscope}%
\end{pgfscope}%
\begin{pgfscope}%
\pgftext[x=0.881391in,y=0.457834in,,top]{\rmfamily\fontsize{10.000000}{12.000000}\selectfont \(\displaystyle 0.00000\)}%
\end{pgfscope}%
\begin{pgfscope}%
\pgfsetbuttcap%
\pgfsetroundjoin%
\definecolor{currentfill}{rgb}{0.000000,0.000000,0.000000}%
\pgfsetfillcolor{currentfill}%
\pgfsetlinewidth{0.803000pt}%
\definecolor{currentstroke}{rgb}{0.000000,0.000000,0.000000}%
\pgfsetstrokecolor{currentstroke}%
\pgfsetdash{}{0pt}%
\pgfsys@defobject{currentmarker}{\pgfqpoint{0.000000in}{-0.048611in}}{\pgfqpoint{0.000000in}{0.000000in}}{%
\pgfpathmoveto{\pgfqpoint{0.000000in}{0.000000in}}%
\pgfpathlineto{\pgfqpoint{0.000000in}{-0.048611in}}%
\pgfusepath{stroke,fill}%
}%
\begin{pgfscope}%
\pgfsys@transformshift{1.833833in}{0.555056in}%
\pgfsys@useobject{currentmarker}{}%
\end{pgfscope}%
\end{pgfscope}%
\begin{pgfscope}%
\pgftext[x=1.833833in,y=0.457834in,,top]{\rmfamily\fontsize{10.000000}{12.000000}\selectfont \(\displaystyle 0.00005\)}%
\end{pgfscope}%
\begin{pgfscope}%
\pgfsetbuttcap%
\pgfsetroundjoin%
\definecolor{currentfill}{rgb}{0.000000,0.000000,0.000000}%
\pgfsetfillcolor{currentfill}%
\pgfsetlinewidth{0.803000pt}%
\definecolor{currentstroke}{rgb}{0.000000,0.000000,0.000000}%
\pgfsetstrokecolor{currentstroke}%
\pgfsetdash{}{0pt}%
\pgfsys@defobject{currentmarker}{\pgfqpoint{0.000000in}{-0.048611in}}{\pgfqpoint{0.000000in}{0.000000in}}{%
\pgfpathmoveto{\pgfqpoint{0.000000in}{0.000000in}}%
\pgfpathlineto{\pgfqpoint{0.000000in}{-0.048611in}}%
\pgfusepath{stroke,fill}%
}%
\begin{pgfscope}%
\pgfsys@transformshift{2.786274in}{0.555056in}%
\pgfsys@useobject{currentmarker}{}%
\end{pgfscope}%
\end{pgfscope}%
\begin{pgfscope}%
\pgftext[x=2.786274in,y=0.457834in,,top]{\rmfamily\fontsize{10.000000}{12.000000}\selectfont \(\displaystyle 0.00010\)}%
\end{pgfscope}%
\begin{pgfscope}%
\pgfsetbuttcap%
\pgfsetroundjoin%
\definecolor{currentfill}{rgb}{0.000000,0.000000,0.000000}%
\pgfsetfillcolor{currentfill}%
\pgfsetlinewidth{0.803000pt}%
\definecolor{currentstroke}{rgb}{0.000000,0.000000,0.000000}%
\pgfsetstrokecolor{currentstroke}%
\pgfsetdash{}{0pt}%
\pgfsys@defobject{currentmarker}{\pgfqpoint{0.000000in}{-0.048611in}}{\pgfqpoint{0.000000in}{0.000000in}}{%
\pgfpathmoveto{\pgfqpoint{0.000000in}{0.000000in}}%
\pgfpathlineto{\pgfqpoint{0.000000in}{-0.048611in}}%
\pgfusepath{stroke,fill}%
}%
\begin{pgfscope}%
\pgfsys@transformshift{3.738716in}{0.555056in}%
\pgfsys@useobject{currentmarker}{}%
\end{pgfscope}%
\end{pgfscope}%
\begin{pgfscope}%
\pgftext[x=3.738716in,y=0.457834in,,top]{\rmfamily\fontsize{10.000000}{12.000000}\selectfont \(\displaystyle 0.00015\)}%
\end{pgfscope}%
\begin{pgfscope}%
\pgfsetbuttcap%
\pgfsetroundjoin%
\definecolor{currentfill}{rgb}{0.000000,0.000000,0.000000}%
\pgfsetfillcolor{currentfill}%
\pgfsetlinewidth{0.803000pt}%
\definecolor{currentstroke}{rgb}{0.000000,0.000000,0.000000}%
\pgfsetstrokecolor{currentstroke}%
\pgfsetdash{}{0pt}%
\pgfsys@defobject{currentmarker}{\pgfqpoint{0.000000in}{-0.048611in}}{\pgfqpoint{0.000000in}{0.000000in}}{%
\pgfpathmoveto{\pgfqpoint{0.000000in}{0.000000in}}%
\pgfpathlineto{\pgfqpoint{0.000000in}{-0.048611in}}%
\pgfusepath{stroke,fill}%
}%
\begin{pgfscope}%
\pgfsys@transformshift{4.691157in}{0.555056in}%
\pgfsys@useobject{currentmarker}{}%
\end{pgfscope}%
\end{pgfscope}%
\begin{pgfscope}%
\pgftext[x=4.691157in,y=0.457834in,,top]{\rmfamily\fontsize{10.000000}{12.000000}\selectfont \(\displaystyle 0.00020\)}%
\end{pgfscope}%
\begin{pgfscope}%
\pgftext[x=3.014860in,y=0.276139in,,top]{\rmfamily\fontsize{10.000000}{12.000000}\selectfont CARA \(\displaystyle \rho\)}%
\end{pgfscope}%
\begin{pgfscope}%
\pgfsetbuttcap%
\pgfsetroundjoin%
\definecolor{currentfill}{rgb}{0.000000,0.000000,0.000000}%
\pgfsetfillcolor{currentfill}%
\pgfsetlinewidth{0.803000pt}%
\definecolor{currentstroke}{rgb}{0.000000,0.000000,0.000000}%
\pgfsetstrokecolor{currentstroke}%
\pgfsetdash{}{0pt}%
\pgfsys@defobject{currentmarker}{\pgfqpoint{-0.048611in}{0.000000in}}{\pgfqpoint{0.000000in}{0.000000in}}{%
\pgfpathmoveto{\pgfqpoint{0.000000in}{0.000000in}}%
\pgfpathlineto{\pgfqpoint{-0.048611in}{0.000000in}}%
\pgfusepath{stroke,fill}%
}%
\begin{pgfscope}%
\pgfsys@transformshift{0.814721in}{1.025952in}%
\pgfsys@useobject{currentmarker}{}%
\end{pgfscope}%
\end{pgfscope}%
\begin{pgfscope}%
\pgftext[x=0.331695in,y=0.977734in,left,base]{\rmfamily\fontsize{10.000000}{12.000000}\selectfont \(\displaystyle 0.0170\)}%
\end{pgfscope}%
\begin{pgfscope}%
\pgfsetbuttcap%
\pgfsetroundjoin%
\definecolor{currentfill}{rgb}{0.000000,0.000000,0.000000}%
\pgfsetfillcolor{currentfill}%
\pgfsetlinewidth{0.803000pt}%
\definecolor{currentstroke}{rgb}{0.000000,0.000000,0.000000}%
\pgfsetstrokecolor{currentstroke}%
\pgfsetdash{}{0pt}%
\pgfsys@defobject{currentmarker}{\pgfqpoint{-0.048611in}{0.000000in}}{\pgfqpoint{0.000000in}{0.000000in}}{%
\pgfpathmoveto{\pgfqpoint{0.000000in}{0.000000in}}%
\pgfpathlineto{\pgfqpoint{-0.048611in}{0.000000in}}%
\pgfusepath{stroke,fill}%
}%
\begin{pgfscope}%
\pgfsys@transformshift{0.814721in}{1.535028in}%
\pgfsys@useobject{currentmarker}{}%
\end{pgfscope}%
\end{pgfscope}%
\begin{pgfscope}%
\pgftext[x=0.331695in,y=1.486810in,left,base]{\rmfamily\fontsize{10.000000}{12.000000}\selectfont \(\displaystyle 0.0175\)}%
\end{pgfscope}%
\begin{pgfscope}%
\pgfsetbuttcap%
\pgfsetroundjoin%
\definecolor{currentfill}{rgb}{0.000000,0.000000,0.000000}%
\pgfsetfillcolor{currentfill}%
\pgfsetlinewidth{0.803000pt}%
\definecolor{currentstroke}{rgb}{0.000000,0.000000,0.000000}%
\pgfsetstrokecolor{currentstroke}%
\pgfsetdash{}{0pt}%
\pgfsys@defobject{currentmarker}{\pgfqpoint{-0.048611in}{0.000000in}}{\pgfqpoint{0.000000in}{0.000000in}}{%
\pgfpathmoveto{\pgfqpoint{0.000000in}{0.000000in}}%
\pgfpathlineto{\pgfqpoint{-0.048611in}{0.000000in}}%
\pgfusepath{stroke,fill}%
}%
\begin{pgfscope}%
\pgfsys@transformshift{0.814721in}{2.044104in}%
\pgfsys@useobject{currentmarker}{}%
\end{pgfscope}%
\end{pgfscope}%
\begin{pgfscope}%
\pgftext[x=0.331695in,y=1.995887in,left,base]{\rmfamily\fontsize{10.000000}{12.000000}\selectfont \(\displaystyle 0.0180\)}%
\end{pgfscope}%
\begin{pgfscope}%
\definecolor{textcolor}{rgb}{0.000000,0.000000,1.000000}%
\pgfsetstrokecolor{textcolor}%
\pgfsetfillcolor{textcolor}%
\pgftext[x=0.276139in,y=1.535028in,,bottom,rotate=90.000000]{\color{textcolor}\rmfamily\fontsize{10.000000}{12.000000}\selectfont Sharpe ratio}%
\end{pgfscope}%
\begin{pgfscope}%
\pgfpathrectangle{\pgfqpoint{0.814721in}{0.555056in}}{\pgfqpoint{4.400279in}{1.959944in}}%
\pgfusepath{clip}%
\pgfsetrectcap%
\pgfsetroundjoin%
\pgfsetlinewidth{1.003750pt}%
\definecolor{currentstroke}{rgb}{0.000000,0.000000,1.000000}%
\pgfsetstrokecolor{currentstroke}%
\pgfsetdash{}{0pt}%
\pgfpathmoveto{\pgfqpoint{1.014733in}{1.535028in}}%
\pgfpathlineto{\pgfqpoint{1.129026in}{1.535028in}}%
\pgfpathlineto{\pgfqpoint{1.243319in}{1.535028in}}%
\pgfpathlineto{\pgfqpoint{1.357612in}{1.535028in}}%
\pgfpathlineto{\pgfqpoint{1.471905in}{1.535028in}}%
\pgfpathlineto{\pgfqpoint{1.586198in}{1.535028in}}%
\pgfpathlineto{\pgfqpoint{1.700491in}{1.535028in}}%
\pgfpathlineto{\pgfqpoint{1.814784in}{1.535028in}}%
\pgfpathlineto{\pgfqpoint{1.929077in}{1.535028in}}%
\pgfpathlineto{\pgfqpoint{2.043370in}{1.535028in}}%
\pgfpathlineto{\pgfqpoint{2.157663in}{1.535028in}}%
\pgfpathlineto{\pgfqpoint{2.271956in}{1.535028in}}%
\pgfpathlineto{\pgfqpoint{2.386249in}{1.535028in}}%
\pgfpathlineto{\pgfqpoint{2.500542in}{1.535028in}}%
\pgfpathlineto{\pgfqpoint{2.614835in}{1.535028in}}%
\pgfpathlineto{\pgfqpoint{2.729128in}{1.535028in}}%
\pgfpathlineto{\pgfqpoint{2.843421in}{1.535028in}}%
\pgfpathlineto{\pgfqpoint{2.957714in}{1.535028in}}%
\pgfpathlineto{\pgfqpoint{3.072007in}{1.535028in}}%
\pgfpathlineto{\pgfqpoint{3.186300in}{1.535028in}}%
\pgfpathlineto{\pgfqpoint{3.300593in}{1.535028in}}%
\pgfpathlineto{\pgfqpoint{3.414886in}{1.535028in}}%
\pgfpathlineto{\pgfqpoint{3.529179in}{1.535028in}}%
\pgfpathlineto{\pgfqpoint{3.643472in}{1.535028in}}%
\pgfpathlineto{\pgfqpoint{3.757765in}{1.535028in}}%
\pgfpathlineto{\pgfqpoint{3.872058in}{1.535028in}}%
\pgfpathlineto{\pgfqpoint{3.986351in}{1.535028in}}%
\pgfpathlineto{\pgfqpoint{4.100644in}{1.535028in}}%
\pgfpathlineto{\pgfqpoint{4.214936in}{1.535028in}}%
\pgfpathlineto{\pgfqpoint{4.329229in}{1.535028in}}%
\pgfpathlineto{\pgfqpoint{4.443522in}{1.535028in}}%
\pgfpathlineto{\pgfqpoint{4.557815in}{1.535028in}}%
\pgfpathlineto{\pgfqpoint{4.672108in}{1.535028in}}%
\pgfpathlineto{\pgfqpoint{4.786401in}{1.535028in}}%
\pgfpathlineto{\pgfqpoint{4.900694in}{1.535028in}}%
\pgfpathlineto{\pgfqpoint{5.014987in}{1.535028in}}%
\pgfusepath{stroke}%
\end{pgfscope}%
\begin{pgfscope}%
\pgfsetrectcap%
\pgfsetmiterjoin%
\pgfsetlinewidth{0.803000pt}%
\definecolor{currentstroke}{rgb}{0.000000,0.000000,0.000000}%
\pgfsetstrokecolor{currentstroke}%
\pgfsetdash{}{0pt}%
\pgfpathmoveto{\pgfqpoint{0.814721in}{0.555056in}}%
\pgfpathlineto{\pgfqpoint{0.814721in}{2.515000in}}%
\pgfusepath{stroke}%
\end{pgfscope}%
\begin{pgfscope}%
\pgfsetrectcap%
\pgfsetmiterjoin%
\pgfsetlinewidth{0.803000pt}%
\definecolor{currentstroke}{rgb}{0.000000,0.000000,0.000000}%
\pgfsetstrokecolor{currentstroke}%
\pgfsetdash{}{0pt}%
\pgfpathmoveto{\pgfqpoint{5.215000in}{0.555056in}}%
\pgfpathlineto{\pgfqpoint{5.215000in}{2.515000in}}%
\pgfusepath{stroke}%
\end{pgfscope}%
\begin{pgfscope}%
\pgfsetrectcap%
\pgfsetmiterjoin%
\pgfsetlinewidth{0.803000pt}%
\definecolor{currentstroke}{rgb}{0.000000,0.000000,0.000000}%
\pgfsetstrokecolor{currentstroke}%
\pgfsetdash{}{0pt}%
\pgfpathmoveto{\pgfqpoint{0.814721in}{0.555056in}}%
\pgfpathlineto{\pgfqpoint{5.215000in}{0.555056in}}%
\pgfusepath{stroke}%
\end{pgfscope}%
\begin{pgfscope}%
\pgfsetrectcap%
\pgfsetmiterjoin%
\pgfsetlinewidth{0.803000pt}%
\definecolor{currentstroke}{rgb}{0.000000,0.000000,0.000000}%
\pgfsetstrokecolor{currentstroke}%
\pgfsetdash{}{0pt}%
\pgfpathmoveto{\pgfqpoint{0.814721in}{2.515000in}}%
\pgfpathlineto{\pgfqpoint{5.215000in}{2.515000in}}%
\pgfusepath{stroke}%
\end{pgfscope}%
\end{pgfpicture}%
\makeatother%
\endgroup%

%% file: sharpe-rewards-vs-miningpower-dpool.pgf
\begingroup%
\makeatletter%
\begin{pgfpicture}%
\pgfpathrectangle{\pgfpointorigin}{\pgfqpoint{5.400000in}{2.700000in}}%
\pgfusepath{use as bounding box, clip}%
\begin{pgfscope}%
\pgfsetbuttcap%
\pgfsetmiterjoin%
\definecolor{currentfill}{rgb}{1.000000,1.000000,1.000000}%
\pgfsetfillcolor{currentfill}%
\pgfsetlinewidth{0.000000pt}%
\definecolor{currentstroke}{rgb}{1.000000,1.000000,1.000000}%
\pgfsetstrokecolor{currentstroke}%
\pgfsetdash{}{0pt}%
\pgfpathmoveto{\pgfqpoint{0.000000in}{0.000000in}}%
\pgfpathlineto{\pgfqpoint{5.400000in}{0.000000in}}%
\pgfpathlineto{\pgfqpoint{5.400000in}{2.700000in}}%
\pgfpathlineto{\pgfqpoint{0.000000in}{2.700000in}}%
\pgfpathclose%
\pgfusepath{fill}%
\end{pgfscope}%
\begin{pgfscope}%
\pgfsetbuttcap%
\pgfsetmiterjoin%
\definecolor{currentfill}{rgb}{1.000000,1.000000,1.000000}%
\pgfsetfillcolor{currentfill}%
\pgfsetlinewidth{0.000000pt}%
\definecolor{currentstroke}{rgb}{0.000000,0.000000,0.000000}%
\pgfsetstrokecolor{currentstroke}%
\pgfsetstrokeopacity{0.000000}%
\pgfsetdash{}{0pt}%
\pgfpathmoveto{\pgfqpoint{0.745276in}{0.555327in}}%
\pgfpathlineto{\pgfqpoint{5.215000in}{0.555327in}}%
\pgfpathlineto{\pgfqpoint{5.215000in}{2.515000in}}%
\pgfpathlineto{\pgfqpoint{0.745276in}{2.515000in}}%
\pgfpathclose%
\pgfusepath{fill}%
\end{pgfscope}%
\begin{pgfscope}%
\pgfsetbuttcap%
\pgfsetroundjoin%
\definecolor{currentfill}{rgb}{0.000000,0.000000,0.000000}%
\pgfsetfillcolor{currentfill}%
\pgfsetlinewidth{0.803000pt}%
\definecolor{currentstroke}{rgb}{0.000000,0.000000,0.000000}%
\pgfsetstrokecolor{currentstroke}%
\pgfsetdash{}{0pt}%
\pgfsys@defobject{currentmarker}{\pgfqpoint{0.000000in}{-0.048611in}}{\pgfqpoint{0.000000in}{0.000000in}}{%
\pgfpathmoveto{\pgfqpoint{0.000000in}{0.000000in}}%
\pgfpathlineto{\pgfqpoint{0.000000in}{-0.048611in}}%
\pgfusepath{stroke,fill}%
}%
\begin{pgfscope}%
\pgfsys@transformshift{2.537546in}{0.555327in}%
\pgfsys@useobject{currentmarker}{}%
\end{pgfscope}%
\end{pgfscope}%
\begin{pgfscope}%
\pgftext[x=2.537546in,y=0.458105in,,top]{\rmfamily\fontsize{10.000000}{12.000000}\selectfont \(\displaystyle 10^{2}\)}%
\end{pgfscope}%
\begin{pgfscope}%
\pgfsetbuttcap%
\pgfsetroundjoin%
\definecolor{currentfill}{rgb}{0.000000,0.000000,0.000000}%
\pgfsetfillcolor{currentfill}%
\pgfsetlinewidth{0.803000pt}%
\definecolor{currentstroke}{rgb}{0.000000,0.000000,0.000000}%
\pgfsetstrokecolor{currentstroke}%
\pgfsetdash{}{0pt}%
\pgfsys@defobject{currentmarker}{\pgfqpoint{0.000000in}{-0.048611in}}{\pgfqpoint{0.000000in}{0.000000in}}{%
\pgfpathmoveto{\pgfqpoint{0.000000in}{0.000000in}}%
\pgfpathlineto{\pgfqpoint{0.000000in}{-0.048611in}}%
\pgfusepath{stroke,fill}%
}%
\begin{pgfscope}%
\pgfsys@transformshift{5.101485in}{0.555327in}%
\pgfsys@useobject{currentmarker}{}%
\end{pgfscope}%
\end{pgfscope}%
\begin{pgfscope}%
\pgftext[x=5.101485in,y=0.458105in,,top]{\rmfamily\fontsize{10.000000}{12.000000}\selectfont \(\displaystyle 10^{3}\)}%
\end{pgfscope}%
\begin{pgfscope}%
\pgfsetbuttcap%
\pgfsetroundjoin%
\definecolor{currentfill}{rgb}{0.000000,0.000000,0.000000}%
\pgfsetfillcolor{currentfill}%
\pgfsetlinewidth{0.602250pt}%
\definecolor{currentstroke}{rgb}{0.000000,0.000000,0.000000}%
\pgfsetstrokecolor{currentstroke}%
\pgfsetdash{}{0pt}%
\pgfsys@defobject{currentmarker}{\pgfqpoint{0.000000in}{-0.027778in}}{\pgfqpoint{0.000000in}{0.000000in}}{%
\pgfpathmoveto{\pgfqpoint{0.000000in}{0.000000in}}%
\pgfpathlineto{\pgfqpoint{0.000000in}{-0.027778in}}%
\pgfusepath{stroke,fill}%
}%
\begin{pgfscope}%
\pgfsys@transformshift{0.745429in}{0.555327in}%
\pgfsys@useobject{currentmarker}{}%
\end{pgfscope}%
\end{pgfscope}%
\begin{pgfscope}%
\pgfsetbuttcap%
\pgfsetroundjoin%
\definecolor{currentfill}{rgb}{0.000000,0.000000,0.000000}%
\pgfsetfillcolor{currentfill}%
\pgfsetlinewidth{0.602250pt}%
\definecolor{currentstroke}{rgb}{0.000000,0.000000,0.000000}%
\pgfsetstrokecolor{currentstroke}%
\pgfsetdash{}{0pt}%
\pgfsys@defobject{currentmarker}{\pgfqpoint{0.000000in}{-0.027778in}}{\pgfqpoint{0.000000in}{0.000000in}}{%
\pgfpathmoveto{\pgfqpoint{0.000000in}{0.000000in}}%
\pgfpathlineto{\pgfqpoint{0.000000in}{-0.027778in}}%
\pgfusepath{stroke,fill}%
}%
\begin{pgfscope}%
\pgfsys@transformshift{1.196917in}{0.555327in}%
\pgfsys@useobject{currentmarker}{}%
\end{pgfscope}%
\end{pgfscope}%
\begin{pgfscope}%
\pgfsetbuttcap%
\pgfsetroundjoin%
\definecolor{currentfill}{rgb}{0.000000,0.000000,0.000000}%
\pgfsetfillcolor{currentfill}%
\pgfsetlinewidth{0.602250pt}%
\definecolor{currentstroke}{rgb}{0.000000,0.000000,0.000000}%
\pgfsetstrokecolor{currentstroke}%
\pgfsetdash{}{0pt}%
\pgfsys@defobject{currentmarker}{\pgfqpoint{0.000000in}{-0.027778in}}{\pgfqpoint{0.000000in}{0.000000in}}{%
\pgfpathmoveto{\pgfqpoint{0.000000in}{0.000000in}}%
\pgfpathlineto{\pgfqpoint{0.000000in}{-0.027778in}}%
\pgfusepath{stroke,fill}%
}%
\begin{pgfscope}%
\pgfsys@transformshift{1.517252in}{0.555327in}%
\pgfsys@useobject{currentmarker}{}%
\end{pgfscope}%
\end{pgfscope}%
\begin{pgfscope}%
\pgfsetbuttcap%
\pgfsetroundjoin%
\definecolor{currentfill}{rgb}{0.000000,0.000000,0.000000}%
\pgfsetfillcolor{currentfill}%
\pgfsetlinewidth{0.602250pt}%
\definecolor{currentstroke}{rgb}{0.000000,0.000000,0.000000}%
\pgfsetstrokecolor{currentstroke}%
\pgfsetdash{}{0pt}%
\pgfsys@defobject{currentmarker}{\pgfqpoint{0.000000in}{-0.027778in}}{\pgfqpoint{0.000000in}{0.000000in}}{%
\pgfpathmoveto{\pgfqpoint{0.000000in}{0.000000in}}%
\pgfpathlineto{\pgfqpoint{0.000000in}{-0.027778in}}%
\pgfusepath{stroke,fill}%
}%
\begin{pgfscope}%
\pgfsys@transformshift{1.765723in}{0.555327in}%
\pgfsys@useobject{currentmarker}{}%
\end{pgfscope}%
\end{pgfscope}%
\begin{pgfscope}%
\pgfsetbuttcap%
\pgfsetroundjoin%
\definecolor{currentfill}{rgb}{0.000000,0.000000,0.000000}%
\pgfsetfillcolor{currentfill}%
\pgfsetlinewidth{0.602250pt}%
\definecolor{currentstroke}{rgb}{0.000000,0.000000,0.000000}%
\pgfsetstrokecolor{currentstroke}%
\pgfsetdash{}{0pt}%
\pgfsys@defobject{currentmarker}{\pgfqpoint{0.000000in}{-0.027778in}}{\pgfqpoint{0.000000in}{0.000000in}}{%
\pgfpathmoveto{\pgfqpoint{0.000000in}{0.000000in}}%
\pgfpathlineto{\pgfqpoint{0.000000in}{-0.027778in}}%
\pgfusepath{stroke,fill}%
}%
\begin{pgfscope}%
\pgfsys@transformshift{1.968739in}{0.555327in}%
\pgfsys@useobject{currentmarker}{}%
\end{pgfscope}%
\end{pgfscope}%
\begin{pgfscope}%
\pgfsetbuttcap%
\pgfsetroundjoin%
\definecolor{currentfill}{rgb}{0.000000,0.000000,0.000000}%
\pgfsetfillcolor{currentfill}%
\pgfsetlinewidth{0.602250pt}%
\definecolor{currentstroke}{rgb}{0.000000,0.000000,0.000000}%
\pgfsetstrokecolor{currentstroke}%
\pgfsetdash{}{0pt}%
\pgfsys@defobject{currentmarker}{\pgfqpoint{0.000000in}{-0.027778in}}{\pgfqpoint{0.000000in}{0.000000in}}{%
\pgfpathmoveto{\pgfqpoint{0.000000in}{0.000000in}}%
\pgfpathlineto{\pgfqpoint{0.000000in}{-0.027778in}}%
\pgfusepath{stroke,fill}%
}%
\begin{pgfscope}%
\pgfsys@transformshift{2.140387in}{0.555327in}%
\pgfsys@useobject{currentmarker}{}%
\end{pgfscope}%
\end{pgfscope}%
\begin{pgfscope}%
\pgfsetbuttcap%
\pgfsetroundjoin%
\definecolor{currentfill}{rgb}{0.000000,0.000000,0.000000}%
\pgfsetfillcolor{currentfill}%
\pgfsetlinewidth{0.602250pt}%
\definecolor{currentstroke}{rgb}{0.000000,0.000000,0.000000}%
\pgfsetstrokecolor{currentstroke}%
\pgfsetdash{}{0pt}%
\pgfsys@defobject{currentmarker}{\pgfqpoint{0.000000in}{-0.027778in}}{\pgfqpoint{0.000000in}{0.000000in}}{%
\pgfpathmoveto{\pgfqpoint{0.000000in}{0.000000in}}%
\pgfpathlineto{\pgfqpoint{0.000000in}{-0.027778in}}%
\pgfusepath{stroke,fill}%
}%
\begin{pgfscope}%
\pgfsys@transformshift{2.289075in}{0.555327in}%
\pgfsys@useobject{currentmarker}{}%
\end{pgfscope}%
\end{pgfscope}%
\begin{pgfscope}%
\pgfsetbuttcap%
\pgfsetroundjoin%
\definecolor{currentfill}{rgb}{0.000000,0.000000,0.000000}%
\pgfsetfillcolor{currentfill}%
\pgfsetlinewidth{0.602250pt}%
\definecolor{currentstroke}{rgb}{0.000000,0.000000,0.000000}%
\pgfsetstrokecolor{currentstroke}%
\pgfsetdash{}{0pt}%
\pgfsys@defobject{currentmarker}{\pgfqpoint{0.000000in}{-0.027778in}}{\pgfqpoint{0.000000in}{0.000000in}}{%
\pgfpathmoveto{\pgfqpoint{0.000000in}{0.000000in}}%
\pgfpathlineto{\pgfqpoint{0.000000in}{-0.027778in}}%
\pgfusepath{stroke,fill}%
}%
\begin{pgfscope}%
\pgfsys@transformshift{2.420226in}{0.555327in}%
\pgfsys@useobject{currentmarker}{}%
\end{pgfscope}%
\end{pgfscope}%
\begin{pgfscope}%
\pgfsetbuttcap%
\pgfsetroundjoin%
\definecolor{currentfill}{rgb}{0.000000,0.000000,0.000000}%
\pgfsetfillcolor{currentfill}%
\pgfsetlinewidth{0.602250pt}%
\definecolor{currentstroke}{rgb}{0.000000,0.000000,0.000000}%
\pgfsetstrokecolor{currentstroke}%
\pgfsetdash{}{0pt}%
\pgfsys@defobject{currentmarker}{\pgfqpoint{0.000000in}{-0.027778in}}{\pgfqpoint{0.000000in}{0.000000in}}{%
\pgfpathmoveto{\pgfqpoint{0.000000in}{0.000000in}}%
\pgfpathlineto{\pgfqpoint{0.000000in}{-0.027778in}}%
\pgfusepath{stroke,fill}%
}%
\begin{pgfscope}%
\pgfsys@transformshift{3.309369in}{0.555327in}%
\pgfsys@useobject{currentmarker}{}%
\end{pgfscope}%
\end{pgfscope}%
\begin{pgfscope}%
\pgfsetbuttcap%
\pgfsetroundjoin%
\definecolor{currentfill}{rgb}{0.000000,0.000000,0.000000}%
\pgfsetfillcolor{currentfill}%
\pgfsetlinewidth{0.602250pt}%
\definecolor{currentstroke}{rgb}{0.000000,0.000000,0.000000}%
\pgfsetstrokecolor{currentstroke}%
\pgfsetdash{}{0pt}%
\pgfsys@defobject{currentmarker}{\pgfqpoint{0.000000in}{-0.027778in}}{\pgfqpoint{0.000000in}{0.000000in}}{%
\pgfpathmoveto{\pgfqpoint{0.000000in}{0.000000in}}%
\pgfpathlineto{\pgfqpoint{0.000000in}{-0.027778in}}%
\pgfusepath{stroke,fill}%
}%
\begin{pgfscope}%
\pgfsys@transformshift{3.760856in}{0.555327in}%
\pgfsys@useobject{currentmarker}{}%
\end{pgfscope}%
\end{pgfscope}%
\begin{pgfscope}%
\pgfsetbuttcap%
\pgfsetroundjoin%
\definecolor{currentfill}{rgb}{0.000000,0.000000,0.000000}%
\pgfsetfillcolor{currentfill}%
\pgfsetlinewidth{0.602250pt}%
\definecolor{currentstroke}{rgb}{0.000000,0.000000,0.000000}%
\pgfsetstrokecolor{currentstroke}%
\pgfsetdash{}{0pt}%
\pgfsys@defobject{currentmarker}{\pgfqpoint{0.000000in}{-0.027778in}}{\pgfqpoint{0.000000in}{0.000000in}}{%
\pgfpathmoveto{\pgfqpoint{0.000000in}{0.000000in}}%
\pgfpathlineto{\pgfqpoint{0.000000in}{-0.027778in}}%
\pgfusepath{stroke,fill}%
}%
\begin{pgfscope}%
\pgfsys@transformshift{4.081191in}{0.555327in}%
\pgfsys@useobject{currentmarker}{}%
\end{pgfscope}%
\end{pgfscope}%
\begin{pgfscope}%
\pgfsetbuttcap%
\pgfsetroundjoin%
\definecolor{currentfill}{rgb}{0.000000,0.000000,0.000000}%
\pgfsetfillcolor{currentfill}%
\pgfsetlinewidth{0.602250pt}%
\definecolor{currentstroke}{rgb}{0.000000,0.000000,0.000000}%
\pgfsetstrokecolor{currentstroke}%
\pgfsetdash{}{0pt}%
\pgfsys@defobject{currentmarker}{\pgfqpoint{0.000000in}{-0.027778in}}{\pgfqpoint{0.000000in}{0.000000in}}{%
\pgfpathmoveto{\pgfqpoint{0.000000in}{0.000000in}}%
\pgfpathlineto{\pgfqpoint{0.000000in}{-0.027778in}}%
\pgfusepath{stroke,fill}%
}%
\begin{pgfscope}%
\pgfsys@transformshift{4.329662in}{0.555327in}%
\pgfsys@useobject{currentmarker}{}%
\end{pgfscope}%
\end{pgfscope}%
\begin{pgfscope}%
\pgfsetbuttcap%
\pgfsetroundjoin%
\definecolor{currentfill}{rgb}{0.000000,0.000000,0.000000}%
\pgfsetfillcolor{currentfill}%
\pgfsetlinewidth{0.602250pt}%
\definecolor{currentstroke}{rgb}{0.000000,0.000000,0.000000}%
\pgfsetstrokecolor{currentstroke}%
\pgfsetdash{}{0pt}%
\pgfsys@defobject{currentmarker}{\pgfqpoint{0.000000in}{-0.027778in}}{\pgfqpoint{0.000000in}{0.000000in}}{%
\pgfpathmoveto{\pgfqpoint{0.000000in}{0.000000in}}%
\pgfpathlineto{\pgfqpoint{0.000000in}{-0.027778in}}%
\pgfusepath{stroke,fill}%
}%
\begin{pgfscope}%
\pgfsys@transformshift{4.532678in}{0.555327in}%
\pgfsys@useobject{currentmarker}{}%
\end{pgfscope}%
\end{pgfscope}%
\begin{pgfscope}%
\pgfsetbuttcap%
\pgfsetroundjoin%
\definecolor{currentfill}{rgb}{0.000000,0.000000,0.000000}%
\pgfsetfillcolor{currentfill}%
\pgfsetlinewidth{0.602250pt}%
\definecolor{currentstroke}{rgb}{0.000000,0.000000,0.000000}%
\pgfsetstrokecolor{currentstroke}%
\pgfsetdash{}{0pt}%
\pgfsys@defobject{currentmarker}{\pgfqpoint{0.000000in}{-0.027778in}}{\pgfqpoint{0.000000in}{0.000000in}}{%
\pgfpathmoveto{\pgfqpoint{0.000000in}{0.000000in}}%
\pgfpathlineto{\pgfqpoint{0.000000in}{-0.027778in}}%
\pgfusepath{stroke,fill}%
}%
\begin{pgfscope}%
\pgfsys@transformshift{4.704326in}{0.555327in}%
\pgfsys@useobject{currentmarker}{}%
\end{pgfscope}%
\end{pgfscope}%
\begin{pgfscope}%
\pgfsetbuttcap%
\pgfsetroundjoin%
\definecolor{currentfill}{rgb}{0.000000,0.000000,0.000000}%
\pgfsetfillcolor{currentfill}%
\pgfsetlinewidth{0.602250pt}%
\definecolor{currentstroke}{rgb}{0.000000,0.000000,0.000000}%
\pgfsetstrokecolor{currentstroke}%
\pgfsetdash{}{0pt}%
\pgfsys@defobject{currentmarker}{\pgfqpoint{0.000000in}{-0.027778in}}{\pgfqpoint{0.000000in}{0.000000in}}{%
\pgfpathmoveto{\pgfqpoint{0.000000in}{0.000000in}}%
\pgfpathlineto{\pgfqpoint{0.000000in}{-0.027778in}}%
\pgfusepath{stroke,fill}%
}%
\begin{pgfscope}%
\pgfsys@transformshift{4.853014in}{0.555327in}%
\pgfsys@useobject{currentmarker}{}%
\end{pgfscope}%
\end{pgfscope}%
\begin{pgfscope}%
\pgfsetbuttcap%
\pgfsetroundjoin%
\definecolor{currentfill}{rgb}{0.000000,0.000000,0.000000}%
\pgfsetfillcolor{currentfill}%
\pgfsetlinewidth{0.602250pt}%
\definecolor{currentstroke}{rgb}{0.000000,0.000000,0.000000}%
\pgfsetstrokecolor{currentstroke}%
\pgfsetdash{}{0pt}%
\pgfsys@defobject{currentmarker}{\pgfqpoint{0.000000in}{-0.027778in}}{\pgfqpoint{0.000000in}{0.000000in}}{%
\pgfpathmoveto{\pgfqpoint{0.000000in}{0.000000in}}%
\pgfpathlineto{\pgfqpoint{0.000000in}{-0.027778in}}%
\pgfusepath{stroke,fill}%
}%
\begin{pgfscope}%
\pgfsys@transformshift{4.984166in}{0.555327in}%
\pgfsys@useobject{currentmarker}{}%
\end{pgfscope}%
\end{pgfscope}%
\begin{pgfscope}%
\pgftext[x=2.980138in,y=0.276410in,,top]{\rmfamily\fontsize{10.000000}{12.000000}\selectfont Mining power (TH/s)}%
\end{pgfscope}%
\begin{pgfscope}%
\pgfsetbuttcap%
\pgfsetroundjoin%
\definecolor{currentfill}{rgb}{0.000000,0.000000,0.000000}%
\pgfsetfillcolor{currentfill}%
\pgfsetlinewidth{0.803000pt}%
\definecolor{currentstroke}{rgb}{0.000000,0.000000,0.000000}%
\pgfsetstrokecolor{currentstroke}%
\pgfsetdash{}{0pt}%
\pgfsys@defobject{currentmarker}{\pgfqpoint{-0.048611in}{0.000000in}}{\pgfqpoint{0.000000in}{0.000000in}}{%
\pgfpathmoveto{\pgfqpoint{0.000000in}{0.000000in}}%
\pgfpathlineto{\pgfqpoint{-0.048611in}{0.000000in}}%
\pgfusepath{stroke,fill}%
}%
\begin{pgfscope}%
\pgfsys@transformshift{0.745276in}{0.794426in}%
\pgfsys@useobject{currentmarker}{}%
\end{pgfscope}%
\end{pgfscope}%
\begin{pgfscope}%
\pgftext[x=0.331695in,y=0.746208in,left,base]{\rmfamily\fontsize{10.000000}{12.000000}\selectfont \(\displaystyle 0.175\)}%
\end{pgfscope}%
\begin{pgfscope}%
\pgfsetbuttcap%
\pgfsetroundjoin%
\definecolor{currentfill}{rgb}{0.000000,0.000000,0.000000}%
\pgfsetfillcolor{currentfill}%
\pgfsetlinewidth{0.803000pt}%
\definecolor{currentstroke}{rgb}{0.000000,0.000000,0.000000}%
\pgfsetstrokecolor{currentstroke}%
\pgfsetdash{}{0pt}%
\pgfsys@defobject{currentmarker}{\pgfqpoint{-0.048611in}{0.000000in}}{\pgfqpoint{0.000000in}{0.000000in}}{%
\pgfpathmoveto{\pgfqpoint{0.000000in}{0.000000in}}%
\pgfpathlineto{\pgfqpoint{-0.048611in}{0.000000in}}%
\pgfusepath{stroke,fill}%
}%
\begin{pgfscope}%
\pgfsys@transformshift{0.745276in}{1.106974in}%
\pgfsys@useobject{currentmarker}{}%
\end{pgfscope}%
\end{pgfscope}%
\begin{pgfscope}%
\pgftext[x=0.331695in,y=1.058756in,left,base]{\rmfamily\fontsize{10.000000}{12.000000}\selectfont \(\displaystyle 0.180\)}%
\end{pgfscope}%
\begin{pgfscope}%
\pgfsetbuttcap%
\pgfsetroundjoin%
\definecolor{currentfill}{rgb}{0.000000,0.000000,0.000000}%
\pgfsetfillcolor{currentfill}%
\pgfsetlinewidth{0.803000pt}%
\definecolor{currentstroke}{rgb}{0.000000,0.000000,0.000000}%
\pgfsetstrokecolor{currentstroke}%
\pgfsetdash{}{0pt}%
\pgfsys@defobject{currentmarker}{\pgfqpoint{-0.048611in}{0.000000in}}{\pgfqpoint{0.000000in}{0.000000in}}{%
\pgfpathmoveto{\pgfqpoint{0.000000in}{0.000000in}}%
\pgfpathlineto{\pgfqpoint{-0.048611in}{0.000000in}}%
\pgfusepath{stroke,fill}%
}%
\begin{pgfscope}%
\pgfsys@transformshift{0.745276in}{1.419521in}%
\pgfsys@useobject{currentmarker}{}%
\end{pgfscope}%
\end{pgfscope}%
\begin{pgfscope}%
\pgftext[x=0.331695in,y=1.371303in,left,base]{\rmfamily\fontsize{10.000000}{12.000000}\selectfont \(\displaystyle 0.185\)}%
\end{pgfscope}%
\begin{pgfscope}%
\pgfsetbuttcap%
\pgfsetroundjoin%
\definecolor{currentfill}{rgb}{0.000000,0.000000,0.000000}%
\pgfsetfillcolor{currentfill}%
\pgfsetlinewidth{0.803000pt}%
\definecolor{currentstroke}{rgb}{0.000000,0.000000,0.000000}%
\pgfsetstrokecolor{currentstroke}%
\pgfsetdash{}{0pt}%
\pgfsys@defobject{currentmarker}{\pgfqpoint{-0.048611in}{0.000000in}}{\pgfqpoint{0.000000in}{0.000000in}}{%
\pgfpathmoveto{\pgfqpoint{0.000000in}{0.000000in}}%
\pgfpathlineto{\pgfqpoint{-0.048611in}{0.000000in}}%
\pgfusepath{stroke,fill}%
}%
\begin{pgfscope}%
\pgfsys@transformshift{0.745276in}{1.732069in}%
\pgfsys@useobject{currentmarker}{}%
\end{pgfscope}%
\end{pgfscope}%
\begin{pgfscope}%
\pgftext[x=0.331695in,y=1.683851in,left,base]{\rmfamily\fontsize{10.000000}{12.000000}\selectfont \(\displaystyle 0.190\)}%
\end{pgfscope}%
\begin{pgfscope}%
\pgfsetbuttcap%
\pgfsetroundjoin%
\definecolor{currentfill}{rgb}{0.000000,0.000000,0.000000}%
\pgfsetfillcolor{currentfill}%
\pgfsetlinewidth{0.803000pt}%
\definecolor{currentstroke}{rgb}{0.000000,0.000000,0.000000}%
\pgfsetstrokecolor{currentstroke}%
\pgfsetdash{}{0pt}%
\pgfsys@defobject{currentmarker}{\pgfqpoint{-0.048611in}{0.000000in}}{\pgfqpoint{0.000000in}{0.000000in}}{%
\pgfpathmoveto{\pgfqpoint{0.000000in}{0.000000in}}%
\pgfpathlineto{\pgfqpoint{-0.048611in}{0.000000in}}%
\pgfusepath{stroke,fill}%
}%
\begin{pgfscope}%
\pgfsys@transformshift{0.745276in}{2.044616in}%
\pgfsys@useobject{currentmarker}{}%
\end{pgfscope}%
\end{pgfscope}%
\begin{pgfscope}%
\pgftext[x=0.331695in,y=1.996398in,left,base]{\rmfamily\fontsize{10.000000}{12.000000}\selectfont \(\displaystyle 0.195\)}%
\end{pgfscope}%
\begin{pgfscope}%
\pgfsetbuttcap%
\pgfsetroundjoin%
\definecolor{currentfill}{rgb}{0.000000,0.000000,0.000000}%
\pgfsetfillcolor{currentfill}%
\pgfsetlinewidth{0.803000pt}%
\definecolor{currentstroke}{rgb}{0.000000,0.000000,0.000000}%
\pgfsetstrokecolor{currentstroke}%
\pgfsetdash{}{0pt}%
\pgfsys@defobject{currentmarker}{\pgfqpoint{-0.048611in}{0.000000in}}{\pgfqpoint{0.000000in}{0.000000in}}{%
\pgfpathmoveto{\pgfqpoint{0.000000in}{0.000000in}}%
\pgfpathlineto{\pgfqpoint{-0.048611in}{0.000000in}}%
\pgfusepath{stroke,fill}%
}%
\begin{pgfscope}%
\pgfsys@transformshift{0.745276in}{2.357164in}%
\pgfsys@useobject{currentmarker}{}%
\end{pgfscope}%
\end{pgfscope}%
\begin{pgfscope}%
\pgftext[x=0.331695in,y=2.308946in,left,base]{\rmfamily\fontsize{10.000000}{12.000000}\selectfont \(\displaystyle 0.200\)}%
\end{pgfscope}%
\begin{pgfscope}%
\definecolor{textcolor}{rgb}{0.000000,0.000000,1.000000}%
\pgfsetstrokecolor{textcolor}%
\pgfsetfillcolor{textcolor}%
\pgftext[x=0.276139in,y=1.535164in,,bottom,rotate=90.000000]{\color{textcolor}\rmfamily\fontsize{10.000000}{12.000000}\selectfont Sharpe ratio}%
\end{pgfscope}%
\begin{pgfscope}%
\pgfpathrectangle{\pgfqpoint{0.745276in}{0.555327in}}{\pgfqpoint{4.469724in}{1.959673in}}%
\pgfusepath{clip}%
\pgfsetrectcap%
\pgfsetroundjoin%
\pgfsetlinewidth{1.003750pt}%
\definecolor{currentstroke}{rgb}{0.000000,0.000000,1.000000}%
\pgfsetstrokecolor{currentstroke}%
\pgfsetdash{}{0pt}%
\pgfpathmoveto{\pgfqpoint{0.948445in}{2.425924in}}%
\pgfpathlineto{\pgfqpoint{1.399933in}{2.407171in}}%
\pgfpathlineto{\pgfqpoint{1.851420in}{2.407171in}}%
\pgfpathlineto{\pgfqpoint{2.302907in}{2.388418in}}%
\pgfpathlineto{\pgfqpoint{2.754394in}{2.388418in}}%
\pgfpathlineto{\pgfqpoint{3.205882in}{2.388418in}}%
\pgfpathlineto{\pgfqpoint{3.657369in}{2.388418in}}%
\pgfpathlineto{\pgfqpoint{4.108856in}{2.225894in}}%
\pgfpathlineto{\pgfqpoint{4.560343in}{1.600799in}}%
\pgfpathlineto{\pgfqpoint{5.011831in}{0.644403in}}%
\pgfusepath{stroke}%
\end{pgfscope}%
\begin{pgfscope}%
\pgfsetrectcap%
\pgfsetmiterjoin%
\pgfsetlinewidth{0.803000pt}%
\definecolor{currentstroke}{rgb}{0.000000,0.000000,0.000000}%
\pgfsetstrokecolor{currentstroke}%
\pgfsetdash{}{0pt}%
\pgfpathmoveto{\pgfqpoint{0.745276in}{0.555327in}}%
\pgfpathlineto{\pgfqpoint{0.745276in}{2.515000in}}%
\pgfusepath{stroke}%
\end{pgfscope}%
\begin{pgfscope}%
\pgfsetrectcap%
\pgfsetmiterjoin%
\pgfsetlinewidth{0.803000pt}%
\definecolor{currentstroke}{rgb}{0.000000,0.000000,0.000000}%
\pgfsetstrokecolor{currentstroke}%
\pgfsetdash{}{0pt}%
\pgfpathmoveto{\pgfqpoint{5.215000in}{0.555327in}}%
\pgfpathlineto{\pgfqpoint{5.215000in}{2.515000in}}%
\pgfusepath{stroke}%
\end{pgfscope}%
\begin{pgfscope}%
\pgfsetrectcap%
\pgfsetmiterjoin%
\pgfsetlinewidth{0.803000pt}%
\definecolor{currentstroke}{rgb}{0.000000,0.000000,0.000000}%
\pgfsetstrokecolor{currentstroke}%
\pgfsetdash{}{0pt}%
\pgfpathmoveto{\pgfqpoint{0.745276in}{0.555327in}}%
\pgfpathlineto{\pgfqpoint{5.215000in}{0.555327in}}%
\pgfusepath{stroke}%
\end{pgfscope}%
\begin{pgfscope}%
\pgfsetrectcap%
\pgfsetmiterjoin%
\pgfsetlinewidth{0.803000pt}%
\definecolor{currentstroke}{rgb}{0.000000,0.000000,0.000000}%
\pgfsetstrokecolor{currentstroke}%
\pgfsetdash{}{0pt}%
\pgfpathmoveto{\pgfqpoint{0.745276in}{2.515000in}}%
\pgfpathlineto{\pgfqpoint{5.215000in}{2.515000in}}%
\pgfusepath{stroke}%
\end{pgfscope}%
\end{pgfpicture}%
\makeatother%
\endgroup%

%% file: sharpe-rewards-vs-miningpower-large.pgf
\begingroup%
\makeatletter%
\begin{pgfpicture}%
\pgfpathrectangle{\pgfpointorigin}{\pgfqpoint{5.400000in}{2.700000in}}%
\pgfusepath{use as bounding box, clip}%
\begin{pgfscope}%
\pgfsetbuttcap%
\pgfsetmiterjoin%
\definecolor{currentfill}{rgb}{1.000000,1.000000,1.000000}%
\pgfsetfillcolor{currentfill}%
\pgfsetlinewidth{0.000000pt}%
\definecolor{currentstroke}{rgb}{1.000000,1.000000,1.000000}%
\pgfsetstrokecolor{currentstroke}%
\pgfsetdash{}{0pt}%
\pgfpathmoveto{\pgfqpoint{0.000000in}{0.000000in}}%
\pgfpathlineto{\pgfqpoint{5.400000in}{0.000000in}}%
\pgfpathlineto{\pgfqpoint{5.400000in}{2.700000in}}%
\pgfpathlineto{\pgfqpoint{0.000000in}{2.700000in}}%
\pgfpathclose%
\pgfusepath{fill}%
\end{pgfscope}%
\begin{pgfscope}%
\pgfsetbuttcap%
\pgfsetmiterjoin%
\definecolor{currentfill}{rgb}{1.000000,1.000000,1.000000}%
\pgfsetfillcolor{currentfill}%
\pgfsetlinewidth{0.000000pt}%
\definecolor{currentstroke}{rgb}{0.000000,0.000000,0.000000}%
\pgfsetstrokecolor{currentstroke}%
\pgfsetstrokeopacity{0.000000}%
\pgfsetdash{}{0pt}%
\pgfpathmoveto{\pgfqpoint{0.745276in}{0.555327in}}%
\pgfpathlineto{\pgfqpoint{5.215000in}{0.555327in}}%
\pgfpathlineto{\pgfqpoint{5.215000in}{2.515000in}}%
\pgfpathlineto{\pgfqpoint{0.745276in}{2.515000in}}%
\pgfpathclose%
\pgfusepath{fill}%
\end{pgfscope}%
\begin{pgfscope}%
\pgfsetbuttcap%
\pgfsetroundjoin%
\definecolor{currentfill}{rgb}{0.000000,0.000000,0.000000}%
\pgfsetfillcolor{currentfill}%
\pgfsetlinewidth{0.803000pt}%
\definecolor{currentstroke}{rgb}{0.000000,0.000000,0.000000}%
\pgfsetstrokecolor{currentstroke}%
\pgfsetdash{}{0pt}%
\pgfsys@defobject{currentmarker}{\pgfqpoint{0.000000in}{-0.048611in}}{\pgfqpoint{0.000000in}{0.000000in}}{%
\pgfpathmoveto{\pgfqpoint{0.000000in}{0.000000in}}%
\pgfpathlineto{\pgfqpoint{0.000000in}{-0.048611in}}%
\pgfusepath{stroke,fill}%
}%
\begin{pgfscope}%
\pgfsys@transformshift{2.537546in}{0.555327in}%
\pgfsys@useobject{currentmarker}{}%
\end{pgfscope}%
\end{pgfscope}%
\begin{pgfscope}%
\pgftext[x=2.537546in,y=0.458105in,,top]{\rmfamily\fontsize{10.000000}{12.000000}\selectfont \(\displaystyle 10^{2}\)}%
\end{pgfscope}%
\begin{pgfscope}%
\pgfsetbuttcap%
\pgfsetroundjoin%
\definecolor{currentfill}{rgb}{0.000000,0.000000,0.000000}%
\pgfsetfillcolor{currentfill}%
\pgfsetlinewidth{0.803000pt}%
\definecolor{currentstroke}{rgb}{0.000000,0.000000,0.000000}%
\pgfsetstrokecolor{currentstroke}%
\pgfsetdash{}{0pt}%
\pgfsys@defobject{currentmarker}{\pgfqpoint{0.000000in}{-0.048611in}}{\pgfqpoint{0.000000in}{0.000000in}}{%
\pgfpathmoveto{\pgfqpoint{0.000000in}{0.000000in}}%
\pgfpathlineto{\pgfqpoint{0.000000in}{-0.048611in}}%
\pgfusepath{stroke,fill}%
}%
\begin{pgfscope}%
\pgfsys@transformshift{5.101485in}{0.555327in}%
\pgfsys@useobject{currentmarker}{}%
\end{pgfscope}%
\end{pgfscope}%
\begin{pgfscope}%
\pgftext[x=5.101485in,y=0.458105in,,top]{\rmfamily\fontsize{10.000000}{12.000000}\selectfont \(\displaystyle 10^{3}\)}%
\end{pgfscope}%
\begin{pgfscope}%
\pgfsetbuttcap%
\pgfsetroundjoin%
\definecolor{currentfill}{rgb}{0.000000,0.000000,0.000000}%
\pgfsetfillcolor{currentfill}%
\pgfsetlinewidth{0.602250pt}%
\definecolor{currentstroke}{rgb}{0.000000,0.000000,0.000000}%
\pgfsetstrokecolor{currentstroke}%
\pgfsetdash{}{0pt}%
\pgfsys@defobject{currentmarker}{\pgfqpoint{0.000000in}{-0.027778in}}{\pgfqpoint{0.000000in}{0.000000in}}{%
\pgfpathmoveto{\pgfqpoint{0.000000in}{0.000000in}}%
\pgfpathlineto{\pgfqpoint{0.000000in}{-0.027778in}}%
\pgfusepath{stroke,fill}%
}%
\begin{pgfscope}%
\pgfsys@transformshift{0.745429in}{0.555327in}%
\pgfsys@useobject{currentmarker}{}%
\end{pgfscope}%
\end{pgfscope}%
\begin{pgfscope}%
\pgfsetbuttcap%
\pgfsetroundjoin%
\definecolor{currentfill}{rgb}{0.000000,0.000000,0.000000}%
\pgfsetfillcolor{currentfill}%
\pgfsetlinewidth{0.602250pt}%
\definecolor{currentstroke}{rgb}{0.000000,0.000000,0.000000}%
\pgfsetstrokecolor{currentstroke}%
\pgfsetdash{}{0pt}%
\pgfsys@defobject{currentmarker}{\pgfqpoint{0.000000in}{-0.027778in}}{\pgfqpoint{0.000000in}{0.000000in}}{%
\pgfpathmoveto{\pgfqpoint{0.000000in}{0.000000in}}%
\pgfpathlineto{\pgfqpoint{0.000000in}{-0.027778in}}%
\pgfusepath{stroke,fill}%
}%
\begin{pgfscope}%
\pgfsys@transformshift{1.196917in}{0.555327in}%
\pgfsys@useobject{currentmarker}{}%
\end{pgfscope}%
\end{pgfscope}%
\begin{pgfscope}%
\pgfsetbuttcap%
\pgfsetroundjoin%
\definecolor{currentfill}{rgb}{0.000000,0.000000,0.000000}%
\pgfsetfillcolor{currentfill}%
\pgfsetlinewidth{0.602250pt}%
\definecolor{currentstroke}{rgb}{0.000000,0.000000,0.000000}%
\pgfsetstrokecolor{currentstroke}%
\pgfsetdash{}{0pt}%
\pgfsys@defobject{currentmarker}{\pgfqpoint{0.000000in}{-0.027778in}}{\pgfqpoint{0.000000in}{0.000000in}}{%
\pgfpathmoveto{\pgfqpoint{0.000000in}{0.000000in}}%
\pgfpathlineto{\pgfqpoint{0.000000in}{-0.027778in}}%
\pgfusepath{stroke,fill}%
}%
\begin{pgfscope}%
\pgfsys@transformshift{1.517252in}{0.555327in}%
\pgfsys@useobject{currentmarker}{}%
\end{pgfscope}%
\end{pgfscope}%
\begin{pgfscope}%
\pgfsetbuttcap%
\pgfsetroundjoin%
\definecolor{currentfill}{rgb}{0.000000,0.000000,0.000000}%
\pgfsetfillcolor{currentfill}%
\pgfsetlinewidth{0.602250pt}%
\definecolor{currentstroke}{rgb}{0.000000,0.000000,0.000000}%
\pgfsetstrokecolor{currentstroke}%
\pgfsetdash{}{0pt}%
\pgfsys@defobject{currentmarker}{\pgfqpoint{0.000000in}{-0.027778in}}{\pgfqpoint{0.000000in}{0.000000in}}{%
\pgfpathmoveto{\pgfqpoint{0.000000in}{0.000000in}}%
\pgfpathlineto{\pgfqpoint{0.000000in}{-0.027778in}}%
\pgfusepath{stroke,fill}%
}%
\begin{pgfscope}%
\pgfsys@transformshift{1.765723in}{0.555327in}%
\pgfsys@useobject{currentmarker}{}%
\end{pgfscope}%
\end{pgfscope}%
\begin{pgfscope}%
\pgfsetbuttcap%
\pgfsetroundjoin%
\definecolor{currentfill}{rgb}{0.000000,0.000000,0.000000}%
\pgfsetfillcolor{currentfill}%
\pgfsetlinewidth{0.602250pt}%
\definecolor{currentstroke}{rgb}{0.000000,0.000000,0.000000}%
\pgfsetstrokecolor{currentstroke}%
\pgfsetdash{}{0pt}%
\pgfsys@defobject{currentmarker}{\pgfqpoint{0.000000in}{-0.027778in}}{\pgfqpoint{0.000000in}{0.000000in}}{%
\pgfpathmoveto{\pgfqpoint{0.000000in}{0.000000in}}%
\pgfpathlineto{\pgfqpoint{0.000000in}{-0.027778in}}%
\pgfusepath{stroke,fill}%
}%
\begin{pgfscope}%
\pgfsys@transformshift{1.968739in}{0.555327in}%
\pgfsys@useobject{currentmarker}{}%
\end{pgfscope}%
\end{pgfscope}%
\begin{pgfscope}%
\pgfsetbuttcap%
\pgfsetroundjoin%
\definecolor{currentfill}{rgb}{0.000000,0.000000,0.000000}%
\pgfsetfillcolor{currentfill}%
\pgfsetlinewidth{0.602250pt}%
\definecolor{currentstroke}{rgb}{0.000000,0.000000,0.000000}%
\pgfsetstrokecolor{currentstroke}%
\pgfsetdash{}{0pt}%
\pgfsys@defobject{currentmarker}{\pgfqpoint{0.000000in}{-0.027778in}}{\pgfqpoint{0.000000in}{0.000000in}}{%
\pgfpathmoveto{\pgfqpoint{0.000000in}{0.000000in}}%
\pgfpathlineto{\pgfqpoint{0.000000in}{-0.027778in}}%
\pgfusepath{stroke,fill}%
}%
\begin{pgfscope}%
\pgfsys@transformshift{2.140387in}{0.555327in}%
\pgfsys@useobject{currentmarker}{}%
\end{pgfscope}%
\end{pgfscope}%
\begin{pgfscope}%
\pgfsetbuttcap%
\pgfsetroundjoin%
\definecolor{currentfill}{rgb}{0.000000,0.000000,0.000000}%
\pgfsetfillcolor{currentfill}%
\pgfsetlinewidth{0.602250pt}%
\definecolor{currentstroke}{rgb}{0.000000,0.000000,0.000000}%
\pgfsetstrokecolor{currentstroke}%
\pgfsetdash{}{0pt}%
\pgfsys@defobject{currentmarker}{\pgfqpoint{0.000000in}{-0.027778in}}{\pgfqpoint{0.000000in}{0.000000in}}{%
\pgfpathmoveto{\pgfqpoint{0.000000in}{0.000000in}}%
\pgfpathlineto{\pgfqpoint{0.000000in}{-0.027778in}}%
\pgfusepath{stroke,fill}%
}%
\begin{pgfscope}%
\pgfsys@transformshift{2.289075in}{0.555327in}%
\pgfsys@useobject{currentmarker}{}%
\end{pgfscope}%
\end{pgfscope}%
\begin{pgfscope}%
\pgfsetbuttcap%
\pgfsetroundjoin%
\definecolor{currentfill}{rgb}{0.000000,0.000000,0.000000}%
\pgfsetfillcolor{currentfill}%
\pgfsetlinewidth{0.602250pt}%
\definecolor{currentstroke}{rgb}{0.000000,0.000000,0.000000}%
\pgfsetstrokecolor{currentstroke}%
\pgfsetdash{}{0pt}%
\pgfsys@defobject{currentmarker}{\pgfqpoint{0.000000in}{-0.027778in}}{\pgfqpoint{0.000000in}{0.000000in}}{%
\pgfpathmoveto{\pgfqpoint{0.000000in}{0.000000in}}%
\pgfpathlineto{\pgfqpoint{0.000000in}{-0.027778in}}%
\pgfusepath{stroke,fill}%
}%
\begin{pgfscope}%
\pgfsys@transformshift{2.420226in}{0.555327in}%
\pgfsys@useobject{currentmarker}{}%
\end{pgfscope}%
\end{pgfscope}%
\begin{pgfscope}%
\pgfsetbuttcap%
\pgfsetroundjoin%
\definecolor{currentfill}{rgb}{0.000000,0.000000,0.000000}%
\pgfsetfillcolor{currentfill}%
\pgfsetlinewidth{0.602250pt}%
\definecolor{currentstroke}{rgb}{0.000000,0.000000,0.000000}%
\pgfsetstrokecolor{currentstroke}%
\pgfsetdash{}{0pt}%
\pgfsys@defobject{currentmarker}{\pgfqpoint{0.000000in}{-0.027778in}}{\pgfqpoint{0.000000in}{0.000000in}}{%
\pgfpathmoveto{\pgfqpoint{0.000000in}{0.000000in}}%
\pgfpathlineto{\pgfqpoint{0.000000in}{-0.027778in}}%
\pgfusepath{stroke,fill}%
}%
\begin{pgfscope}%
\pgfsys@transformshift{3.309369in}{0.555327in}%
\pgfsys@useobject{currentmarker}{}%
\end{pgfscope}%
\end{pgfscope}%
\begin{pgfscope}%
\pgfsetbuttcap%
\pgfsetroundjoin%
\definecolor{currentfill}{rgb}{0.000000,0.000000,0.000000}%
\pgfsetfillcolor{currentfill}%
\pgfsetlinewidth{0.602250pt}%
\definecolor{currentstroke}{rgb}{0.000000,0.000000,0.000000}%
\pgfsetstrokecolor{currentstroke}%
\pgfsetdash{}{0pt}%
\pgfsys@defobject{currentmarker}{\pgfqpoint{0.000000in}{-0.027778in}}{\pgfqpoint{0.000000in}{0.000000in}}{%
\pgfpathmoveto{\pgfqpoint{0.000000in}{0.000000in}}%
\pgfpathlineto{\pgfqpoint{0.000000in}{-0.027778in}}%
\pgfusepath{stroke,fill}%
}%
\begin{pgfscope}%
\pgfsys@transformshift{3.760856in}{0.555327in}%
\pgfsys@useobject{currentmarker}{}%
\end{pgfscope}%
\end{pgfscope}%
\begin{pgfscope}%
\pgfsetbuttcap%
\pgfsetroundjoin%
\definecolor{currentfill}{rgb}{0.000000,0.000000,0.000000}%
\pgfsetfillcolor{currentfill}%
\pgfsetlinewidth{0.602250pt}%
\definecolor{currentstroke}{rgb}{0.000000,0.000000,0.000000}%
\pgfsetstrokecolor{currentstroke}%
\pgfsetdash{}{0pt}%
\pgfsys@defobject{currentmarker}{\pgfqpoint{0.000000in}{-0.027778in}}{\pgfqpoint{0.000000in}{0.000000in}}{%
\pgfpathmoveto{\pgfqpoint{0.000000in}{0.000000in}}%
\pgfpathlineto{\pgfqpoint{0.000000in}{-0.027778in}}%
\pgfusepath{stroke,fill}%
}%
\begin{pgfscope}%
\pgfsys@transformshift{4.081191in}{0.555327in}%
\pgfsys@useobject{currentmarker}{}%
\end{pgfscope}%
\end{pgfscope}%
\begin{pgfscope}%
\pgfsetbuttcap%
\pgfsetroundjoin%
\definecolor{currentfill}{rgb}{0.000000,0.000000,0.000000}%
\pgfsetfillcolor{currentfill}%
\pgfsetlinewidth{0.602250pt}%
\definecolor{currentstroke}{rgb}{0.000000,0.000000,0.000000}%
\pgfsetstrokecolor{currentstroke}%
\pgfsetdash{}{0pt}%
\pgfsys@defobject{currentmarker}{\pgfqpoint{0.000000in}{-0.027778in}}{\pgfqpoint{0.000000in}{0.000000in}}{%
\pgfpathmoveto{\pgfqpoint{0.000000in}{0.000000in}}%
\pgfpathlineto{\pgfqpoint{0.000000in}{-0.027778in}}%
\pgfusepath{stroke,fill}%
}%
\begin{pgfscope}%
\pgfsys@transformshift{4.329662in}{0.555327in}%
\pgfsys@useobject{currentmarker}{}%
\end{pgfscope}%
\end{pgfscope}%
\begin{pgfscope}%
\pgfsetbuttcap%
\pgfsetroundjoin%
\definecolor{currentfill}{rgb}{0.000000,0.000000,0.000000}%
\pgfsetfillcolor{currentfill}%
\pgfsetlinewidth{0.602250pt}%
\definecolor{currentstroke}{rgb}{0.000000,0.000000,0.000000}%
\pgfsetstrokecolor{currentstroke}%
\pgfsetdash{}{0pt}%
\pgfsys@defobject{currentmarker}{\pgfqpoint{0.000000in}{-0.027778in}}{\pgfqpoint{0.000000in}{0.000000in}}{%
\pgfpathmoveto{\pgfqpoint{0.000000in}{0.000000in}}%
\pgfpathlineto{\pgfqpoint{0.000000in}{-0.027778in}}%
\pgfusepath{stroke,fill}%
}%
\begin{pgfscope}%
\pgfsys@transformshift{4.532678in}{0.555327in}%
\pgfsys@useobject{currentmarker}{}%
\end{pgfscope}%
\end{pgfscope}%
\begin{pgfscope}%
\pgfsetbuttcap%
\pgfsetroundjoin%
\definecolor{currentfill}{rgb}{0.000000,0.000000,0.000000}%
\pgfsetfillcolor{currentfill}%
\pgfsetlinewidth{0.602250pt}%
\definecolor{currentstroke}{rgb}{0.000000,0.000000,0.000000}%
\pgfsetstrokecolor{currentstroke}%
\pgfsetdash{}{0pt}%
\pgfsys@defobject{currentmarker}{\pgfqpoint{0.000000in}{-0.027778in}}{\pgfqpoint{0.000000in}{0.000000in}}{%
\pgfpathmoveto{\pgfqpoint{0.000000in}{0.000000in}}%
\pgfpathlineto{\pgfqpoint{0.000000in}{-0.027778in}}%
\pgfusepath{stroke,fill}%
}%
\begin{pgfscope}%
\pgfsys@transformshift{4.704326in}{0.555327in}%
\pgfsys@useobject{currentmarker}{}%
\end{pgfscope}%
\end{pgfscope}%
\begin{pgfscope}%
\pgfsetbuttcap%
\pgfsetroundjoin%
\definecolor{currentfill}{rgb}{0.000000,0.000000,0.000000}%
\pgfsetfillcolor{currentfill}%
\pgfsetlinewidth{0.602250pt}%
\definecolor{currentstroke}{rgb}{0.000000,0.000000,0.000000}%
\pgfsetstrokecolor{currentstroke}%
\pgfsetdash{}{0pt}%
\pgfsys@defobject{currentmarker}{\pgfqpoint{0.000000in}{-0.027778in}}{\pgfqpoint{0.000000in}{0.000000in}}{%
\pgfpathmoveto{\pgfqpoint{0.000000in}{0.000000in}}%
\pgfpathlineto{\pgfqpoint{0.000000in}{-0.027778in}}%
\pgfusepath{stroke,fill}%
}%
\begin{pgfscope}%
\pgfsys@transformshift{4.853014in}{0.555327in}%
\pgfsys@useobject{currentmarker}{}%
\end{pgfscope}%
\end{pgfscope}%
\begin{pgfscope}%
\pgfsetbuttcap%
\pgfsetroundjoin%
\definecolor{currentfill}{rgb}{0.000000,0.000000,0.000000}%
\pgfsetfillcolor{currentfill}%
\pgfsetlinewidth{0.602250pt}%
\definecolor{currentstroke}{rgb}{0.000000,0.000000,0.000000}%
\pgfsetstrokecolor{currentstroke}%
\pgfsetdash{}{0pt}%
\pgfsys@defobject{currentmarker}{\pgfqpoint{0.000000in}{-0.027778in}}{\pgfqpoint{0.000000in}{0.000000in}}{%
\pgfpathmoveto{\pgfqpoint{0.000000in}{0.000000in}}%
\pgfpathlineto{\pgfqpoint{0.000000in}{-0.027778in}}%
\pgfusepath{stroke,fill}%
}%
\begin{pgfscope}%
\pgfsys@transformshift{4.984166in}{0.555327in}%
\pgfsys@useobject{currentmarker}{}%
\end{pgfscope}%
\end{pgfscope}%
\begin{pgfscope}%
\pgftext[x=2.980138in,y=0.276410in,,top]{\rmfamily\fontsize{10.000000}{12.000000}\selectfont Mining power (TH/s)}%
\end{pgfscope}%
\begin{pgfscope}%
\pgfsetbuttcap%
\pgfsetroundjoin%
\definecolor{currentfill}{rgb}{0.000000,0.000000,0.000000}%
\pgfsetfillcolor{currentfill}%
\pgfsetlinewidth{0.803000pt}%
\definecolor{currentstroke}{rgb}{0.000000,0.000000,0.000000}%
\pgfsetstrokecolor{currentstroke}%
\pgfsetdash{}{0pt}%
\pgfsys@defobject{currentmarker}{\pgfqpoint{-0.048611in}{0.000000in}}{\pgfqpoint{0.000000in}{0.000000in}}{%
\pgfpathmoveto{\pgfqpoint{0.000000in}{0.000000in}}%
\pgfpathlineto{\pgfqpoint{-0.048611in}{0.000000in}}%
\pgfusepath{stroke,fill}%
}%
\begin{pgfscope}%
\pgfsys@transformshift{0.745276in}{1.015554in}%
\pgfsys@useobject{currentmarker}{}%
\end{pgfscope}%
\end{pgfscope}%
\begin{pgfscope}%
\pgftext[x=0.331695in,y=0.967336in,left,base]{\rmfamily\fontsize{10.000000}{12.000000}\selectfont \(\displaystyle 0.017\)}%
\end{pgfscope}%
\begin{pgfscope}%
\pgfsetbuttcap%
\pgfsetroundjoin%
\definecolor{currentfill}{rgb}{0.000000,0.000000,0.000000}%
\pgfsetfillcolor{currentfill}%
\pgfsetlinewidth{0.803000pt}%
\definecolor{currentstroke}{rgb}{0.000000,0.000000,0.000000}%
\pgfsetstrokecolor{currentstroke}%
\pgfsetdash{}{0pt}%
\pgfsys@defobject{currentmarker}{\pgfqpoint{-0.048611in}{0.000000in}}{\pgfqpoint{0.000000in}{0.000000in}}{%
\pgfpathmoveto{\pgfqpoint{0.000000in}{0.000000in}}%
\pgfpathlineto{\pgfqpoint{-0.048611in}{0.000000in}}%
\pgfusepath{stroke,fill}%
}%
\begin{pgfscope}%
\pgfsys@transformshift{0.745276in}{1.510420in}%
\pgfsys@useobject{currentmarker}{}%
\end{pgfscope}%
\end{pgfscope}%
\begin{pgfscope}%
\pgftext[x=0.331695in,y=1.462203in,left,base]{\rmfamily\fontsize{10.000000}{12.000000}\selectfont \(\displaystyle 0.018\)}%
\end{pgfscope}%
\begin{pgfscope}%
\pgfsetbuttcap%
\pgfsetroundjoin%
\definecolor{currentfill}{rgb}{0.000000,0.000000,0.000000}%
\pgfsetfillcolor{currentfill}%
\pgfsetlinewidth{0.803000pt}%
\definecolor{currentstroke}{rgb}{0.000000,0.000000,0.000000}%
\pgfsetstrokecolor{currentstroke}%
\pgfsetdash{}{0pt}%
\pgfsys@defobject{currentmarker}{\pgfqpoint{-0.048611in}{0.000000in}}{\pgfqpoint{0.000000in}{0.000000in}}{%
\pgfpathmoveto{\pgfqpoint{0.000000in}{0.000000in}}%
\pgfpathlineto{\pgfqpoint{-0.048611in}{0.000000in}}%
\pgfusepath{stroke,fill}%
}%
\begin{pgfscope}%
\pgfsys@transformshift{0.745276in}{2.005287in}%
\pgfsys@useobject{currentmarker}{}%
\end{pgfscope}%
\end{pgfscope}%
\begin{pgfscope}%
\pgftext[x=0.331695in,y=1.957069in,left,base]{\rmfamily\fontsize{10.000000}{12.000000}\selectfont \(\displaystyle 0.019\)}%
\end{pgfscope}%
\begin{pgfscope}%
\pgfsetbuttcap%
\pgfsetroundjoin%
\definecolor{currentfill}{rgb}{0.000000,0.000000,0.000000}%
\pgfsetfillcolor{currentfill}%
\pgfsetlinewidth{0.803000pt}%
\definecolor{currentstroke}{rgb}{0.000000,0.000000,0.000000}%
\pgfsetstrokecolor{currentstroke}%
\pgfsetdash{}{0pt}%
\pgfsys@defobject{currentmarker}{\pgfqpoint{-0.048611in}{0.000000in}}{\pgfqpoint{0.000000in}{0.000000in}}{%
\pgfpathmoveto{\pgfqpoint{0.000000in}{0.000000in}}%
\pgfpathlineto{\pgfqpoint{-0.048611in}{0.000000in}}%
\pgfusepath{stroke,fill}%
}%
\begin{pgfscope}%
\pgfsys@transformshift{0.745276in}{2.500154in}%
\pgfsys@useobject{currentmarker}{}%
\end{pgfscope}%
\end{pgfscope}%
\begin{pgfscope}%
\pgftext[x=0.331695in,y=2.451936in,left,base]{\rmfamily\fontsize{10.000000}{12.000000}\selectfont \(\displaystyle 0.020\)}%
\end{pgfscope}%
\begin{pgfscope}%
\definecolor{textcolor}{rgb}{0.000000,0.000000,1.000000}%
\pgfsetstrokecolor{textcolor}%
\pgfsetfillcolor{textcolor}%
\pgftext[x=0.276139in,y=1.535164in,,bottom,rotate=90.000000]{\color{textcolor}\rmfamily\fontsize{10.000000}{12.000000}\selectfont Sharpe ratio}%
\end{pgfscope}%
\begin{pgfscope}%
\pgfpathrectangle{\pgfqpoint{0.745276in}{0.555327in}}{\pgfqpoint{4.469724in}{1.959673in}}%
\pgfusepath{clip}%
\pgfsetrectcap%
\pgfsetroundjoin%
\pgfsetlinewidth{1.003750pt}%
\definecolor{currentstroke}{rgb}{0.000000,0.000000,1.000000}%
\pgfsetstrokecolor{currentstroke}%
\pgfsetdash{}{0pt}%
\pgfpathmoveto{\pgfqpoint{0.948445in}{1.807340in}}%
\pgfpathlineto{\pgfqpoint{1.399933in}{1.510420in}}%
\pgfpathlineto{\pgfqpoint{1.851420in}{1.510420in}}%
\pgfpathlineto{\pgfqpoint{2.302907in}{1.262987in}}%
\pgfpathlineto{\pgfqpoint{2.754394in}{1.262987in}}%
\pgfpathlineto{\pgfqpoint{3.205882in}{1.262987in}}%
\pgfpathlineto{\pgfqpoint{3.657369in}{1.262987in}}%
\pgfpathlineto{\pgfqpoint{4.108856in}{1.262987in}}%
\pgfpathlineto{\pgfqpoint{4.560343in}{1.262987in}}%
\pgfpathlineto{\pgfqpoint{5.011831in}{1.262987in}}%
\pgfusepath{stroke}%
\end{pgfscope}%
\begin{pgfscope}%
\pgfsetrectcap%
\pgfsetmiterjoin%
\pgfsetlinewidth{0.803000pt}%
\definecolor{currentstroke}{rgb}{0.000000,0.000000,0.000000}%
\pgfsetstrokecolor{currentstroke}%
\pgfsetdash{}{0pt}%
\pgfpathmoveto{\pgfqpoint{0.745276in}{0.555327in}}%
\pgfpathlineto{\pgfqpoint{0.745276in}{2.515000in}}%
\pgfusepath{stroke}%
\end{pgfscope}%
\begin{pgfscope}%
\pgfsetrectcap%
\pgfsetmiterjoin%
\pgfsetlinewidth{0.803000pt}%
\definecolor{currentstroke}{rgb}{0.000000,0.000000,0.000000}%
\pgfsetstrokecolor{currentstroke}%
\pgfsetdash{}{0pt}%
\pgfpathmoveto{\pgfqpoint{5.215000in}{0.555327in}}%
\pgfpathlineto{\pgfqpoint{5.215000in}{2.515000in}}%
\pgfusepath{stroke}%
\end{pgfscope}%
\begin{pgfscope}%
\pgfsetrectcap%
\pgfsetmiterjoin%
\pgfsetlinewidth{0.803000pt}%
\definecolor{currentstroke}{rgb}{0.000000,0.000000,0.000000}%
\pgfsetstrokecolor{currentstroke}%
\pgfsetdash{}{0pt}%
\pgfpathmoveto{\pgfqpoint{0.745276in}{0.555327in}}%
\pgfpathlineto{\pgfqpoint{5.215000in}{0.555327in}}%
\pgfusepath{stroke}%
\end{pgfscope}%
\begin{pgfscope}%
\pgfsetrectcap%
\pgfsetmiterjoin%
\pgfsetlinewidth{0.803000pt}%
\definecolor{currentstroke}{rgb}{0.000000,0.000000,0.000000}%
\pgfsetstrokecolor{currentstroke}%
\pgfsetdash{}{0pt}%
\pgfpathmoveto{\pgfqpoint{0.745276in}{2.515000in}}%
\pgfpathlineto{\pgfqpoint{5.215000in}{2.515000in}}%
\pgfusepath{stroke}%
\end{pgfscope}%
\end{pgfpicture}%
\makeatother%
\endgroup%

%% file: sharpe-rewards-vs-interval.pgf
\begingroup%
\makeatletter%
\begin{pgfpicture}%
\pgfpathrectangle{\pgfpointorigin}{\pgfqpoint{5.400000in}{2.700000in}}%
\pgfusepath{use as bounding box, clip}%
\begin{pgfscope}%
\pgfsetbuttcap%
\pgfsetmiterjoin%
\definecolor{currentfill}{rgb}{1.000000,1.000000,1.000000}%
\pgfsetfillcolor{currentfill}%
\pgfsetlinewidth{0.000000pt}%
\definecolor{currentstroke}{rgb}{1.000000,1.000000,1.000000}%
\pgfsetstrokecolor{currentstroke}%
\pgfsetdash{}{0pt}%
\pgfpathmoveto{\pgfqpoint{0.000000in}{0.000000in}}%
\pgfpathlineto{\pgfqpoint{5.400000in}{0.000000in}}%
\pgfpathlineto{\pgfqpoint{5.400000in}{2.700000in}}%
\pgfpathlineto{\pgfqpoint{0.000000in}{2.700000in}}%
\pgfpathclose%
\pgfusepath{fill}%
\end{pgfscope}%
\begin{pgfscope}%
\pgfsetbuttcap%
\pgfsetmiterjoin%
\definecolor{currentfill}{rgb}{1.000000,1.000000,1.000000}%
\pgfsetfillcolor{currentfill}%
\pgfsetlinewidth{0.000000pt}%
\definecolor{currentstroke}{rgb}{0.000000,0.000000,0.000000}%
\pgfsetstrokecolor{currentstroke}%
\pgfsetstrokeopacity{0.000000}%
\pgfsetdash{}{0pt}%
\pgfpathmoveto{\pgfqpoint{0.675831in}{0.555327in}}%
\pgfpathlineto{\pgfqpoint{5.215000in}{0.555327in}}%
\pgfpathlineto{\pgfqpoint{5.215000in}{2.515000in}}%
\pgfpathlineto{\pgfqpoint{0.675831in}{2.515000in}}%
\pgfpathclose%
\pgfusepath{fill}%
\end{pgfscope}%
\begin{pgfscope}%
\pgfsetbuttcap%
\pgfsetroundjoin%
\definecolor{currentfill}{rgb}{0.000000,0.000000,0.000000}%
\pgfsetfillcolor{currentfill}%
\pgfsetlinewidth{0.803000pt}%
\definecolor{currentstroke}{rgb}{0.000000,0.000000,0.000000}%
\pgfsetstrokecolor{currentstroke}%
\pgfsetdash{}{0pt}%
\pgfsys@defobject{currentmarker}{\pgfqpoint{0.000000in}{-0.048611in}}{\pgfqpoint{0.000000in}{0.000000in}}{%
\pgfpathmoveto{\pgfqpoint{0.000000in}{0.000000in}}%
\pgfpathlineto{\pgfqpoint{0.000000in}{-0.048611in}}%
\pgfusepath{stroke,fill}%
}%
\begin{pgfscope}%
\pgfsys@transformshift{0.796188in}{0.555327in}%
\pgfsys@useobject{currentmarker}{}%
\end{pgfscope}%
\end{pgfscope}%
\begin{pgfscope}%
\pgftext[x=0.796188in,y=0.458105in,,top]{\rmfamily\fontsize{10.000000}{12.000000}\selectfont \(\displaystyle 0\)}%
\end{pgfscope}%
\begin{pgfscope}%
\pgfsetbuttcap%
\pgfsetroundjoin%
\definecolor{currentfill}{rgb}{0.000000,0.000000,0.000000}%
\pgfsetfillcolor{currentfill}%
\pgfsetlinewidth{0.803000pt}%
\definecolor{currentstroke}{rgb}{0.000000,0.000000,0.000000}%
\pgfsetstrokecolor{currentstroke}%
\pgfsetdash{}{0pt}%
\pgfsys@defobject{currentmarker}{\pgfqpoint{0.000000in}{-0.048611in}}{\pgfqpoint{0.000000in}{0.000000in}}{%
\pgfpathmoveto{\pgfqpoint{0.000000in}{0.000000in}}%
\pgfpathlineto{\pgfqpoint{0.000000in}{-0.048611in}}%
\pgfusepath{stroke,fill}%
}%
\begin{pgfscope}%
\pgfsys@transformshift{1.655879in}{0.555327in}%
\pgfsys@useobject{currentmarker}{}%
\end{pgfscope}%
\end{pgfscope}%
\begin{pgfscope}%
\pgftext[x=1.655879in,y=0.458105in,,top]{\rmfamily\fontsize{10.000000}{12.000000}\selectfont \(\displaystyle 10\)}%
\end{pgfscope}%
\begin{pgfscope}%
\pgfsetbuttcap%
\pgfsetroundjoin%
\definecolor{currentfill}{rgb}{0.000000,0.000000,0.000000}%
\pgfsetfillcolor{currentfill}%
\pgfsetlinewidth{0.803000pt}%
\definecolor{currentstroke}{rgb}{0.000000,0.000000,0.000000}%
\pgfsetstrokecolor{currentstroke}%
\pgfsetdash{}{0pt}%
\pgfsys@defobject{currentmarker}{\pgfqpoint{0.000000in}{-0.048611in}}{\pgfqpoint{0.000000in}{0.000000in}}{%
\pgfpathmoveto{\pgfqpoint{0.000000in}{0.000000in}}%
\pgfpathlineto{\pgfqpoint{0.000000in}{-0.048611in}}%
\pgfusepath{stroke,fill}%
}%
\begin{pgfscope}%
\pgfsys@transformshift{2.515570in}{0.555327in}%
\pgfsys@useobject{currentmarker}{}%
\end{pgfscope}%
\end{pgfscope}%
\begin{pgfscope}%
\pgftext[x=2.515570in,y=0.458105in,,top]{\rmfamily\fontsize{10.000000}{12.000000}\selectfont \(\displaystyle 20\)}%
\end{pgfscope}%
\begin{pgfscope}%
\pgfsetbuttcap%
\pgfsetroundjoin%
\definecolor{currentfill}{rgb}{0.000000,0.000000,0.000000}%
\pgfsetfillcolor{currentfill}%
\pgfsetlinewidth{0.803000pt}%
\definecolor{currentstroke}{rgb}{0.000000,0.000000,0.000000}%
\pgfsetstrokecolor{currentstroke}%
\pgfsetdash{}{0pt}%
\pgfsys@defobject{currentmarker}{\pgfqpoint{0.000000in}{-0.048611in}}{\pgfqpoint{0.000000in}{0.000000in}}{%
\pgfpathmoveto{\pgfqpoint{0.000000in}{0.000000in}}%
\pgfpathlineto{\pgfqpoint{0.000000in}{-0.048611in}}%
\pgfusepath{stroke,fill}%
}%
\begin{pgfscope}%
\pgfsys@transformshift{3.375261in}{0.555327in}%
\pgfsys@useobject{currentmarker}{}%
\end{pgfscope}%
\end{pgfscope}%
\begin{pgfscope}%
\pgftext[x=3.375261in,y=0.458105in,,top]{\rmfamily\fontsize{10.000000}{12.000000}\selectfont \(\displaystyle 30\)}%
\end{pgfscope}%
\begin{pgfscope}%
\pgfsetbuttcap%
\pgfsetroundjoin%
\definecolor{currentfill}{rgb}{0.000000,0.000000,0.000000}%
\pgfsetfillcolor{currentfill}%
\pgfsetlinewidth{0.803000pt}%
\definecolor{currentstroke}{rgb}{0.000000,0.000000,0.000000}%
\pgfsetstrokecolor{currentstroke}%
\pgfsetdash{}{0pt}%
\pgfsys@defobject{currentmarker}{\pgfqpoint{0.000000in}{-0.048611in}}{\pgfqpoint{0.000000in}{0.000000in}}{%
\pgfpathmoveto{\pgfqpoint{0.000000in}{0.000000in}}%
\pgfpathlineto{\pgfqpoint{0.000000in}{-0.048611in}}%
\pgfusepath{stroke,fill}%
}%
\begin{pgfscope}%
\pgfsys@transformshift{4.234952in}{0.555327in}%
\pgfsys@useobject{currentmarker}{}%
\end{pgfscope}%
\end{pgfscope}%
\begin{pgfscope}%
\pgftext[x=4.234952in,y=0.458105in,,top]{\rmfamily\fontsize{10.000000}{12.000000}\selectfont \(\displaystyle 40\)}%
\end{pgfscope}%
\begin{pgfscope}%
\pgfsetbuttcap%
\pgfsetroundjoin%
\definecolor{currentfill}{rgb}{0.000000,0.000000,0.000000}%
\pgfsetfillcolor{currentfill}%
\pgfsetlinewidth{0.803000pt}%
\definecolor{currentstroke}{rgb}{0.000000,0.000000,0.000000}%
\pgfsetstrokecolor{currentstroke}%
\pgfsetdash{}{0pt}%
\pgfsys@defobject{currentmarker}{\pgfqpoint{0.000000in}{-0.048611in}}{\pgfqpoint{0.000000in}{0.000000in}}{%
\pgfpathmoveto{\pgfqpoint{0.000000in}{0.000000in}}%
\pgfpathlineto{\pgfqpoint{0.000000in}{-0.048611in}}%
\pgfusepath{stroke,fill}%
}%
\begin{pgfscope}%
\pgfsys@transformshift{5.094643in}{0.555327in}%
\pgfsys@useobject{currentmarker}{}%
\end{pgfscope}%
\end{pgfscope}%
\begin{pgfscope}%
\pgftext[x=5.094643in,y=0.458105in,,top]{\rmfamily\fontsize{10.000000}{12.000000}\selectfont \(\displaystyle 50\)}%
\end{pgfscope}%
\begin{pgfscope}%
\pgftext[x=2.945416in,y=0.276410in,,top]{\rmfamily\fontsize{10.000000}{12.000000}\selectfont Diversify interval (days)}%
\end{pgfscope}%
\begin{pgfscope}%
\pgfsetbuttcap%
\pgfsetroundjoin%
\definecolor{currentfill}{rgb}{0.000000,0.000000,0.000000}%
\pgfsetfillcolor{currentfill}%
\pgfsetlinewidth{0.803000pt}%
\definecolor{currentstroke}{rgb}{0.000000,0.000000,0.000000}%
\pgfsetstrokecolor{currentstroke}%
\pgfsetdash{}{0pt}%
\pgfsys@defobject{currentmarker}{\pgfqpoint{-0.048611in}{0.000000in}}{\pgfqpoint{0.000000in}{0.000000in}}{%
\pgfpathmoveto{\pgfqpoint{0.000000in}{0.000000in}}%
\pgfpathlineto{\pgfqpoint{-0.048611in}{0.000000in}}%
\pgfusepath{stroke,fill}%
}%
\begin{pgfscope}%
\pgfsys@transformshift{0.675831in}{0.858459in}%
\pgfsys@useobject{currentmarker}{}%
\end{pgfscope}%
\end{pgfscope}%
\begin{pgfscope}%
\pgftext[x=0.331695in,y=0.810241in,left,base]{\rmfamily\fontsize{10.000000}{12.000000}\selectfont \(\displaystyle 0.00\)}%
\end{pgfscope}%
\begin{pgfscope}%
\pgfsetbuttcap%
\pgfsetroundjoin%
\definecolor{currentfill}{rgb}{0.000000,0.000000,0.000000}%
\pgfsetfillcolor{currentfill}%
\pgfsetlinewidth{0.803000pt}%
\definecolor{currentstroke}{rgb}{0.000000,0.000000,0.000000}%
\pgfsetstrokecolor{currentstroke}%
\pgfsetdash{}{0pt}%
\pgfsys@defobject{currentmarker}{\pgfqpoint{-0.048611in}{0.000000in}}{\pgfqpoint{0.000000in}{0.000000in}}{%
\pgfpathmoveto{\pgfqpoint{0.000000in}{0.000000in}}%
\pgfpathlineto{\pgfqpoint{-0.048611in}{0.000000in}}%
\pgfusepath{stroke,fill}%
}%
\begin{pgfscope}%
\pgfsys@transformshift{0.675831in}{1.237991in}%
\pgfsys@useobject{currentmarker}{}%
\end{pgfscope}%
\end{pgfscope}%
\begin{pgfscope}%
\pgftext[x=0.331695in,y=1.189773in,left,base]{\rmfamily\fontsize{10.000000}{12.000000}\selectfont \(\displaystyle 0.05\)}%
\end{pgfscope}%
\begin{pgfscope}%
\pgfsetbuttcap%
\pgfsetroundjoin%
\definecolor{currentfill}{rgb}{0.000000,0.000000,0.000000}%
\pgfsetfillcolor{currentfill}%
\pgfsetlinewidth{0.803000pt}%
\definecolor{currentstroke}{rgb}{0.000000,0.000000,0.000000}%
\pgfsetstrokecolor{currentstroke}%
\pgfsetdash{}{0pt}%
\pgfsys@defobject{currentmarker}{\pgfqpoint{-0.048611in}{0.000000in}}{\pgfqpoint{0.000000in}{0.000000in}}{%
\pgfpathmoveto{\pgfqpoint{0.000000in}{0.000000in}}%
\pgfpathlineto{\pgfqpoint{-0.048611in}{0.000000in}}%
\pgfusepath{stroke,fill}%
}%
\begin{pgfscope}%
\pgfsys@transformshift{0.675831in}{1.617522in}%
\pgfsys@useobject{currentmarker}{}%
\end{pgfscope}%
\end{pgfscope}%
\begin{pgfscope}%
\pgftext[x=0.331695in,y=1.569304in,left,base]{\rmfamily\fontsize{10.000000}{12.000000}\selectfont \(\displaystyle 0.10\)}%
\end{pgfscope}%
\begin{pgfscope}%
\pgfsetbuttcap%
\pgfsetroundjoin%
\definecolor{currentfill}{rgb}{0.000000,0.000000,0.000000}%
\pgfsetfillcolor{currentfill}%
\pgfsetlinewidth{0.803000pt}%
\definecolor{currentstroke}{rgb}{0.000000,0.000000,0.000000}%
\pgfsetstrokecolor{currentstroke}%
\pgfsetdash{}{0pt}%
\pgfsys@defobject{currentmarker}{\pgfqpoint{-0.048611in}{0.000000in}}{\pgfqpoint{0.000000in}{0.000000in}}{%
\pgfpathmoveto{\pgfqpoint{0.000000in}{0.000000in}}%
\pgfpathlineto{\pgfqpoint{-0.048611in}{0.000000in}}%
\pgfusepath{stroke,fill}%
}%
\begin{pgfscope}%
\pgfsys@transformshift{0.675831in}{1.997053in}%
\pgfsys@useobject{currentmarker}{}%
\end{pgfscope}%
\end{pgfscope}%
\begin{pgfscope}%
\pgftext[x=0.331695in,y=1.948836in,left,base]{\rmfamily\fontsize{10.000000}{12.000000}\selectfont \(\displaystyle 0.15\)}%
\end{pgfscope}%
\begin{pgfscope}%
\pgfsetbuttcap%
\pgfsetroundjoin%
\definecolor{currentfill}{rgb}{0.000000,0.000000,0.000000}%
\pgfsetfillcolor{currentfill}%
\pgfsetlinewidth{0.803000pt}%
\definecolor{currentstroke}{rgb}{0.000000,0.000000,0.000000}%
\pgfsetstrokecolor{currentstroke}%
\pgfsetdash{}{0pt}%
\pgfsys@defobject{currentmarker}{\pgfqpoint{-0.048611in}{0.000000in}}{\pgfqpoint{0.000000in}{0.000000in}}{%
\pgfpathmoveto{\pgfqpoint{0.000000in}{0.000000in}}%
\pgfpathlineto{\pgfqpoint{-0.048611in}{0.000000in}}%
\pgfusepath{stroke,fill}%
}%
\begin{pgfscope}%
\pgfsys@transformshift{0.675831in}{2.376585in}%
\pgfsys@useobject{currentmarker}{}%
\end{pgfscope}%
\end{pgfscope}%
\begin{pgfscope}%
\pgftext[x=0.331695in,y=2.328367in,left,base]{\rmfamily\fontsize{10.000000}{12.000000}\selectfont \(\displaystyle 0.20\)}%
\end{pgfscope}%
\begin{pgfscope}%
\definecolor{textcolor}{rgb}{0.000000,0.000000,1.000000}%
\pgfsetstrokecolor{textcolor}%
\pgfsetfillcolor{textcolor}%
\pgftext[x=0.276139in,y=1.535164in,,bottom,rotate=90.000000]{\color{textcolor}\rmfamily\fontsize{10.000000}{12.000000}\selectfont Sharpe ratio}%
\end{pgfscope}%
\begin{pgfscope}%
\pgfpathrectangle{\pgfqpoint{0.675831in}{0.555327in}}{\pgfqpoint{4.539169in}{1.959673in}}%
\pgfusepath{clip}%
\pgfsetrectcap%
\pgfsetroundjoin%
\pgfsetlinewidth{1.003750pt}%
\definecolor{currentstroke}{rgb}{0.000000,0.000000,1.000000}%
\pgfsetstrokecolor{currentstroke}%
\pgfsetdash{}{0pt}%
\pgfpathmoveto{\pgfqpoint{0.882157in}{2.425924in}}%
\pgfpathlineto{\pgfqpoint{0.968126in}{2.092695in}}%
\pgfpathlineto{\pgfqpoint{1.054095in}{2.043356in}}%
\pgfpathlineto{\pgfqpoint{1.140064in}{1.956823in}}%
\pgfpathlineto{\pgfqpoint{1.226034in}{1.940124in}}%
\pgfpathlineto{\pgfqpoint{1.312003in}{1.868772in}}%
\pgfpathlineto{\pgfqpoint{1.397972in}{1.711646in}}%
\pgfpathlineto{\pgfqpoint{1.483941in}{1.553002in}}%
\pgfpathlineto{\pgfqpoint{1.569910in}{1.697983in}}%
\pgfpathlineto{\pgfqpoint{1.655879in}{1.795902in}}%
\pgfpathlineto{\pgfqpoint{1.741848in}{1.411816in}}%
\pgfpathlineto{\pgfqpoint{1.827817in}{1.748081in}}%
\pgfpathlineto{\pgfqpoint{1.913786in}{1.514289in}}%
\pgfpathlineto{\pgfqpoint{1.999755in}{1.285812in}}%
\pgfpathlineto{\pgfqpoint{2.085725in}{1.625113in}}%
\pgfpathlineto{\pgfqpoint{2.171694in}{1.539339in}}%
\pgfpathlineto{\pgfqpoint{2.257663in}{1.225846in}}%
\pgfpathlineto{\pgfqpoint{2.343632in}{1.193206in}}%
\pgfpathlineto{\pgfqpoint{2.429601in}{1.014067in}}%
\pgfpathlineto{\pgfqpoint{2.515570in}{1.662307in}}%
\pgfpathlineto{\pgfqpoint{2.601539in}{1.679006in}}%
\pgfpathlineto{\pgfqpoint{2.687508in}{1.410298in}}%
\pgfpathlineto{\pgfqpoint{2.773477in}{1.269112in}}%
\pgfpathlineto{\pgfqpoint{2.859447in}{1.187892in}}%
\pgfpathlineto{\pgfqpoint{2.945416in}{1.007235in}}%
\pgfpathlineto{\pgfqpoint{3.031385in}{1.119577in}}%
\pgfpathlineto{\pgfqpoint{3.117354in}{1.105914in}}%
\pgfpathlineto{\pgfqpoint{3.203323in}{1.069479in}}%
\pgfpathlineto{\pgfqpoint{3.289292in}{1.056575in}}%
\pgfpathlineto{\pgfqpoint{3.375261in}{1.748840in}}%
\pgfpathlineto{\pgfqpoint{3.461230in}{1.644089in}}%
\pgfpathlineto{\pgfqpoint{3.547199in}{1.508976in}}%
\pgfpathlineto{\pgfqpoint{3.633168in}{1.416370in}}%
\pgfpathlineto{\pgfqpoint{3.719138in}{1.225846in}}%
\pgfpathlineto{\pgfqpoint{3.805107in}{1.285052in}}%
\pgfpathlineto{\pgfqpoint{3.891076in}{1.197001in}}%
\pgfpathlineto{\pgfqpoint{3.977045in}{1.070997in}}%
\pgfpathlineto{\pgfqpoint{4.063014in}{1.021658in}}%
\pgfpathlineto{\pgfqpoint{4.148983in}{1.124890in}}%
\pgfpathlineto{\pgfqpoint{4.234952in}{1.054297in}}%
\pgfpathlineto{\pgfqpoint{4.320921in}{1.194724in}}%
\pgfpathlineto{\pgfqpoint{4.406890in}{1.070238in}}%
\pgfpathlineto{\pgfqpoint{4.492860in}{1.145385in}}%
\pgfpathlineto{\pgfqpoint{4.578829in}{1.071756in}}%
\pgfpathlineto{\pgfqpoint{4.664798in}{1.130204in}}%
\pgfpathlineto{\pgfqpoint{4.750767in}{1.052020in}}%
\pgfpathlineto{\pgfqpoint{4.836736in}{0.900208in}}%
\pgfpathlineto{\pgfqpoint{4.922705in}{0.752949in}}%
\pgfpathlineto{\pgfqpoint{5.008674in}{0.644403in}}%
\pgfusepath{stroke}%
\end{pgfscope}%
\begin{pgfscope}%
\pgfsetrectcap%
\pgfsetmiterjoin%
\pgfsetlinewidth{0.803000pt}%
\definecolor{currentstroke}{rgb}{0.000000,0.000000,0.000000}%
\pgfsetstrokecolor{currentstroke}%
\pgfsetdash{}{0pt}%
\pgfpathmoveto{\pgfqpoint{0.675831in}{0.555327in}}%
\pgfpathlineto{\pgfqpoint{0.675831in}{2.515000in}}%
\pgfusepath{stroke}%
\end{pgfscope}%
\begin{pgfscope}%
\pgfsetrectcap%
\pgfsetmiterjoin%
\pgfsetlinewidth{0.803000pt}%
\definecolor{currentstroke}{rgb}{0.000000,0.000000,0.000000}%
\pgfsetstrokecolor{currentstroke}%
\pgfsetdash{}{0pt}%
\pgfpathmoveto{\pgfqpoint{5.215000in}{0.555327in}}%
\pgfpathlineto{\pgfqpoint{5.215000in}{2.515000in}}%
\pgfusepath{stroke}%
\end{pgfscope}%
\begin{pgfscope}%
\pgfsetrectcap%
\pgfsetmiterjoin%
\pgfsetlinewidth{0.803000pt}%
\definecolor{currentstroke}{rgb}{0.000000,0.000000,0.000000}%
\pgfsetstrokecolor{currentstroke}%
\pgfsetdash{}{0pt}%
\pgfpathmoveto{\pgfqpoint{0.675831in}{0.555327in}}%
\pgfpathlineto{\pgfqpoint{5.215000in}{0.555327in}}%
\pgfusepath{stroke}%
\end{pgfscope}%
\begin{pgfscope}%
\pgfsetrectcap%
\pgfsetmiterjoin%
\pgfsetlinewidth{0.803000pt}%
\definecolor{currentstroke}{rgb}{0.000000,0.000000,0.000000}%
\pgfsetstrokecolor{currentstroke}%
\pgfsetdash{}{0pt}%
\pgfpathmoveto{\pgfqpoint{0.675831in}{2.515000in}}%
\pgfpathlineto{\pgfqpoint{5.215000in}{2.515000in}}%
\pgfusepath{stroke}%
\end{pgfscope}%
\end{pgfpicture}%
\makeatother%
\endgroup%

%% file: sharpe-rewards-vs-pools1.pgf
\begingroup%
\makeatletter%
\begin{pgfpicture}%
\pgfpathrectangle{\pgfpointorigin}{\pgfqpoint{5.400000in}{2.700000in}}%
\pgfusepath{use as bounding box, clip}%
\begin{pgfscope}%
\pgfsetbuttcap%
\pgfsetmiterjoin%
\definecolor{currentfill}{rgb}{1.000000,1.000000,1.000000}%
\pgfsetfillcolor{currentfill}%
\pgfsetlinewidth{0.000000pt}%
\definecolor{currentstroke}{rgb}{1.000000,1.000000,1.000000}%
\pgfsetstrokecolor{currentstroke}%
\pgfsetdash{}{0pt}%
\pgfpathmoveto{\pgfqpoint{0.000000in}{0.000000in}}%
\pgfpathlineto{\pgfqpoint{5.400000in}{0.000000in}}%
\pgfpathlineto{\pgfqpoint{5.400000in}{2.700000in}}%
\pgfpathlineto{\pgfqpoint{0.000000in}{2.700000in}}%
\pgfpathclose%
\pgfusepath{fill}%
\end{pgfscope}%
\begin{pgfscope}%
\pgfsetbuttcap%
\pgfsetmiterjoin%
\definecolor{currentfill}{rgb}{1.000000,1.000000,1.000000}%
\pgfsetfillcolor{currentfill}%
\pgfsetlinewidth{0.000000pt}%
\definecolor{currentstroke}{rgb}{0.000000,0.000000,0.000000}%
\pgfsetstrokecolor{currentstroke}%
\pgfsetstrokeopacity{0.000000}%
\pgfsetdash{}{0pt}%
\pgfpathmoveto{\pgfqpoint{0.853301in}{0.949216in}}%
\pgfpathlineto{\pgfqpoint{5.161821in}{0.949216in}}%
\pgfpathlineto{\pgfqpoint{5.161821in}{2.515000in}}%
\pgfpathlineto{\pgfqpoint{0.853301in}{2.515000in}}%
\pgfpathclose%
\pgfusepath{fill}%
\end{pgfscope}%
\begin{pgfscope}%
\pgfsetbuttcap%
\pgfsetroundjoin%
\definecolor{currentfill}{rgb}{0.000000,0.000000,0.000000}%
\pgfsetfillcolor{currentfill}%
\pgfsetlinewidth{0.803000pt}%
\definecolor{currentstroke}{rgb}{0.000000,0.000000,0.000000}%
\pgfsetstrokecolor{currentstroke}%
\pgfsetdash{}{0pt}%
\pgfsys@defobject{currentmarker}{\pgfqpoint{0.000000in}{-0.048611in}}{\pgfqpoint{0.000000in}{0.000000in}}{%
\pgfpathmoveto{\pgfqpoint{0.000000in}{0.000000in}}%
\pgfpathlineto{\pgfqpoint{0.000000in}{-0.048611in}}%
\pgfusepath{stroke,fill}%
}%
\begin{pgfscope}%
\pgfsys@transformshift{1.049143in}{0.949216in}%
\pgfsys@useobject{currentmarker}{}%
\end{pgfscope}%
\end{pgfscope}%
\begin{pgfscope}%
\pgftext[x=0.853747in,y=0.569214in,left,base,rotate=25.000000]{\rmfamily\fontsize{10.000000}{12.000000}\selectfont ViaBTC}%
\end{pgfscope}%
\begin{pgfscope}%
\pgfsetbuttcap%
\pgfsetroundjoin%
\definecolor{currentfill}{rgb}{0.000000,0.000000,0.000000}%
\pgfsetfillcolor{currentfill}%
\pgfsetlinewidth{0.803000pt}%
\definecolor{currentstroke}{rgb}{0.000000,0.000000,0.000000}%
\pgfsetstrokecolor{currentstroke}%
\pgfsetdash{}{0pt}%
\pgfsys@defobject{currentmarker}{\pgfqpoint{0.000000in}{-0.048611in}}{\pgfqpoint{0.000000in}{0.000000in}}{%
\pgfpathmoveto{\pgfqpoint{0.000000in}{0.000000in}}%
\pgfpathlineto{\pgfqpoint{0.000000in}{-0.048611in}}%
\pgfusepath{stroke,fill}%
}%
\begin{pgfscope}%
\pgfsys@transformshift{1.832510in}{0.949216in}%
\pgfsys@useobject{currentmarker}{}%
\end{pgfscope}%
\end{pgfscope}%
\begin{pgfscope}%
\pgftext[x=1.636777in,y=0.568898in,left,base,rotate=25.000000]{\rmfamily\fontsize{10.000000}{12.000000}\selectfont AntPool}%
\end{pgfscope}%
\begin{pgfscope}%
\pgfsetbuttcap%
\pgfsetroundjoin%
\definecolor{currentfill}{rgb}{0.000000,0.000000,0.000000}%
\pgfsetfillcolor{currentfill}%
\pgfsetlinewidth{0.803000pt}%
\definecolor{currentstroke}{rgb}{0.000000,0.000000,0.000000}%
\pgfsetstrokecolor{currentstroke}%
\pgfsetdash{}{0pt}%
\pgfsys@defobject{currentmarker}{\pgfqpoint{0.000000in}{-0.048611in}}{\pgfqpoint{0.000000in}{0.000000in}}{%
\pgfpathmoveto{\pgfqpoint{0.000000in}{0.000000in}}%
\pgfpathlineto{\pgfqpoint{0.000000in}{-0.048611in}}%
\pgfusepath{stroke,fill}%
}%
\begin{pgfscope}%
\pgfsys@transformshift{2.615877in}{0.949216in}%
\pgfsys@useobject{currentmarker}{}%
\end{pgfscope}%
\end{pgfscope}%
\begin{pgfscope}%
\pgftext[x=2.374631in,y=0.526452in,left,base,rotate=25.000000]{\rmfamily\fontsize{10.000000}{12.000000}\selectfont SlushPool}%
\end{pgfscope}%
\begin{pgfscope}%
\pgfsetbuttcap%
\pgfsetroundjoin%
\definecolor{currentfill}{rgb}{0.000000,0.000000,0.000000}%
\pgfsetfillcolor{currentfill}%
\pgfsetlinewidth{0.803000pt}%
\definecolor{currentstroke}{rgb}{0.000000,0.000000,0.000000}%
\pgfsetstrokecolor{currentstroke}%
\pgfsetdash{}{0pt}%
\pgfsys@defobject{currentmarker}{\pgfqpoint{0.000000in}{-0.048611in}}{\pgfqpoint{0.000000in}{0.000000in}}{%
\pgfpathmoveto{\pgfqpoint{0.000000in}{0.000000in}}%
\pgfpathlineto{\pgfqpoint{0.000000in}{-0.048611in}}%
\pgfusepath{stroke,fill}%
}%
\begin{pgfscope}%
\pgfsys@transformshift{3.399244in}{0.949216in}%
\pgfsys@useobject{currentmarker}{}%
\end{pgfscope}%
\end{pgfscope}%
\begin{pgfscope}%
\pgftext[x=3.203542in,y=0.568927in,left,base,rotate=25.000000]{\rmfamily\fontsize{10.000000}{12.000000}\selectfont DPOOL}%
\end{pgfscope}%
\begin{pgfscope}%
\pgfsetbuttcap%
\pgfsetroundjoin%
\definecolor{currentfill}{rgb}{0.000000,0.000000,0.000000}%
\pgfsetfillcolor{currentfill}%
\pgfsetlinewidth{0.803000pt}%
\definecolor{currentstroke}{rgb}{0.000000,0.000000,0.000000}%
\pgfsetstrokecolor{currentstroke}%
\pgfsetdash{}{0pt}%
\pgfsys@defobject{currentmarker}{\pgfqpoint{0.000000in}{-0.048611in}}{\pgfqpoint{0.000000in}{0.000000in}}{%
\pgfpathmoveto{\pgfqpoint{0.000000in}{0.000000in}}%
\pgfpathlineto{\pgfqpoint{0.000000in}{-0.048611in}}%
\pgfusepath{stroke,fill}%
}%
\begin{pgfscope}%
\pgfsys@transformshift{4.182612in}{0.949216in}%
\pgfsys@useobject{currentmarker}{}%
\end{pgfscope}%
\end{pgfscope}%
\begin{pgfscope}%
\pgftext[x=3.761402in,y=0.358615in,left,base,rotate=25.000000]{\rmfamily\fontsize{10.000000}{12.000000}\selectfont BitClub Network}%
\end{pgfscope}%
\begin{pgfscope}%
\pgfsetbuttcap%
\pgfsetroundjoin%
\definecolor{currentfill}{rgb}{0.000000,0.000000,0.000000}%
\pgfsetfillcolor{currentfill}%
\pgfsetlinewidth{0.803000pt}%
\definecolor{currentstroke}{rgb}{0.000000,0.000000,0.000000}%
\pgfsetstrokecolor{currentstroke}%
\pgfsetdash{}{0pt}%
\pgfsys@defobject{currentmarker}{\pgfqpoint{0.000000in}{-0.048611in}}{\pgfqpoint{0.000000in}{0.000000in}}{%
\pgfpathmoveto{\pgfqpoint{0.000000in}{0.000000in}}%
\pgfpathlineto{\pgfqpoint{0.000000in}{-0.048611in}}%
\pgfusepath{stroke,fill}%
}%
\begin{pgfscope}%
\pgfsys@transformshift{4.965979in}{0.949216in}%
\pgfsys@useobject{currentmarker}{}%
\end{pgfscope}%
\end{pgfscope}%
\begin{pgfscope}%
\pgftext[x=4.728328in,y=0.529805in,left,base,rotate=25.000000]{\rmfamily\fontsize{10.000000}{12.000000}\selectfont KanoPool}%
\end{pgfscope}%
\begin{pgfscope}%
\pgftext[x=3.007561in,y=0.276139in,,top]{\rmfamily\fontsize{10.000000}{12.000000}\selectfont Pools set}%
\end{pgfscope}%
\begin{pgfscope}%
\pgfsetbuttcap%
\pgfsetroundjoin%
\definecolor{currentfill}{rgb}{0.000000,0.000000,0.000000}%
\pgfsetfillcolor{currentfill}%
\pgfsetlinewidth{0.803000pt}%
\definecolor{currentstroke}{rgb}{0.000000,0.000000,0.000000}%
\pgfsetstrokecolor{currentstroke}%
\pgfsetdash{}{0pt}%
\pgfsys@defobject{currentmarker}{\pgfqpoint{-0.048611in}{0.000000in}}{\pgfqpoint{0.000000in}{0.000000in}}{%
\pgfpathmoveto{\pgfqpoint{0.000000in}{0.000000in}}%
\pgfpathlineto{\pgfqpoint{-0.048611in}{0.000000in}}%
\pgfusepath{stroke,fill}%
}%
\begin{pgfscope}%
\pgfsys@transformshift{0.853301in}{1.220941in}%
\pgfsys@useobject{currentmarker}{}%
\end{pgfscope}%
\end{pgfscope}%
\begin{pgfscope}%
\pgftext[x=0.509164in,y=1.172723in,left,base]{\rmfamily\fontsize{10.000000}{12.000000}\selectfont \(\displaystyle 0.00\)}%
\end{pgfscope}%
\begin{pgfscope}%
\pgfsetbuttcap%
\pgfsetroundjoin%
\definecolor{currentfill}{rgb}{0.000000,0.000000,0.000000}%
\pgfsetfillcolor{currentfill}%
\pgfsetlinewidth{0.803000pt}%
\definecolor{currentstroke}{rgb}{0.000000,0.000000,0.000000}%
\pgfsetstrokecolor{currentstroke}%
\pgfsetdash{}{0pt}%
\pgfsys@defobject{currentmarker}{\pgfqpoint{-0.048611in}{0.000000in}}{\pgfqpoint{0.000000in}{0.000000in}}{%
\pgfpathmoveto{\pgfqpoint{0.000000in}{0.000000in}}%
\pgfpathlineto{\pgfqpoint{-0.048611in}{0.000000in}}%
\pgfusepath{stroke,fill}%
}%
\begin{pgfscope}%
\pgfsys@transformshift{0.853301in}{1.710096in}%
\pgfsys@useobject{currentmarker}{}%
\end{pgfscope}%
\end{pgfscope}%
\begin{pgfscope}%
\pgftext[x=0.509164in,y=1.661878in,left,base]{\rmfamily\fontsize{10.000000}{12.000000}\selectfont \(\displaystyle 0.01\)}%
\end{pgfscope}%
\begin{pgfscope}%
\pgfsetbuttcap%
\pgfsetroundjoin%
\definecolor{currentfill}{rgb}{0.000000,0.000000,0.000000}%
\pgfsetfillcolor{currentfill}%
\pgfsetlinewidth{0.803000pt}%
\definecolor{currentstroke}{rgb}{0.000000,0.000000,0.000000}%
\pgfsetstrokecolor{currentstroke}%
\pgfsetdash{}{0pt}%
\pgfsys@defobject{currentmarker}{\pgfqpoint{-0.048611in}{0.000000in}}{\pgfqpoint{0.000000in}{0.000000in}}{%
\pgfpathmoveto{\pgfqpoint{0.000000in}{0.000000in}}%
\pgfpathlineto{\pgfqpoint{-0.048611in}{0.000000in}}%
\pgfusepath{stroke,fill}%
}%
\begin{pgfscope}%
\pgfsys@transformshift{0.853301in}{2.199251in}%
\pgfsys@useobject{currentmarker}{}%
\end{pgfscope}%
\end{pgfscope}%
\begin{pgfscope}%
\pgftext[x=0.509164in,y=2.151033in,left,base]{\rmfamily\fontsize{10.000000}{12.000000}\selectfont \(\displaystyle 0.02\)}%
\end{pgfscope}%
\begin{pgfscope}%
\definecolor{textcolor}{rgb}{0.000000,0.000000,1.000000}%
\pgfsetstrokecolor{textcolor}%
\pgfsetfillcolor{textcolor}%
\pgftext[x=0.453609in,y=1.732108in,,bottom,rotate=90.000000]{\color{textcolor}\rmfamily\fontsize{10.000000}{12.000000}\selectfont Sharpe ratio}%
\end{pgfscope}%
\begin{pgfscope}%
\pgfpathrectangle{\pgfqpoint{0.853301in}{0.949216in}}{\pgfqpoint{4.308520in}{1.565784in}}%
\pgfusepath{clip}%
\pgfsetrectcap%
\pgfsetroundjoin%
\pgfsetlinewidth{1.003750pt}%
\definecolor{currentstroke}{rgb}{0.000000,0.000000,1.000000}%
\pgfsetstrokecolor{currentstroke}%
\pgfsetdash{}{0pt}%
\pgfpathmoveto{\pgfqpoint{1.049143in}{1.020388in}}%
\pgfpathlineto{\pgfqpoint{1.832510in}{2.306865in}}%
\pgfpathlineto{\pgfqpoint{2.615877in}{2.306865in}}%
\pgfpathlineto{\pgfqpoint{3.399244in}{2.306865in}}%
\pgfpathlineto{\pgfqpoint{4.182612in}{2.306865in}}%
\pgfpathlineto{\pgfqpoint{4.965979in}{2.443828in}}%
\pgfusepath{stroke}%
\end{pgfscope}%
\begin{pgfscope}%
\pgfsetrectcap%
\pgfsetmiterjoin%
\pgfsetlinewidth{0.803000pt}%
\definecolor{currentstroke}{rgb}{0.000000,0.000000,0.000000}%
\pgfsetstrokecolor{currentstroke}%
\pgfsetdash{}{0pt}%
\pgfpathmoveto{\pgfqpoint{0.853301in}{0.949216in}}%
\pgfpathlineto{\pgfqpoint{0.853301in}{2.515000in}}%
\pgfusepath{stroke}%
\end{pgfscope}%
\begin{pgfscope}%
\pgfsetrectcap%
\pgfsetmiterjoin%
\pgfsetlinewidth{0.803000pt}%
\definecolor{currentstroke}{rgb}{0.000000,0.000000,0.000000}%
\pgfsetstrokecolor{currentstroke}%
\pgfsetdash{}{0pt}%
\pgfpathmoveto{\pgfqpoint{5.161821in}{0.949216in}}%
\pgfpathlineto{\pgfqpoint{5.161821in}{2.515000in}}%
\pgfusepath{stroke}%
\end{pgfscope}%
\begin{pgfscope}%
\pgfsetrectcap%
\pgfsetmiterjoin%
\pgfsetlinewidth{0.803000pt}%
\definecolor{currentstroke}{rgb}{0.000000,0.000000,0.000000}%
\pgfsetstrokecolor{currentstroke}%
\pgfsetdash{}{0pt}%
\pgfpathmoveto{\pgfqpoint{0.853301in}{0.949216in}}%
\pgfpathlineto{\pgfqpoint{5.161821in}{0.949216in}}%
\pgfusepath{stroke}%
\end{pgfscope}%
\begin{pgfscope}%
\pgfsetrectcap%
\pgfsetmiterjoin%
\pgfsetlinewidth{0.803000pt}%
\definecolor{currentstroke}{rgb}{0.000000,0.000000,0.000000}%
\pgfsetstrokecolor{currentstroke}%
\pgfsetdash{}{0pt}%
\pgfpathmoveto{\pgfqpoint{0.853301in}{2.515000in}}%
\pgfpathlineto{\pgfqpoint{5.161821in}{2.515000in}}%
\pgfusepath{stroke}%
\end{pgfscope}%
\end{pgfpicture}%
\makeatother%
\endgroup%

%% file: conclusion.tex
\section{Conclusions}
We present an analytical tool that allows risk-averse miners to optimally create a mining portfolio that maximizes their risk-adjusted rewards, using a theoretical model that optimally allocates miner's resources over mining pools based on their risk aversion levels. We provide multiple extensions of the base model to enable miners to optimally distribute their power between mining pools of different cryptocurrencies, which might even use different PoW algorithms. Then, we develop an analytical tool publicly available (as provided in Section \ref{sec:evaluation}) for miners to compute their optimal hash power allocation based on their inputs, and we present both time-static and historical-retroactive evaluations of our tool. The retroactive evaluation results show a direct benefit for the individual miner in terms of reward amount over reward standard deviation ratio (expressed by the Sharpe ratio). 

As a final note, it is often argued that the massive participation on mining pools has lead to blockchain centralization (e.g. in bitcoin at the time of writing, over 50\% of mining is done by 4 mining pools).
Lack of decentralization can lead to various types of attacks, including double-spending, reversing confirmations of previous transactions or transaction censoring~\cite{SP:Eyal15,FC:EyaSir14,FC:SapSomZoh16,Sompolinsky:2017:BUI:3155112.3168362}. Mining pools (especially mining pools of larger sizes) have been criticized for leading to a high rate of centralization. A number of academic works have studied the level and concerning effects of the centralization trend~\cite{jiasun,DBLP:journals/corr/abs-1801-03998,Sompolinsky:2017:BUI:3155112.3168362} while a number of solutions have been proposed spanning from decentralized mining pools (P2Pool~\cite{p2pool} for Bitcoin), to alternative, non-outsourceable PoW mechanisms~\cite{CCS:MKKS15}.  We believe that tools like the one we present here, are positive steps towards the ``centralization'' problem of PoW systems. Our tool provides incentives to miners in order for them to actively diversify among different pools and cryptocurrencies, potentially increasing power of a large number of smaller pools, while at the same time could also provide insights to mining pool managers in terms of how would rational miners behave.
\section*{Acknowledgement}
Foteini Baldimtsi and Jiasun Li would like to thank the George Mason Multidisciplinary Research (MDR) Initiative for funding this project.

%% file: appendix.tex
\section{Mining Pools for Major Cryptocurrencies}
	\label{app:PoolData}
	
In Table \ref{mining-pools} we summarize a list of mining pools for major cryptocurrencies and the reward type each offers, as of February 15 2019. PPS, PPLNS and proportional we already discussed in Section~\ref{reward-methods}. In the following table we also come across  some variants of the standard reward types. In particular,  FPPS (Full Pay Per Share) is similar to PPS but payments take average transaction fees into account, Score is based on the proportional reward method weighed by time the share was submitted, and Exponential is a PPLNS variant with exponential decay of share values. We manually collected the data using the respective mining pool websites and the cryptocurrencies' block explorers.

\ifllncs
\else
\ifieee
\begin{table}
	\caption{Mining pools for major cryptocurrencies}
	\label{mining-pools}
			\begin{tabular}{|l|l|l|l|}
				\hline
				\textbf{Pool Name}&\textbf{Coin}&\textbf{Reward type}&\textbf{Hash power}\\
				\hline
				BTC.com&BTC&FPPS&8.13 EH/s\\
				&BCH&FPPS&207.97 PH/s \\
				
				\hline
				Antpool&BTC&PPLNS,PPS&5.06 EH/s\\
				\cline{2-4}
				&BCH&PPLNS,PPS&149.48 PH/s\\
				&ETH&PPLNS,PPS&323 GH/s\\
				&LTC&PPLNS,PPS&21.2 TH/s\\
				&ETC&PPLNS,PPS&6.26 GH/s\\
				&ZEC&PPLNS,PPS&258 MSol/s\\
				&DASH&PPLNS,PPS&138 TH/S\\
				&SIA&PPLNS,PPS&816.92 TH/s\\
				
				\hline
				Slush&BTC&Score&3.27 EH/s\\
				
				\hline
				ViaBTC&BTC&PPLNS,PPS&2.87 EH/s\\
				\cline{2-4}
				&BCH&PPLNS,PPS&84.49 PH/s\\
				\cline{2-4}
				&ETH&PPLNS,PPS&878 GH/s\\
				\cline{2-4}
				&LTC&PPLNS,PPS&26.52 TH/s\\
				\cline{2-4}
				&ETC&PPLNS,PPS&6.032 GH/s\\
				\cline{2-4}
				&ZEC&PPLNS,PPS&89.63 MSol/s\\
				\cline{2-4}
				&DASH&PPLNS,PPS&117.36 TH/s\\
				
				\hline
				Miningpoolhub&ETH&PPLNS&8.33 TH/s\\
				\cline{2-4}
				&LTC&PPLNS&277 GH/s\\
				\cline{2-4}
				&ETC&PPLNS&1.494 TH/s\\
				\cline{2-4}
				&ZEC&PPLNS&29.378 MH/s\\
				\cline{2-4}
				&DASH&PPLNS&27.69 TH/s\\
				\cline{2-4}
				&XMR&PPLNS&5.74 MH/s\\
				\cline{2-4}
				&DGB&PPLNS&11.85 TH/s\\
				\hline
				Bitcoin.com&BTC&PPS&892.16 PH/s\\
				\cline{2-4}
				&BCH&PPS&191.72 PH/s\\		
				\hline
				Nanopool&ETH&PPLNS&20.96 TH/s\\
				\cline{2-4}
				&ETC&PPLNS&1.65 TH/s\\
				\cline{2-4}
				&ZEC&PPLNS&53.57 MSol/s\\
				\cline{2-4}
				&GRIN&PPLNS&17.1 Kgp/s\\
				\cline{2-4}
				&XMR&PPLNS&57.488 MH/s\\
				\hline
				Litecoinpool.org&LTC&PPS&33.49 TH/s\\
				\hline
				Ethermine&ETH&PPLNS&41.4 TH/s\\
				\cline{2-4}
				&ETC&PPLNS&3.5 TH/s\\
				\cline{2-4}
				&ZEC&PPLNS&396.9 MSol/s\\
				\hline
				f2pool&BTC&PPS&4.46 EH/s\\
				\cline{2-4}
				&ETH&PPS&17.7 TH/s\\
				\cline{2-4}
				&LTC&PPS&40.94 TH/s\\
				\cline{2-4}
				&ETC&PPS&127.37 GH/s\\
				\cline{2-4}
				&ZEC&PPS&379.73 MSol/s\\
				\cline{2-4}
				&DASH&PPS&74.44 TH/s\\
				\cline{2-4}
				&SIA&PPS&3.59 TH/s\\
				\cline{2-4}
				&XMR&PPS&54.50 MH/s\\
				\hline
				Multipool&BTC&Exponential&0.66 PH/s\\
				\cline{2-4}
				&BCH&PPLNS&3.658 PH/s\\
				\cline{2-4}
				&LTC&PPLNS&122.22 GH/s\\
				\cline{2-4}
				&DGB&Proportional&3.836 PH/s\\
				\hline
				Minergate&BTG&PPLNS,PPS&2.006 KSol/s\\
				\cline{2-4}
				&ETH&PPLNS&14.68 GH/s\\
				\cline{2-4}
				&ETC&PPLNS&4.374 GH/s\\
				\cline{2-4}
				&ZEC&PPLNS&16.12 KSol/s\\
				\cline{2-4}
				&XMR&PPLNS,PPS&3.776 MH/s\\
				\cline{2-4}
				&BCN&PPLNS,PPS&1.641 MH/s\\
				\hline
				Suprnova&BTG&Proportional&0.25 MSol/s\\
				\cline{2-4}
				&LTC&Proportional&24.09 GH/\\
				\cline{2-4}
				&ZEC&Proportional&19.03 KSol/s\\
				\cline{2-4}
				&DGB&Proportional&9.45 TH/s\\
				\cline{2-4}
				&DASH&Proportional&9.26 TH/s\\
				\hline
				Coinotron&ETH&PPLNS,RBPPS&806.6 GH/s\\
				\cline{2-4}
				&LTC&PPLNS,PPS&37.8 GH/s\\
				\cline{2-4}
				&ETC&PPLNS,RBPPS&53.8 GH/s\\
				\cline{2-4}
				&ZEC&PPLNS,PPS&491.5 MSol/s\\
				\cline{2-4}
				&BTG&PPLNS,PPS&332.6 KH/s\\
				\cline{2-4}
				&DASH&PPLNS,PPS&2.4 TH/s\\
				\hline
				
			\end{tabular}
\end{table}
\else
\begin{table}[b]
	\caption{Mining pools for major cryptocurrencies}
	\label{mining-pools}
	\begin{tabular}{|l|l|l|l|}
		\hline
		\textbf{Pool Name}&\textbf{Coin}&\textbf{Reward type}&\textbf{Hash power}\\
		\hline
		BTC.com&BTC&FPPS&8.13 EH/s\\
		&BCH&FPPS&207.97 PH/s \\
		
		\hline
		Antpool&BTC&PPLNS,PPS&5.06 EH/s\\
		\cline{2-4}
		&BCH&PPLNS,PPS&149.48 PH/s\\
		&ETH&PPLNS,PPS&323 GH/s\\
		&LTC&PPLNS,PPS&21.2 TH/s\\
		&ETC&PPLNS,PPS&6.26 GH/s\\
		&ZEC&PPLNS,PPS&258 MSol/s\\
		&DASH&PPLNS,PPS&138 TH/S\\
		&SIA&PPLNS,PPS&816.92 TH/s\\
		
		\hline
		Slush&BTC&Score&3.27 EH/s\\
		
		\hline
		ViaBTC&BTC&PPLNS,PPS&2.87 EH/s\\
		\cline{2-4}
		&BCH&PPLNS,PPS&84.49 PH/s\\
		\cline{2-4}
		&ETH&PPLNS,PPS&878 GH/s\\
		\cline{2-4}
		&LTC&PPLNS,PPS&26.52 TH/s\\
		\cline{2-4}
		&ETC&PPLNS,PPS&6.032 GH/s\\
		\cline{2-4}
		&ZEC&PPLNS,PPS&89.63 MSol/s\\
		\cline{2-4}
		&DASH&PPLNS,PPS&117.36 TH/s\\
		
		\hline
		Miningpoolhub&ETH&PPLNS&8.33 TH/s\\
		\cline{2-4}
		&LTC&PPLNS&277 GH/s\\
		\cline{2-4}
		&ETC&PPLNS&1.494 TH/s\\
		\cline{2-4}
		&ZEC&PPLNS&29.378 MH/s\\
		\cline{2-4}
		&DASH&PPLNS&27.69 TH/s\\
		\cline{2-4}
		&XMR&PPLNS&5.74 MH/s\\
		\cline{2-4}
		&DGB&PPLNS&11.85 TH/s\\
		\hline
		Bitcoin.com&BTC&PPS&892.16 PH/s\\
		\cline{2-4}
		&BCH&PPS&191.72 PH/s\\		
		\hline
		Nanopool&ETH&PPLNS&20.96 TH/s\\
		\cline{2-4}
		&ETC&PPLNS&1.65 TH/s\\
		\cline{2-4}
		&ZEC&PPLNS&53.57 MSol/s\\
		\cline{2-4}
		&GRIN&PPLNS&17.1 Kgp/s\\
		\cline{2-4}
		&XMR&PPLNS&57.488 MH/s\\
		\hline
		Litecoinpool.org&LTC&PPS&33.49 TH/s\\
		\hline
		\end{tabular}
\end{table}

\begin{table}
		\begin{tabular}{|l|l|l|l|}
		\hline
		Ethermine&ETH&PPLNS&41.4 TH/s\\
		\cline{2-4}
		&ETC&PPLNS&3.5 TH/s\\
		\cline{2-4}
		&ZEC&PPLNS&396.9 MSol/s\\
		\hline
		f2pool&BTC&PPS&4.46 EH/s\\
		\cline{2-4}
		&ETH&PPS&17.7 TH/s\\
		\cline{2-4}
		&LTC&PPS&40.94 TH/s\\
		\cline{2-4}
		&ETC&PPS&127.37 GH/s\\
		\cline{2-4}
		&ZEC&PPS&379.73 MSol/s\\
		\cline{2-4}
		&DASH&PPS&74.44 TH/s\\
		\cline{2-4}
		&SIA&PPS&3.59 TH/s\\
		\cline{2-4}
		&XMR&PPS&54.50 MH/s\\
		\hline
		Multipool&BTC&Exponential&0.66 PH/s\\
		\cline{2-4}
		&BCH&PPLNS&3.658 PH/s\\
		\cline{2-4}
		&LTC&PPLNS&122.22 GH/s\\
		\cline{2-4}
		&DGB&Proportional&3.836 PH/s\\
		\hline
		Minergate&BTG&PPLNS,PPS&2.006 KSol/s\\
		\cline{2-4}
		&ETH&PPLNS&14.68 GH/s\\
		\cline{2-4}
		&ETC&PPLNS&4.374 GH/s\\
		\cline{2-4}
		&ZEC&PPLNS&16.12 KSol/s\\
		\cline{2-4}
		&XMR&PPLNS,PPS&3.776 MH/s\\
		\cline{2-4}
		&BCN&PPLNS,PPS&1.641 MH/s\\
		\hline
		Suprnova&BTG&Proportional&0.25 MSol/s\\
		\cline{2-4}
		&LTC&Proportional&24.09 GH/\\
		\cline{2-4}
		&ZEC&Proportional&19.03 KSol/s\\
		\cline{2-4}
		&DGB&Proportional&9.45 TH/s\\
		\cline{2-4}
		&DASH&Proportional&9.26 TH/s\\
		\hline
		Coinotron&ETH&PPLNS,RBPPS&806.6 GH/s\\
		\cline{2-4}
		&LTC&PPLNS,PPS&37.8 GH/s\\
		\cline{2-4}
		&ETC&PPLNS,RBPPS&53.8 GH/s\\
		\cline{2-4}
		&ZEC&PPLNS,PPS&491.5 MSol/s\\
		\cline{2-4}
		&BTG&PPLNS,PPS&332.6 KH/s\\
		\cline{2-4}
		&DASH&PPLNS,PPS&2.4 TH/s\\
		\hline
		
	\end{tabular}
\end{table}
\fi
\fi